\documentclass[
 reprint,
superscriptaddress,
 amsmath,amssymb,
 aps,
 journal,
 floatfix,
 physrev
]{revtex4-2}

\usepackage{graphicx}
\usepackage{dcolumn}
\usepackage{bm}
\usepackage{hyperref}
\usepackage{url}

\usepackage{newtxtext,newtxmath}
\usepackage{bm}
\usepackage[T1]{fontenc}
\usepackage{xcolor}
\usepackage{orcidlink}

\newcommand{\mnras}{Mon.~Not.~Roy.~Astron.~Soc.}
\newcommand{\physrep}{Phys.~Rep.}
\newcommand{\jcap}{JCAP}
\newcommand{\pasj}{Publ.~Astron.~Soc.~Japan}
\newcommand{\apjl}{\apj~Lett.}
\newcommand{\apjs}{\apj~Suppl.}
\newcommand{\aap}{Astronomy~\&~Astrophysics}
\newcommand{\aj}{The~Astronomical~J.}

\newcommand{\dnnz}{{\sc DNNz}~}
\newcommand{\mizuki}{{\sc Mizuki}~}
\newcommand{\dempz}{{\sc DEmPz}~}
\newcommand{\etal}{{\it et al.}~}

\usepackage{graphicx}
\usepackage{amsmath}
\usepackage{booktabs}
\newcommand{\dSigma}{\Delta\!\Sigma}
\newcommand{\wproj}{w_{\rm p}}

\newcommand{\mpch}{h^{-1}{\rm Mpc}}

\begin{document}

\title{Hyper Suprime-Cam Year 3 Results: Measurements of Clustering of SDSS-BOSS Galaxies, Galaxy-Galaxy Lensing and Cosmic Shear}

\author{Surhud~More\orcidlink{0000-0002-2986-2371}}
\affiliation{The Inter-University Centre for Astronomy and Astrophysics, Post bag 4, Ganeshkhind, Pune 411007, India}
\affiliation{Kavli Institute for the Physics and Mathematics of the Universe
(WPI), The University of Tokyo Institutes for Advanced Study (UTIAS),
The University of Tokyo, Chiba 277-8583, Japan}

\author{Sunao~Sugiyama\orcidlink{0000-0003-1153-6735}}
\affiliation{Kavli Institute for the Physics and Mathematics of the Universe (WPI), The University of Tokyo Institutes for Advanced Study (UTIAS), The University of Tokyo, Chiba 277-8583, Japan}
\affiliation{Department of Physics, The University of Tokyo, Bunkyo, Tokyo 113-0031, Japan}

\author{Hironao~Miyatake\orcidlink{0000-0001-7964-9766}}
\affiliation{Kobayashi-Maskawa Institute for the Origin of Particles and the Universe (KMI),
Nagoya University, Nagoya, 464-8602, Japan}
\affiliation{Institute for Advanced Research, Nagoya University, Nagoya 464-8601, Japan}
\affiliation{Kavli Institute for the Physics and Mathematics of the Universe (WPI), The University of Tokyo Institutes for Advanced Study (UTIAS), The University of Tokyo, Chiba 277-8583, Japan}

\author{Markus~Michael~Rau\orcidlink{0000-0003-3709-1324}}
\affiliation{High Energy Physics Division, Argonne National Laboratory, Lemont, IL 60439, USA}
\affiliation{McWilliams Center for Cosmology, Department of Physics, Carnegie Mellon University, Pittsburgh, PA 15213}

\author{Masato~Shirasaki\orcidlink{0000-0002-1706-5797}}
\affiliation{National Astronomical Observatory of Japan, Mitaka, Tokyo 181-8588, Japan}
\affiliation{The Institute of Statistical Mathematics,
Tachikawa, Tokyo 190-8562, Japan}

\author{Xiangchong~Li\orcidlink{0000-0003-2880-5102}}
\affiliation{McWilliams Center for Cosmology, Department of Physics, Carnegie Mellon University, Pittsburgh, PA 15213}
\affiliation{Kavli Institute for the Physics and Mathematics of the Universe
(WPI), The University of Tokyo Institutes for Advanced Study (UTIAS),
The University of Tokyo, Chiba 277-8583, Japan}

\author{Atsushi~J.~Nishizawa\orcidlink{0000-0002-6109-2397}}
\affiliation{Gifu Shotoku Gakuen University, Gifu 501-6194, Japan}
\affiliation{Institute for Advanced Research/Kobayashi Maskawa Institute, Nagoya University, Nagoya 464-8602, Japan}

\author{Ken~Osato\orcidlink{0000-0002-7934-2569}}
\affiliation{Center for Frontier Science, Chiba University, Chiba 263-8522, Japan}
\affiliation{Department of Physics, Graduate School of Science, Chiba University, Chiba 263-8522, Japan}

\author{Tianqing~Zhang\orcidlink{0000-0002-5596-198X}}
\affiliation{McWilliams Center for Cosmology, Department of Physics, Carnegie Mellon University, Pittsburgh, PA 15213}

\author{Masahiro~Takada\orcidlink{0000-0002-5578-6472}}
\affiliation{Kavli Institute for the Physics and Mathematics of the Universe (WPI), The University of Tokyo Institutes for Advanced Study (UTIAS), The University of Tokyo, Chiba 277-8583, Japan}

\author{Takashi~Hamana}
\affiliation{National Astronomical Observatory of Japan, National Institutes of Natural Sciences, Mitaka, Tokyo 181-8588, Japan}

\author{Ryuichi~Takahashi}
\affiliation{Faculty of Science and Technology, Hirosaki University, 3 Bunkyo-cho, Hirosaki, Aomori 036-8561, Japan}

\author{Roohi~Dalal\orcidlink{0000-0002-7998-9899}}
\affiliation{Department of Astrophysical Sciences, Princeton University, Princeton, NJ 08544, USA}

\author{Rachel~Mandelbaum\orcidlink{0000-0003-2271-1527}}
\affiliation{McWilliams Center for Cosmology, Department of Physics, Carnegie Mellon University, Pittsburgh, PA 15213}

\author{Michael~A.~Strauss\orcidlink{0000-0002-0106-7755}}
\affiliation{Department of Astrophysical Sciences, Princeton University, Princeton, NJ 08544, USA}

\author{Yosuke~Kobayashi\orcidlink{0000-0002-6633-5036}}
\affiliation{Department of Astronomy/Steward Observatory, University of Arizona, 933 North Cherry Avenue, Tucson, AZ 85721-0065, USA}
\affiliation{Kavli Institute for the Physics and Mathematics of the Universe
(WPI), The University of Tokyo Institutes for Advanced Study (UTIAS),
The University of Tokyo, Chiba 277-8583, Japan}

\author{Takahiro~Nishimichi\orcidlink{0000-0002-9664-0760}}
\affiliation{Center for Gravitational Physics and Quantum Information, Yukawa Institute for Theoretical Physics, Kyoto University, Kyoto 606-8502, Japan}
\affiliation{Kavli Institute for the Physics and Mathematics of the Universe
(WPI), The University of Tokyo Institutes for Advanced Study (UTIAS),
The University of Tokyo, Chiba 277-8583, Japan}
\affiliation{Department of Astrophysics and Atmospheric Sciences, Faculty of Science, Kyoto Sangyo University, Motoyama, Kamigamo, Kita-ku, Kyoto 603-8555, Japan}

\author{Masamune~Oguri\orcidlink{0000-0003-3484-399X}}
\affiliation{Center for Frontier Science, Chiba University, Chiba 263-8522, Japan}
\affiliation{Department of Physics, Graduate School of Science, Chiba University, Chiba 263-8522, Japan}

\author{Wentao~Luo\orcidlink{0000-0003-1297-6142}}
\affiliation{School of Physical Sciences, University of Science and Technology of China, Hefei, Anhui 230026, China}
\affiliation{CAS Key Laboratory for Researches in Galaxies and Cosmology/Department of Astronomy, School of Astronomy and Space Science, University of Science and Technology of China, Hefei, Anhui 230026, China}

\author{Arun~Kannawadi\orcidlink{0000-0001-8783-6529}}
\affiliation{Department of Astrophysical Sciences, Princeton University, Princeton, NJ 08544, USA}

\author{Bau-Ching~Hsieh\orcidlink{0000-0001-5615-4904}}
\affiliation{Academia Sinica Institute of Astronomy and Astrophysics, No.~1, Sec.~4, Roosevelt Rd., Taipei 10617, Taiwan}

\author{Robert~Armstrong}
\affiliation{Lawrence Livermore National Laboratory, Livermore, CA 94551, USA}

\author{James~Bosch\orcidlink{0000-0003-2759-5764}}
\affiliation{Department of Astrophysical Sciences, Princeton University, Princeton, NJ 08544, USA}

\author{Yutaka~Komiyama\orcidlink{0000-0002-3852-6329}}
\affiliation{Department of Advanced Sciences, Faculty of Science and Engineering, Hosei University, 3-7-2 Kajino-cho, Koganei-shi, Tokyo 184-8584, Japan}

\author{Robert~H.~Lupton\orcidlink{0000-0003-1666-0962}}
\affiliation{Department of Astrophysical Sciences, Princeton University, Princeton, NJ 08544, USA}

\author{Nate~B.~Lust\orcidlink{0000-0002-4122-9384}}
\affiliation{Department of Astrophysical Sciences, Princeton University, Princeton, NJ 08544, USA}

\author{Lauren~A.~MacArthur}
\affiliation{Department of Astrophysical Sciences, Princeton University, Princeton, NJ 08544, USA}

\author{Satoshi Miyazaki\orcidlink{0000-0002-1962-904X}}
\affiliation{Subaru Telescope,  National Astronomical Observatory of Japan, 650 N Aohoku Place Hilo HI 96720 USA}

\author{Hitoshi~Murayama\orcidlink{0000-0001-5769-9471}}
\affiliation{Berkeley Center for Theoretical Physics, University of California, Berkeley, CA 94720, USA}
\affiliation{Theory Group, Lawrence Berkeley National Laboratory, Berkeley, CA 94720, USA}
\affiliation{Kavli Institute for the Physics and Mathematics of the Universe (WPI), The University of Tokyo Institutes for Advanced Study (UTIAS), The University of Tokyo, Chiba 277-8583, Japan}

\author{Yuki~Okura\orcidlink{0000-0001-6623-4190}}
\affiliation{National Astronomical Observatory of Japan, National Institutes of Natural Sciences, Mitaka, Tokyo 181-8588, Japan}

\author{Paul~A.~Price\orcidlink{0000-0003-0511-0228}}
\affiliation{Department of Astrophysical Sciences, Princeton University, Princeton, NJ 08544, USA}

\author{Philip~J.~Tait}
\affiliation{Subaru Telescope,  National Astronomical Observatory of Japan, 650 N Aohoku Place Hilo HI 96720 USA}

\author{Masayuki~Tanaka}
\affiliation{National Astronomical Observatory of Japan, National Institutes of Natural Sciences, Mitaka, Tokyo 181-8588, Japan}

\author{Shiang-Yu~Wang}
\affiliation{Academia Sinica Institute of Astronomy and Astrophysics, No.~1, Sec.~4, Roosevelt Rd., Taipei 10617, Taiwan}

\date{\today}

\begin{abstract}
We utilize the Sloan Digital Sky Survey Baryon Oscillation Spectroscopic Survey (SDSS-BOSS) galaxies and its overlap with approximately 416 sq.~degree of deep $grizy$-band imaging from the Subaru Hyper Suprime-Cam Survey (HSC). We  perform measurements of three two-point correlations which form the basis of the cosmological inference presented in our companion papers, Miyatake \etal and Sugiyama \etal We use three approximately volume limited subsamples of spectroscopic galaxies by their $i$-band magnitude from the SDSS-BOSS: LOWZ ($0.1<z<0.35$), CMASS1 ($0.43<z<0.55$) and CMASS2 ($0.55<z<0.7$), respectively. We present high signal-to-noise ratio measurements of the projected correlation functions of these galaxies, which is expected to be proportional to the projected matter correlation function on large scales with a proportionality constant dependent on the bias of galaxies. In order to help break the degeneracy between the amplitude of the matter correlation and the bias of these spectroscopic galaxies, we use the distortions of the shapes of fainter galaxies in HSC due to weak gravitational lensing, to measure the galaxy-galaxy lensing signal, which probes the projected galaxy-matter cross-correlation function of the SDSS-BOSS galaxies. We also measure the cosmic shear correlation functions from HSC galaxies which is related to the projected matter correlation function. We demonstrate the robustness of our measurements by subjecting each of them to a variety of systematic tests. Our use of a single sample of HSC source galaxies is crucial to calibrate any residual systematic biases in the inferred redshifts of our galaxies. We also describe the construction of a suite of mocks: i) spectroscopic galaxy catalogs which obey the clustering and abundance of each of the three SDSS-BOSS subsamples, and ii) galaxy shape catalogs which obey the footprint of the HSC survey and have been appropriately sheared by the large-scale structure expected in a $\Lambda$CDM model. We use these mock catalogs to compute the covariance of each of our observables. 
\end{abstract}


\maketitle

\section{Introduction}

The concordance cosmological model, $\Lambda$CDM, is rooted in its ability to accurately describe a variety of cosmological observations. Primary among these are the statistics of anisotropies in the cosmic microwave background (CMB) \cite[e.g.,][]{PlanckCosmology:16, 2020A&A...641A...6P}, the distance-redshift relation obtained using type-Ia supernovae \cite[e.g.,][]{Scolnic:2018} and baryon acoustic oscillations \cite[e.g.,][]{Ross:2017, Beutler:2017, Alam:2021}, the abundance of galaxy clusters \cite[e.g,][]{Aiola:2020, Chaubal:2022}, the redshift space clustering of galaxies \cite[e.g.,][]{Sanchez:2017, Grieb:2017,Kobayashi:2021}, and the weak gravitational lensing signal \cite[e.g.,][]{Hikage:2019, Hamana:2020, Heymans:2021, Pandey:2021, Porredon:2021, Sugiyama:2022, Miyatake:2022, Amon:2022, DES-Y3}. The presence of dark matter and dark energy is essential to the success of the model, in which dark matter causes density fluctuations to grow due to gravitational instabilities, while dark energy causes an accelerated expansion at late times \cite[e.g.,][]{MovdBWhite:2020}. Although the evidence for the presence of dark matter and dark energy in the Universe is quite compelling, their existence is a grand challenge to our current understanding of the physics of the Universe. The empirical characterization of the abundance of dark matter, the growth in its density fluctuations, and the behavior of the equation of state for dark energy can aid in the understanding of these components.

The measurements of density fluctuations in the early Universe have been mapped by CMB experiments such as WMAP \cite{Hinshaw:2013} and Planck \cite{2020A&A...641A...6P}. These precise observations suggest that our Universe can be described as a simple flat $\Lambda$ cold dark matter model with only a handful of parameters. Under the assumptions of this simple model, the CMB observations predict the value of $\sigma_8$, which characterizes the root mean square dispersion in the density fluctuations when averaged in spheres of radius 8 $\mpch$ today. This quantity can be directly inferred from observables in the Universe at late times. 

The current observational frontier in this area is driven by Stage III dark energy experiments which are designed to probe dark energy and dark matter. The Kilo-Degree Survey (KiDS) \cite{deJong:2013} \footnote{\url{https://kids.strw.leidenuniv.nl/}}, the Dark Energy Survey (DES) \cite{Frieman:2005} \footnote{\url{https://www.darkenergysurvey.org}} and the Hyper Suprime-Cam survey Subaru Strategic Program (HSC) \cite{HSCoverview:17} \footnote{\url{https://hsc.mtk.nao.ac.jp/ssp/}} have all conducted complementary galaxy imaging surveys that target multiple probes in order to address key questions related to dark matter and dark energy. Although these surveys use a combination of probes to address cosmology, gravitational lensing is the primary tool of interest. The images of background galaxies get sheared in a coherent manner due to the presence of intervening matter distribution between them and us \cite[e.g.,][]{GravitationalLensing:SaasFee}. The correlation of these coherent distortion patterns, commonly called the cosmic shear signal, is related to the projected matter density distribution and is therefore sensitive to the cosmological parameters. The amplitude of this signal and its variation with redshift can be used to infer the growth of structure in dark matter and constrain the parameter combination $S_8=\sigma_8\Omega_{\rm m}^{0.5}$, where $\Omega_{\rm m}$ is the matter density parameter.

The clustering of galaxies can also be used to probe the large-scale structure of the Universe, as galaxies trace the matter distribution \cite[e.g.,][]{Tegmarketal:04}. Galaxies form within dark matter halos which are biased tracers of the matter distribution \cite[e.g.,][]{CooraySheth:02}, and the galaxy bias, $b$, is expected to depend upon the halo bias of their parent halos \cite[e.g.][]{Kaiser:1984}. On large scales this bias is expected to be linear, and the clustering of galaxies thus reflects the shape of the matter correlation function, $\xi_{\rm mm}$. Such observations can therefore constrain cosmological parameters that determine the shape of $\xi_{\rm mm}$, but its amplitude is entirely degenerate with galaxy bias \cite[e.g.,][]{Seljak:2005}. The dependence of the halo bias on the mass of the dark matter halo is further a function of the cosmological parameters, especially $\Omega_{\rm m}$ and $\sigma_8$ \cite[e.g.,][]{Tinker:2010}. Therefore, the amplitude of the matter correlation function can be inferred if dark matter halo masses can be measured for galaxies using the galaxy-galaxy lensing \cite{Seljak:2005, Cacciato:2009, vandenBosch:2013}. The galaxy-galaxy lensing signal is the cross correlation of lens galaxy positions with shapes of background galaxies. Thus together a combination of the clustering measurements and small-scale lensing measurements can help infer cosmological constraints \cite[e.g.,][]{Seljak:2005, Cacciato:2009, More:2013b, Cacciato:2013, More:2015, Miyatake:2022}. Another avenue to use galaxy-galaxy lensing is to only focus on large scales so that the cosmological inference is not affected by issues related to galaxy assembly bias \cite[e.g.,][]{Mandelbaumetal:13}. The galaxy-galaxy lensing signal depends upon the halo matter cross-correlation, which on large scales is proportional to $b\xi_{\rm mm}$. The large-scale clustering of of galaxies is sensitive to $b^2\xi_{\rm mm}$. Thus together they can help determine both the amplitude and shape of the matter correlation function.

The cosmological parameter dependences of cosmic shear or the combination of galaxy clustering and galaxy-galaxy lensing signal are expected to be different and thus complementary. Thus a combination of all the three two-point correlations can further reduce the uncertainties on the inferred cosmological parameters \cite[see e.g.][]{DES-Y3, Heymans:2021}. The first step in the careful inference of the cosmological parameters from these observables is the reliable measurement of the observables and their covariance. There are a variety of systematics that can affect each of these measurements, decisions need to be taken regarding the scales over which the signals can be reliably modelled as well as a demonstration that the measurements pass a variety of null tests. Inference of the redshift distribution of source galaxies is yet another step before the weak lensing measurements can be reliably modelled.

This is the first paper in a series of the 3$\times$2pt cosmology analyses of the Subaru HSC Year 3 data (hereafter HSC-Y3). In this paper (Paper I), we define the lens and source galaxy samples to be used for the 3$\times$2pt cosmological analyses. We present measurements of the three two-point functions as well as the results of various systematics tests which allow us to narrow down the scales to be used for cosmological analysis. We also present mock galaxy catalogs and shear catalogs that were used to obtain the covariance matrix of our measurements. The inference of the cosmological parameters were performed in a blind manner using these measurements, and those results will be presented in Miyatake \etal \cite[][Paper II]{Miyatake_hscy3} and Sugiyama \etal \cite[][Paper III]{Sugiyama_hscy3}, respectively. Paper II will use information on quasi-linear scales by using the emulator based halo model, while the analysis in Paper III is based on linear scales where a model based on perturbation theory can be used reliably. The measurement and the cosmological analysis of the cosmic shear tomography in real space and Fourier space will be presented in companion papers, Li \etal \cite{Li_hscy3} and Dalal \etal \cite{Dalal_hscy3}, respectively.

This paper is organized as follows. In Section~\ref{sec:data}, we describe the HSC three-year shear catalog that we use, our blinding strategy, the construction of the lens galaxy sample, as well as the construction of the mock galaxy catalogs which are used to determine the covariance of our measurements. In Section~\ref{sec:measurement_codes}, we present a brief description of our pipelines that we use to measure the galaxy clustering, the galaxy-galaxy lensing signal and the cosmic shear signal from our data. In Section~\ref{sec:clustering}, we present the measurements of the galaxy clustering signal, its covariance and a variety of systematics tests designed to assess the robustness of our measurements. In Section~\ref{sec:lensing}, we present the measurements of the galaxy-galaxy lensing signal, its covariance, and a variety of null and systematics tests. In Section~\ref{sec:cosmic_shear}, we similarly present the measurements of the cosmic shear correlation functions and its covariance, the systematic tests as well as various PSF related systematic tests. Finally, we summarize our results in Section~\ref{sec:summary}.

When performing the galaxy clustering and galaxy-galaxy lensing analyses, we adopt a fiducial flat $\Lambda$CDM cosmological model with parameters consistent with the cosmic microwave background analysis from WMAP9. The cosmological model is specified by the CDM density parameter $\Omega_{\rm cdm}=0.233$, the baryon density $\Omega_{\rm b0}=0.046$, the matter density $\Omega_{\rm m}=\Omega_{\rm cdm}+\Omega_{\rm b} = 0.279$, the cosmological constant $\Omega_{\Lambda}=0.721$, the Hubble parameter $h= 0.7$, the amplitude of density fluctuations $\sigma_8= 0.82$, and the spectral index $n_s= 0.97$. The choice of these cosmological parameters is dictated by the cosmological parameters adopted for the simulations that form the basis of the mock catalogs that we use for our covariance calculations \cite{2017ApJ...850...24T}. While carrying out our cosmological analyses in Paper II and III, we will account for the cosmological dependence of our observables, so that our choice of this fiducial cosmological model has no impact on our cosmological inference.

\section{Data}
\label{sec:data}

\begin{figure*}
    \centering
    \includegraphics[width=\textwidth]{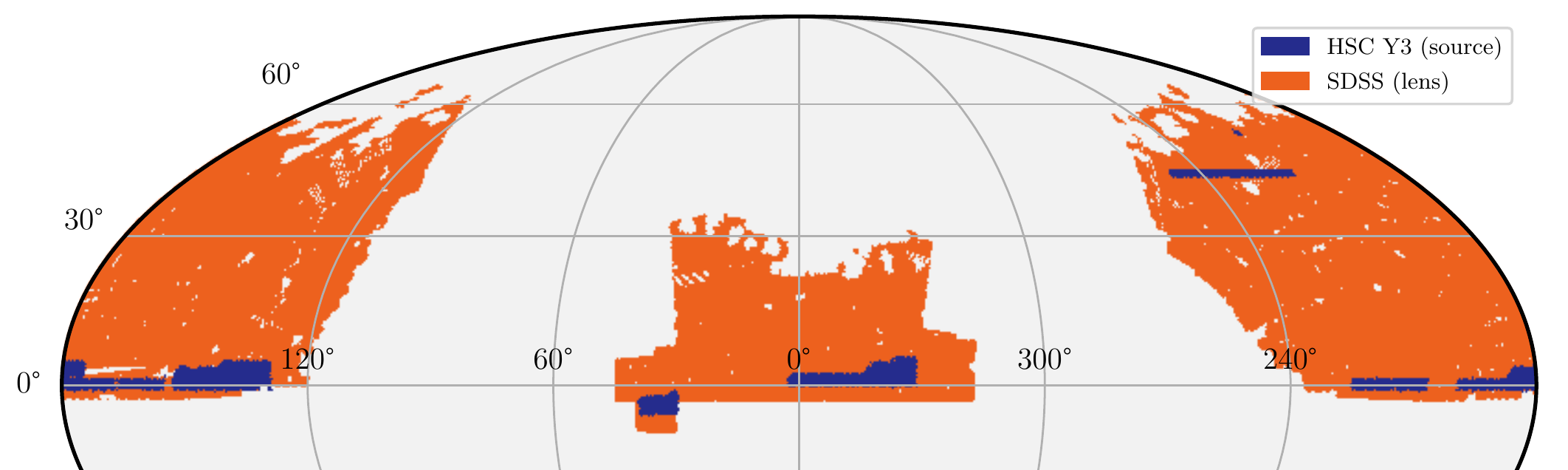}
    \caption{The area coverage of the data used in this paper for performing
3$\times$2pt
measurements. The catalog of HSC galaxies used for the weak lensing
measurements are shown using the purple shaded region, while the  catalog of SDSS
galaxies is shown using yellow. The overlap between the two catalogs is $\sim
416$~sq. degree. Our measurement of the clustering signal uses the entire SDSS
region, while the measurements of the galaxy-galaxy lensing signal and the
cosmic shear signal utilize source galaxies in the overlapping area.}
    \label{fig:hscs19a_sdss}
\end{figure*}

The 3$\times$2pt cosmological analyses for HSC-Y3 data will focus on a combination of probes: the cosmic shear signal, the clustering of galaxies and their galaxy-galaxy lensing signal. In this section, we describe the data we use for the gravitational lensing measurements (Section~\ref{sec:data_shape_catalog}), our blinding strategy (Section~\ref{sec:blinding}),  the galaxy samples used for the clustering and the galaxy-galaxy lensing analyses (Section~\ref{sec:data_lens_catalog}) as well as the mock catalogs that are used to estimate the covariance of these measurements (Section~\ref{sec:data_sims}).

\subsection{Shape Catalog}
\label{sec:data_shape_catalog}
The Hyper Suprime-Cam instrument (HSC) is a wide-field optical imaging camera mounted at the prime focus of the 8.2-meter Subaru Telescope \cite{2018PASJ...70S...1M,2018PASJ...70S...2K,2018PASJ...70S...3F,2018PASJ...70...66K}. The wide field of view combined with the excellent seeing conditions at Maunakea and the large aperture make it an ideal telescope for surveys targetting weak lensing measurements of large-scale structure.
The HSC-Subaru Strategic Program was allocated 330 nights to carry out a
three-layered imaging survey in multiple bands with different depths \cite{HSCoverview:17}. The Wide
layer, which is designed for weak lensing cosmology,  
consists of multi-band $grizy$ imaging that will cover an area of
approximately 1100 deg$^2$, upon completion. The galaxy shape measurements are
carried out in the $i$-band which is observed at a $5-\sigma$ depth of $i\sim
26$ ($2^{\prime\prime}$ aperture for a point source). The images in the $i$ band were preferentially taken 
under good seeing conditions, which resulted in a median seeing of about $0.6$~arcsec.

The data from the survey has been processed using a fork of Rubin’s LSST
Science Pipelines catered to images taken with HSC \cite{2018PASJ...70S...5B},
and which is updated from time to time to include new features, any bug fixes,
and improvements to deal with the Subaru HSC data. The Subaru HSC survey has
made three public data releases thus far \cite{HSCDR1:17, HSCDR2:2019,
HSCDR3:2022}.  In this paper, we use the shape catalog from the S19a internal
data release intermediate to PDR2 and PDR3, which was processed with hscPipe~v7
\cite{Li2021}. There were a number of improvements to the PSF modeling, image
warping kernel, background subtraction and bright star masks, which have
improved the quality of the shape catalog in HSC-Y3 compared to the Year~1
shape catalog  \cite{HSCDR1_shear:17,2018MNRAS.481.3170M}. The detailed
selection of galaxies that form the shape catalog is presented in
\citet{Li2021}. Briefly, the shape catalog consists of galaxies selected from
the full-depth and full-color region in all five filters. Apart from some basic
quality cuts related to pixel level information, we select extended objects
with an extinction corrected cmodel magnitude $i<24.5$, $i$-band SNR$\ge 10$,
resolution $>0.3$, $>5\sigma$ detection in at least two bands other than $i$, a
1 arcsec diameter aperture magnitude cut of $i<25.5$, and a blendedness cut in
the $i$-band of $10^{-3.8}$.

The shape catalog consists of a total of 35.7 million galaxies spanning an area
of about 433~sq.~deg, an effective number density of 19.9 arcmin$^{-2}$. It is
divided into six disjoint regions: XMM, VVDS, GAMA09H, WIDE12H, GAMA15H and
HECTOMAP fields \cite[see Fig.~2 in][]{Li2021}. The shape measurements in the
catalog were calibrated using detailed image simulations, such that the galaxy
property dependent multiplicative shear estimation bias is less than $\sim
10^{-2}$. \citet{Li2021} also presented a number of systematics tests and null
tests, and quantified the level of residual systematics in the shape catalog
that could affect the cosmological science analyses carried out using the data.
Given that \citet{Li2021} flag residual additive biases due to PSF model shape
residual correlations and star galaxy shape correlations as systematics
requiring special attention and marginalization, we will also investigate the
effect of these systematics on the cosmic shear measurements.

As shown in Appendix~\ref{sec:gama09h_bmodes_app}, we exclude a $\sim 20$~sq.
degree patch of the sky in the GAMA09H region from our analysis. This region
was a significant source of $B$-mode systematics in the cosmic shear analysis
presented in our companion paper Li \etal \cite{Li_hscy3}. The resultant area of our shape
catalog is $\sim 416$~sq. degree. The on-sky projection of this area is shown in
Fig.~\ref{fig:hscs19a_sdss}, as the purple shaded region.

The HSC-Y3 shape catalog is accompanied by a photometric redshift catalog of
galaxies based on three different methods
\cite{2020arXiv200301511N}\footnote{However, the photo-$z$ catalog released for
PDR2 is different from the catalog we use in this analysis. We will make the
shape catalog, with the photo-$z$ information, publicly available after the
cosmology papers are published.}. The software \mizuki is a template fitting
based photometric redshift estimate code, while \dempz and \dnnz provide
machine learning based estimates of the photometric redshifts of galaxies\footnote{Self-organizing map is yet another way of characterizing the photometric redshift distribution similar to \dempz/\dnnz~methods. The differences between these methods lie in the machine learning architecture used for the calibration and inference. Please see \citet{Rau:2023} for detailed discussion about the differences in the calibration strategies adopted in the current surveys.}. Each
of these methods provides an estimate of the posterior distribution function of
the redshift for each galaxy, denoted as $P(z_{\rm s})$. Photometric redshift
uncertainties are one of most important systematic effects in weak lensing
cosmology, and can cause significant biases in the cosmological parameters if
they are affected by unknown residual systematic errors. To minimize the impact
of such errors, we will adopt the method in
\citet{OguriTakada:11} that allows to self-consistently calibrate such
photo-$z$ errors, using a single sample of photometric source galaxies and
multiple samples of spectroscopic lens galaxies in galaxy-galaxy weak lensing
measurements.

For this purpose, we define a sample of background galaxies that satisfies 
\begin{equation}
\int_{z_{\rm l, max}+z_{\rm diff}}^{7} P(z_{\rm s}) dz_{\rm s} \ge p_{\rm thresh} \,,\label{eq:source-selection}
\end{equation} 
where $z_{\rm l, max}$ was chosen to be equal to $0.70$, the
maximum redshift of the lens samples that we will use for the galaxy-galaxy
lensing measurements (see below), $z_{\rm diff}=0.05$ and $p_{\rm
thresh}=0.99$. Such cuts significantly reduce the contamination of source
galaxies which are physically associated with the lens galaxies which can
dilute the weak lensing signal at small separations. As our default choice, we
use the $P(z)$ estimates for each galaxy provided by \dempz.

The selection (Eq.~\ref{eq:source-selection}) reduces the total number of
galaxies in our weak lensing sample to be just $24$ percent of the original
shape catalog,\footnote{Although a cut that retains only 24 percent of source galaxies is quite severe, in these initial round of cosmological analyses, we stick to a single source bin at redshifts larger than any of our lens samples to conservatively constrain any systematics in the redshift distribution of the source galaxies and avoid any source galaxies that may be physically correlated with our source galaxies.} with an effective number density of $4.9$ galaxies per square
arcmin. Instead, if we use the posterior distributions of the redshifts given
by \dnnz or \mizuki, the number of galaxies is $9$ or $35$ percent of the
entirety of the shape catalog, respectively. These would correspond to an
effective number density of $1.9$ and $6.8$ galaxies per square arcmin for
\dnnz and \mizuki, respectively. These differences in the number density are
entirely driven by the differences in the widths of the individual $P(z)$'s
inferred by the different codes. The inferred $P(z)$ estimates for individual
galaxies obtained using \dnnz are on average broader than those obtained in
\mizuki and \dempz. The broader widths result in a smaller number of galaxies
that satisfy the cut shown in Eq.~(\ref{eq:source-selection}).

\begin{figure}
    \centering
    \includegraphics[width=\columnwidth]{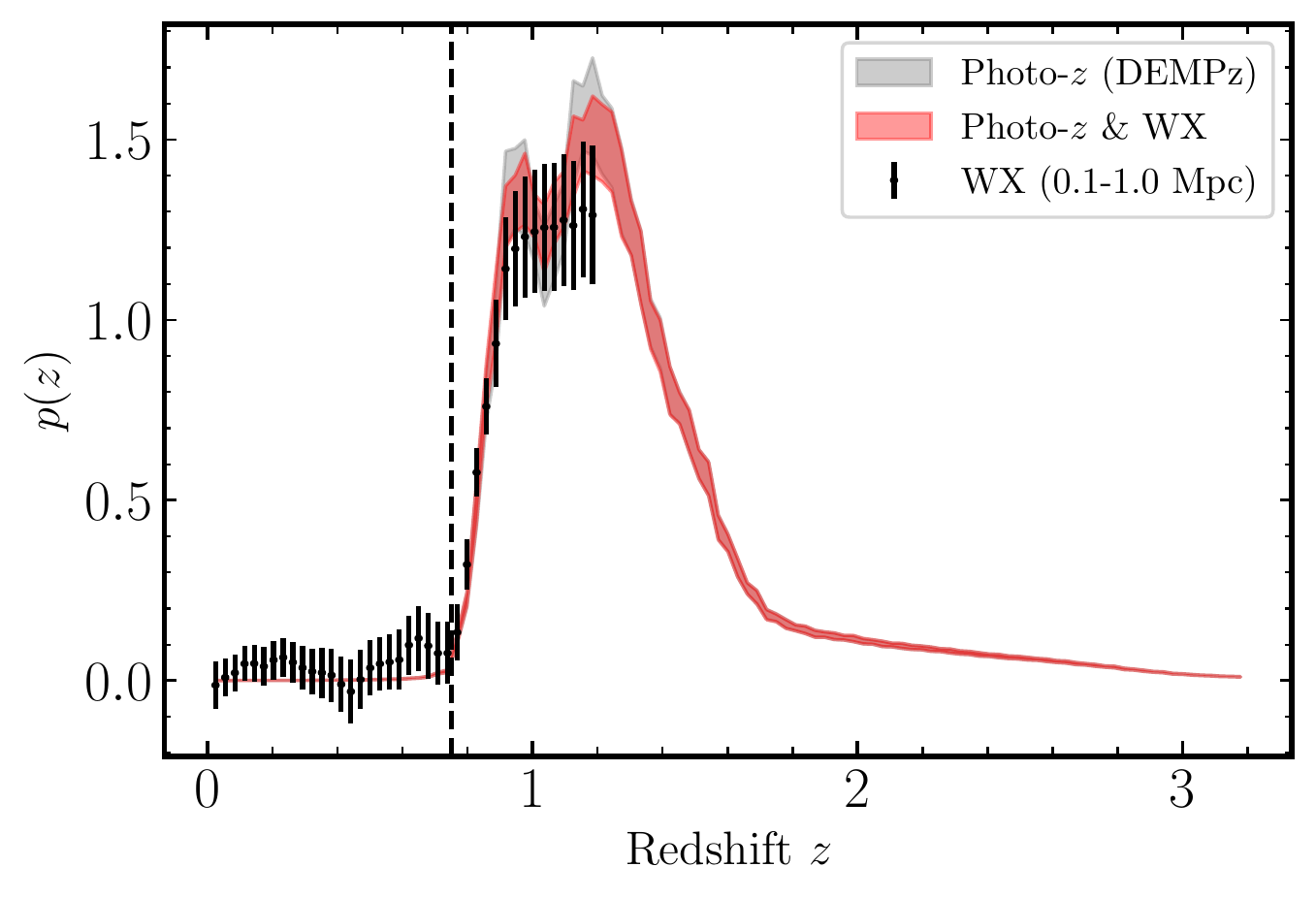}
    \includegraphics[width=\columnwidth]{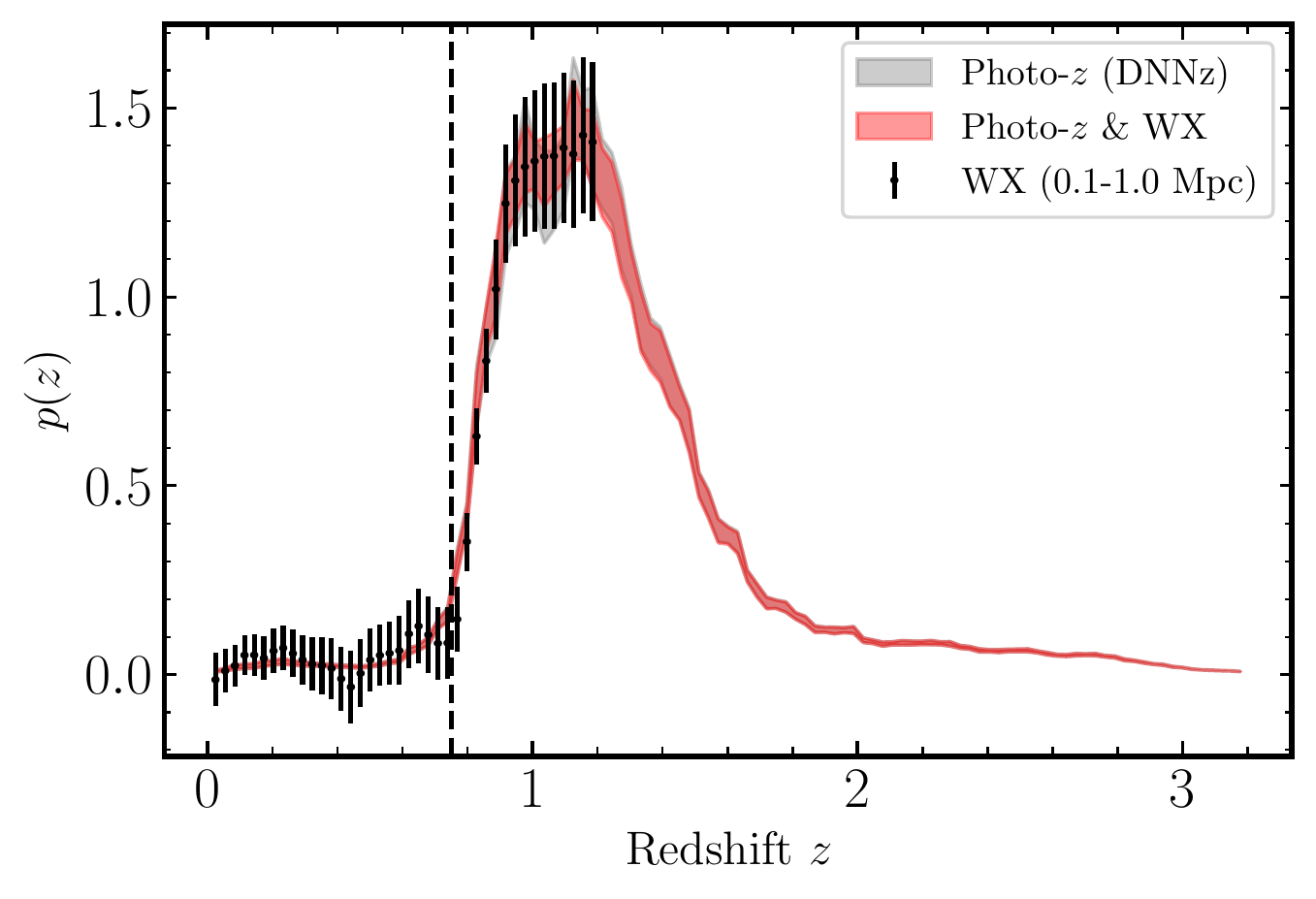}
    \caption{The inferred redshift distribution of the source galaxies used in
our weak lensing analysis obtained using the techniques described in
\cite{Rau:2023}. The upper and lower panels use the photometric redshift PDFs
for the source sample as inferred by the photo-$z$ methods \dempz and \dnnz,
respectively. The grey shaded region shows the posterior based on deconvolving
the photometric redshift errors from these PDFs, the black points with errors
correspond to the redshift inference based purely on the clustering of our
source galaxies with the CAMIRA red galaxy sample, while the red shaded region
corresponds to the posterior combining the two measurements.}
    \label{fig:dndz}
\end{figure}

The inferred redshift distributions of the sample of our fiducial sources are
shown in Fig.~\ref{fig:dndz} based on the Bayesian Hierarchical Inference
presented in
\citet{Rau:2023}. These redshift distribution inferences (90 percent credible
region shown in gray) use both the individual redshift PDFs from photometric
redshift estimation codes, as well as the measurements of clustering of the
galaxies in our source sample, with that of red galaxies selected by the {\sc
CAMIRA} algorithm \citep{2014MNRAS.444..147O,2018PASJ...70S..20O}. The $90$
percent credible intervals on the redshift inference from each of these
techniques individually are shown as the red shaded region and black points
with errors, respectively. As mentioned before, we have multiple choices of
photometric redshift estimates for our sample of galaxies. Even though we
select galaxies using the \dempz redshift PDFs, the same source galaxy sample
also has redshift estimates characterized by the other codes, which help in
pinning down any systematic uncertainties. The two panels in
Fig.~\ref{fig:dndz} correspond to the use of individual redshift PDFs from
\dempz and \dnnz, respectively, but for the same set of source galaxies. The
clustering method does not extend to the entire range of redshifts as we run
out of galaxies with well calibrated redshifts from {\sc CAMIRA} beyond a
redshift of $1.2$. Note that the cross-correlation results (marked by WX) differ in each of the panels by their a-posteriori normalization factor from the joint likelihood inference between WX and \dnnz/\dempz, as can be seen upon a close inspection of the two panels. The redshift axes in each panel are aligned vertically for ease of comparison.

In the upper panel, we see a broad agreement between the redshift inference
based on just the redshift PDFs from \dempz and the clustering redshifts.
However, it is interesting to note that \dempz predicts a bimodal feature in
the redshift distribution with a dip in the number of galaxies with redshifts
at $z\sim 1.0$. The clustering redshift inferences have large and correlated
errors, but do not show any indication of such a bimodal distribution. The
redshift inference based on \dempz shows very small support at redshifts lower
than $0.75$ (less than a percent), by construction.  In the bottom panel, which
uses \dnnz estimates of the redshift PDFs, we do not see a
substantial evidence of a dip at redshift of unity in this inference. However,
we notice that the inference using \dnnz does show some non-zero support
even below the maximum lens redshifts we use of $0.75$, as seen in the
clustering measurements (although at low significance). 

Any systematic differences in the redshifts of our sources could translate into
biases in the measured weak lensing signals. In
Appendix~\ref{sec:nz_differences_app}, we present our estimates of such
potential biases and suggest that shifting the inferred redshift PDFs by a free
parameter $\Delta z$ is sufficient in order to marginalize over such
uncertainties with percent level accuracy. 

\subsection{Blinding strategy}\label{sec:blinding}

All of our cosmological analyses are carried out in a two-tiered blind manner
similar to our strategy in our Year~1 analysis \cite{2019PASJ...71...43H}. As
described in the HSC-Y3 shape catalog paper \citet[][]{Li2021}, we use
different multiplicative bias factors in order to blind our analyses in the
first tier, which provide a convenient way to change the data vectors related
to the weak lensing observables. We obtain three blind catalogs
$i=0,1,2$ where the $j$-th galaxy has a multiplicative bias equal to
\begin{equation}
    m^{ij} =  m^{j} + \mathrm{d}m_1^i + \mathrm{d}m_2^i\,,
\end{equation}
where $\mathrm{d}m_1^i$ is a multiplicative bias known to the analysis lead
designed to prevent unblinding due to accidental comparisons of multiple
versions of the catalog. This value is removed in the measurement codes before
performing any measurements with any of the catalogs. The three values
$\mathrm{d}m_2^i$ are one of the three choices $(-0.1, -0.05, 0.)$, $(-0.05,
0., 0.05)$, $(0., 0.05, 0.1)$. These amplitude offsets are motivated by the
differences in the values of amplitude of density fluctuations, $\sigma_8$, as
obtained by the CMB analysis by the Planck collaboration
\cite[][]{2020A&A...641A...6P} and other large-scale structure probes
\cite[e.g.,][]{Heymans:2021, DES-Y3}. All of our lensing related measurements
were performed with all three blinded catalogs at the same time. The
cosmological analyses and all systematics tests were performed on each of these
catalogs, separately. As we will describe in sections \ref{subsec:gglens-meas}
and \ref{subsec:cosmicshear-meas} in this paper, we modify our covariances to
account for the fact that each of our catalogs has a different value of
$\mathrm{d}m_2$ (a procedure that can be performed without knowledge of
$\mathrm{d}m_2$). In this manner, the $\chi^2$ of the best fit models to
measurements performed using the three catalogs are not able to accidentally
distinguish the true catalog.

The second tier of blinding is performed at the analysis level. The results of
our analyses of all the three catalogs prior to unblinding are presented by
masking the value of the inferred cosmological parameters. Once all the
analyses are performed and
the relevant systematics checks are passed (see Paper II and III for a detailed
checklist of tests to be passed prior to unblinding), we unblind each of the
tiers, first starting with the second tier. The first tier of the catalog level
unblinding is performed by a HSC team member external to the analysis team
right at the end.  Throughout this paper, we will show the measurements from
our analysis of the blind catalog id = $2$, which was found to be the true
catalog post unblinding. No changes have been made to the analysis post
unblinding. The other blinded catalogs also similarly passed each of the
systematic tests as we show here for the blind catalog id $2$.

\subsection{Lens Galaxy Sample}
\label{sec:data_lens_catalog}

We use the large-scale structure sample compiled as part of the Data Release 11
(DR11) \footnote{\url{https://www.sdss.org/dr11/}} \cite{Alam:2015} of the
SDSS-III (Baryon Oscillation Spectroscopic Survey) project
\cite{2013AJ....145...10D} for measurements of the clustering of galaxies and
as lens galaxies for the galaxy-galaxy weak lensing measurements. The lens
galaxy sample used in this paper is the same as that used in the first year
analysis of HSC data (\citet{2021arXiv210100113M,Sugiyama:2021}). The
methodology used to construct the catalog is the same as that described in
\citet{Miyatakeetal:15}. We briefly describe the catalog here. 

The BOSS is a spectroscopic follow-up survey of galaxies and quasars
selected from the imaging data obtained by the SDSS-I/II and covers an area of
approximately 11,000 deg$^2$ \cite{2009ApJS..182..543A} using the dedicated
2.5m SDSS Telescope \cite{2006AJ....131.2332G}.  Imaging data obtained in five
photometric bands ($ugriz$) as part of the SDSS I/II surveys
\cite{1996AJ....111.1748F, 2002AJ....123.2121S, 2010AJ....139.1628D} were
augmented with an additional 3,000 deg$^2$ in SDSS DR9 to cover a larger
portion of the sky in the southern region \cite{2011AJ....142...72E,
2012ApJS..203...21A, 2013AJ....145...10D, 2011ApJS..193...29A}. These data were
processed by a series of the photometric processing pipelines
\cite{2001ASPC..238..269L, 2003AJ....125.1559P, 2008ApJ...674.1217P}, and
corrected for Galactic extinction \cite{1998ApJ...500..525S} to obtain a
reliable photometric catalog which serves as an input to select targets for
spectroscopy \cite{2013AJ....145...10D}. The resulting spectra were processed
by an automated pipeline to perform redshift  determination and spectral
classification \cite{2012AJ....144..144B}. The BOSS large-scale structure (LSS)
samples are selected using algorithms focused on galaxies in different
redshifts:  $0.15<z<0.35$ (LOWZ) and $0.43<z<0.70$ (CMASS). In addition to the
galaxies targeted by the BOSS project, these samples also include galaxies
which pass the target selection but have already been observed as part of the
SDSS-I/II project (legacy galaxies). These legacy galaxies were subsampled in
each SDSS sector \cite{Reid:2016} on the sky so that they obey the same
completeness as that of the LOWZ/CMASS targets in their respective redshift
ranges \cite{2014MNRAS.441...24A}.

Various color-magnitude selections guarantee a population of massive galaxies
spanning a redshift range  $z\in[0.15,0.70]$ in the  spectroscopic survey.
The resultant sample, however, is not entirely a volume or flux limited sample
of galaxies.

The SDSS spectrograph can assign at most 1000 fibers on the sky at one time in
a circular tile region (for the BOSS survey). Therefore an adaptive tiling
algorithm was used to maximize the completeness of the survey. The spectra
obtained from the observations were processed with {\sc specpipe}, the
spectroscopic pipeline for determination of redshift and spectral
classification. Despite all optimizations, there exist galaxies which obey the
target selection, but cannot be assigned a fiber due to limitations of how
close fibers can be assigned on a given tile. In such cases, the galaxy is
assigned the redshift of its nearest neighbour galaxy which was assigned a
fiber. A similar procedure is used to assign redshifts to galaxies for which
{\sc specpipe} failed to determine a redshift. In the fiducial DR11 LSS
catalog, the nearest neighbouring galaxy gets an additional weight to account
for the fiber-collided or redshift failure galaxy. We instead assign the
nearest neighbour redshift to photometric galaxies with fiber collisions or
redshift failures. This should be equivalent when the entire sample is used.
However, our method allows for making further sub-samples based on absolute
magnitude.  \citet{Guo:2012} have shown using detailed tests on mock galaxy
catalogs that the nearest neighbor redshift correction achieves sub-percent
accuracy in the projected galaxy auto-correlation function for scales used in
this paper. 

This modified DR11 LSS sample described above forms our parent catalog. We
obtain the k+e-corrected $i$-band absolute magnitudes for individual SDSS
galaxies using the k and e-corrections tabulated in \citet{2006MNRAS.372..537W}
relying on the ``passive plus star-forming galaxies'' spectral templates
constructed using the stellar population synthesis model in
\citet{Bruzual_Charlot:2003}) of individual galaxies based on {\tt cmodel}
photometry. In order to minimize the effect of k-corrections, we k-correct the
magnitudes of the LOWZ galaxies to a redshift of 0.20 and that of CMASS
galaxies to a redshift of 0.55. As we will see below, these magnitudes allow us
to define subsamples from this parent catalog. 

We use weights provided for all galaxies to account for the inverse correlation
between the number density of galaxies and that of stars \cite{Ross:2012}, and
that of seeing ($w_\ast$) as provided in the SDSS DR11 large
scale structure catalogs \cite{2014MNRAS.441...24A}.

\begin{figure}
    \centering
    \includegraphics[width=\columnwidth]{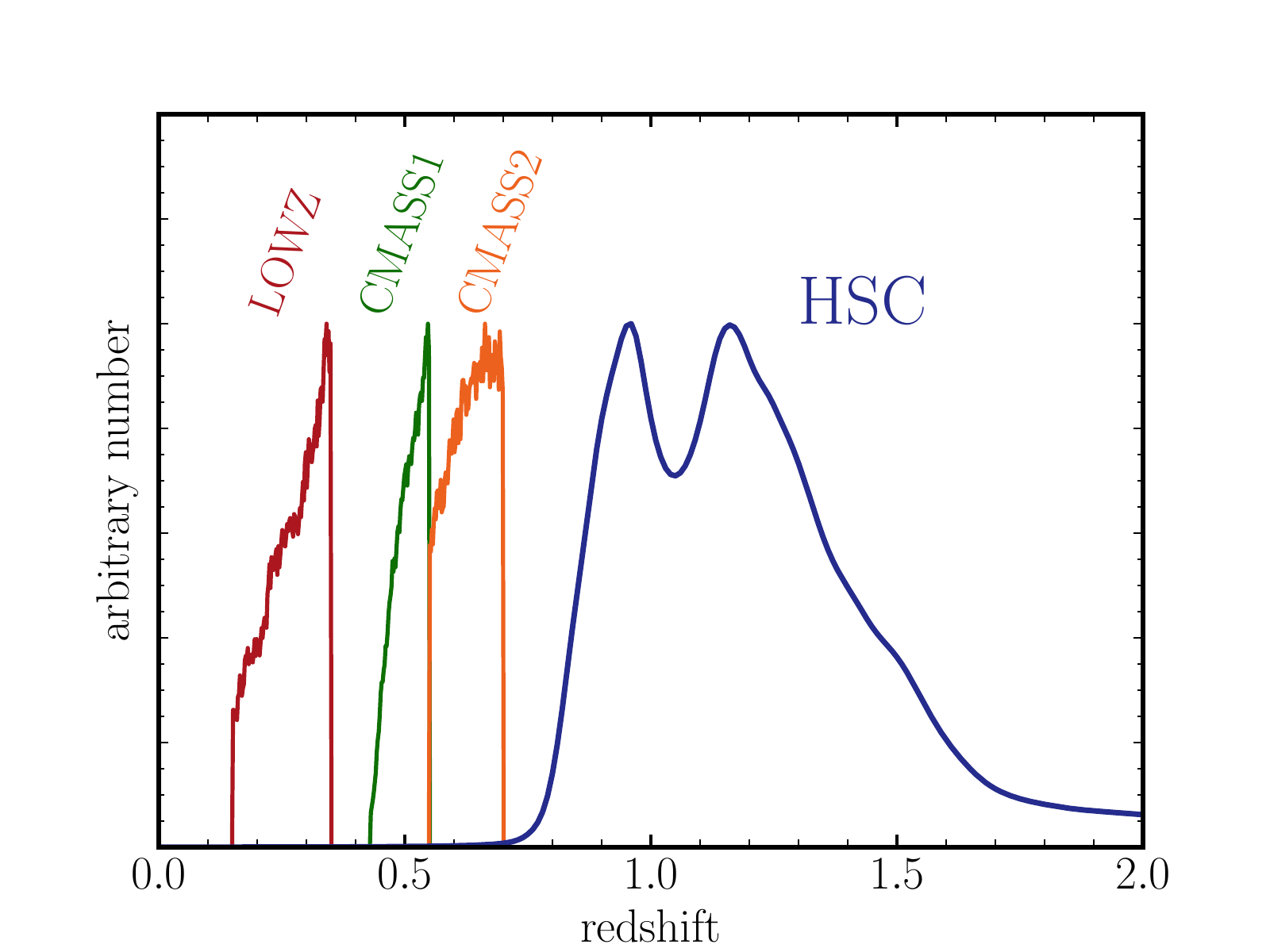}
    \caption{The redshift distribution of the three spectroscopic lens samples
are shown in red, green, and orange, respectively. These distributions can be
compared to the redshift distribution of the HSC source galaxy sample used in
our analysis shown in blue. In this figure the latter is estimated using stacked
$P(z)$ distribution estimated by \dempz for the source galaxies used in our
analysis.}
    \label{fig:pofz_sdss_hsc}
\end{figure}

We define three subsamples which are approximately volume-limited by
luminosity. The ``LOWZ'' subsample consists of galaxies with $z \in
[0.15,0.35]$, and the ``CMASS1'' and ``CMASS2'' subsamples consists of galaxies
with $z \in [0.43,0.55]$ and $z \in [0.55,0.70]$, respectively. We apply
further cuts on the absolute magnitude of galaxies of $M_i-5\log {\rm
h}<-21.5$, $-21.9$ and $-22.2$ for the three subsamples, respectively. This
results in subsamples which have a number density equal to $\bar{n}_{\rm
g}/[10^{-4}\,(h^{-1}{\rm Mpc})^{-3}] \simeq 1.8, 0.74$ and $0.45$,
respectively, which are a few times smaller than those of the entire parent
(color-cut and flux-limited) LOWZ and CMASS samples.  The redshift distribution
of the three lens subsamples compared to the stacked $P(z)$ distribution for
the HSC source galaxy sample we use in this paper is shown in
Fig.~\ref{fig:pofz_sdss_hsc} which shows the well separated redshift
distributions of our lens subsamples compared to the source galaxies.\footnote{The gap in the redshift distributions between LOWZ and CMASS samples is due to our use of the SDSS DR11 parent sample. Using the latest DR12 catalog would have allowed us to add in a few more galaxies at intermediate redshifts. However, that would have required us to modify the BOSS large scale structure sample to include the fiber collided and redshift failure galaxies. Since this procedure was already performed for DR11, we decided to stick with this sample.} Please
see Fig.~1 in \citet{2021arXiv210100113M} for the redshift distribution of the
three luminosity-limited subsamples, compared to that of the flux-limited sample. As we will show in Sections~\ref{subsec:clustering-systematics} and \ref{subsec:lensing_systematics}, use of such subsamples helps reduce systematics related to the variation of the measurements within each of the three redshift bins.

We will use random points in order to measure the galaxy clustering signal from
these subsamples, as well as testing for systematics in the lensing signal. The
random points are also useful to infer the potential dilution of the weak
lensing signal due to the use of galaxies physically clustered with our
subsamples, but are inferred to be at a higher redshift due to systematic and
statistical errors in the photometric redshift distribution. The random points
for each of our galaxy samples were constructed by downsampling the DR11 LSS
random catalogs which follow the angular mask and redshift distributions of all
galaxies in the LOWZ and CMASS samples, respectively, such that they follow the
redshift distributions of our galaxy subsamples.

We will show that the clustering and lensing observables within the redshift
bin of each sample do not evolve significantly given the approximate
volume-limited nature of our catalogs, compared to that for the flux-limited
sample.  This is similar to the findings in \citet[][ their
fig.~3]{Miyatakeetal:15}, but for a sample of stellar mass selected galaxies.
Our samples also allow a simpler treatment of the magnification bias effect on
the galaxy-galaxy weak lensing than the flux-limited sample
\cite[see][]{2021arXiv210100113M,Sugiyama:2021}.

\subsection{Simulations For Covariance Measurements}
\label{sec:data_sims}
In order to carry out the Bayesian inference of the cosmological parameters, we
will compare the measurements of the galaxy clustering and galaxy-galaxy
lensing signal of the lens subsamples described in the subsection above with
theoretical models. The covariance matrix of these measurements provides a
metric to compare the theoretical predictions to these observations, and thus a
robust determination of uncertainties in the inferred parameters, which is a
crucial component of a Bayesian analysis.
We use mock catalogs of SDSS and HSC galaxies, generated from the N-body
simulation based full-sky light-cone simulations \cite{2017ApJ...850...24T}, to
estimate the covariance matrix for our observables \cite[also see Appendix~B of
Ref.][for details]{2021arXiv210100113M}. We produce a large number of synthetic
SDSS galaxy mocks which are consistent with the measured clustering properties
of SDSS galaxies. The HSC galaxy mocks also include properties of the galaxies
such as the angular positions of each galaxy, their galaxy shapes and the
simulated lensing effect. We perform mock measurements of galaxy clustering and
galaxy-galaxy lensing signal with such synthetic galaxy data, in order to
estimate the statistical uncertainties.

\subsubsection{Generation of mock shape catalogs}
\label{subsubsec:mock-shape-catalogs}

We present the creation of the mock catalogs of galaxy shapes for the HSC-Y3
data.  We use the $108$ full sky lensing simulations in
\citet{2017ApJ...850...24T} in order to construct the mock catalogs. These
simulations adopt a flat $\Lambda$CDM cosmology consistent with the 9-year WMAP
cosmology (WMAP9) \cite{WMAP9}. To produce a set of galaxy shape catalogs, we
follow the same methodology as adopted in \citet{2019MNRAS.486...52S}.

We use 13 different rotations of the HSC-Y3 sky area on the full sky to extract
13 approximately non-overlapping HSC-Y3 survey footprints out of each full sky
lensing map. The full sky lensing map yields the shear for a given source
redshift at the location of each healpix pixel within the mock survey
footprints. In total, we thus have $108 \times 13 = 1404$ realizations of shear
in a number of lensing planes on a footprint equivalent to that of HSC-Y3 data.
These shear maps are then rotated back so that they occupy the same sky coordinates
as the HSC-Y3 footprint.

We use these shear maps to construct mock shape catalogs by using the observed
photometric redshifts and angular positions of real galaxies from the HSC-Y3
shape catalog. We rotate the shape of each source galaxy at random to erase the
real lensing signal imprinted on the HSC galaxies. In practice, we first rotate
the distortion of individual galaxies $\bm{\epsilon}^{\rm obs}$ and obtain the
rotated distortion as $\bm{\epsilon}^{\rm ran}=\bm{\epsilon}^{\rm
obs}e^{i\phi}$, where $\phi$ is a random number between 0 and $2\pi$. We have
to be careful that the observed distortion of a galaxy is different from the
intrinsic distortion due to the measurement error. We model the intrinsic shape
$\bm{\epsilon}^{\rm int}$ and measurement error $\bm{\epsilon}^{\rm mea}$ in
the following manner: 
\begin{eqnarray}
\bm{\epsilon}^{\rm int} &=& 
\left(\frac{\epsilon_{\rm rms}}{\sqrt{\epsilon_{\rm rms}^2+\sigma_{\rm e}^2}}\right)\bm{\epsilon}^{\rm ran}, \label{eq:e_int} \\ 
\bm{\epsilon}^{\rm mea} &=& N_1 +i\, N_2, \label{eq:e_mea}
\end{eqnarray}
where $N_{i}$ is a random number drawn from a normal distribution with a
standard deviation of $\sigma_{\rm e}$. The first equation guarantees that the
root-mean-squared (rms) ellipticity of the rotated intrinsic ellipticity is
equal to $\epsilon_{\rm rms}$.  We use $\epsilon_{\rm rms}$ (parameter {\tt
i\verb|_|shapehsmregauss\verb|_|derived\verb|_|rms\verb|_|e}) and $\sigma_{\rm
e}$ (parameter {\tt
i\verb|_|shapehsmregauss\verb|_|derived\verb|_|sigma\verb|_|e}) that are
provided on an object-by-object basis in the HSC-Y3 shape catalog.

We draw a true redshift for the galaxy randomly from its individual photometric
redshift PDF based on the photo-$z$ code {\sc DNNZ}. The choice for the use of
\dnnz was due to the availability of the redshift PDF estimates at an early
stage after the internal data release on which the shape catalog is based. We
add the lensing shear on the source galaxy from the shear map on a lensing
plane closest to the true redshift drawn for the galaxy thus assigning it an
ellipticity given by
\begin{eqnarray}
\epsilon^{\rm mock}_{1} &=& 
\frac{\epsilon^{\rm int}_{1}+
\delta_{1}+(\delta_{2}/\delta^2)
[1-(1-\delta^2)^{1/2}] (\delta_{1}\epsilon^{\rm int}_
{2}-\delta_{2}\epsilon^{\rm int}_{1})} {1+{\bm \delta}\cdot{\bm \epsilon}^{\rm int}} 
\nonumber \\ 
&& +\,\,\epsilon^{\rm mea}_{1}, \label{eq:emock1}\\ 
\epsilon^{\rm mock}_{2} &=& 
\frac{\epsilon^{\rm int}_{2}+\delta_{2}+(\delta_{1}/\delta^2) [1-(1-\delta^2)^{1/2}]
(\delta_{2}\epsilon^{\rm int}_{1}-\delta_{1}\epsilon^{\rm int}_{2})}
{1+{\bm \delta}\cdot{\bm \epsilon}^{\rm int}}
\nonumber \\ 
&& +\,\,\epsilon^{\rm mea}_{2}, \label{eq:emock2}
\end{eqnarray}
where ${\bm \delta}\equiv 2(1-\kappa){\bm
\gamma}/[(1-\kappa)^2+\gamma^2]$ and 
$\kappa$ and $\gamma$ are 
simulated lensing convergence and shear at the galaxy position, taken from the
ray-tracing simulation.  Note $\delta\simeq 2\gamma$ in the weak lensing regime
and we do not include any multiplicative and additive biases in mock catalogs
in Eqs.~(\ref{eq:emock1}) and (\ref{eq:emock2}). Our method maintains the
observed properties of the source galaxies on the sky including the lensing
weights.

\subsubsection{Generation of mock SDSS galaxy catalogs (lens and clustering samples)}
\label{subsubsec:mock-galaxy-catalogs}

\begin{figure*}
    \centering
    \includegraphics[width=2\columnwidth]{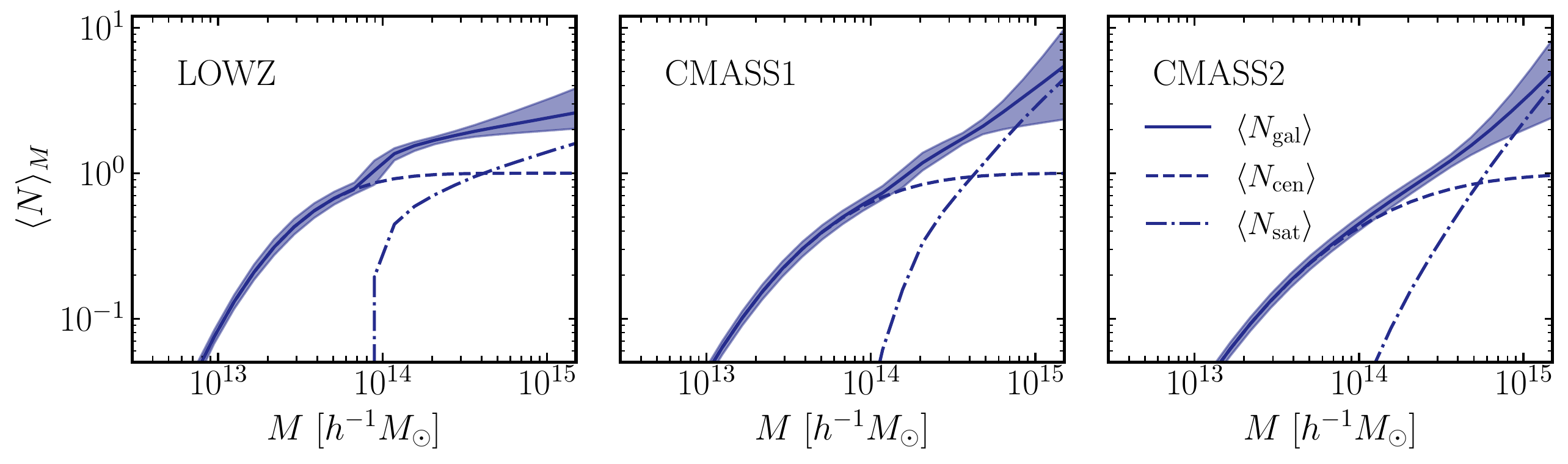}
    \caption{The number of galaxies in a halo of mass $M$ predicted by the HOD
model with HOD parameters obtained from the emulator based model fitting to the
clustering signal for each sample of the SDSS-BOSS galaxies. This HOD model is
used to populate galaxies in mock simulations, which in turn are used to
compute the covariance matrix of our observables.}
    \label{fig:NofM-HOD}
\end{figure*}

We populate galaxies in the dark matter halos in each of our realizations using
a halo occupation distribution (HOD) framework. The HOD $\langle N\rangle_{M}$
defines the number of galaxies in a halo of mass $M$. As is standard practice,
we divide the HOD into a central and a satellite galaxy component, where
\begin{eqnarray}
\langle N_{\mathrm{gal}}\rangle_M &=&  \langle N_{\mathrm{cen}}\rangle_M + 
\langle N_{\mathrm{sat}}\rangle_M\,, \label{eq:HOD} 
\end{eqnarray}
with
\begin{eqnarray}
\langle N_{\mathrm{cen}}\rangle_M &=& \frac{1}{2}\left[1+\mathrm{erf}\left(\frac{\log_{10} M - \log_{10} M_{\mathrm{min}}}{\sigma_{\log_{10} M}}\right)\right]\,, \label{eq:HOD_cen}\\
\langle N_{\mathrm{sat}}\rangle_M &=& \langle N_{\mathrm{cen}}\rangle_M 
\left(\frac{M - \kappa_M M_{\mathrm{min}}}{M_1}\right)^{\alpha_M}\,.\label{eq:HOD_sat}
\end{eqnarray}
To estimate HOD parameters, we first measure the clustering abundance of LOWZ,
CMASS1, and CMASS2 sample using the clustering measurement pipeline described
in Section~\ref{subsec:clustering_pipeline}. The clustering covariance is
obtained using 192 approximately equal area jackknife regions of the SDSS
footprint. We also measure abundance of each sample. To avoid an
over-dependence of the HOD constraints on the abundances, we use a more
conservative 10\% error estimate on these measurements. We model these
measurements by emulator-based halo model \cite{2021arXiv210100113M} to obtain
the HOD parameters, where we adopt the fitting range $0.5 < R/[\mpch] < 80$.
Note that we assume WMAP9 flat-$\Lambda$CDM cosmology \cite{WMAP9} for the
measurements and fitting. The resultant halo occupation distribution fits are
shown in Fig.~\ref{fig:NofM-HOD} and the corresponding best fit parameters for
the three subsamples are noted in Table~\ref{tab:HODmock}. Given that we are
only using the clustering signal to fit the halo occupation distribution
parameters, we expect a large number of degeneracies to be unresolved.

\begin{table}
    \centering
    \caption{HOD parameters that best fit the clustering measurements for the
three different subsamples along with the median redshift are noted in the
different columns of the table. The mock galaxy catalogs are populated with
these halo occupation distribution parameters. We provide the exact numbers we
use for the population of the catalog, the number of significant digits
mentioned are entirely for completeness and should not be taken to indicate the
error on the determination of these HOD parameters.}
\begin{tabular}{rrrr}
\toprule
       &    LOWZ    &  CMASS1 &   CMASS2 \\
Median z & 0.279 & 0.5206 & 0.6264 \\
\midrule
$\log M_{\rm min}$ & 13.502510 &  13.835350 &  14.134250 \\
$\sigma^2$ & 0.271474 & 0.473399 & 0.699328\\
$\log M_1$ & 14.435000 & 14.614390 & 14.686320 \\
$\alpha$ & 0.334696 & 1.868055 & 1.803775 \\
$\kappa$ & 2.620329 & 0.017179 & 0.016590 \\
\bottomrule
\end{tabular}
    \label{tab:HODmock}
\end{table}

We use these parameters to populate galaxies in dark matter halos in
the following manner. For every halo in the simulation, we compute the
central halo occupation $\langle N_{\mathrm{cen}}\rangle_M$, which
acts as a probability for it to host a central galaxy within the halo.
The central galaxy is assumed to reside at the center of each halo and
is at rest with respect to its host halo. For halos which have central
galaxies, we populate satellite galaxies with a Poisson deviate with
mean $\lambda_{M} =\left[(M - \kappa_M M_{\mathrm{min}} ) / M_1
\right]^{\alpha_M}$. The satellite galaxies are assumed to have a
radial distribution that follows the Navarro–Frenk–White (NFW) profile
\citep{Navarroetal:97} with a concentration parameter as measured
by the halo finder {\tt ROCKSTAR} \citep{Behroozi:2013}.
The satellite galaxies are assigned a velocity with respect to the halo
which is drawn from a Gaussian with zero mean and variance equal to
$\sigma^2_{\rm vir} = (1+z) \, G M / (2 R_{\rm 200m})$. The $(1+z)$
factor in this equation accounts for the fact that the halo radius is
measured in comoving units.

\section{Measurement codes}
\label{sec:measurement_codes}

In this section we describe the various analysis codes we use for
carrying out the galaxy clustering, galaxy-galaxy lensing and the
cosmic shear measurements as well as their covariances using the mock
catalogs we have described in the previous section.

\subsection{Clustering Pipeline}
\label{subsec:clustering_pipeline}
The measurement of the clustering signal requires fast computation of pair
counts at a variety of separations. We use a custom `clustering\_pipeline`
specifically developed for the measurement of weighted paircounts at a given
projected separation $R$ and line-of-sight separation $\pi$ starting
from catalogs of angular positions, redshifts and weights. The core functions
of the pipeline are written in C++, which are then exposed to python using the
Simplified Wrapper and Interface Generator (SWIG).

The nearest neighbour searches are carried out with the help of a
kd-tree\footnote{The original code was written in C++ by Matthew B Kennel, the
first author of this paper has carried out some bug fixes, memory leaks and
optimizations.} optimized to utilize the cache and enable fast searches for
potential neighbours in the galaxy catalog in the plane of the sky. The value
of $R_{\rm max}$ and $\pi_{\rm max}$ are used to leave out galaxy pairs that
do not lie within the required projected and line-of-sight distances from each
other.

The python functions allow input catalogs to be read in a variety of formats
such as plain text, fits and csv files with information about the angular
positions, redshifts, and weights of galaxies. These are then passed to the C++
code in order to generate the kd-tree for the paircounts. We use the
Landy-Szalay estimator in order to measure the 3-d correlation function,
\begin{equation}
\xi(R, \pi) = \frac{DD - 2DR + RR}{RR}\,.
\end{equation}
where $DD$ denotes the number of pairs of galaxies in the data, $DR$ the number
of normalized galaxy random pairs, and $RR$ the number of normalized random
random pairs at a given separation ($R, \pi$). We use 50 times more
randoms than the data points in order to reduce the shot noise in the
determination of $DR$ and $RR$. 

The computation of the pair counts are done using the Message Passing Interface
(MPI) using a division of tasks for each of these pair counts. The load
balancing is achieved by dividing the randoms into 50 separate units and
computing the pair counts separately. The C++ code can also compute jackknife
covariances on the fly if every galaxy is associated with a jackknife index.

We compute the 3 dimensional correlation function $\xi(R, \pi)$ and integrate
it along the line of sight direction to obtain the projected correlation
function $\wproj$,
\begin{equation}
\wproj(R) = 2 \int_0^{\pi_{\rm max}} \xi(R, \pi) d\pi\,.
\end{equation}
We use $\pi_{\rm max}=100\mpch$. This length is smaller than the line-of-sight
direction width of our subsamples and larger than the maximum projected
distance to which we compute $\wproj$.

The measurements of the clustering signal $\wproj(R)$ are dependent
upon the fiducial cosmological model used to compute these measurements. In the
modeling analysis we account for the cosmology dependence of the measurements
using the formalism presented in \citet{More:2013b} during the modeling phase. Briefly, the fiducial cosmological model used to measure the clustering enters the measurements when the angular separations between the galaxies get converted into comoving separations and the redshift differences in to line-of-sight comoving separations. The ratio of the comoving distance to the median redshift of each sample between the fiducial model and the cosmological model under consideration, as well as a similar ratio between the Hubble parameters in the two models can correct for the clustering measurements. On the other hand, for the weak lensing signal, we additionally have to consider the dependence of the critical surface density on the fiducial cosmological parameters and the cosmological model under consideration. 

The various inputs to the
clustering pipeline such as the cosmological model to be used for the
measurements, the minimum and maximum projected distances, the number of radial
bins, the number of bins in projected radius, the types of galaxies used to
measure the clustering (cross or auto) are passed via a simple yaml file.

The pipeline has been validated by comparing the correlation function
measurements against those measured by the BOSS survey collaboration using the
entire BOSS-LOWZ and BOSS-CMASS samples.

\subsection{Galaxy-galaxy Lensing Pipeline}
\label{subsec:gglens-pipeline}
The gravitational lensing due to the presence of mass between us and the far
away source galaxies results in coherent distortions in their shapes. Such
distortions can be inferred from the observed distribution of ellipticities of
these galaxies. As galaxies have intrinsic shapes, this signal needs to be
measured as a statistical average of the shapes of a large number of galaxies.

The HSC-Y3 shape catalog provides the ellipticity components $(e_1,
e_2)$ along with the weights for every galaxy $w_{\rm s}$, the additive and
multiplicative biases $(c_{\rm 1}, c_{\rm 2}, m)$ and the variance of the
ellipticity $e_{\rm RMS}$. Given $\phi$, the angle between the line joining the
lens and the source galaxy in the plane of the sky and the x-axis of the
coordinate system, the tangential distortion in shape is given by
\begin{equation}
e_{\rm t}= -e_{\rm 1} \cos(2\phi) - e_{\rm 2} \sin(2\phi)\,.
\label{eq:e_t}
\end{equation}
The average of the tangential ellipticity is proportional to the tangential
shear induced on the galaxy by the matter distribution correlated with the lens
galaxy. The tangential shear depends upon the surface mass density,
$\Sigma(R)$, and the critical surface density for lensing $\Sigma_{\rm crit}$
such that
\begin{equation}
\gamma_{\rm t}(R)=\frac{\Sigma(<R) -\Sigma(R)}{\Sigma_{\rm cr}(z_{\rm l},z_{\rm s})}=\frac{\Delta\Sigma(R)}{\Sigma_{\rm cr}(z_{\rm l},z_{\rm s})}\,.
\label{eq:esd}
\end{equation}
Here, the symbol $\Sigma(<R)$ denotes the average surface mass density within a
projected distance $R$ from the lens, and $\Sigma(R)$ is the surface mass
density at the distance $R$ after performing an azimuthal average. The critical
surface density, $\Sigma_{\rm cr}(z_{\rm l},z_{\rm s})$, is a geometrical factor
dependent on the angular diameter distances to the lens $D_{\rm A}(z_{\rm l})$,
the source $D_{\rm A}(z_{\rm s})$ and between the two, $D_{\rm A}(z_{\rm
l},z_{\rm s})$. When expressed in comoving units, the critical surface density
is given by
\begin{equation}
\Sigma_{\rm cr}(z_{\rm l},z_{\rm s})= \frac{c^2}{4\pi G}\frac{D_{\rm A}(z_{\rm s})}{D_{\rm A}(z_{\rm l}) D_{\rm A}(z_{\rm l},z_{\rm s})(1+z_{\rm l}^2)}\,.
\label{eq:sigma_crit}
\end{equation}
As the galaxy-galaxy lensing signal, we measure the excess surface density
$\Delta\Sigma(R)=\Sigma(<R) -\Sigma(R)$ as a minimum variance weighted
statistical average,
\begin{equation}
    \Delta\Sigma(R)=\frac{1}{(1+\hat{m})}\left(\frac{\sum_{\rm ls}w_{\rm ls} e_{\rm t,ls}\left\langle\Sigma_{\rm cr}^{-1}\right\rangle^{-1}_{\rm ls}}{2\mathcal{R}\sum_{\rm ls}w_{\rm ls} }\right)\,.
\label{eq:stacked_signal}
\end{equation}

The average inverse critical surface density is computed from the full
photometric redshift posterior distribution, $p(z_{\rm s})$,
\begin{equation}
    \langle\Sigma_{\rm cr}^{-1}\rangle_{\rm ls}=\frac{4\pi G(1+z_{\rm l})^2}{c^2}\int_{z_{\rm l}}^{\infty}\frac{D_{\rm A}(z_{\rm l}) D_{\rm A}(z_{\rm l},z_{\rm s})}{D_{\rm A}(z_{\rm s})} p(z_{\rm s})dz_{\rm s} \,,
\end{equation}
where the lower limit of the integration is from the lens redshift $z_{\rm l}$.
Given the photometric redshift errors, one can expect galaxies in the
foreground or at the lens redshifts to be in our source sample. To reduce the
effects of such galaxies we only choose those sources that have a large
probability to lie at a specified redshift according to Eq.~\ref{eq:source-selection}.
The parameter $z_{\rm max}$ can be chosen to correspond to the maximum
redshifts of the sample, and the tunable parameters $z_{\rm diff}$ and $p_{\rm
cut}$ which can be changed to allow the cuts to be more stringent.
Alternatively, the pipeline also allows for the possibility to carry out the
source selection by using the best estimate of the photometric redshift.

The weight used in Eq.~(\ref{eq:stacked_signal}), $w_{\rm ls}=w_{\rm l}w_{\rm s}\langle\Sigma_{\rm crit}^{-1}\rangle^2_{\rm ls}$, is the minimum variance weight for the estimator of the signal $\Delta\Sigma$, and it down weights those lens-source
pairs that are close to each other in redshift. The shear responsivity ${\cal
R}$ is given by 
\begin{equation}
    \mathcal{R}=1- \frac{\sum_{\rm ls}w_{\rm ls }e_{\rm RMS}^2}{\sum _{\rm ls} w_{\rm ls}}\,,
\end{equation}
and can be estimated from $e_{\rm RMS}$. Finally, the term $\hat{m}$ is the
average multiplicative bias and is defined as as $\hat{m}=\Sigma_{\rm ls}w_{\rm
ls}m_s/\Sigma_{\rm ls}w_{\rm ls}$.

In addition, we consider the effect on the multiplicative and the additive
selection bias. \citet{Li2021} found that the multiplicative bias and selection
bias are proportional to the fraction of galaxies at the sharp boundary of
selection cuts on resolution and aperture magnitude as follows:
\begin{align}
    \hat{m}_{\rm sel} = -0.05854\,\hat{P}({\rm mag}_A=25.5) + 0.01919\,\hat{P}(R_2=0.3), \\
    \hat{a}_{\rm sel} = 0.00635\,\hat{P}({\rm mag}_A=25.5) + 0.00627\,\hat{P}(R_2=0.3),
\end{align}
respectively. Here, $\hat{P}(X)$ is the fraction estimate of galaxies at the
boundary of selection cut on $X$. The galaxy-galaxy lensing estimate is
corrected as
\begin{align}
    &\dSigma \rightarrow \frac{1}{1+\hat{m}_{\rm sel}} (\dSigma - \hat{a}_{\rm sel}\dSigma^{\rm psf}),\label{eq:dsig_add_bias}\\
    &\dSigma^{\rm psf} = \frac{\sum_{\rm ls}w_{\rm ls}e_{\rm t, ls}^{\rm psf}}{\sum_{\rm ls}w_{\rm ls}},
\end{align}
where $e_i^{\rm psf}$ is the PSF shape.

Finally we also repeat the same exact measurement procedure mentioned above for the weak lensing signal but
around random points which are 40 times larger than the number of lenses. We
use random catalogs which occupy the same area on the sky and follow the same
redshift distributions as our lens subsamples. This allows us to subtract out
any large-scale systematic effects that may exist, especially on scales
comparable to the field of view of HSC. This changes our final estimator from the one written in Eq.~\ref{eq:dsig_add_bias},
\begin{align}
    \dSigma \rightarrow \dSigma - \dSigma_{\rm rand}\,.
\end{align}

The weak lensing pipeline we have developed implements the above calculations
in order to measure the galaxy-galaxy lensing signal. At its core, it needs to
perform nearest neighbour calculations on the plane of the sky between lens and
source galaxies. It uses the the same kd-tree code as used in the galaxy
clustering pipeline in order to speed up these calculations. The core functions
of the pipeline are again written in C++ for speed, with interfaces provided
via python using SWIG. The input and output is handled on the python side
making it easier to swap in and out different catalogs for the lensing
measurements while preserving the core functionality.

In order to make the code memory efficient, we search for lens galaxies around
source galaxies rather than the other way round. The computational complexity
of the algorithm to generate a tree of $N$ points and to perform $n$ searches
with it is ${\cal O}[(N+n)\log N)]$, thus it is better to generate the tree out
of the smaller of the two numbers. Given the order of magnitude difference
between the number of lens and the source samples, we generate the tree using
the lens catalogs. This in turn also makes the code memory efficient, because
the entirety of the source catalog need not be read right at the beginning in
order to generate the tree.

The computation of the weak lensing signal and the covariance is parallelized
using a division of tasks. Given that the code requires to compute the weak
lensing signal around random points, we usually divide the task into
computation of the weak lensing signal around the lens galaxy samples and the
randoms into separate tasks. The covariance computations are also performed by
running the weak lensing pipeline over a large number of mock catalogs. This
allows us to massively parallelize the computation of the weak lensing signal
and covariances. The code also allows for the computation of the weak lensing
signal in different fields separately. The output of the weak lensing pipeline
has all the relevant outputs which can then later be combined in the
post-processing.

The various inputs to the weak lensing pipeline such as the cosmological model
to be used for the measurements, the minimum and maximum projected distances,
the number of radial bins, the lens and source selection parameters can all be
passed via simple yaml files. The pipeline has been validated by comparing the
weak lensing measurements with a number of independent codes written by other
authors in the HSC survey collaboration. 

\subsection{Cosmic Shear Measurement Code}
\label{subsec:cosmicshear-code}
In order to compute the cosmic shear, we use the estimator in \citet{Li2021}
for the shear estimate for each galaxy,
\begin{equation}
    \hat\gamma_i = \frac{1}{(1+\hat{m})}\left[\frac{e_i}{2{\cal R}} - c_i \right]\,.
\end{equation}
In this case the weighted average multiplicative bias factor is given by
\begin{equation}
    \hat{m} = \frac{\sum_i w_i m_i}{\sum_i w_i}\,,
\end{equation}
and the weight in this equation corresponds to the shape weight $w_{\rm s}$
given in the HSC shape catalog \cite{Li2021}. The shear responsivity in this
case is estimated using
\begin{equation}
    {\cal R} = 1 - \frac{\sum_i w_i e_{\rm rms, i}^2}{\sum_i w_i}\,.
\end{equation}

As in the galaxy-galaxy lensing measurement, we take account of the
multiplicative and the additive selection biases in the cosmic shear
measurement. 
\begin{align}
    \hat\gamma_i \rightarrow \frac{1}{1+m_{\rm sel}}(\hat\gamma_i - a_{\rm sel}e_i^{\rm psf})\,,
\end{align}
where $m_{\rm sel}$ and $a_{\rm sel}$ are evaluated using the entire sample of our source galaxies.

We use the software {\sc treecorr}, in order to compute the two point shear
correlation functions $\xi_+$ and $\xi_-$.  In our case we use a single source
sample which would allow us to self-calibrate any residual uncertainties in the
determination of the true redshift distribution given the photometric redshifts
of our galaxies. We estimate $\xi_{\pm}$ as
\begin{equation}
    \xi_{\pm}(\theta) = \frac{\sum_{ij} w_i w_j \left[ \hat\gamma_{i, t}\hat\gamma_{j, t} \pm \hat\gamma_{i, \times}\hat\gamma_{j, \times} \right]}{\sum_{ij} w_i w_j} \,,
\end{equation}
and the summation runs over all pairs of galaxies $i, j$ whose angular
separation falls within a bin of given width around $\theta$.

We also diagnose any contamination of the above correlations due to the
presence of point spread function (PSF) anisotropies. Such residuals are
expected to be present due to the imperfections in the modeling and the
measurements of the PSF. We continue to follow the prescription presented in
\citet{Hikageetal:11} and \citet{Troxel:2018}, such that the systematics due to
the PSF errors are added in a linear fashion and arise from two sources,
\begin{equation}
    \gamma^{\rm sys} = \alpha_{\rm psf} \gamma^{\rm p} + \beta_{\rm psf} \gamma^{\rm q}\,, 
\end{equation}
where $\gamma^{\rm p}$ is the shape of the PSF model, and $\gamma^{\rm q}$ is
the difference between the true PSF of stars and the model PSF at their
locations. These two terms account for the error in the deconvolution of the
PSF from the galaxy shapes and the error in the shapes due to the imperfect
modeling of the PSF. These terms give rise to a spurious component in the
measured cosmic shear signal, $\xi_{\rm psf}$ given by,
\begin{equation}
    \hat\xi_{\rm psf, \pm}(\theta) = \alpha_{\rm psf}^2 \xi_{\pm}^{\rm pp} + 2\alpha_{\rm psf}\beta_{\rm psf} \xi_{\pm}^{\rm pq} +    \beta_{\rm psf}^2\xi_{\pm}^{qq}\,.
    \label{eq:psf-contamination}
\end{equation}
Here the quantities $\xi_{\pm}^{\rm pp}$ and $\xi_{\pm}^{\rm qq}$ are the
auto-correlations of $\gamma^{\rm p}$ and $\gamma^{\rm q}$, respectively and
$\xi_{\pm}^{\rm pq}$ corresponds to the cross-correlation of the two
quantities. These correlations can be measured directly from the data by using
the shapes of stars, that were reserved for modeling the PSF over the entire
field of view. In HSC, the PSF measurement and modeling is performed over each
exposure, rather than the coadd and 20 percent of the stars are randomly
reserved for PSF testing. Since these stars are randomly selected during each
exposure, the number of stars that are never used in the PSF determination is
quite small. The HSC pipeline assigns a flag {\sc i\_calib\_psf\_used} to
denote stars that have been used for the PSF determination at least in $80$
percent of the visits they belong to. We use stars with flags {\sc
i\_calib\_psf\_used=True} to compute these correlations, because it is less
noisy. We also take into account the uncertainty of the predicted $\hat\xi_{\rm
psf, \pm}$ whether we use {\sc i\_calib\_psf\_used=True} or {\sc False}. In
Section~\ref{sec:cosmic-shear-systematics}, we will present our constraints on
the parameters $\alpha_{\rm PSF}$ and $\beta_{\rm PSF}$ and our suggested
priors for the cosmological analyses.

Note that our treatment of the PSF systematics in the 3$\times$2pt
analyses differs
from the default setup used in the cosmic shear analyses alone in our companion
papers Li \etal \cite{Li_hscy3} and Dalal \etal \cite{Dalal_hscy3}, where we account for the errors in the PSF
moments upto fourth order following the methodology developed in \cite{TQ:2023}.
In Sugiyama \etal \cite{Sugiyama_hscy3} and Miyatake \etal \cite{Miyatake_hscy3}, we will test the impact of including such
effect for the HSC source galaxies used in our analyses and quantify its impact
on the cosmological parameters.

\begin{figure*}
    \centering
    \includegraphics[width=2\columnwidth]{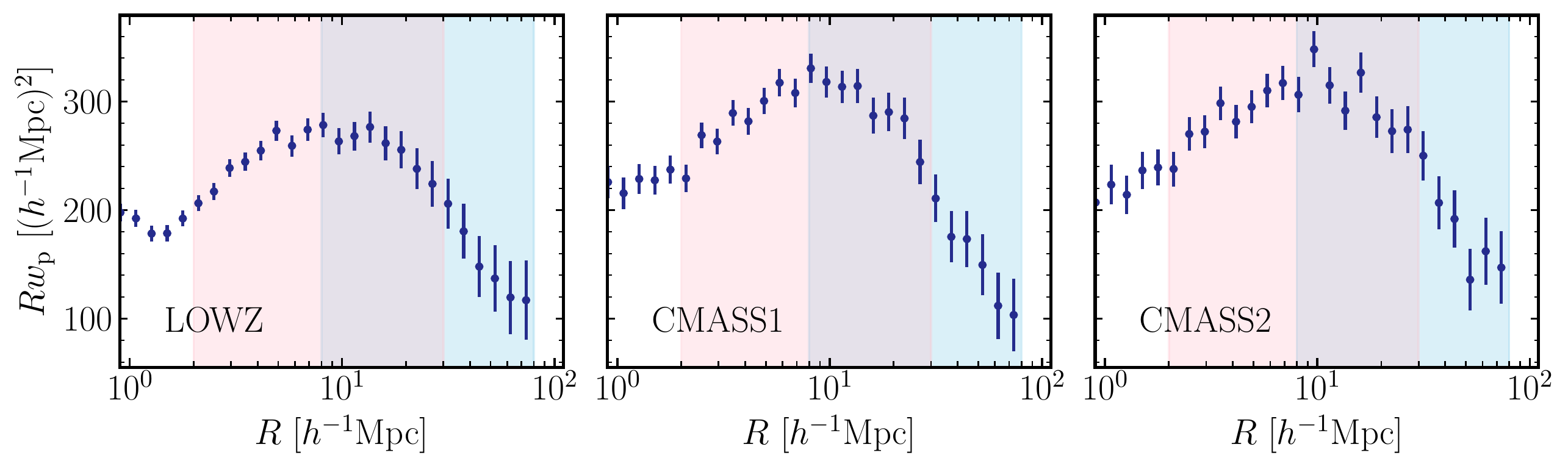}
    \includegraphics[width=2\columnwidth]{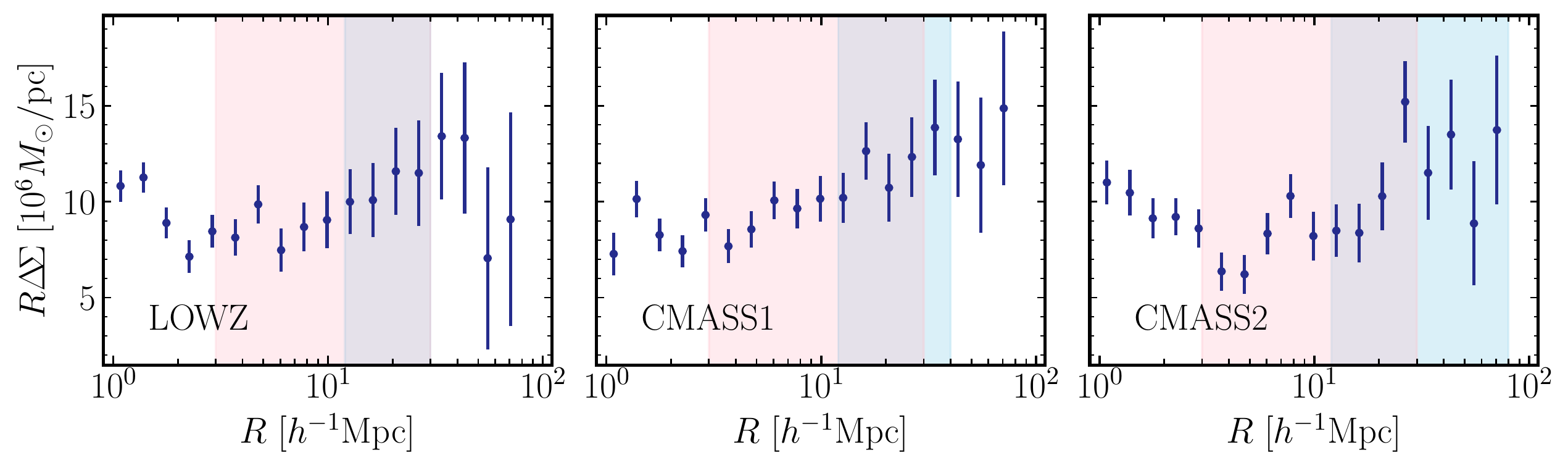}
    \includegraphics[width=1.4\columnwidth]{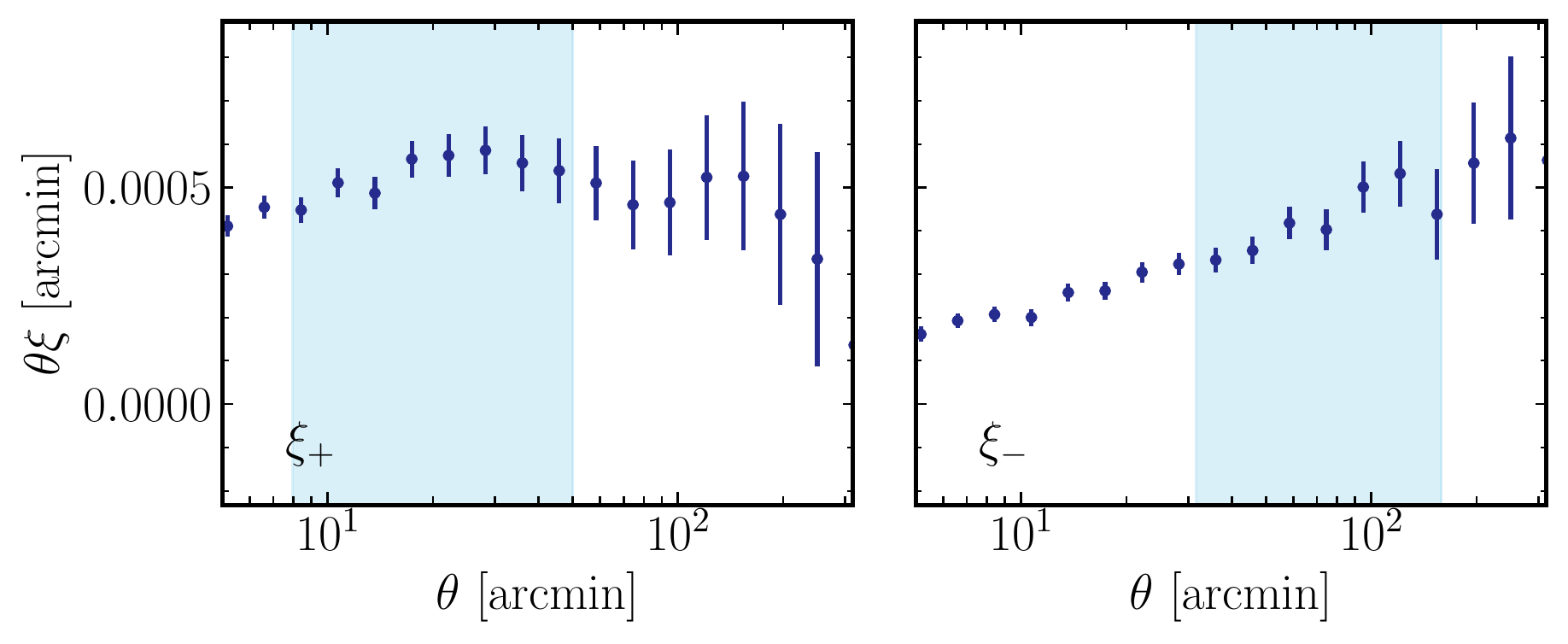}
    \caption{The measurements of the 3$\times$2pt
    functions that will be used for the
cosmology parameter inference in companion papers, Miyatake {\it et al.} and
Sugiyama {\it et al.} {\it Top panels}: The clustering signals of spectroscopic
SDSS galaxies are shown for LOWZ, CMASS1 and CMASS2 in redshift bin
$z\in[0.15,0.35]$, $[0.43, 0.55]$ and $[0.55,0.70]$ from left to right panels,
respectively. Although the signal is identical to the Year~1 analysis
\citep{Miyatake:2022,Sugiyama:2021}, the updated covariance is measured
from mock catalogs described in this paper. The shaded regions indicate the
scales used for cosmology analyses in the companion papers: blue for the
large-scale analysis with minimal bias model by Sugiyama {\it et al.}, and red
for the small-scale analysis by Miyatake {\it et al}.  {\it Middle panels}: The
galaxy-galaxy weak lensing signals measured by combining the spectroscopic SDSS
galaxies and HSC-Y3 source galaxies. For the measurement of clustering and
galaxy-galaxy lensing signals, we used a flat $\Lambda$CDM model with
$\Omega_{\rm m}=0.279$ to convert angular separation $\theta$ to physical
separation $R$ and to compute $\langle\Sigma_{\rm cr}\rangle$. {\it Bottom
panels}: The cosmic shear correlation functions for the plus and minus modes
are shown in the left and the right panels, respectively. The blue shaded
regions indicate the scales used for both of the large-scale and the
small-scale cosmology analyses. 
    }
    \label{fig:signals}
\end{figure*}

\begin{figure*}
    \centering
    \includegraphics[width=\columnwidth]{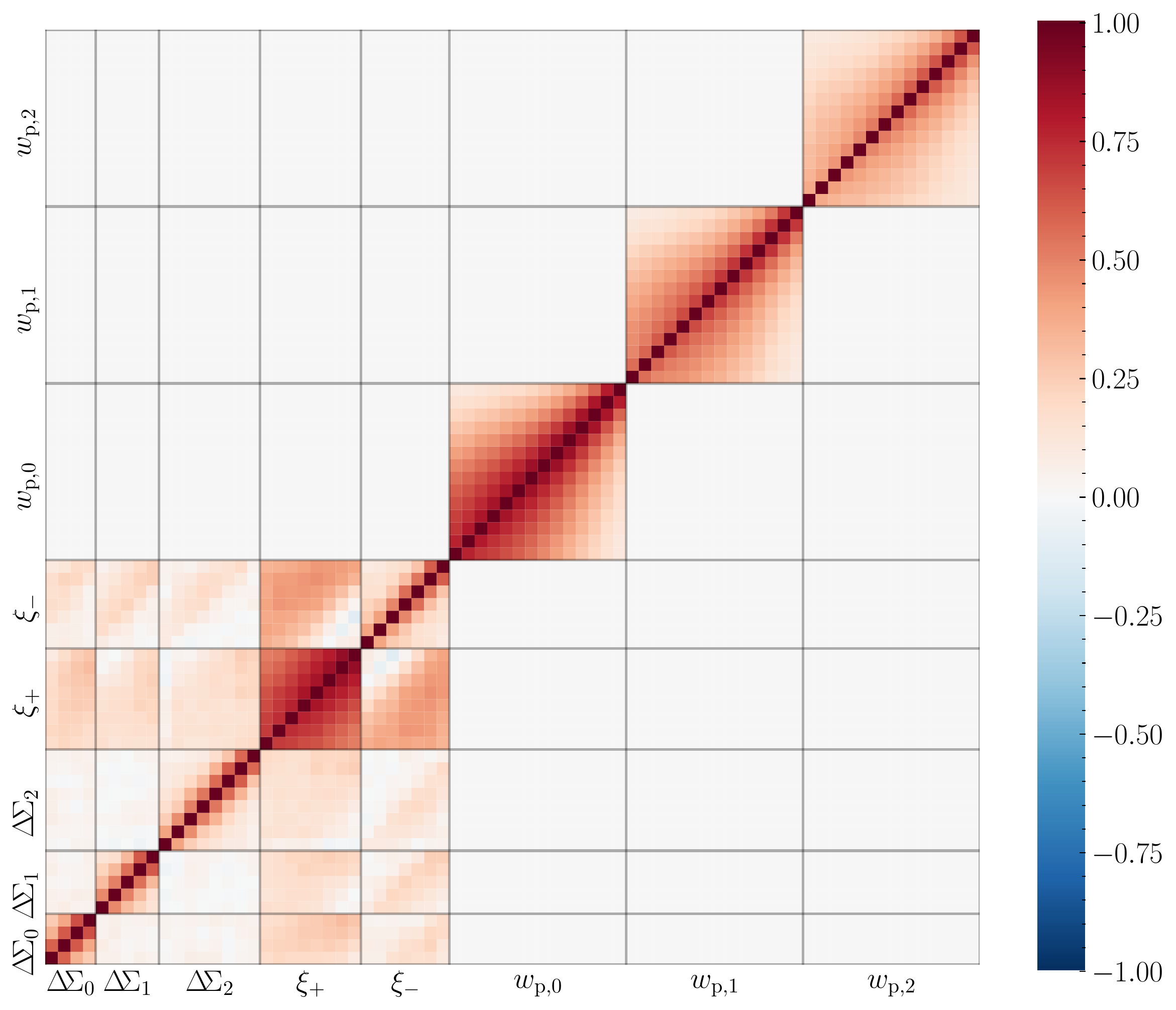}
    \includegraphics[width=\columnwidth]{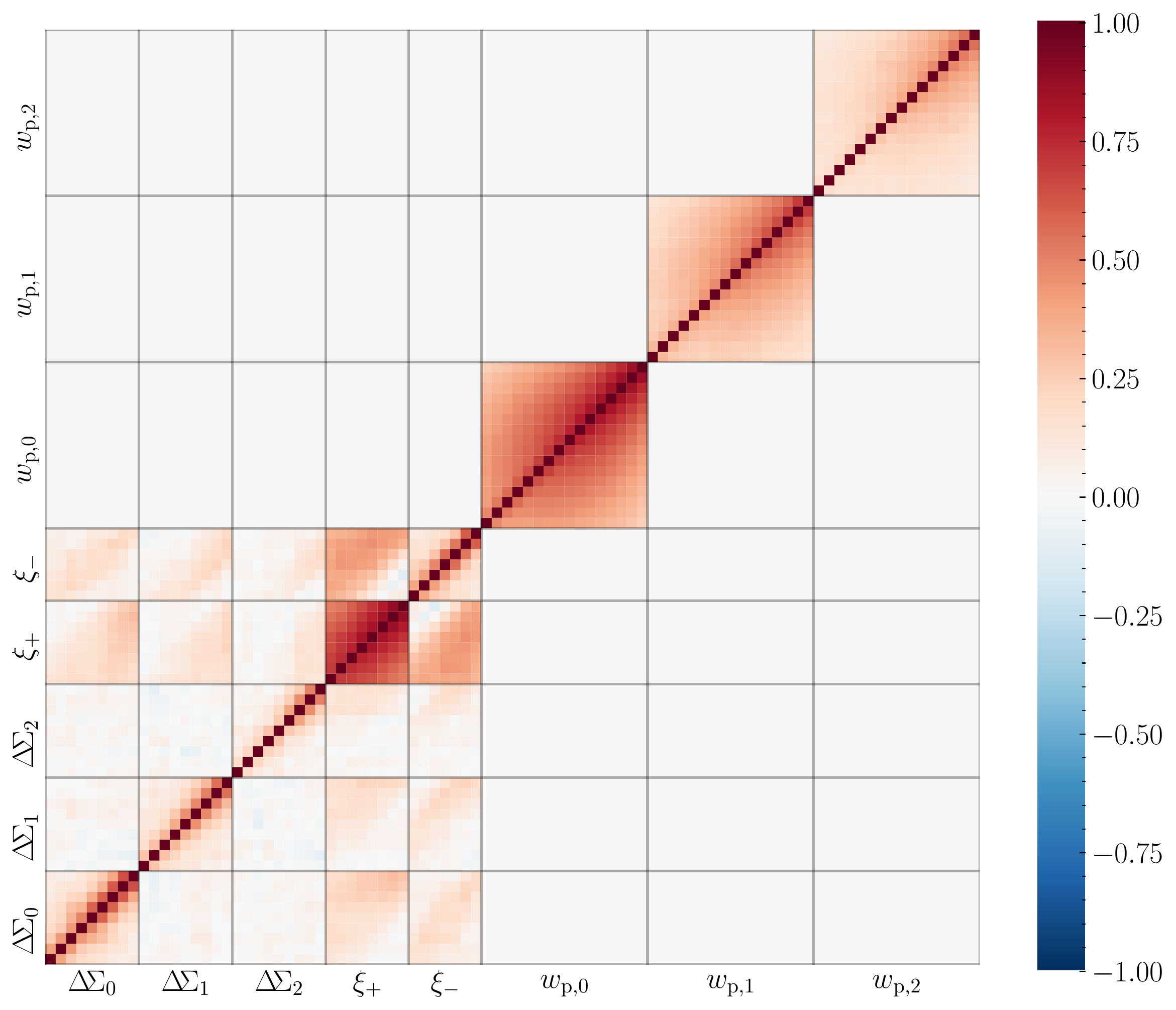}
    \caption{The correlation coefficient of the full covariance matrix. {\it Left}: For the large-scale cosmology analysis with minimal bias model. The scales shown in this figure are indicated by the shaded regions in Fig.~\ref{fig:signals}. Subscripts of 0, 1, and 2 for $\dSigma$ and $\wproj$ stand for the LOWZ, CMASS1 and CMASS2 subsamples, respectively. The contribution of magnification bias is evaluated analytically and added on the covariance estimated from mock measurements. {\it Right}: Similar to the left panel, but for the small-scale analysis.
    }
    \label{fig:covariance}
\end{figure*}

Finally, we also will perform an $E$/$B$ mode decomposition of the measured
cosmic shear two point correlation functions, such that,
\begin{eqnarray}
    \xi_{\rm E}(\theta) &=& \frac{\xi_{+}(\theta) + \xi'(\theta)}{2} \,,\\
    \xi_{\rm B}(\theta) &=& \frac{\xi_{+}(\theta) - \xi'(\theta)}{2} \,,    
\end{eqnarray}
where, $\xi'$ is given by 
\begin{eqnarray}
    \xi'(\theta) = \xi_{-}(\theta) + 4\int_{\theta}^{\infty}\frac{d\theta'}{\theta'}\xi_{-}(\theta') - 12 \theta^2\int_{\theta}^{\infty} \frac{d\theta'}{\theta'^3} \xi_{-}(\theta') \,.\nonumber\\
\end{eqnarray}
We replace the integral with a Riemann sum over the range $\theta'\in [2, 420]$
and with logarithmic bins of size $\Delta \log \theta=0.0222$. Larger bin size
would smooth out the angle dependence of $\xi_{-}$, and affect the accuracy of
the Riemann sum as a consequence. Therefore, we use narrower bins in $\theta$
for the measurement of $\xi_\pm$, decompose them into $E$/$B$ modes using above
equations and then finally smooth the signal to obtain $E$/$B$ signals over
similar bins as the cosmology analysis. We have carefully checked that the
results converge with the choice of the bins we use.

\section{Clustering measurements}
\label{sec:clustering}
\subsection{Measurements}
\label{subsec:clustering_meas}

We show the clustering signals we measure for the three subsamples in the three
different panels of Fig.~\ref{fig:signals}, respectively. Although we have
measured the signals on small scales, we restrict to scales beyond 1 $\mpch$.
The measured projected correlation functions falls off approximately as $1/R$.
The deviation from this power law can be seen clearly when we plot $w_{\rm
p}R$. 

The clustering of LOWZ galaxies clearly shows the transition between the
one-halo term and the two-halo term on scales around $1.5\mpch$. This
transition is
less pronounced for the CMASS1 and CMASS2 subsamples. In both our cosmological
analyses we avoid modeling below the transition scale due to uncertainties in the
accuracy of their modeling. We have tested the accuracy of both our modeling
schemes against different systematic uncertainties and decided on the scale
cuts that we will use for our analysis in Miyatake \etal \cite{Miyatake_hscy3} and Sugiyama \etal \cite{Sugiyama_hscy3}. The validation of the scale cuts include tests on different models for how satellite galaxies populate dark matter halos, whether the central galaxies are off-centered within their respective halos, the presence of incompleteness in the halo occupation distribution of central galaxies, or that of halo assembly bias, in addition to the choice of halo definitions in simulations. Based on the validation analyses, we use a range of $[2, 30]\mpch$ and $[8, 80]\mpch$ for the small-scale and large-scale analyses, respectively.

We use the 108 realizations of the full sky mock catalogs imprinted with the
BOSS LOWZ, CMASS1 and CMASS2 footprints. Within each realization we utilize 192
jackknife regions of the SDSS survey footprint \citep[see][for
details]{Miyatakeetal:15}, measure $\wproj$ for each jackknife region, and
estimate the covariance matrix from the measured $\wproj$'s from all the
jackknife regions. We then average the covariances for the 108 realizations to
estimate the covariance matrix for each of the LOWZ, CMASS1, and CMASS2
subsamples. The covariance matrix estimated in this way is used for our
cosmology analysis. Since the covariance matrix of $\wproj$ is estimated
effectively from a larger number of samples $20,736=(108\times 192)$, the
Hartlap factor \citep{2007A&A...464..399H} can be ignored in evaluating the
inverse of the covariance matrix for the clustering sector. This is different
from our analysis with Y1 data, where we just used the jackknife estimate of
the covariance from the real data. Our new method allows the determination of
the covariance with reduced noise properties.

One can also expect some cross-covariance between the measurements as the
CMASS1 and CMASS2 subsamples share a boundary and the galaxies near the
boundary will share the same large-scale structure, although we expect this to
be a small effect. Finally, we have also analytically checked that the
magnification bias effect on the observed galaxy number density by
gravitational lensing by foreground matter fluctuation has negligible effect on
the clustering covariance, thus we do not include the cross-covariance between
$\wproj$'s for different subsamples.

The cross-correlation coefficients of the covariance result is shown in
Figs.~\ref{fig:signals} and \ref{fig:covariance}. The errorbars in
Fig.~\ref{fig:signals} show the diagonal elements of estimates covariance.
Fig.~\ref{fig:covariance} shows the correlation coefficients of the covariance,
over the scale used for cosmology analyses: the large-scale analysis with
minimal bias model by Sugiyama {\it et al.}, and red for the small-scale
analysis with the emulator based halo model by Miyatake {\it et al},
respectively.

\begin{figure*}
    \centering
    \includegraphics[width=2\columnwidth]{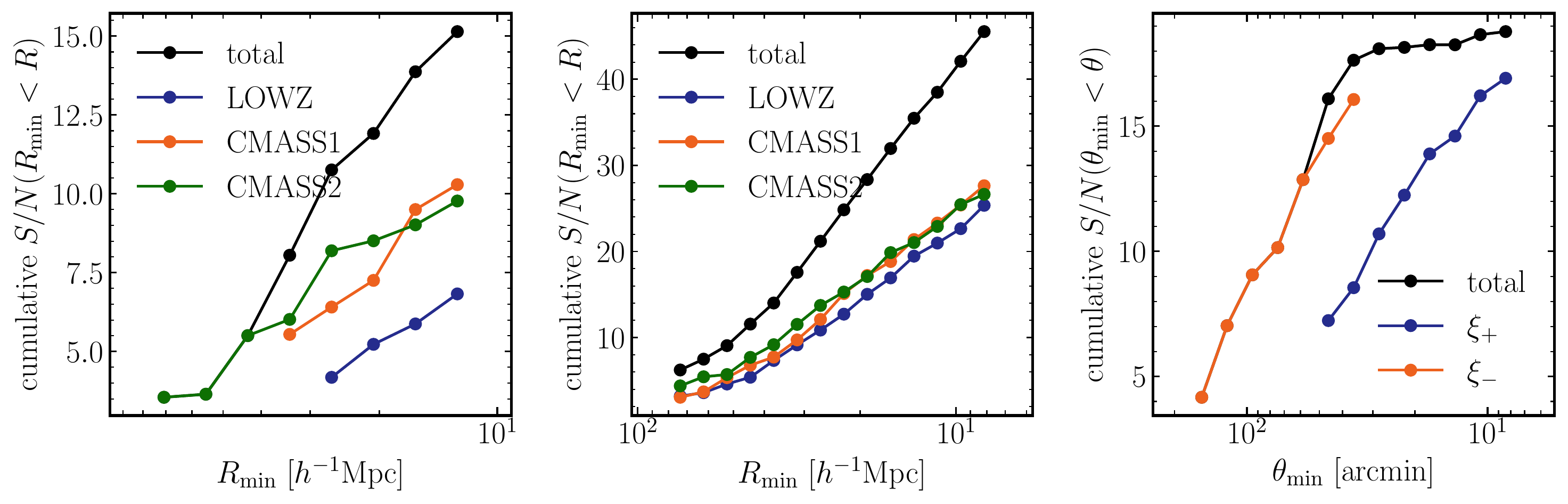}
    \includegraphics[width=2\columnwidth]{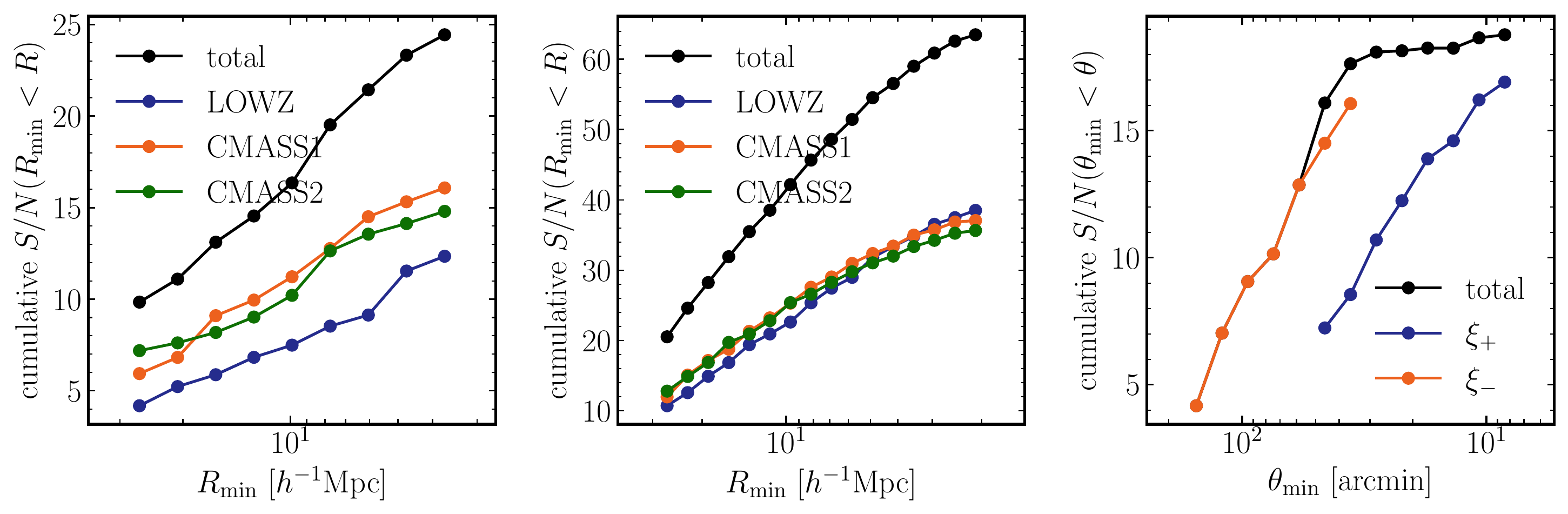}
    \caption{The cumulative signal-to-noise ratios as a function of minimum
scale cuts for clustering, galaxy-galaxy lensing and cosmic shear are shown in
the different columns, respectively. The top and the bottom panels show the
cumulative signal-to-noise ratio for large and small-scale analysis
respectively.}
    \label{fig:cumsumSN}
\end{figure*}

Fig.~\ref{fig:cumsumSN} shows the cumulative signal-to-noise ratio as the
function of minimum scale cut. The upper and lower panels are for large and
small-scale analysis, respectively. The total signal-to-noise ratios of our
measurements in the LOWZ, CMASS1 and CMASS2 subsamples are 25.6, 27.6, 26.6,
respectively for our large-scale analysis. When including the scales that
will be modelled in our small-scale analysis, these signal-to-noise ratios
increase to 38.5, 37.1, 35.7, respectively.

\subsection{Systematic Tests}\label{subsec:clustering-systematics}

We carry out a couple of tests in order to understand the possible systematics
in our analysis. The first is the dependence of the galaxy number density in
our lens sample on seeing and the number density of stars. In
\citet{Ross:2017}, these relations were calibrated based on the entire large
scale structure sample. Since we are including a cut on the luminosity which is
dependent on luminosity we are preferentially selecting the brighter galaxies.
If the dependence is weaker for brighter galaxies then we may end up giving
extra weight to brighter galaxies.

\begin{figure*}
    \centering
    \includegraphics[width=2\columnwidth]{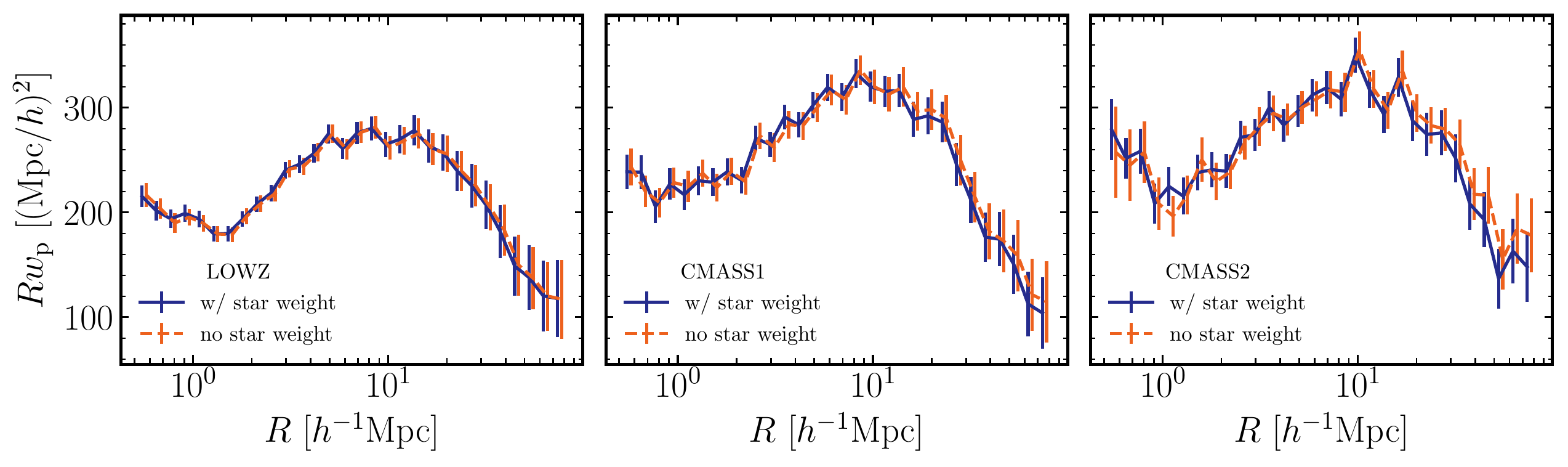}
\caption{The dependence of the clustering signals on the systematic weights
related to seeing and stellar number density. We find a very weak dependence of
these measurements on such systematic weights within the statiscal errors.}
    \label{fig:sys_clustering_weights}
\end{figure*}

To account for the reduction in the number density of galaxies in regions where
there is a large stellar number density, the large-scale structure catalogs
have been assigned a stellar density weight. Since this weight is also
dependent on the magnitude of galaxies, we carry over the weights assigned to
these galaxies even in our subsamples. In order to test how such systematic
stellar weights affect the clustering signal, we show the variation of the
clustering signal with and without the systematic star weights in
Fig~\ref{fig:sys_clustering_weights}. The effects on the clustering signal all
appear within the statistical error budgets, especially considering the fact
that the measurements are correlated on large scales.

Next, we also test out how much the clustering signal varies as a function of
redshift. In our cosmological analyses we will consider the average redshift of
each of the subsamples as the representative redshift for our measurements and
compare it to the theoretical predictions for cosmological inference. The
cosmological ingredients such as the halo mass function, the halo bias, and the
clustering of matter with itself and the halos varies as a function of
redshift. At fixed halo occupation distribution, these changes are at the
sub-percent level. Further, we also expect the clustering signal to vary with
redshift if the halo occupation distribution of galaxies changes as a function
of redshift.

\begin{figure*}
    \centering
    \includegraphics[width=2\columnwidth]{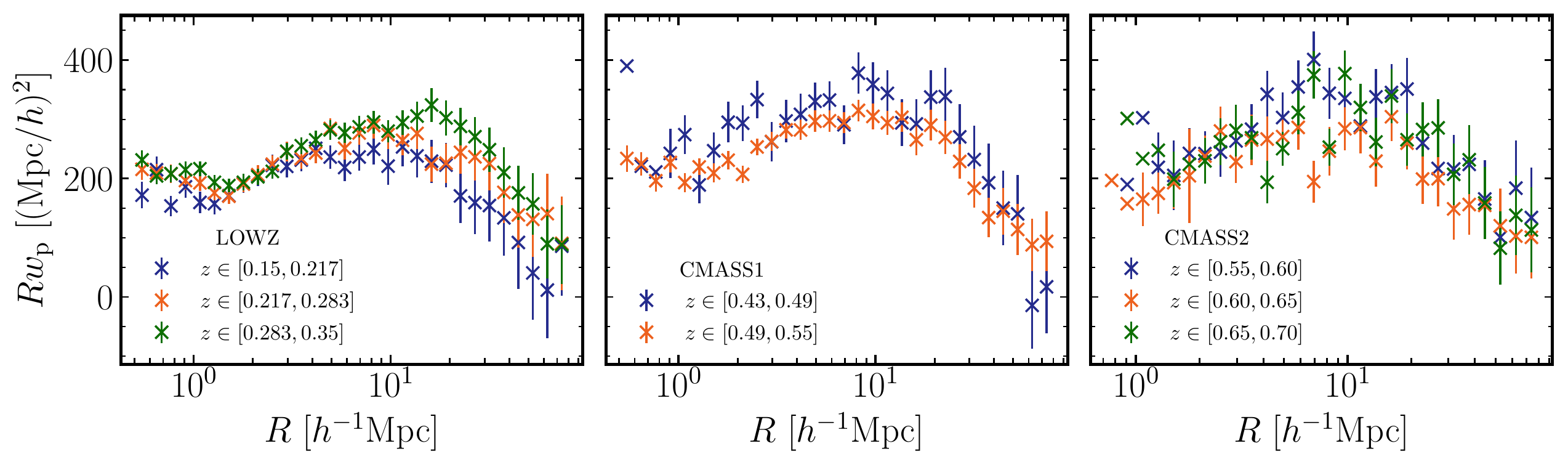}
    \caption{The variation of the clustering signal within each lens redshift
bin. Here, we divide the lens galaxy samples into two or three subsamples in
each redshift bin. This should be compared to
Fig.~\ref{fig:clustering_signal_finer_bin_full_sample}, where the BOSS full
sample, i.e the flux limited sample, is used for $\wproj$ measurement rather
than the luminosity-limited samples presented in this paper.}
    \label{fig:clustering_signal_finer_bin}
\end{figure*}

We divide each of the LOWZ, CMASS1 and CMASS2 subsamples into three bins each
and remeasure the clustering signal. We show the variation of the clustering
signal as a function of redshift for the subsamples in each of the sub-panels
of Fig.~\ref{fig:clustering_signal_finer_bin}. To assess the variation of the
clustering amplitudes of the signal in each of the bins, we fit the signal with
a fiducial model with our fiducial cosmological parameters at the median
redshift of the sample with a free amplitude parameterized as
\begin{align}
A(z) &= A_0 + \alpha(z-z_{\rm med})\,.
\label{eq:amplitude_variation}
\end{align}
When we fit the clustering measurements on the large or small scales we obtain
values of $\alpha$ all consistent within 2-$\sigma$ (see Table
~\ref{tab:dwp_zred_varn}). This shows that the clustering signal does not show
an appreciable variation with redshift.

\begin{table}
    \centering
    \caption{The slope of the systematic variation of the clustering signal
with redshift for each of the subsamples and the small and large-scale analyses
are presented in the top and bottom sections, respectively. We find no
significant evidence of variation with redshift.}
    \begin{tabular}{rrcr}
\toprule
Analysis & Sample & $\alpha$ \\
\midrule
Small scale & LOWZ &  $0.0 \pm 0.5$  \\
            & CMASS1 &  $-2.5 \pm 1.4$ \\
            & CMASS2 &  $-1.1 \pm 0.8$  \\
Large scale & LOWZ &  $0.2 \pm 0.7$  \\
            & CMASS1 &  $-2.7 \pm 1.8$  \\
            & CMASS2 &  $-0.6 \pm 1.1$  \\
\bottomrule
\end{tabular}
    \label{tab:dwp_zred_varn}
\end{table}

\section{Lensing measurements}
\label{sec:lensing}

We use the weak lensing pipeline described in
Section~\ref{subsec:gglens-pipeline} in order to measure the galaxy-galaxy
lensing signal around the three subsamples of lenses that we consider. In this
section, we present these lensing measurements as well as our estimates of the
covariance of these measurements using the mock catalogs.

We measure $\Delta\Sigma$ in 30 logarithmic radial bins in projected distance
in the range $[0.05, 80.0] \mpch$. As described before, We use the source catalog of HSC galaxies
satisfying Eq.~(\ref{eq:source-selection}) with $z_{\rm min}=0.7, z_{\rm
diff}=0.05, p_{\rm cut}=0.99$.  We carry out our measurements of the weak
lensing signal in the six different fields separately and later on combine them
in post-processing.

The measurement of the galaxy-galaxy lensing signal includes the conversion of
galaxy shear $\gamma_{\rm t}$ to the matter surface density $\dSigma$ and the
conversion of the angular radial bin $\theta$ to the projected radial bin $R$.
To do these conversion, we assume a fiducial cosmology for measurement. Because
we use the radial bin in the unit of $\mpch$ and it is independent to the value
of Hubble parameter $h$, the only relevant cosmological parameter is the matter
density parameter $\Omega_{\rm m}$ in the flat $\Lambda$CDM model that we focus
in the companion cosmology papers. We use a flat $\Lambda$CDM model with
$\Omega_{\rm m}=0.279$ as the fiducial cosmology for measurement.

\subsection{Measurements And Covariance}\label{subsec:gglens-meas}

We show the weak lensing signal $\Delta\Sigma$ for the three subsamples used in
our analysis in the three different panels in the middle row of
Fig.~\ref{fig:signals}, respectively. Although we have measured the signals on
small scales, we restrict to scales beyond 1 $\mpch$ in our figure. The
measured weak lensing signal falls off approximately as $1/R$. In the figure, we
plot the quantity $R\times\Delta\Sigma$ as this shows the deviations from such a
power law.  In general we see similar amplitudes for the weak lensing signals
across the three different subsamples.

For the covariance estimation, we measure the galaxy-galaxy lensing signals
using mock lens galaxy catalogs described in Section
\ref{subsubsec:mock-galaxy-catalogs} and the corresponding $1404$ mock shape
catalogs in Section \ref{subsubsec:mock-shape-catalogs}. We construct a total
data vector $\Delta\Sigma$ by concatenating the measurements of all our three
subsamples together. We obtain the covariance between these measurements as 
\begin{equation}
    C_{ij} = \langle ( \Delta\Sigma_{i} - \bar{\Delta\Sigma}_{i}) (\Delta\Sigma_{j} - \bar{\Delta\Sigma}_{j})\rangle\,.
\end{equation}
Our blinding scheme described in detail in Section~\ref{sec:blinding} relies on
the use of different multiplicative bias factors for three different catalogs,
only one of which is correct. However, our mock simulations do not include any
multiplicative bias factors. The covariance for the galaxy-galaxy lensing
measurements computed in this manner cannot therefore be directly used with the
three different blind catalogs.

To estimate the covariance for each of the blinded catalogs with non-zero
multiplicative biases, we include the multiplicative bias in
Eqs.~(\ref{eq:emock1}) and (\ref{eq:emock2}) by changing $\gamma\rightarrow
(1+\hat{m})(1+\hat{m}_{\rm sel})\gamma$ at the catalog level. The value of the
multiplicative bias $\hat{m}$ is taken from the corresponding blind catalog.
However, we have three different blind catalogs with different multiplicative
bias, and iterating mock measurements for all the blind catalogs is
computationally expensive. To avoid this iteration, we develop a rescaling
method in order to obtain the galaxy-galaxy lensing signal with non-zero
multiplicative bias from one without multiplicative bias. We first note that
the observed galaxy ellipticity in Eqs.~(\ref{eq:emock1}) and (\ref{eq:emock2})
in the presence of multiplicative biases can be expanded as 
\begin{align}
    {\bm e}^{\rm mock} = 2{\cal R}(1+\hat{m})(1+\hat{m}_{\rm sel}){\bm \gamma} + {\bm \epsilon}^{\rm n}\,,
\end{align}
to the lowest order of the intrinsic shape and the lensing shear, where ${\bm
\epsilon}^{\rm n} = {\bm \epsilon}^{\rm int}+{\bm \epsilon}^{\rm meas}$. Using
this expansion, the estimator in Eq.~(\ref{eq:stacked_signal}) and
Eq.~(\ref{eq:dsig_add_bias}) is 
\begin{align}
    \hat\dSigma(\hat{m}) \sim \hat\dSigma^{\rm sim}+
    \frac{1}{(1+\hat{m})}\frac{\sum_{\rm ls} w_{\rm ls} \epsilon^{\rm n}_{\rm t,ls}\langle\Sigma_{\rm cr}^{-1}\rangle^{-1}}{(1+\hat{m}_{\rm sel})2{\cal R}\sum_{\rm ls}w_{\rm ls}}
    \label{eq:dSigma-mbias-expansion}\,,
\end{align}
where the first term is
\begin{align}
    \hat\dSigma^{\rm sim} = \frac{\sum_{\rm ls}w_{\rm ls}\gamma_{\rm t, ls} \langle\Sigma_{\rm cr}^{-1}\rangle^{-1}}{\sum_{\rm ls}w_{\rm ls}}-\frac{\hat{a}_{\rm sel}}{1+\hat{m}_{\rm sel}}\dSigma^{\rm psf}\,,
\end{align}
and $\epsilon^{\rm int}_{\rm t, ls}$, $\epsilon^{\rm meas}_{\rm t,ls}$ and
$\gamma_{\rm t, ls}$ are defined similarly as in Eq.~(\ref{eq:e_t}) but with
${\bm e}\rightarrow \epsilon^{\rm int}$, $\epsilon^{\rm meas}$ and ${\bm
\gamma}$. Note that $\hat\dSigma^{\rm sim}$ can be measured from the mock
catalog because we know $\gamma$, but this is an unknown in the real
measurement. The measurement of $\hat\dSigma^{\rm sim}$ can be done at the same
time by using the same lens-source pair stacking as $\hat\dSigma$, and does not
require additional measurement. Using the dependence of $\hat\dSigma(\hat{m})$
on $\hat{m}$ in Eq.~(\ref{eq:dSigma-mbias-expansion}), we obtain an equation
for rescaling the $\dSigma$ measurements in the presence of multiplicative
bias,
\begin{align}
    \hat\dSigma(\hat{m}) = \hat\dSigma^{\rm sim} + \frac{\hat\dSigma(m=0)-\hat\dSigma^{\rm sim}}{1+\hat{m}}.
\end{align}
This method relies on the expansion of the estimator with respect to the
intrinsic shape and the lensing shear in Eq.~(\ref{eq:dSigma-mbias-expansion}),
which may lead to inaccuracy in the covariance estimate. We have checked that
the rescaling method works at the 1\% level in the covariance amplitude, by
using 100 mock measurements and comparing the measurements made without the
multiplicative bias but corrected using our formalism to those made with the
correct multiplicative biases.

The mock galaxy catalogs represent the intrinsic galaxy distribution at the
redshift of galaxy, but does not include the magnification bias effect on the
observed galaxy distribution by gravitational lensing by foreground matter
fluctuation. We evaluate the contribution of magnification bias to the
auto-covariance of galaxy-galaxy lensing signals analytically as in
\citet{Sugiyama:2021} (see Appendix A of the paper). In this paper, we also
evaluate the contribution of magnification bias to cross-covariance between
galaxy-galaxy lensing and cosmic shear signals. The formulation is summarized
in Appendix~\ref{sec:cov-mag}.

We note that the redshifts in the mock catalogs were assigned based on the
\dnnz estimates due to their availability in time when the covariance
calibrations were started. Therefore the mock measurements for the covariance
were also performed with a source sample selected based on using the redshift
PDFs from \dnnz used in Eq.~(\ref{eq:source-selection}). The covariance in the
galaxy-galaxy lensing signal consists of two terms, the shape noise and the
covariance due to large-scale structure. The shape noise depends upon the
number density of the source galaxy sample, while the large-scale structure
term arises independent of the source galaxy sample. The shape noise term
dominates on small scales but decreases as we consider larger and larger
separations owing to the larger number of lens-source pairs, where the large
scale structure term then can play a dominant part. We measure each of these
terms separately.\footnote{Note that unlike cosmic shear covariance, there is no mixed term between large structure and shape noise in the galaxy-galaxy lensing measurements.}

Given that our fiducial sample consists of galaxies selected using redshift
PDFs from \dempz which results in a source sample with a higher source galaxy
number density, we scale the shape noise term by the square root of the ratio
of the source galaxy number density when selected using \dnnz to that obtained
using a selection in \dempz. We maintain the large-scale structure term as is.

Finally, we also note that the mock galaxy-galaxy lensing signals are used
further to estimate the cross-covariance with the cosmic shear measurements by
combining these measurements with the cosmic shear signals measured from the
same set of mock shape catalogs in Section~\ref{subsec:cosmicshear-meas}.
 
Fig.~\ref{fig:cumsumSN} shows the cumulative signal-to-noise ratio as the
function of minimum scale cut. The upper and lower panels are for large and
small-scale analysis, respectively. The total signal-to-noise ratios of our
measurements in the LOWZ, CMASS1 and CMASS2 subsamples are 7.6, 11.3, 10.6,
respectively for our large-scale analysis. When including the scales that will
be modelled in our small-scale analysis, these signal-to-noise ratios increase
to 13.4, 17.3, 15.8, respectively.

\subsection{Systematic Tests}
\label{subsec:lensing_systematics}

\begin{table}
    \centering
    \caption{The slope of the systematic variation of the lensing signal with
redshift for each of the subsamples and the small and large-scale analyses are
presented in the top and bottom sections, respectively. We find no significant
evidence of variation with redshift.}
    \begin{tabular}{rrcr}
\toprule
Analysis & Sample & $\alpha$ \\
\midrule
Small scale & LOWZ &  $1.6 \pm 1.9$  \\
            & CMASS1 &  $-5.9 \pm 3.1$ \\
            & CMASS2 &  $-4.4 \pm 2.2$  \\
Large scale & LOWZ &  $4.3 \pm 3.0$  \\
            & CMASS1 &  $-2.3 \pm 4.8$  \\
            & CMASS2 &  $-2.8 \pm 3.3$  \\
\bottomrule
\end{tabular}
    \label{tab:esd_zred_varn}
\end{table}

\begin{figure*}
    \centering
    \includegraphics[width=2\columnwidth]{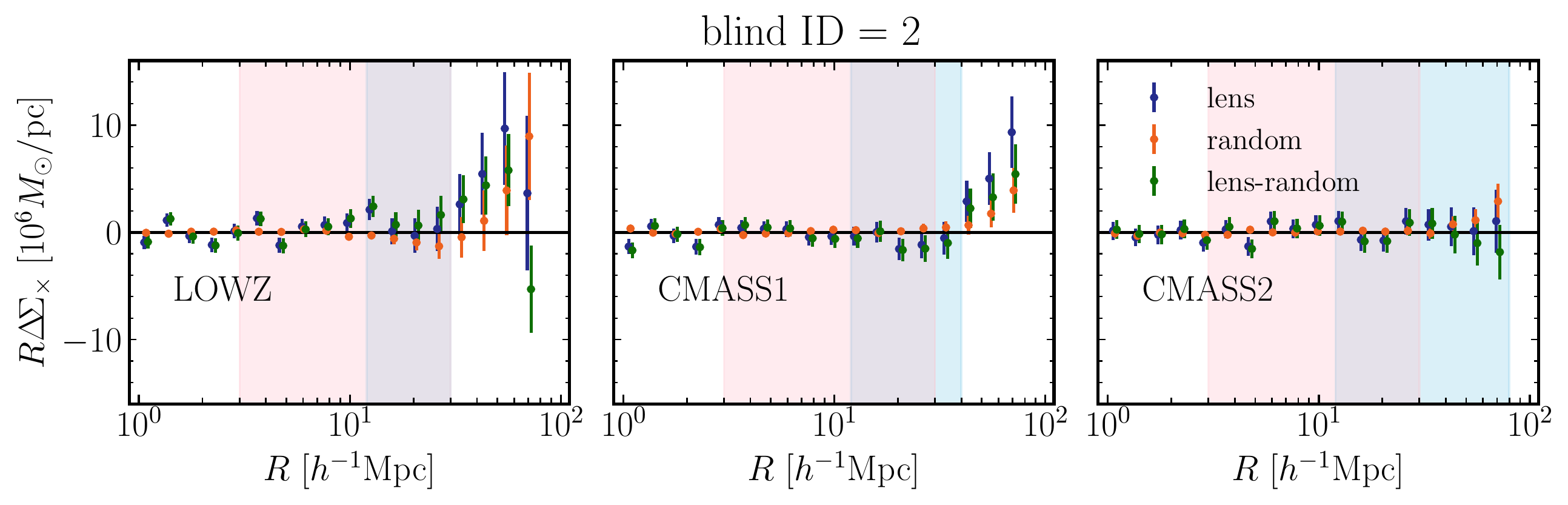}
    \caption{Systematics test of galaxy-galaxy lensing signal, i.e null test of
$\dSigma_{\times}$. From left to right panels, the cross signal from LOWZ,
CMASS1 and CMASS2 subsamples are shown. The blue points shows lensing signal
around lens galaxy, the orange points shows signals around random points, and
the green points shows the subtracted signal. The skyblue and pink shaded
region indicates the scale which the large-scale only analysis and small-scale
only analysis uses for cosmology inference.}
    \label{fig:dsigmacross}
\end{figure*}

\begin{table}
    \centering
    \caption{Summary of systematics test of galaxy-galaxy lensing: lensing
cross mode. The chi-square, degree of freedom (dof) and p-value are shown in
the format of $\chi^2/{\rm dof}~(p)$. Different columns show the results for
large-scale analysis and small-scale analysis, where $\chi^2$ is computed over
the scale of cosmology analyses indicated by blue and red shaded regions in
Fig.~\ref{fig:dsigmacross}.}
    \begin{tabular}{rrrr}
\toprule
       &        &     large scale &     small scale \\
photo-$z$ & lens sample &                 &                 \\
\midrule
\dempz & LOWZ &   7.57/4 (0.11) &  16.65/9 (0.05) \\
       & CMASS1 &   4.87/5 (0.43) &   6.86/9 (0.65) \\
       & CMASS2 &   3.84/8 (0.87) &   7.78/9 (0.56) \\
\dnnz & LOWZ &   2.28/4 (0.68) &  10.47/9 (0.31) \\
       & CMASS1 &   7.14/5 (0.21) &  12.56/9 (0.18) \\
       & CMASS2 &   5.48/8 (0.71) &   4.98/9 (0.84) \\
\mizuki & LOWZ &   7.47/4 (0.11) &  25.05/9 (0.003) \\
       & CMASS1 &   4.69/5 (0.45) &  20.47/9 (0.02) \\
       & CMASS2 &  11.03/8 (0.20) &  15.56/9 (0.08) \\
\bottomrule
\end{tabular}

    \label{tab:dsigmacross}
\end{table}

\begin{figure*}
    \centering
    \includegraphics[width=2\columnwidth]{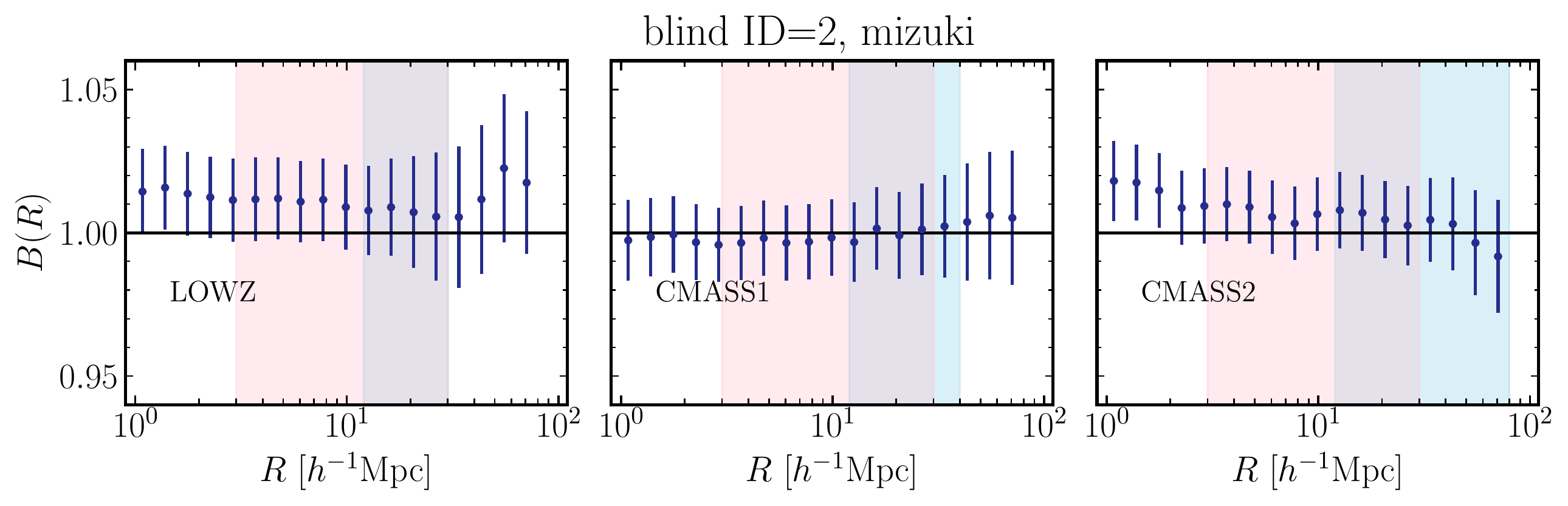}
    \caption{Systematics test of galaxy-galaxy lensing signal: boost factor for
each of the lens subsamples. The boost factor is consistent with unity in the
scales of interest for LOWZ, CMASS1 and CMASS2 subsamples with p-values greater
than $0.10$.}
    \label{fig:boost}
\end{figure*}

\begin{table}
    \centering
    \caption{Summary of systematics of galaxy-galaxy lensing: boost factor. The
format is similar to Table \ref{tab:dsigmacross}}
    \begin{tabular}{rrrr}
\toprule
       &        &     large scale &     small scale \\
photo-$z$ & lens sample &                 &                 \\
\midrule
\dempz & LOWZ &   0.89/4 (0.93) &   2.85/9 (0.97) \\
       & CMASS1 &  10.80/5 (0.06) &  15.41/9 (0.08) \\
       & CMASS2 &   6.61/8 (0.58) &   8.16/9 (0.52) \\
\dnnz & LOWZ &   0.76/4 (0.94) &   3.28/9 (0.95) \\
       & CMASS1 &  12.22/5 (0.03) &  14.83/9 (0.10) \\
       & CMASS2 &   8.49/8 (0.39) &   8.77/9 (0.46) \\
\mizuki & LOWZ &   0.52/4 (0.97) &   3.26/9 (0.95) \\
       & CMASS1 &  11.21/5 (0.05) &  13.83/9 (0.13) \\
       & CMASS2 &   6.22/8 (0.62) &   8.06/9 (0.53) \\
\bottomrule
\end{tabular}

    \label{tab:boost}
\end{table}
We carry out a variety of systematic tests in order to validate the measured
weak lensing signal. In Fig.~\ref{fig:dsigmacross}, we show the cross component
of the weak lensing signal for the three subsamples for the fiducial
photometric redshift estimates from \dnnz in the three different columns.
Fig.~\ref{fig:dsigmacross-dnnz-mizuki} in
Appendix~\ref{sec:dempz_mizuki_meas_app} corresponds to the same measurements
but using the photometric redshift codes \mizuki and \dempz.  The blue color
points correspond to the measurement around the lens samples, the red ones
correspond to the measurement around the random points, and the green points
correspond to the subtraction of the two. 

The cross-signal is expected to be zero apart from the presence of any
systematics. On large scales, we observe a significant deviation from zero in
the LOWZ and the CMASS1 subsamples, but we do not see such effect for the
CMASS2 subsample. Similarly the lensing signal around random points is also
non-zero, but it is not large enough to explain the deviation seen around the
lenses. Therefore the cross-systematic around the lenses still survives after
subtracting the cross-signal around random points.

One likely explanation is that the cross systematic appears on a fixed angular
scale, and given the redshift differences in each of our subsamples, this
angular scale corresponds to a different distance in comoving coordinates for
each of our subsample. The scale where this cross systematic appears is
$2.3~{\rm deg}$, which is slightly larger than the HSC field-of-view size,
$1.5~{\rm deg}$, beyond which even the random subtraction does not seem to
help.

The presence of the cross-systematic dictates the large-scale cut we will use
for the cosmological analysis of these measurements. We will use large scale
cuts of $30, 40$ and $80 \mpch$ for the analysis of the signals from the LOWZ,
CMASS1, and CMASS2 subsamples, respectively. In our cosmological analyses, we
will also adopt cuts on the small-scale which are motivated by modeling
uncertainties on small scales. The large-scale only analysis which uses
perturbation theory based techniques and the small-scale only analysis which
uses a halo occupation distribution modeling framework will use small scale
cuts of $12 \mpch$ and $3 \mpch$, respectively. In the small-scale analysis, we
also discard lensing signals over $R>30\mpch$ because they gives negligible
contribution of signal-to-noise ratio compared to smaller scales around
$R\sim3\mpch$.

We compute the reduced chi-squared: $\chi_{\rm red}^2=\chi^2/$ degrees of
freedom (dof) away from a null value of the cross signal around the lens sample
after the subtraction of the cross signal around random points. These values
along with the corresponding p-values to exceed the $\chi^2$ given the dof are
also tabulated in Table~\ref{tab:dsigmacross} computed over the scales used in
the large-scale and small-scale analyses, respectively. These values justify
our choice of the scale cuts for our cosmology analysis.

We note that we do see some evidence of cross-systematic with p-values smaller
than $0.05$ for the LOWZ and CMASS1 subsamples when using the \mizuki source
galaxy sample. This sample of source galaxies has the highest
number density, which results in smaller errors. In the bottom panel of
Fig.~\ref{fig:dsigmacross-dnnz-mizuki}, we do not see a systematic deviation in
one direction, but the cross-points lie above and below the zero line, with a
scatter not consistent with the errors on these points. We flag this issue, in
case, there is an interest in using the \mizuki based source sample for
cosmological analyses. We proceed further by noting that our fiducial analysis
will rely on the \dempz based sample of source galaxies.

Galaxies which are physically correlated with the lens sample could seep into
our source samples due to imperfections in the photometric redshift PDFs
despite our stringent cuts. Such galaxies are not expected to be efficiently
lensed by our sample of lens galaxies. The inclusion of such galaxies could
therefore result in a diluted signal. The presence of correlated galaxies in
the source sample can be inferred by comparing the number of source galaxies
around the lens sample with the number of pairs around random points. If the
source galaxies are correlated, we expect the ratio of the two pair counts,
also called the boost factor, to be consistent with unity.

In Fig.~\ref{fig:boost}, we show the boost factor for all three of our lens
samples. In the scales of our interest shown as the colored shaded regions, we
do not see a significant deviation away from unity. The measurements shown in
the Figure are expected to be correlated. The chi-squared for the expectation
$B(R)=1$, and the corresponding p-values for all three cases for the large
scale and the small-scale analysis are written in Tables~\ref{tab:boost}.  We
find that the boost factors are consistent with unity for the LOWZ and CMASS2
samples, although the CMASS1 sample shows p-values smaller than 10 percent.
This is related to the large covariance in the measurement of the boost factor
and the measurements of boost values which swing from being below unity to
above unity in some cases, which results in a large $\chi^2$. We have verified
that if we ignore the covariance then the $\chi^2$ values are small and result
in a large p-value. Furthermore, the boost factor values in the case of the
CMASS1 subsample are away from unity at the sub-percent level. Therefore we do not apply any boost factor corrections to our signals.

\begin{figure*}
    \centering
    \includegraphics[width=2\columnwidth]{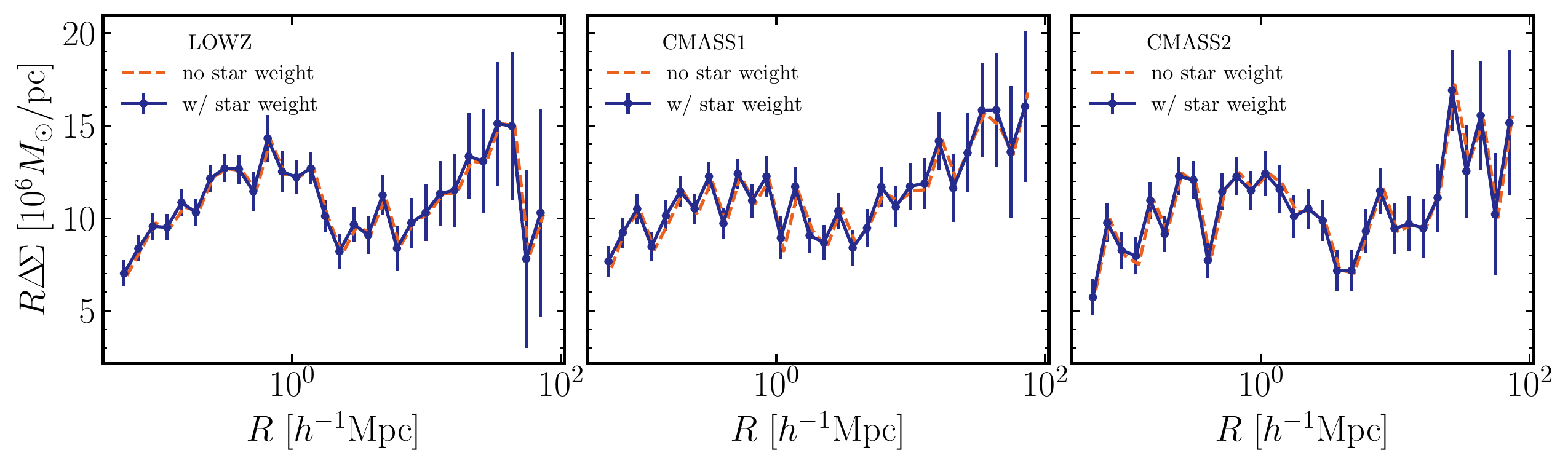}
    \caption{The variation of the galaxy-galaxy lensing signal with and without
systematic weights related to the number density of stars and the seeing. The
difference between these signals is within the statistical errors.}
    \label{fig:systematics_lensing}
\end{figure*}

\begin{figure*}
    \centering
    \includegraphics[width=2\columnwidth]{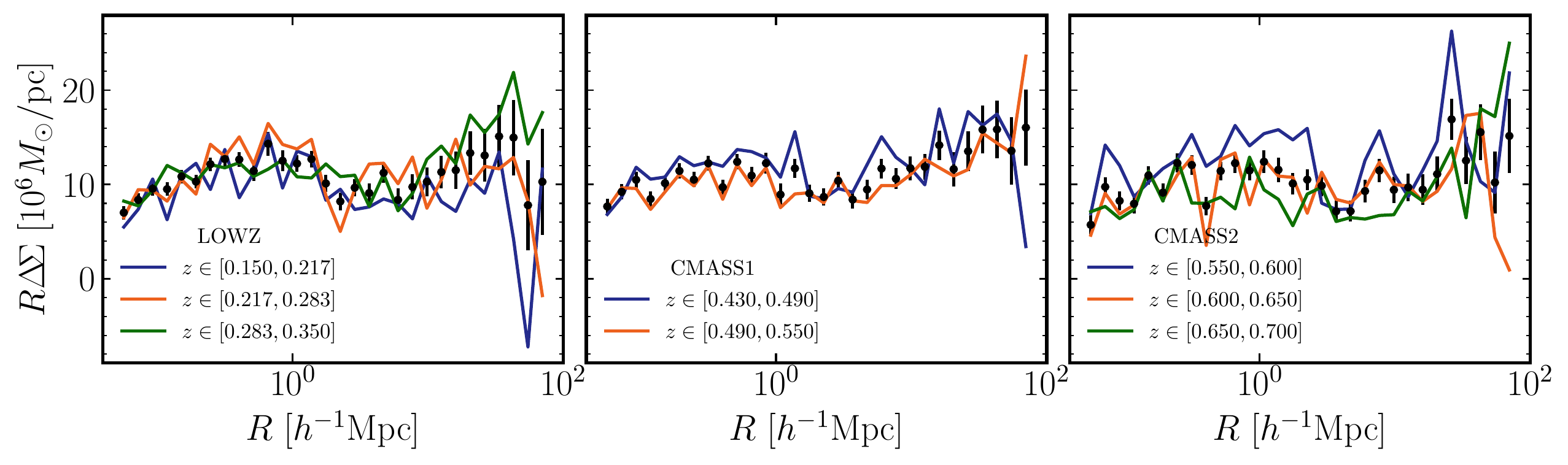}
    \caption{The variation of the galaxy-galaxy lensing signal in each redshift bins is shown in the three panels. In each panel we show the lensing signal in a given redshift bin by points with errors, while the lensing signal when the redshift bin is further subdivided by solid lines of varying colors. We fit a varying amplitude as a function of redshift in each panel to the solid lines and do not find a significant evidence for evolution given the statistical errors.}
    \label{fig:systematics_lensing_finer_bin}
\end{figure*}

Just as in the case of the clustering signal we also quantify the variation of
the galaxy-galaxy lensing signal with redshift. We measure the galaxy-galaxy
lensing signals in subsamples of each redshift bin and present the result in
each of the sub-panels of Fig.~\ref{fig:systematics_lensing_finer_bin}. We fit
these signals with a fiducial model at the median redshift of the sample with
fiducial cosmological parameters with a free amplitude that varies with
redshift according to Eq.~(\ref{eq:amplitude_variation}) and again obtain
results which indicate a result consistent with no variation at the 2-$\sigma$
level (see Table~\ref{tab:esd_zred_varn}). In addition, we also show the very
weak dependence of the weak lensing signal whether we use or do not use the
systematic weights related to the stellar density in the various sub-panels of
Fig.~\ref{fig:systematics_lensing}.

\section{Cosmic shear measurements}
\label{sec:cosmic_shear}

We use the infrastructure described in Section~\ref{subsec:cosmicshear-code} in
order to measure the cosmic shear signal using the common sample of source
galaxies used in the galaxy-galaxy lensing measurement. Note that in 
the 3$\times$2pt
analyses, we do not perform a tomographic measurement of the cosmic shear. In
this section, we will present these cosmic shear measurements as well as our
estimates of the covariance of these measurements using the mock catalogs. We
use a total of $30$ logarithmic bins starting from $0.25$ arcmin to a maximum
distance of $360$ arcmin in order to carry our measurements. As mentioned
previously, we will use a single source sample, and not perform any tomographic
measurement of the cosmic shear signal. A single source sample will allow us to
self-calibrate any residual source redshift uncertainties.

\subsection{Measurements}\label{subsec:cosmicshear-meas}

We show the measurements of the cosmic shear correlation functions, $\xi_+$ and
$\xi_-$, in the bottom panels of Fig.~\ref{fig:signals}. Both the measurements
approximately scale as $1/\theta$. Although we measure the signals on a wide
range of scales, we will only use the signal shown in the shaded regions. The cut on small scales is dictated by our requirement that the modelling uncertainties in the power spectrum due to the impact of baryonic physics do not result in significant biases. To this end, in our companion paper, Miyatake et al., we show that the dark matter only model, compared to various models which account for the uncertain baryonic physics give consistent cosmological inference on scale cuts implemented in this paper.
For $\xi_+$ this corresponds to a small-scale cut of $8$~arcmin, while the
corresponding cut for $\xi_-$ is equal to $30$~arcmin. On large scales, the
scale cuts are dictated by systematics in the measured signals as we will
describe below.

The measurements of the cosmic shear signals were performed on the mock shape
catalogs in order to obtain the covariance of the cosmic shear signals. Note
that the mock catalogs themselves were constructed without the inclusion of any
multiplicative biases. Therefore, in order to estimate the covariances for the
three blind catalogs with non-zero multiplicative biases, we include the
multiplicative bias in Eqs.~(\ref{eq:emock1}) and (\ref{eq:emock2}) by changing
$\gamma\rightarrow (1+\hat{m})\gamma$ at the catalog level. The values of the
multiplicative bias $\hat{m}$ are taken from the corresponding blind catalog.

As mentioned in the previous section, the mocks were constructed based on the
\dnnz photometric redshift estimates. We have measured the mock cosmic shear
signals with galaxies from the \dnnz selected source sample. These measurements
are then used to measure the covariance matrix. Given the difference in the
number density of \dnnz and \dempz selected source samples, we have rescaled
the shot noise related term in the covariance matrix.

The cross-correlation coefficient of the covariance matrix obtained from our
cosmic shear measurements can be seen as two of the blocks in
Fig.~\ref{fig:covariance}. Since the cosmic shear measurements and the
galaxy-galaxy lensing measurements also share the same mock catalogs, we have
also obtained the cross-covariance between these measurements. Although they
are not very large, we do find non-zero cross-correlations between these
measurements, which we will take into account in our analyses. The
cross-correlation between the clustering and the lensing measurements are
considered to be zero, given that the clustering measurements come from the
entire SDSS footprint, while the lensing measurements are restricted to HSC
regions, which are a small fraction of the entire SDSS footprint.

Fig.~\ref{fig:cumsumSN} shows the cumulative signal-to-noise ratio for large
and small-scale analysis as a function of the minimum scale cut. The total
signal-to-noise ratios of our measurements for $\xi_+$ and $\xi_-$ are 20.2 and
19.0, respectively.

\subsection{Systematic Tests}\label{sec:cosmic-shear-systematics}

\begin{figure*}
    \centering
    \includegraphics[width=1.5\columnwidth]{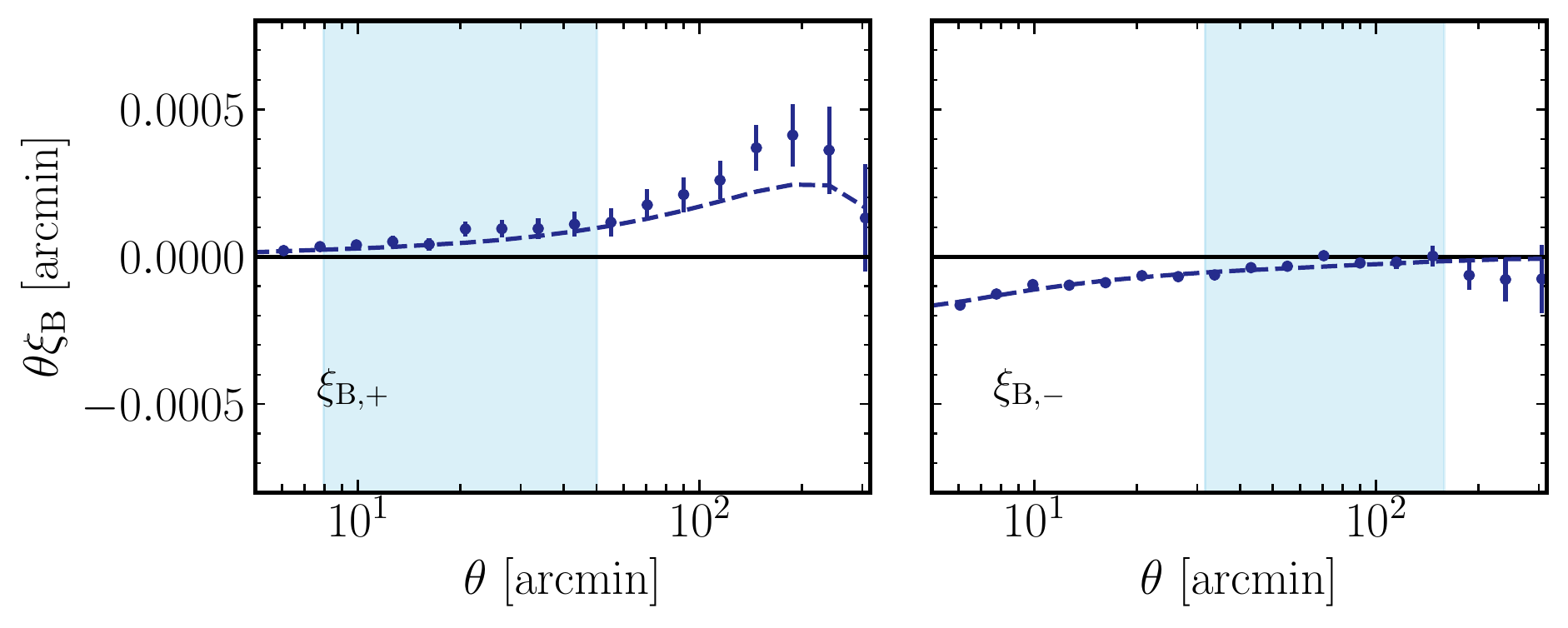}
\caption{Systematics test of cosmic shear signal, i.e. $B$-mode test. The
B-mode expectations are consistent with that due to survey geometry effects and
is consistent with the expectation from mock catalogs shown as the blue dashed
line in each of the panels.}
    \label{fig:Bmode}
\end{figure*}

In Fig.~\ref{fig:Bmode}, we present our measurements of the B-mode signals,
$\xi_{B,+}$ and $\xi_{B,-}$ computed using the measurements of $\xi_+$ and
$\xi_-$, respectively, obtained using finer binning to avoid noise. The finite
field size of our weak lensing shape catalog regions can result in a residual
non-zero B-mode in the cosmic shear signals. We evaluate the presence of such
residual B-modes using our mock shape catalogs (which do not include any
systematics) and present the mean from the mocks as a dashed line. We compute
the $\chi^2$ per degree of freedom and the p-value for the $\chi^2$ to exceed
the one measured in the data. We use the mean mock measurement of the B-modes
with respect to the measured B-modes in our data for the above purpose. We
obtain reasonably good $\chi^2$ values with large p-values if we restrict
ourselves to scales below $50$~arcmin for $\xi_+$ and below $150$~arcmin for
$\xi_-$. 

We note that this result was obtained after excluding a problematic 20 sq. deg
area in the GAMA09H region whose inclusion would result in significantly larger
B-mode signal especially in the cosmic shear tomography analyses presented in
Li \etal \cite{Li_hscy3} and Dalal \etal \cite{Dalal_hscy3}. We comment on this region in
Appendix~\ref{sec:gama09h_bmodes_app}.

\begin{table}
    \centering
    \caption{Summary of cosmic shear $B$ mode null tests. $\chi^2$ is defined
as the deviation from mean signal of mock measurements indicated by the dotted
line in Figs.~\ref{fig:Bmode} and \ref{fig:Bmode-dnnz-mizuki}.}
    \begin{tabular}{rrr}
\toprule
       &                       & $\chi^2/\nu$ $(p)$ \\
photo-$z$ & signal &                    \\
\midrule
\dempz & $\xi_{\mathrm{B}, +}$ &      8.81/7 (0.27) \\
       & $\xi_{\mathrm{B}, -}$ &      6.87/7 (0.44) \\
\dnnz & $\xi_{\mathrm{B}, +}$ &     11.20/7 (0.13) \\
       & $\xi_{\mathrm{B}, -}$ &      9.94/7 (0.19) \\
\mizuki & $\xi_{\mathrm{B}, +}$ &     11.44/7 (0.12) \\
       & $\xi_{\mathrm{B}, -}$ &      6.89/7 (0.44) \\
\bottomrule
\end{tabular}

    \label{tab:Bmode}
\end{table}

As discussed in Section \ref{subsec:cosmicshear-code}, PSF leakage and residual
PSF model error contaminate the measured cosmic shear signal. In our cosmology
analyses, we model this contamination as Eq.~(\ref{eq:psf-contamination}) and
add it onto the model prediction of gravitational lensing signal. Here, we
present the constraint on coefficients $\alpha_{\rm psf}$ and $\beta_{\rm psf}$
to be used as prior during cosmological parameter inference. We follow the
method in \citet{2020PASJ...72...16H}. The coefficients $\alpha_{\rm psf}$ and
$\beta_{\rm psf}$ can be estimated from the cross correlation functions between
$\gamma^{\rm p,q}$ from star catalog and galaxy shears from galaxy shape
catalog. These cross correlation functions can be expressed as
\begin{align}
    \label{eq:xigp}
    \xi^{\rm gp}_\pm = \alpha_{\rm psf}\xi^{\rm pp}_\pm + \beta_{\rm psf}\xi^{\rm pq}_\pm,\\
    \xi^{\rm gq}_\pm = \alpha_{\rm psf}\xi^{\rm pq}_\pm + \beta_{\rm psf}\xi^{\rm qq}_\pm.
    \label{eq:xigq}
\end{align}
All the correlation functions in the above equations are measurable from star
catalog and galaxy catalogs, and hence $\alpha_{\rm psf}$ and $\beta_{\rm psf}$
can be estimated. In the following figures, we  show the measurement result
with stars with flags {\sc i\_calib\_psf\_used=True}. We also perform the same
analysis with the stars with flags {\sc i\_calib\_psf\_used=False}, and finally
take account of the uncertainty of $\hat\xi_{\rm psf, \pm}$ due to the
difference of stars whether {\sc i\_calib\_psf\_used=True} or {\sc False}.

\begin{figure}
    \centering
    \includegraphics[width=1\columnwidth]{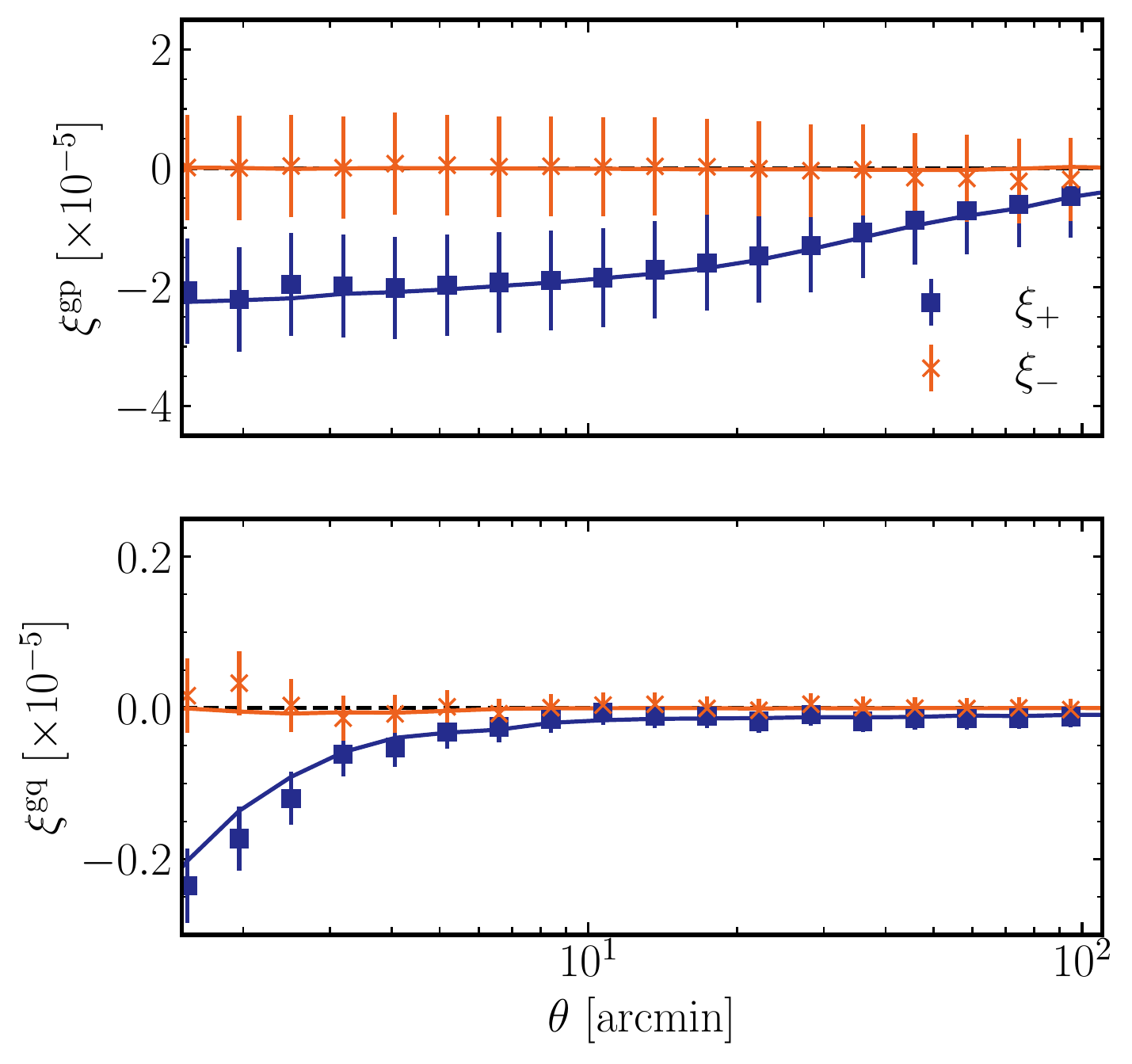}
    \caption{Cross correlation function between galaxies and PSF ellipticity
($\gamma^{\rm p}$) and the PSF model error ($\gamma^{\rm q}$) are shown in
the upper and lower panels, respectively. We model these functions to determine
the values of the PSF systematics model parameters, $\alpha_{\rm psf}$ and
$\beta_{\rm psf}$.}
    \label{fig:psf-gpgq}
\end{figure}

\begin{figure}
    \centering
    \includegraphics[width=1\columnwidth]{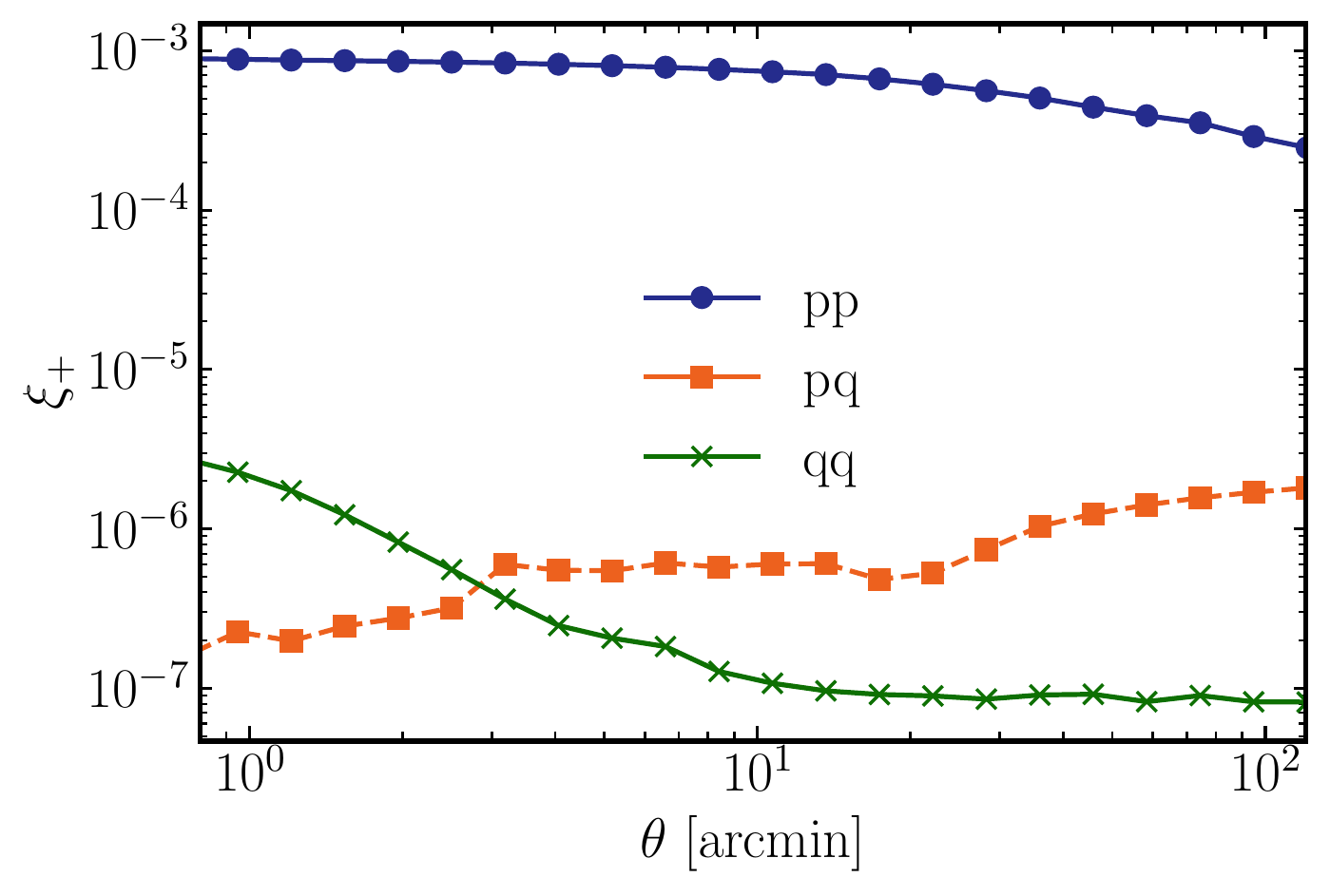}
    \caption{Auto-correlation and cross-correlation functions of the PSF
ellipticity ($\gamma^{\rm p}$) and the PSF model error ($\gamma^{\rm q}$). The
contribution of each of these to the cosmic shear correlation functions are
governed by the parameters $\alpha_{\rm psf}$ and $\beta_{\rm psf}$.}
    \label{fig:psf-pppqqq}
\end{figure}

\begin{figure}
    \centering
    \includegraphics[width=1\columnwidth]{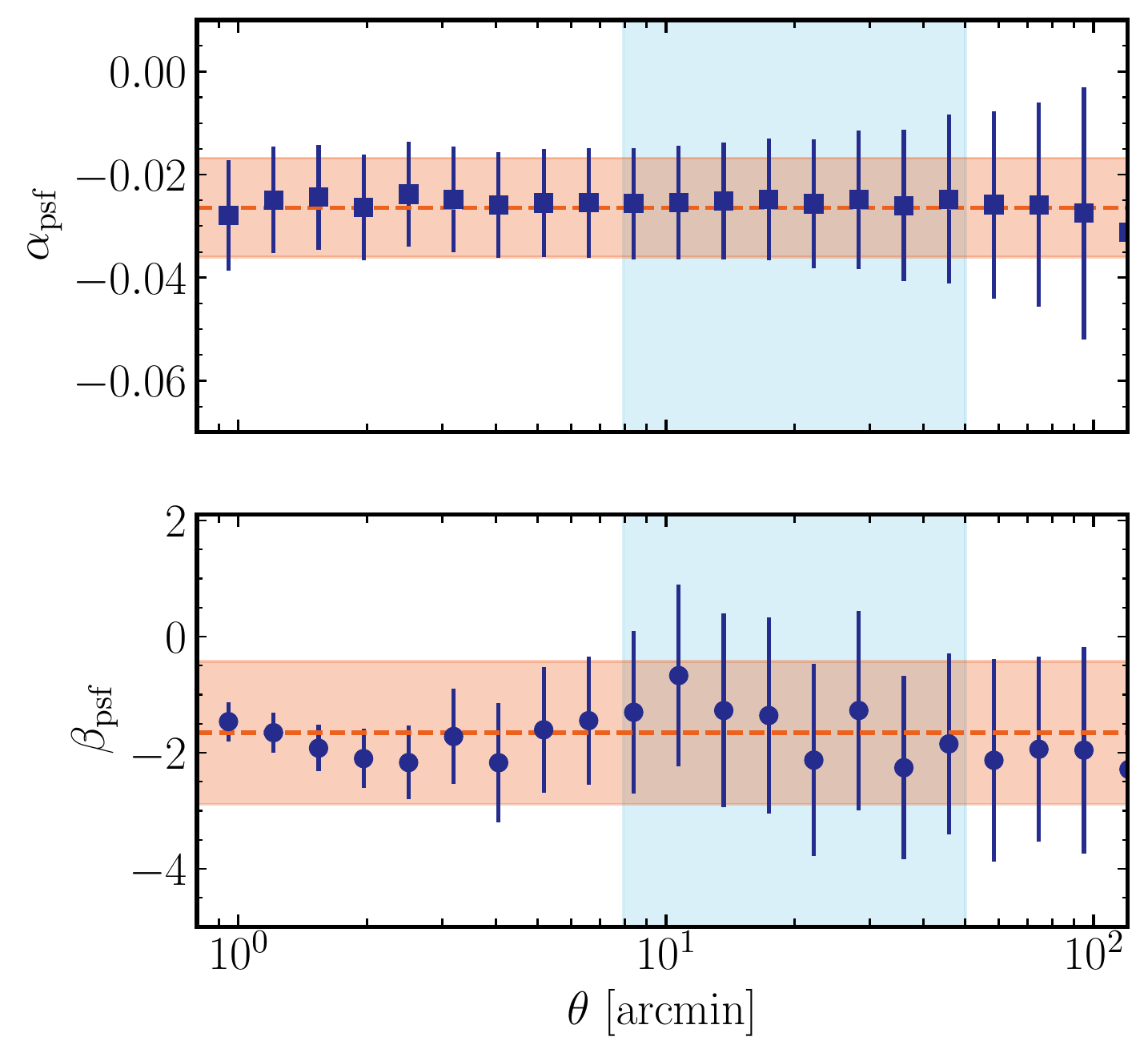}
    \caption{The scale dependence of the PSF systematic coefficients
$\alpha_{\rm psf}$ and $\beta_{\rm psf}$, obtained by fitting the correlation
functions shown in Fig.~\ref{fig:psf-gpgq}. We do not observe any significant
scale dependence in the values of these PSF systematics coefficients.}
    \label{fig:psf-alpha-beta}
\end{figure}

Fig.~\ref{fig:psf-gpgq} shows the cross correlations between $\gamma^{\rm p,q}$
and galaxy shears. The errorbar is estimated from the measurements of mock
galaxy shape catalogs and the real star catalog, i.e. errorbar takes account of
the cosmic variance. We find that $\xi_-^{\rm gp, gq}$ are consistent with zero
within the statistical uncertainty, and hence we will focus on the $\xi_+$ mode
alone in the following analysis. Fig.~\ref{fig:psf-pppqqq} shows the auto
correlations of PSF leakage and PSF model error. We first estimate the
coefficients $\alpha_{\rm PSF}$ and $\beta_{\rm PSF}$ as a function of angular
separation by fitting signals in each bin in order to check the assumption that
coefficients are scale independent. The result is shown in
Fig.~\ref{fig:psf-alpha-beta}, indicating that the coefficients are scale
invariant within the statistical uncertainty. Combining the coefficient
estimates over the scales for cosmology analyses indicated by blue shaded
region, we estimate the mean coefficients and the errors on them, shown in the
horizontal orange line and shaded region.

\begin{figure}
    \centering
    \includegraphics[width=1\columnwidth]{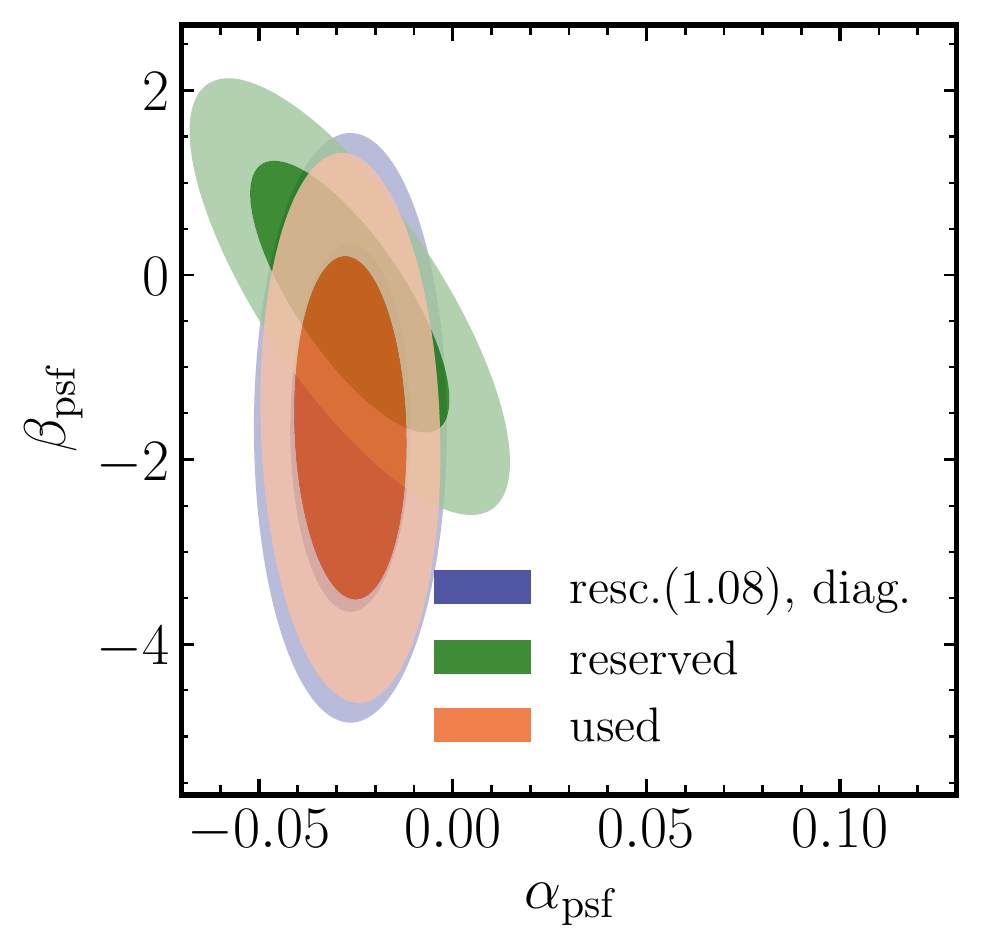}
    \caption{Constraint on $\alpha_{\rm psf}$ and $\beta_{\rm psf}$ parameters.
The orange and green contours show the constraints on $(\alpha_{\rm psf},
\beta_{\rm psf})$ using stars selected by {\sc i\_calib\_PSF\_used=True} and
{\sc False}, respectively. The blue contour is the diagonalized Gaussian
posterior with respect to $\alpha$ and $\beta$, and the posterior size is
rescaled by a factor of $1.08$ to cover the uncertainty $\xi_{\rm psf, +}$ due
to the choice of star selection, i.e. {\sc i\_calib\_PSF\_used=True} or {\sc
False}. The blue contour is used as a conservative prior on $\alpha_{\rm psf}$
and $\beta_{\rm psf}$ in the cosmology analyses.}
\label{fig:psf-alpha-beta-contour}
\end{figure}

The orange contour in Fig.~\ref{fig:psf-alpha-beta-contour} shows the
constraint on the coefficients using stars with flags {\sc
i\_calib\_psf\_used=True}, while the green contour shows with flags {\sc
i\_calib\_psf\_used=False}. These contours are two dimensional Gaussian
distribution, because Eqs.~(\ref{eq:xigp}) and (\ref{eq:xigq}) are linear in
$\alpha_{\rm psf}$ and $\beta_{\rm psf}$. The coefficients from the orange and
green contour predicts slightly different $\xi^{\rm psf}_\pm$. We find that
this uncertainty of $\xi^{\rm psf}_\pm$ due to the difference of stars whether
{\sc i\_calib\_psf\_used=True} or {False} can be covered by rescaling the
statistical uncertainty of orange contour by a factor of $1.08$ \footnote{One
may think that the factor of $1.08$ is not enough because the blue contour does
not cover the green contour in Fig.~\ref{fig:psf-alpha-beta-contour}. Depending
on the flags, {\sc i\_calib\_psf\_used=True} or {False}, we not only use
different coefficients but also different $\xi^{\rm pp,pq,qq}$, and the
difference is not so significant in predicted $\hat\xi^{\rm psf}$ than how it
appears in Fig.~\ref{fig:psf-alpha-beta-contour}}. For simplicity, we also
diagonalize the rescaled constraint with respect to $\alpha_{\rm psf}$ and
$\beta_{\rm psf}$, which will be used in the cosmology analyses.

\begin{figure}
    \centering
    \includegraphics[width=1\columnwidth]{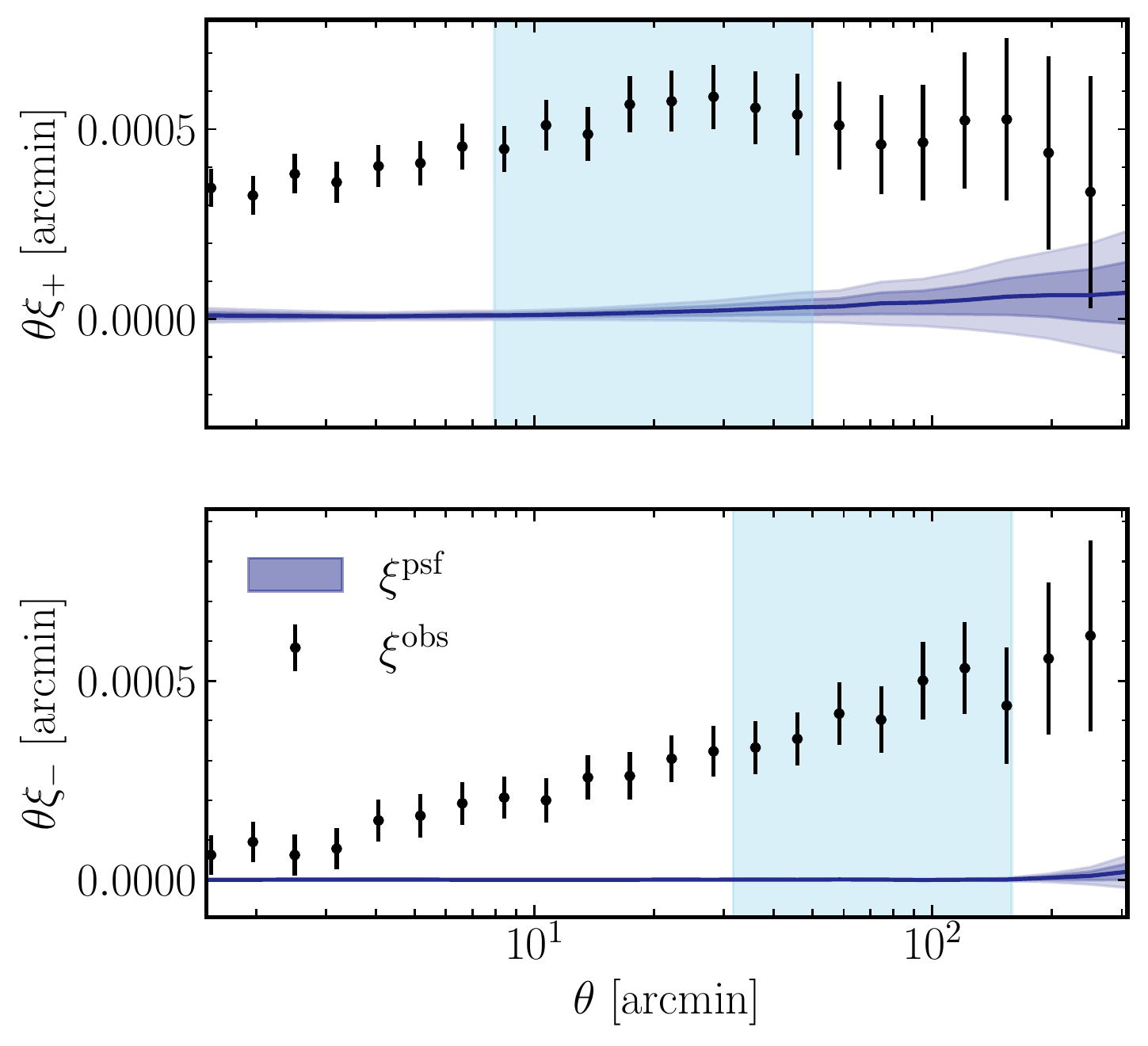}
    \caption{Posterior distribution of PSF correlation term which contaminates the observed cosmic shear correlation function. The measured cosmic shear correlation functions are much larger than the contamination from PSF systematics.}
    \label{fig:psf-xipsf}
\end{figure}

\section{Summary}
\label{sec:summary}

A joint analysis of the three two-point functions: the galaxy clustering
signal, the galaxy-galaxy lensing signal and the cosmic shear signal is a
unique probe of cosmology. In a series of papers, based on imaging data
obtained over a three year period from the HSC survey and the spectroscopic
data from the SDSS, we will present the cosmological constraints on the matter
density parameter as well as the amplitude of density fluctuations. In this
paper, the first in the series, we presents robust measurements and systematic
tests corresponding to the measurement of each of the above signals. A summary
of the results and products made available in this paper are as follows:

\begin{itemize}
    \item We define three different subsamples of spectroscopic lens galaxies
from the large-scale structure samples of SDSS BOSS galaxies. These subsamples
are defined to be approximately volume limited by absolute magnitude (k+e
corrected), the LOWZ, CMASS1 and CMASS2 subsamples of galaxies with $z \in
[0.15,0.35], [0.43,0.55]$ and $[0.55, 0.70]$, respectively, and absolute
magnitude of galaxies of $M_i-5\log {\rm h}<-21.5$, $-21.9$ and $-22.2$ for the
three subsamples, respectively. 
    \item We conservatively define a single subsample of source galaxies from
the HSC-Y3 shape catalog that we use for the weak lensing signals to be
measured in this paper. Our subsample of galaxies has a greater than 99 percent
probability assigned by the photometric redshift assignment algorithm \dempz to
be above a redshift of $0.75$, higher than the redshift of any of our lens
subsamples by more than $\mathrm{d}z=0.05$. We also present the inferred
redshift distribution of our source galaxies based on the method presented in
\cite{Rau:2023}.
    \item We have generated 108 mock catalogs for our subsample of SDSS
galaxies to aid in the computation of the covariance matrix of their clustering
signal. These mock catalogs were created by populating halos with galaxies with
a halo occupation distribution designed to reproduce the abundance and
clustering of the galaxy subsamples we use on scales of $0.5 \mpch > R > 80
\mpch$. Our mock catalogs further mimic the footprint of the SDSS survey.
    \item We have also generated 1404 mock shape catalogs of galaxies by
randomly rotating the ellipticities of galaxies in our HSC-Y3 shape catalog,
and distorting them with shears that would arise from the large-scale structure
distribution in a $\Lambda$CDM universe. These mock catalogs aid in the
computation of the covariance of the two point measurements involving weak
gravitational lensing shears.
    \item We describe the codes and pipelines used to optimally
perform the three two-point correlation function measurements on data as well
as on the mock catalogs in order to estimate the full covariance matrix of our
observables.
    \item Based on the clustering pipeline, we present the measurements of the
projected clustering signal of the three subsamples of galaxies. In the range
of scale cuts over which we will perform our cosmological analyses with the
large-scale perturbation theory-based model (Paper III), the signal-to-noise
ratio of the clustering signals are $25.6, 27.6, 26.6$ for the LOWZ, CMASS1 and
CMASS2 subsamples, respectively. When including the small scales that will be
modelled in Paper II, the corresponding signal-to-noise ratios are, $38.5,
37.1, 35.7$, respectively.
    \item We also show that the clustering signals within each subsample do not
change substantially irrespective of the use of systematic weights, suggested
by the SDSS BOSS team, that are related to the the number density of stars or
the seeing in the SDSS imaging data used to define the spectroscopic targets.
We also show that the clustering signal of our subsample of galaxies does not
show a substantial evidence for variation within different redshift bins.
    \item Using the weak lensing pipeline, we present our measurements of the
galaxy-galaxy lensing signal for each of our lens subsamples using the single
source galaxy sample. For the large-scale analyses, the signal-to-noise ratio
of the galaxy-galaxy lensing signals are $7.6, 11.3$ and $10.6$, for the LOWZ,
CMASS1, and CMASS2 subsamples, respectively. The corresponding signal-to-noise
ratios for the measurements in the scales of interest relevant to small-scale
analysis are $13.4, 17.3$ and $15.8$, respectively.
    \item We presented a number of systematic tests for the weak lensing
signals: the null signals around random points, the null cross-signals,
corresponding to each of our subsamples of galaxies. We also presented that the
contamination of our measured signals due to presence of source galaxies
physically associated with our lens galaxies is negligible due to our
conservative source selection. We have further shown that the measured lensing
signals are not impacted by the use of the systematic weights corresponding to
the SDSS BOSS large-scale structure subsamples. The lensing signal amplitudes
within the redshift bin for each subsamples also show no significant variation
as a function of redshift.
    \item We also presented the measurements of the cosmic shear correlation
functions $\xi_\pm$ for our source galaxy sample. The signal-to-noise ratio of
our measurements of $\xi_+$ and $\xi_-$ are 20.2 and 19.0, respectively. We
also presented systematics tests for the cosmis shear measurements including a
decomposition into $E$ and $B$ modes. The largest scales of interest for the
cosmological analyses were chosen based on the null detection of $B$ modes,
while the smallest scales were chosen based on the accuracy of our theoretical
templates.
    \item Finally, we also presented our best estimates for the PSF systematics
parameters $\alpha_{\rm PSF}$ and $\beta_{\rm PSF}$ which quantify the PSF
leakage and the residual PSF model error. The values we obtain will be used in
the cosmological analyses to model the PSF systematics component of the cosmic
shear signal, in order to marginalize over these nuisance parameters.
\end{itemize}

The measurements presented in this paper, the covariances and the constraints
on parameter systematics have been used in the 3$\times$2pt analysis of the
data to infer the constraints on the cosmological parameters $\Omega_{\rm m}$
and $\sigma_8$, in particular the parameter combination, $S_8$. These results
will be presented in our companion papers Sugiyama \etal \cite{Sugiyama_hscy3} and Miyatake \etal \cite{Miyatake_hscy3}.

The Subaru Hyper Suprime Cam survey has also finished collecting data from its
entirety of operations which spanned 330 nights in total. The wide survey area
which is the most useful in terms of its cosmological constraining power will
span a total of about 1100 sq. deg. Numerous challenges will be involved in the
processing of data with such statistical power, in particular, to keep the
systematic error budget under control. As we will show in the companion papers,
the photometric redshift uncertainties are still our dominant source of errors.
Addressing this challenge using additional measurements, better data as well as
innovative techniques is going to be a necessary task before the arrival of
data from the Rubin LSST. 

\begin{acknowledgments}
We thank the anonymous referee for a careful reading of the manuscript and the constructive inputs on the version of the manuscript submitted for review.
This work was supported in part by World Premier International Research Center Initiative (WPI Initiative), MEXT, Japan, and JSPS KAKENHI Grant Numbers JP18H04350, JP18H04358, JP19H00677, JP19K14767, JP20H00181, JP20H01932, JP20H04723, JP20H05850, JP20H05855, JP20H05856, JP20H05861, JP21J00011, JP21H05456, JP21J10314, JP21H01081, JP21H05456,  JP22H00130, JP22K03634, JP22K03655, JP22K21349, JP23H00108 and JP23H04005, by JSPS Core-to-Core Program (grant number: JPJSCCA20210003), by Japan Science and Technology Agency (JST) CREST JPMHCR1414, by JST AIP Acceleration Research Grant Number JP20317829, Japan, and by Basic Research Grant (Super AI) of Institute for AI and Beyond of the University of Tokyo. SS was supported in part by International Graduate Program for Excellence in Earth-Space Science (IGPEES), WINGS Program, the University of Tokyo. RD acknowledges support from the NSF Graduate Research Fellowship Program under Grant No.\ DGE-2039656. YK is supported in part by the David and Lucile Packard foundation. W.L. acknowledge the support from the National Key R\&D Program of China (2021YFC2203100), the 111 Project for "Observational and Theoretical Research on Dark Matter and Dark Energy'' (B23042), NSFC(NO. 11833005, 12192224) as well as the Fundamental Research Funds for the Central Universities (WK3440000006).

The Hyper Suprime-Cam (HSC) collaboration includes the astronomical
communities of Japan and Taiwan, and Princeton University. The HSC
instrumentation and software were developed by the National Astronomical
Observatory of Japan (NAOJ), the Kavli Institute for the Physics and
Mathematics of the Universe (Kavli IPMU), the University of Tokyo, the
High Energy Accelerator Research Organization (KEK), the Academia Sinica
Institute for Astronomy and Astrophysics in Taiwan (ASIAA), and
Princeton University. Funding was contributed by the FIRST program from
Japanese Cabinet Office, the Ministry of Education, Culture, Sports,
Science and Technology (MEXT), the Japan Society for the Promotion of
Science (JSPS), Japan Science and Technology Agency (JST), the Toray
Science Foundation, NAOJ, Kavli IPMU, KEK, ASIAA, and Princeton
University.
This paper makes use of software developed for the Large Synoptic Survey
Telescope. We thank the LSST Project for making their code available as
free software at \url{http://dm.lsst.org}

The Pan-STARRS1 Surveys (PS1) have been made possible through
contributions of the Institute for Astronomy, the University of Hawaii,
the Pan-STARRS Project Office, the Max-Planck Society and its
participating institutes, the Max Planck Institute for Astronomy,
Heidelberg and the Max Planck Institute for Extraterrestrial Physics,
Garching, The Johns Hopkins University, Durham University, the
University of Edinburgh, Queen's University Belfast, the
Harvard-Smithsonian Center for Astrophysics, the Las Cumbres Observatory
Global Telescope Network Incorporated, the National Central University
of Taiwan, the Space Telescope Science Institute, the National
Aeronautics and Space Administration under Grant No. NNX08AR22G issued
through the Planetary Science Division of the NASA Science Mission
Directorate, the National Science Foundation under Grant
No. AST-1238877, the University of Maryland, and Eotvos Lorand
University (ELTE) and the Los Alamos National Laboratory.

Based in part on data collected at the Subaru Telescope and retrieved
from the HSC data archive system, which is operated by Subaru Telescope
and Astronomy Data Center at National Astronomical Observatory of Japan.
\end{acknowledgments}

\bibliography{refs-short}

%apsrev4-2.bst 2019-01-14 (MD) hand-edited version of apsrev4-1.bst
%Control: key (0)
%Control: author (72) initials jnrlst
%Control: editor formatted (1) identically to author
%Control: production of article title (-1) disabled
%Control: page (0) single
%Control: year (1) truncated
%Control: production of eprint (0) enabled
\begin{thebibliography}{103}%
\makeatletter
\providecommand \@ifxundefined [1]{%
 \@ifx{#1\undefined}
}%
\providecommand \@ifnum [1]{%
 \ifnum #1\expandafter \@firstoftwo
 \else \expandafter \@secondoftwo
 \fi
}%
\providecommand \@ifx [1]{%
 \ifx #1\expandafter \@firstoftwo
 \else \expandafter \@secondoftwo
 \fi
}%
\providecommand \natexlab [1]{#1}%
\providecommand \enquote  [1]{``#1''}%
\providecommand \bibnamefont  [1]{#1}%
\providecommand \bibfnamefont [1]{#1}%
\providecommand \citenamefont [1]{#1}%
\providecommand \href@noop [0]{\@secondoftwo}%
\providecommand \href [0]{\begingroup \@sanitize@url \@href}%
\providecommand \@href[1]{\@@startlink{#1}\@@href}%
\providecommand \@@href[1]{\endgroup#1\@@endlink}%
\providecommand \@sanitize@url [0]{\catcode `\\12\catcode `\$12\catcode
  `\&12\catcode `\#12\catcode `\^12\catcode `\_12\catcode `\%12\relax}%
\providecommand \@@startlink[1]{}%
\providecommand \@@endlink[0]{}%
\providecommand \url  [0]{\begingroup\@sanitize@url \@url }%
\providecommand \@url [1]{\endgroup\@href {#1}{\urlprefix }}%
\providecommand \urlprefix  [0]{URL }%
\providecommand \Eprint [0]{\href }%
\providecommand \doibase [0]{https://doi.org/}%
\providecommand \selectlanguage [0]{\@gobble}%
\providecommand \bibinfo  [0]{\@secondoftwo}%
\providecommand \bibfield  [0]{\@secondoftwo}%
\providecommand \translation [1]{[#1]}%
\providecommand \BibitemOpen [0]{}%
\providecommand \bibitemStop [0]{}%
\providecommand \bibitemNoStop [0]{.\EOS\space}%
\providecommand \EOS [0]{\spacefactor3000\relax}%
\providecommand \BibitemShut  [1]{\csname bibitem#1\endcsname}%
\let\auto@bib@innerbib\@empty
%</preamble>
\bibitem [{\citenamefont {{Planck Collaboration}}\ \emph
  {et~al.}(2016)\citenamefont {{Planck Collaboration}}, \citenamefont {{Ade}},
  \citenamefont {{Aghanim}}, \citenamefont {{Arnaud}}, \citenamefont
  {{Ashdown}}, \citenamefont {{Aumont}}, \citenamefont {{Baccigalupi}},
  \citenamefont {{Banday}}, \citenamefont {{Barreiro}}, \citenamefont
  {{Bartlett}},\ and\ \citenamefont {et~al.}}]{PlanckCosmology:16}%
  \BibitemOpen
  \bibfield  {author} {\bibinfo {author} {\bibnamefont {{Planck
  Collaboration}}}, \bibinfo {author} {\bibfnamefont {P.~A.~R.}\ \bibnamefont
  {{Ade}}}, \bibinfo {author} {\bibfnamefont {N.}~\bibnamefont {{Aghanim}}},
  \bibinfo {author} {\bibfnamefont {M.}~\bibnamefont {{Arnaud}}}, \bibinfo
  {author} {\bibfnamefont {M.}~\bibnamefont {{Ashdown}}}, \bibinfo {author}
  {\bibfnamefont {J.}~\bibnamefont {{Aumont}}}, \bibinfo {author}
  {\bibfnamefont {C.}~\bibnamefont {{Baccigalupi}}}, \bibinfo {author}
  {\bibfnamefont {A.~J.}\ \bibnamefont {{Banday}}}, \bibinfo {author}
  {\bibfnamefont {R.~B.}\ \bibnamefont {{Barreiro}}}, \bibinfo {author}
  {\bibfnamefont {J.~G.}\ \bibnamefont {{Bartlett}}},\ and\ \bibinfo {author}
  {\bibnamefont {et~al.}},\ }\href
  {https://doi.org/10.1051/0004-6361/201525830} {\bibfield  {journal} {\bibinfo
   {journal} {\aap}\ }\textbf {\bibinfo {volume} {594}},\ \bibinfo {eid} {A13}
  (\bibinfo {year} {2016})},\ \Eprint {https://arxiv.org/abs/1502.01589}
  {arXiv:1502.01589} \BibitemShut {NoStop}%
\bibitem [{\citenamefont {{Planck Collaboration}}\ \emph
  {et~al.}(2020)\citenamefont {{Planck Collaboration}}, \citenamefont
  {{Aghanim}}, \citenamefont {{Akrami}}, \citenamefont {{Ashdown}},
  \citenamefont {{Aumont}}, \citenamefont {{Baccigalupi}}, \citenamefont
  {{Ballardini}}, \citenamefont {{Banday}}, \citenamefont {{Barreiro}},
  \citenamefont {{Bartolo}}, \citenamefont {{Basak}}, \citenamefont {{Battye}},
  \citenamefont {{Benabed}}, \citenamefont {{Bernard}}, \citenamefont
  {{Bersanelli}}, \citenamefont {{Bielewicz}}, \citenamefont {{Bock}},
  \citenamefont {{Bond}}, \citenamefont {{Borrill}}, \citenamefont {{Bouchet}},
  \citenamefont {{Boulanger}} \emph {et~al.}}]{2020A&A...641A...6P}%
  \BibitemOpen
  \bibfield  {author} {\bibinfo {author} {\bibnamefont {{Planck
  Collaboration}}}, \bibinfo {author} {\bibfnamefont {N.}~\bibnamefont
  {{Aghanim}}}, \bibinfo {author} {\bibfnamefont {Y.}~\bibnamefont {{Akrami}}},
  \bibinfo {author} {\bibfnamefont {M.}~\bibnamefont {{Ashdown}}}, \bibinfo
  {author} {\bibfnamefont {J.}~\bibnamefont {{Aumont}}}, \bibinfo {author}
  {\bibfnamefont {C.}~\bibnamefont {{Baccigalupi}}}, \bibinfo {author}
  {\bibfnamefont {M.}~\bibnamefont {{Ballardini}}}, \bibinfo {author}
  {\bibfnamefont {A.~J.}\ \bibnamefont {{Banday}}}, \bibinfo {author}
  {\bibfnamefont {R.~B.}\ \bibnamefont {{Barreiro}}}, \bibinfo {author}
  {\bibfnamefont {N.}~\bibnamefont {{Bartolo}}}, \bibinfo {author}
  {\bibfnamefont {S.}~\bibnamefont {{Basak}}}, \bibinfo {author} {\bibfnamefont
  {R.}~\bibnamefont {{Battye}}}, \bibinfo {author} {\bibfnamefont
  {K.}~\bibnamefont {{Benabed}}}, \bibinfo {author} {\bibfnamefont {J.~P.}\
  \bibnamefont {{Bernard}}}, \bibinfo {author} {\bibfnamefont {M.}~\bibnamefont
  {{Bersanelli}}}, \bibinfo {author} {\bibfnamefont {P.}~\bibnamefont
  {{Bielewicz}}}, \bibinfo {author} {\bibfnamefont {J.~J.}\ \bibnamefont
  {{Bock}}}, \bibinfo {author} {\bibfnamefont {J.~R.}\ \bibnamefont {{Bond}}},
  \bibinfo {author} {\bibfnamefont {J.}~\bibnamefont {{Borrill}}}, \bibinfo
  {author} {\bibfnamefont {F.~R.}\ \bibnamefont {{Bouchet}}}, \bibinfo {author}
  {\bibfnamefont {F.}~\bibnamefont {{Boulanger}}}, \emph {et~al.},\ }\href
  {https://doi.org/10.1051/0004-6361/201833910} {\bibfield  {journal} {\bibinfo
   {journal} {\aap}\ }\textbf {\bibinfo {volume} {641}},\ \bibinfo {eid} {A6}
  (\bibinfo {year} {2020})},\ \Eprint {https://arxiv.org/abs/1807.06209}
  {arXiv:1807.06209 [astro-ph.CO]} \BibitemShut {NoStop}%
\bibitem [{\citenamefont {{Scolnic}}\ \emph {et~al.}(2018)\citenamefont
  {{Scolnic}}, \citenamefont {{Jones}}, \citenamefont {{Rest}}, \citenamefont
  {{Pan}}, \citenamefont {{Chornock}}, \citenamefont {{Foley}}, \citenamefont
  {{Huber}}, \citenamefont {{Kessler}}, \citenamefont {{Narayan}},
  \citenamefont {{Riess}}, \citenamefont {{Rodney}}, \citenamefont {{Berger}},
  \citenamefont {{Brout}}, \citenamefont {{Challis}}, \citenamefont {{Drout}}
  \emph {et~al.}}]{Scolnic:2018}%
  \BibitemOpen
  \bibfield  {author} {\bibinfo {author} {\bibfnamefont {D.~M.}\ \bibnamefont
  {{Scolnic}}}, \bibinfo {author} {\bibfnamefont {D.~O.}\ \bibnamefont
  {{Jones}}}, \bibinfo {author} {\bibfnamefont {A.}~\bibnamefont {{Rest}}},
  \bibinfo {author} {\bibfnamefont {Y.~C.}\ \bibnamefont {{Pan}}}, \bibinfo
  {author} {\bibfnamefont {R.}~\bibnamefont {{Chornock}}}, \bibinfo {author}
  {\bibfnamefont {R.~J.}\ \bibnamefont {{Foley}}}, \bibinfo {author}
  {\bibfnamefont {M.~E.}\ \bibnamefont {{Huber}}}, \bibinfo {author}
  {\bibfnamefont {R.}~\bibnamefont {{Kessler}}}, \bibinfo {author}
  {\bibfnamefont {G.}~\bibnamefont {{Narayan}}}, \bibinfo {author}
  {\bibfnamefont {A.~G.}\ \bibnamefont {{Riess}}}, \bibinfo {author}
  {\bibfnamefont {S.}~\bibnamefont {{Rodney}}}, \bibinfo {author}
  {\bibfnamefont {E.}~\bibnamefont {{Berger}}}, \bibinfo {author}
  {\bibfnamefont {D.~J.}\ \bibnamefont {{Brout}}}, \bibinfo {author}
  {\bibfnamefont {P.~J.}\ \bibnamefont {{Challis}}}, \bibinfo {author}
  {\bibfnamefont {M.}~\bibnamefont {{Drout}}}, \emph {et~al.},\ }\href
  {https://doi.org/10.3847/1538-4357/aab9bb} {\bibfield  {journal} {\bibinfo
  {journal} {\apj}\ }\textbf {\bibinfo {volume} {859}},\ \bibinfo {eid} {101}
  (\bibinfo {year} {2018})},\ \Eprint {https://arxiv.org/abs/1710.00845}
  {arXiv:1710.00845 [astro-ph.CO]} \BibitemShut {NoStop}%
\bibitem [{\citenamefont {{Ross}}\ \emph {et~al.}(2017)\citenamefont {{Ross}},
  \citenamefont {{Beutler}}, \citenamefont {{Chuang}}, \citenamefont
  {{Pellejero-Ibanez}}, \citenamefont {{Seo}}, \citenamefont
  {{Vargas-Maga{\~n}a}}, \citenamefont {{Cuesta}}, \citenamefont {{Percival}},
  \citenamefont {{Burden}}, \citenamefont {{S{\'a}nchez}}, \citenamefont
  {{Grieb}}, \citenamefont {{Reid}}, \citenamefont {{Brownstein}},
  \citenamefont {{Dawson}}, \citenamefont {{Eisenstein}} \emph
  {et~al.}}]{Ross:2017}%
  \BibitemOpen
  \bibfield  {author} {\bibinfo {author} {\bibfnamefont {A.~J.}\ \bibnamefont
  {{Ross}}}, \bibinfo {author} {\bibfnamefont {F.}~\bibnamefont {{Beutler}}},
  \bibinfo {author} {\bibfnamefont {C.-H.}\ \bibnamefont {{Chuang}}}, \bibinfo
  {author} {\bibfnamefont {M.}~\bibnamefont {{Pellejero-Ibanez}}}, \bibinfo
  {author} {\bibfnamefont {H.-J.}\ \bibnamefont {{Seo}}}, \bibinfo {author}
  {\bibfnamefont {M.}~\bibnamefont {{Vargas-Maga{\~n}a}}}, \bibinfo {author}
  {\bibfnamefont {A.~J.}\ \bibnamefont {{Cuesta}}}, \bibinfo {author}
  {\bibfnamefont {W.~J.}\ \bibnamefont {{Percival}}}, \bibinfo {author}
  {\bibfnamefont {A.}~\bibnamefont {{Burden}}}, \bibinfo {author}
  {\bibfnamefont {A.~G.}\ \bibnamefont {{S{\'a}nchez}}}, \bibinfo {author}
  {\bibfnamefont {J.~N.}\ \bibnamefont {{Grieb}}}, \bibinfo {author}
  {\bibfnamefont {B.}~\bibnamefont {{Reid}}}, \bibinfo {author} {\bibfnamefont
  {J.~R.}\ \bibnamefont {{Brownstein}}}, \bibinfo {author} {\bibfnamefont
  {K.~S.}\ \bibnamefont {{Dawson}}}, \bibinfo {author} {\bibfnamefont {D.~J.}\
  \bibnamefont {{Eisenstein}}}, \emph {et~al.},\ }\href
  {https://doi.org/10.1093/mnras/stw2372} {\bibfield  {journal} {\bibinfo
  {journal} {\mnras}\ }\textbf {\bibinfo {volume} {464}},\ \bibinfo {pages}
  {1168} (\bibinfo {year} {2017})},\ \Eprint {https://arxiv.org/abs/1607.03145}
  {arXiv:1607.03145 [astro-ph.CO]} \BibitemShut {NoStop}%
\bibitem [{\citenamefont {{Beutler}}\ \emph {et~al.}(2017)\citenamefont
  {{Beutler}}, \citenamefont {{Seo}}, \citenamefont {{Ross}}, \citenamefont
  {{McDonald}}, \citenamefont {{Saito}}, \citenamefont {{Bolton}},
  \citenamefont {{Brownstein}}, \citenamefont {{Chuang}}, \citenamefont
  {{Cuesta}}, \citenamefont {{Eisenstein}}, \citenamefont {{Font-Ribera}},
  \citenamefont {{Grieb}}, \citenamefont {{Hand}}, \citenamefont {{Kitaura}},
  \citenamefont {{Modi}} \emph {et~al.}}]{Beutler:2017}%
  \BibitemOpen
  \bibfield  {author} {\bibinfo {author} {\bibfnamefont {F.}~\bibnamefont
  {{Beutler}}}, \bibinfo {author} {\bibfnamefont {H.-J.}\ \bibnamefont
  {{Seo}}}, \bibinfo {author} {\bibfnamefont {A.~J.}\ \bibnamefont {{Ross}}},
  \bibinfo {author} {\bibfnamefont {P.}~\bibnamefont {{McDonald}}}, \bibinfo
  {author} {\bibfnamefont {S.}~\bibnamefont {{Saito}}}, \bibinfo {author}
  {\bibfnamefont {A.~S.}\ \bibnamefont {{Bolton}}}, \bibinfo {author}
  {\bibfnamefont {J.~R.}\ \bibnamefont {{Brownstein}}}, \bibinfo {author}
  {\bibfnamefont {C.-H.}\ \bibnamefont {{Chuang}}}, \bibinfo {author}
  {\bibfnamefont {A.~J.}\ \bibnamefont {{Cuesta}}}, \bibinfo {author}
  {\bibfnamefont {D.~J.}\ \bibnamefont {{Eisenstein}}}, \bibinfo {author}
  {\bibfnamefont {A.}~\bibnamefont {{Font-Ribera}}}, \bibinfo {author}
  {\bibfnamefont {J.~N.}\ \bibnamefont {{Grieb}}}, \bibinfo {author}
  {\bibfnamefont {N.}~\bibnamefont {{Hand}}}, \bibinfo {author} {\bibfnamefont
  {F.-S.}\ \bibnamefont {{Kitaura}}}, \bibinfo {author} {\bibfnamefont
  {C.}~\bibnamefont {{Modi}}}, \emph {et~al.},\ }\href
  {https://doi.org/10.1093/mnras/stw2373} {\bibfield  {journal} {\bibinfo
  {journal} {\mnras}\ }\textbf {\bibinfo {volume} {464}},\ \bibinfo {pages}
  {3409} (\bibinfo {year} {2017})},\ \Eprint {https://arxiv.org/abs/1607.03149}
  {arXiv:1607.03149 [astro-ph.CO]} \BibitemShut {NoStop}%
\bibitem [{\citenamefont {{Alam}}\ \emph {et~al.}(2021)\citenamefont {{Alam}},
  \citenamefont {{Aubert}}, \citenamefont {{Avila}}, \citenamefont {{Balland}},
  \citenamefont {{Bautista}}, \citenamefont {{Bershady}}, \citenamefont
  {{Bizyaev}}, \citenamefont {{Blanton}}, \citenamefont {{Bolton}},
  \citenamefont {{Bovy}}, \citenamefont {{Brinkmann}}, \citenamefont
  {{Brownstein}}, \citenamefont {{Burtin}}, \citenamefont {{Chabanier}},
  \citenamefont {{Chapman}} \emph {et~al.}}]{Alam:2021}%
  \BibitemOpen
  \bibfield  {author} {\bibinfo {author} {\bibfnamefont {S.}~\bibnamefont
  {{Alam}}}, \bibinfo {author} {\bibfnamefont {M.}~\bibnamefont {{Aubert}}},
  \bibinfo {author} {\bibfnamefont {S.}~\bibnamefont {{Avila}}}, \bibinfo
  {author} {\bibfnamefont {C.}~\bibnamefont {{Balland}}}, \bibinfo {author}
  {\bibfnamefont {J.~E.}\ \bibnamefont {{Bautista}}}, \bibinfo {author}
  {\bibfnamefont {M.~A.}\ \bibnamefont {{Bershady}}}, \bibinfo {author}
  {\bibfnamefont {D.}~\bibnamefont {{Bizyaev}}}, \bibinfo {author}
  {\bibfnamefont {M.~R.}\ \bibnamefont {{Blanton}}}, \bibinfo {author}
  {\bibfnamefont {A.~S.}\ \bibnamefont {{Bolton}}}, \bibinfo {author}
  {\bibfnamefont {J.}~\bibnamefont {{Bovy}}}, \bibinfo {author} {\bibfnamefont
  {J.}~\bibnamefont {{Brinkmann}}}, \bibinfo {author} {\bibfnamefont {J.~R.}\
  \bibnamefont {{Brownstein}}}, \bibinfo {author} {\bibfnamefont
  {E.}~\bibnamefont {{Burtin}}}, \bibinfo {author} {\bibfnamefont
  {S.}~\bibnamefont {{Chabanier}}}, \bibinfo {author} {\bibfnamefont {M.~J.}\
  \bibnamefont {{Chapman}}}, \emph {et~al.},\ }\href
  {https://doi.org/10.1103/PhysRevD.103.083533} {\bibfield  {journal} {\bibinfo
   {journal} {\prd}\ }\textbf {\bibinfo {volume} {103}},\ \bibinfo {eid}
  {083533} (\bibinfo {year} {2021})},\ \Eprint
  {https://arxiv.org/abs/2007.08991} {arXiv:2007.08991 [astro-ph.CO]}
  \BibitemShut {NoStop}%
\bibitem [{\citenamefont {{Aiola}}\ \emph {et~al.}(2020)\citenamefont
  {{Aiola}}, \citenamefont {{Calabrese}}, \citenamefont {{Maurin}},
  \citenamefont {{Naess}}, \citenamefont {{Schmitt}}, \citenamefont
  {{Abitbol}}, \citenamefont {{Addison}}, \citenamefont {{Ade}}, \citenamefont
  {{Alonso}}, \citenamefont {{Amiri}}, \citenamefont {{Amodeo}}, \citenamefont
  {{Angile}}, \citenamefont {{Austermann}}, \citenamefont {{Baildon}},
  \citenamefont {{Battaglia}} \emph {et~al.}}]{Aiola:2020}%
  \BibitemOpen
  \bibfield  {author} {\bibinfo {author} {\bibfnamefont {S.}~\bibnamefont
  {{Aiola}}}, \bibinfo {author} {\bibfnamefont {E.}~\bibnamefont
  {{Calabrese}}}, \bibinfo {author} {\bibfnamefont {L.}~\bibnamefont
  {{Maurin}}}, \bibinfo {author} {\bibfnamefont {S.}~\bibnamefont {{Naess}}},
  \bibinfo {author} {\bibfnamefont {B.~L.}\ \bibnamefont {{Schmitt}}}, \bibinfo
  {author} {\bibfnamefont {M.~H.}\ \bibnamefont {{Abitbol}}}, \bibinfo {author}
  {\bibfnamefont {G.~E.}\ \bibnamefont {{Addison}}}, \bibinfo {author}
  {\bibfnamefont {P.~A.~R.}\ \bibnamefont {{Ade}}}, \bibinfo {author}
  {\bibfnamefont {D.}~\bibnamefont {{Alonso}}}, \bibinfo {author}
  {\bibfnamefont {M.}~\bibnamefont {{Amiri}}}, \bibinfo {author} {\bibfnamefont
  {S.}~\bibnamefont {{Amodeo}}}, \bibinfo {author} {\bibfnamefont
  {E.}~\bibnamefont {{Angile}}}, \bibinfo {author} {\bibfnamefont {J.~E.}\
  \bibnamefont {{Austermann}}}, \bibinfo {author} {\bibfnamefont
  {T.}~\bibnamefont {{Baildon}}}, \bibinfo {author} {\bibfnamefont
  {N.}~\bibnamefont {{Battaglia}}}, \emph {et~al.},\ }\href
  {https://doi.org/10.1088/1475-7516/2020/12/047} {\bibfield  {journal}
  {\bibinfo  {journal} {\jcap}\ }\textbf {\bibinfo {volume} {2020}},\ \bibinfo
  {eid} {047} (\bibinfo {year} {2020})},\ \Eprint
  {https://arxiv.org/abs/2007.07288} {arXiv:2007.07288 [astro-ph.CO]}
  \BibitemShut {NoStop}%
\bibitem [{\citenamefont {{Chaubal}}\ \emph {et~al.}(2022)\citenamefont
  {{Chaubal}}, \citenamefont {{Reichardt}}, \citenamefont {{Gupta}},
  \citenamefont {{Ansarinejad}}, \citenamefont {{Aylor}}, \citenamefont
  {{Balkenhol}}, \citenamefont {{Baxter}}, \citenamefont {{Bianchini}},
  \citenamefont {{Benson}}, \citenamefont {{Bleem}}, \citenamefont {{Bocquet}},
  \citenamefont {{Carlstrom}}, \citenamefont {{Chang}}, \citenamefont
  {{Crawford}}, \citenamefont {{Crites}} \emph {et~al.}}]{Chaubal:2022}%
  \BibitemOpen
  \bibfield  {author} {\bibinfo {author} {\bibfnamefont {P.~S.}\ \bibnamefont
  {{Chaubal}}}, \bibinfo {author} {\bibfnamefont {C.~L.}\ \bibnamefont
  {{Reichardt}}}, \bibinfo {author} {\bibfnamefont {N.}~\bibnamefont
  {{Gupta}}}, \bibinfo {author} {\bibfnamefont {B.}~\bibnamefont
  {{Ansarinejad}}}, \bibinfo {author} {\bibfnamefont {K.}~\bibnamefont
  {{Aylor}}}, \bibinfo {author} {\bibfnamefont {L.}~\bibnamefont
  {{Balkenhol}}}, \bibinfo {author} {\bibfnamefont {E.~J.}\ \bibnamefont
  {{Baxter}}}, \bibinfo {author} {\bibfnamefont {F.}~\bibnamefont
  {{Bianchini}}}, \bibinfo {author} {\bibfnamefont {B.~A.}\ \bibnamefont
  {{Benson}}}, \bibinfo {author} {\bibfnamefont {L.~E.}\ \bibnamefont
  {{Bleem}}}, \bibinfo {author} {\bibfnamefont {S.}~\bibnamefont {{Bocquet}}},
  \bibinfo {author} {\bibfnamefont {J.~E.}\ \bibnamefont {{Carlstrom}}},
  \bibinfo {author} {\bibfnamefont {C.~L.}\ \bibnamefont {{Chang}}}, \bibinfo
  {author} {\bibfnamefont {T.~M.}\ \bibnamefont {{Crawford}}}, \bibinfo
  {author} {\bibfnamefont {A.~T.}\ \bibnamefont {{Crites}}}, \emph {et~al.},\
  }\href {https://doi.org/10.3847/1538-4357/ac6a55} {\bibfield  {journal}
  {\bibinfo  {journal} {\apj}\ }\textbf {\bibinfo {volume} {931}},\ \bibinfo
  {eid} {139} (\bibinfo {year} {2022})},\ \Eprint
  {https://arxiv.org/abs/2111.07491} {arXiv:2111.07491 [astro-ph.CO]}
  \BibitemShut {NoStop}%
\bibitem [{\citenamefont {{S{\'a}nchez}}\ \emph {et~al.}(2017)\citenamefont
  {{S{\'a}nchez}}, \citenamefont {{Scoccimarro}}, \citenamefont {{Crocce}},
  \citenamefont {{Grieb}}, \citenamefont {{Salazar-Albornoz}}, \citenamefont
  {{Dalla Vecchia}}, \citenamefont {{Lippich}}, \citenamefont {{Beutler}},
  \citenamefont {{Brownstein}}, \citenamefont {{Chuang}}, \citenamefont
  {{Eisenstein}}, \citenamefont {{Kitaura}}, \citenamefont {{Olmstead}},
  \citenamefont {{Percival}}, \citenamefont {{Prada}} \emph
  {et~al.}}]{Sanchez:2017}%
  \BibitemOpen
  \bibfield  {author} {\bibinfo {author} {\bibfnamefont {A.~G.}\ \bibnamefont
  {{S{\'a}nchez}}}, \bibinfo {author} {\bibfnamefont {R.}~\bibnamefont
  {{Scoccimarro}}}, \bibinfo {author} {\bibfnamefont {M.}~\bibnamefont
  {{Crocce}}}, \bibinfo {author} {\bibfnamefont {J.~N.}\ \bibnamefont
  {{Grieb}}}, \bibinfo {author} {\bibfnamefont {S.}~\bibnamefont
  {{Salazar-Albornoz}}}, \bibinfo {author} {\bibfnamefont {C.}~\bibnamefont
  {{Dalla Vecchia}}}, \bibinfo {author} {\bibfnamefont {M.}~\bibnamefont
  {{Lippich}}}, \bibinfo {author} {\bibfnamefont {F.}~\bibnamefont
  {{Beutler}}}, \bibinfo {author} {\bibfnamefont {J.~R.}\ \bibnamefont
  {{Brownstein}}}, \bibinfo {author} {\bibfnamefont {C.-H.}\ \bibnamefont
  {{Chuang}}}, \bibinfo {author} {\bibfnamefont {D.~J.}\ \bibnamefont
  {{Eisenstein}}}, \bibinfo {author} {\bibfnamefont {F.-S.}\ \bibnamefont
  {{Kitaura}}}, \bibinfo {author} {\bibfnamefont {M.~D.}\ \bibnamefont
  {{Olmstead}}}, \bibinfo {author} {\bibfnamefont {W.~J.}\ \bibnamefont
  {{Percival}}}, \bibinfo {author} {\bibfnamefont {F.}~\bibnamefont {{Prada}}},
  \emph {et~al.},\ }\href {https://doi.org/10.1093/mnras/stw2443} {\bibfield
  {journal} {\bibinfo  {journal} {\mnras}\ }\textbf {\bibinfo {volume} {464}},\
  \bibinfo {pages} {1640} (\bibinfo {year} {2017})},\ \Eprint
  {https://arxiv.org/abs/1607.03147} {arXiv:1607.03147 [astro-ph.CO]}
  \BibitemShut {NoStop}%
\bibitem [{\citenamefont {{Grieb}}\ \emph {et~al.}(2017)\citenamefont
  {{Grieb}}, \citenamefont {{S{\'a}nchez}}, \citenamefont {{Salazar-Albornoz}},
  \citenamefont {{Scoccimarro}}, \citenamefont {{Crocce}}, \citenamefont
  {{Dalla Vecchia}}, \citenamefont {{Montesano}}, \citenamefont
  {{Gil-Mar{\'\i}n}}, \citenamefont {{Ross}}, \citenamefont {{Beutler}},
  \citenamefont {{Rodr{\'\i}guez-Torres}}, \citenamefont {{Chuang}},
  \citenamefont {{Prada}}, \citenamefont {{Kitaura}}, \citenamefont {{Cuesta}}
  \emph {et~al.}}]{Grieb:2017}%
  \BibitemOpen
  \bibfield  {author} {\bibinfo {author} {\bibfnamefont {J.~N.}\ \bibnamefont
  {{Grieb}}}, \bibinfo {author} {\bibfnamefont {A.~G.}\ \bibnamefont
  {{S{\'a}nchez}}}, \bibinfo {author} {\bibfnamefont {S.}~\bibnamefont
  {{Salazar-Albornoz}}}, \bibinfo {author} {\bibfnamefont {R.}~\bibnamefont
  {{Scoccimarro}}}, \bibinfo {author} {\bibfnamefont {M.}~\bibnamefont
  {{Crocce}}}, \bibinfo {author} {\bibfnamefont {C.}~\bibnamefont {{Dalla
  Vecchia}}}, \bibinfo {author} {\bibfnamefont {F.}~\bibnamefont
  {{Montesano}}}, \bibinfo {author} {\bibfnamefont {H.}~\bibnamefont
  {{Gil-Mar{\'\i}n}}}, \bibinfo {author} {\bibfnamefont {A.~J.}\ \bibnamefont
  {{Ross}}}, \bibinfo {author} {\bibfnamefont {F.}~\bibnamefont {{Beutler}}},
  \bibinfo {author} {\bibfnamefont {S.}~\bibnamefont
  {{Rodr{\'\i}guez-Torres}}}, \bibinfo {author} {\bibfnamefont {C.-H.}\
  \bibnamefont {{Chuang}}}, \bibinfo {author} {\bibfnamefont {F.}~\bibnamefont
  {{Prada}}}, \bibinfo {author} {\bibfnamefont {F.-S.}\ \bibnamefont
  {{Kitaura}}}, \bibinfo {author} {\bibfnamefont {A.~J.}\ \bibnamefont
  {{Cuesta}}}, \emph {et~al.},\ }\href {https://doi.org/10.1093/mnras/stw3384}
  {\bibfield  {journal} {\bibinfo  {journal} {\mnras}\ }\textbf {\bibinfo
  {volume} {467}},\ \bibinfo {pages} {2085} (\bibinfo {year} {2017})},\ \Eprint
  {https://arxiv.org/abs/1607.03143} {arXiv:1607.03143 [astro-ph.CO]}
  \BibitemShut {NoStop}%
\bibitem [{\citenamefont {{Kobayashi}}\ \emph {et~al.}(2022)\citenamefont
  {{Kobayashi}}, \citenamefont {{Nishimichi}}, \citenamefont {{Takada}},\ and\
  \citenamefont {{Miyatake}}}]{Kobayashi:2021}%
  \BibitemOpen
  \bibfield  {author} {\bibinfo {author} {\bibfnamefont {Y.}~\bibnamefont
  {{Kobayashi}}}, \bibinfo {author} {\bibfnamefont {T.}~\bibnamefont
  {{Nishimichi}}}, \bibinfo {author} {\bibfnamefont {M.}~\bibnamefont
  {{Takada}}},\ and\ \bibinfo {author} {\bibfnamefont {H.}~\bibnamefont
  {{Miyatake}}},\ }\href {https://doi.org/10.1103/PhysRevD.105.083517}
  {\bibfield  {journal} {\bibinfo  {journal} {\prd}\ }\textbf {\bibinfo
  {volume} {105}},\ \bibinfo {eid} {083517} (\bibinfo {year} {2022})},\ \Eprint
  {https://arxiv.org/abs/2110.06969} {arXiv:2110.06969 [astro-ph.CO]}
  \BibitemShut {NoStop}%
\bibitem [{\citenamefont {{Hikage}}\ \emph
  {et~al.}(2019{\natexlab{a}})\citenamefont {{Hikage}}, \citenamefont
  {{Oguri}}, \citenamefont {{Hamana}}, \citenamefont {{More}}, \citenamefont
  {{Mandelbaum}}, \citenamefont {{Takada}}, \citenamefont {{K{\"o}hlinger}},
  \citenamefont {{Miyatake}}, \citenamefont {{Nishizawa}}, \citenamefont
  {{Aihara}}, \citenamefont {{Armstrong}}, \citenamefont {{Bosch}},
  \citenamefont {{Coupon}}, \citenamefont {{Ducout}}, \citenamefont {{Ho}}
  \emph {et~al.}}]{Hikage:2019}%
  \BibitemOpen
  \bibfield  {author} {\bibinfo {author} {\bibfnamefont {C.}~\bibnamefont
  {{Hikage}}}, \bibinfo {author} {\bibfnamefont {M.}~\bibnamefont {{Oguri}}},
  \bibinfo {author} {\bibfnamefont {T.}~\bibnamefont {{Hamana}}}, \bibinfo
  {author} {\bibfnamefont {S.}~\bibnamefont {{More}}}, \bibinfo {author}
  {\bibfnamefont {R.}~\bibnamefont {{Mandelbaum}}}, \bibinfo {author}
  {\bibfnamefont {M.}~\bibnamefont {{Takada}}}, \bibinfo {author}
  {\bibfnamefont {F.}~\bibnamefont {{K{\"o}hlinger}}}, \bibinfo {author}
  {\bibfnamefont {H.}~\bibnamefont {{Miyatake}}}, \bibinfo {author}
  {\bibfnamefont {A.~J.}\ \bibnamefont {{Nishizawa}}}, \bibinfo {author}
  {\bibfnamefont {H.}~\bibnamefont {{Aihara}}}, \bibinfo {author}
  {\bibfnamefont {R.}~\bibnamefont {{Armstrong}}}, \bibinfo {author}
  {\bibfnamefont {J.}~\bibnamefont {{Bosch}}}, \bibinfo {author} {\bibfnamefont
  {J.}~\bibnamefont {{Coupon}}}, \bibinfo {author} {\bibfnamefont
  {A.}~\bibnamefont {{Ducout}}}, \bibinfo {author} {\bibfnamefont
  {P.}~\bibnamefont {{Ho}}}, \emph {et~al.},\ }\href
  {https://doi.org/10.1093/pasj/psz010} {\bibfield  {journal} {\bibinfo
  {journal} {\pasj}\ }\textbf {\bibinfo {volume} {71}},\ \bibinfo {eid} {43}
  (\bibinfo {year} {2019}{\natexlab{a}})},\ \Eprint
  {https://arxiv.org/abs/1809.09148} {arXiv:1809.09148 [astro-ph.CO]}
  \BibitemShut {NoStop}%
\bibitem [{\citenamefont {{Hamana}}\ \emph
  {et~al.}(2020{\natexlab{a}})\citenamefont {{Hamana}}, \citenamefont
  {{Shirasaki}}, \citenamefont {{Miyazaki}}, \citenamefont {{Hikage}},
  \citenamefont {{Oguri}}, \citenamefont {{More}}, \citenamefont {{Armstrong}},
  \citenamefont {{Leauthaud}}, \citenamefont {{Mandelbaum}}, \citenamefont
  {{Miyatake}}, \citenamefont {{Nishizawa}}, \citenamefont {{Simet}},
  \citenamefont {{Takada}}, \citenamefont {{Aihara}}, \citenamefont {{Bosch}}
  \emph {et~al.}}]{Hamana:2020}%
  \BibitemOpen
  \bibfield  {author} {\bibinfo {author} {\bibfnamefont {T.}~\bibnamefont
  {{Hamana}}}, \bibinfo {author} {\bibfnamefont {M.}~\bibnamefont
  {{Shirasaki}}}, \bibinfo {author} {\bibfnamefont {S.}~\bibnamefont
  {{Miyazaki}}}, \bibinfo {author} {\bibfnamefont {C.}~\bibnamefont
  {{Hikage}}}, \bibinfo {author} {\bibfnamefont {M.}~\bibnamefont {{Oguri}}},
  \bibinfo {author} {\bibfnamefont {S.}~\bibnamefont {{More}}}, \bibinfo
  {author} {\bibfnamefont {R.}~\bibnamefont {{Armstrong}}}, \bibinfo {author}
  {\bibfnamefont {A.}~\bibnamefont {{Leauthaud}}}, \bibinfo {author}
  {\bibfnamefont {R.}~\bibnamefont {{Mandelbaum}}}, \bibinfo {author}
  {\bibfnamefont {H.}~\bibnamefont {{Miyatake}}}, \bibinfo {author}
  {\bibfnamefont {A.~J.}\ \bibnamefont {{Nishizawa}}}, \bibinfo {author}
  {\bibfnamefont {M.}~\bibnamefont {{Simet}}}, \bibinfo {author} {\bibfnamefont
  {M.}~\bibnamefont {{Takada}}}, \bibinfo {author} {\bibfnamefont
  {H.}~\bibnamefont {{Aihara}}}, \bibinfo {author} {\bibfnamefont
  {J.}~\bibnamefont {{Bosch}}}, \emph {et~al.},\ }\href
  {https://doi.org/10.1093/pasj/psz138} {\bibfield  {journal} {\bibinfo
  {journal} {\pasj}\ }\textbf {\bibinfo {volume} {72}},\ \bibinfo {eid} {16}
  (\bibinfo {year} {2020}{\natexlab{a}})},\ \Eprint
  {https://arxiv.org/abs/1906.06041} {arXiv:1906.06041 [astro-ph.CO]}
  \BibitemShut {NoStop}%
\bibitem [{\citenamefont {{Heymans}}\ \emph {et~al.}(2021)\citenamefont
  {{Heymans}}, \citenamefont {{Tr{\"o}ster}}, \citenamefont {{Asgari}},
  \citenamefont {{Blake}}, \citenamefont {{Hildebrandt}}, \citenamefont
  {{Joachimi}}, \citenamefont {{Kuijken}}, \citenamefont {{Lin}}, \citenamefont
  {{S{\'a}nchez}}, \citenamefont {{van den Busch}}, \citenamefont {{Wright}},
  \citenamefont {{Amon}}, \citenamefont {{Bilicki}}, \citenamefont {{de Jong}},
  \citenamefont {{Crocce}} \emph {et~al.}}]{Heymans:2021}%
  \BibitemOpen
  \bibfield  {author} {\bibinfo {author} {\bibfnamefont {C.}~\bibnamefont
  {{Heymans}}}, \bibinfo {author} {\bibfnamefont {T.}~\bibnamefont
  {{Tr{\"o}ster}}}, \bibinfo {author} {\bibfnamefont {M.}~\bibnamefont
  {{Asgari}}}, \bibinfo {author} {\bibfnamefont {C.}~\bibnamefont {{Blake}}},
  \bibinfo {author} {\bibfnamefont {H.}~\bibnamefont {{Hildebrandt}}}, \bibinfo
  {author} {\bibfnamefont {B.}~\bibnamefont {{Joachimi}}}, \bibinfo {author}
  {\bibfnamefont {K.}~\bibnamefont {{Kuijken}}}, \bibinfo {author}
  {\bibfnamefont {C.-A.}\ \bibnamefont {{Lin}}}, \bibinfo {author}
  {\bibfnamefont {A.~G.}\ \bibnamefont {{S{\'a}nchez}}}, \bibinfo {author}
  {\bibfnamefont {J.~L.}\ \bibnamefont {{van den Busch}}}, \bibinfo {author}
  {\bibfnamefont {A.~H.}\ \bibnamefont {{Wright}}}, \bibinfo {author}
  {\bibfnamefont {A.}~\bibnamefont {{Amon}}}, \bibinfo {author} {\bibfnamefont
  {M.}~\bibnamefont {{Bilicki}}}, \bibinfo {author} {\bibfnamefont
  {J.}~\bibnamefont {{de Jong}}}, \bibinfo {author} {\bibfnamefont
  {M.}~\bibnamefont {{Crocce}}}, \emph {et~al.},\ }\href
  {https://doi.org/10.1051/0004-6361/202039063} {\bibfield  {journal} {\bibinfo
   {journal} {\aap}\ }\textbf {\bibinfo {volume} {646}},\ \bibinfo {eid} {A140}
  (\bibinfo {year} {2021})},\ \Eprint {https://arxiv.org/abs/2007.15632}
  {arXiv:2007.15632 [astro-ph.CO]} \BibitemShut {NoStop}%
\bibitem [{\citenamefont {{Pandey}}\ \emph {et~al.}(2021)\citenamefont
  {{Pandey}}, \citenamefont {{Krause}}, \citenamefont {{DeRose}}, \citenamefont
  {{MacCrann}}, \citenamefont {{Jain}}, \citenamefont {{Crocce}}, \citenamefont
  {{Blazek}}, \citenamefont {{Choi}}, \citenamefont {{Huang}}, \citenamefont
  {{To}}, \citenamefont {{Fang}}, \citenamefont {{Elvin-Poole}}, \citenamefont
  {{Prat}}, \citenamefont {{Porredon}}, \citenamefont {{Secco}} \emph
  {et~al.}}]{Pandey:2021}%
  \BibitemOpen
  \bibfield  {author} {\bibinfo {author} {\bibfnamefont {S.}~\bibnamefont
  {{Pandey}}}, \bibinfo {author} {\bibfnamefont {E.}~\bibnamefont {{Krause}}},
  \bibinfo {author} {\bibfnamefont {J.}~\bibnamefont {{DeRose}}}, \bibinfo
  {author} {\bibfnamefont {N.}~\bibnamefont {{MacCrann}}}, \bibinfo {author}
  {\bibfnamefont {B.}~\bibnamefont {{Jain}}}, \bibinfo {author} {\bibfnamefont
  {M.}~\bibnamefont {{Crocce}}}, \bibinfo {author} {\bibfnamefont
  {J.}~\bibnamefont {{Blazek}}}, \bibinfo {author} {\bibfnamefont
  {A.}~\bibnamefont {{Choi}}}, \bibinfo {author} {\bibfnamefont
  {H.}~\bibnamefont {{Huang}}}, \bibinfo {author} {\bibfnamefont
  {C.}~\bibnamefont {{To}}}, \bibinfo {author} {\bibfnamefont {X.}~\bibnamefont
  {{Fang}}}, \bibinfo {author} {\bibfnamefont {J.}~\bibnamefont
  {{Elvin-Poole}}}, \bibinfo {author} {\bibfnamefont {J.}~\bibnamefont
  {{Prat}}}, \bibinfo {author} {\bibfnamefont {A.}~\bibnamefont {{Porredon}}},
  \bibinfo {author} {\bibfnamefont {L.~F.}\ \bibnamefont {{Secco}}}, \emph
  {et~al.},\ }\href@noop {} {\bibfield  {journal} {\bibinfo  {journal} {arXiv
  e-prints}\ ,\ \bibinfo {eid} {arXiv:2105.13545}} (\bibinfo {year} {2021})},\
  \Eprint {https://arxiv.org/abs/2105.13545} {arXiv:2105.13545 [astro-ph.CO]}
  \BibitemShut {NoStop}%
\bibitem [{\citenamefont {{Porredon}}\ \emph {et~al.}(2021)\citenamefont
  {{Porredon}}, \citenamefont {{Crocce}}, \citenamefont {{Elvin-Poole}},
  \citenamefont {{Cawthon}}, \citenamefont {{Giannini}}, \citenamefont {{De
  Vicente}}, \citenamefont {{Carnero Rosell}}, \citenamefont {{Ferrero}},
  \citenamefont {{Krause}}, \citenamefont {{Fang}}, \citenamefont {{Prat}},
  \citenamefont {{Rodriguez-Monroy}}, \citenamefont {{Pandey}}, \citenamefont
  {{Pocino}}, \citenamefont {{Castander}} \emph {et~al.}}]{Porredon:2021}%
  \BibitemOpen
  \bibfield  {author} {\bibinfo {author} {\bibfnamefont {A.}~\bibnamefont
  {{Porredon}}}, \bibinfo {author} {\bibfnamefont {M.}~\bibnamefont
  {{Crocce}}}, \bibinfo {author} {\bibfnamefont {J.}~\bibnamefont
  {{Elvin-Poole}}}, \bibinfo {author} {\bibfnamefont {R.}~\bibnamefont
  {{Cawthon}}}, \bibinfo {author} {\bibfnamefont {G.}~\bibnamefont
  {{Giannini}}}, \bibinfo {author} {\bibfnamefont {J.}~\bibnamefont {{De
  Vicente}}}, \bibinfo {author} {\bibfnamefont {A.}~\bibnamefont {{Carnero
  Rosell}}}, \bibinfo {author} {\bibfnamefont {I.}~\bibnamefont {{Ferrero}}},
  \bibinfo {author} {\bibfnamefont {E.}~\bibnamefont {{Krause}}}, \bibinfo
  {author} {\bibfnamefont {X.}~\bibnamefont {{Fang}}}, \bibinfo {author}
  {\bibfnamefont {J.}~\bibnamefont {{Prat}}}, \bibinfo {author} {\bibfnamefont
  {M.}~\bibnamefont {{Rodriguez-Monroy}}}, \bibinfo {author} {\bibfnamefont
  {S.}~\bibnamefont {{Pandey}}}, \bibinfo {author} {\bibfnamefont
  {A.}~\bibnamefont {{Pocino}}}, \bibinfo {author} {\bibfnamefont {F.~J.}\
  \bibnamefont {{Castander}}}, \emph {et~al.},\ }\href@noop {} {\bibfield
  {journal} {\bibinfo  {journal} {arXiv e-prints}\ ,\ \bibinfo {eid}
  {arXiv:2105.13546}} (\bibinfo {year} {2021})},\ \Eprint
  {https://arxiv.org/abs/2105.13546} {arXiv:2105.13546 [astro-ph.CO]}
  \BibitemShut {NoStop}%
\bibitem [{\citenamefont {{Sugiyama}}\ \emph {et~al.}(2022)\citenamefont
  {{Sugiyama}}, \citenamefont {{Takada}}, \citenamefont {{Miyatake}},
  \citenamefont {{Nishimichi}}, \citenamefont {{Shirasaki}}, \citenamefont
  {{Kobayashi}}, \citenamefont {{Mandelbaum}}, \citenamefont {{More}},
  \citenamefont {{Takahashi}}, \citenamefont {{Osato}}, \citenamefont
  {{Oguri}}, \citenamefont {{Coupon}}, \citenamefont {{Hikage}}, \citenamefont
  {{Hsieh}}, \citenamefont {{Komiyama}} \emph {et~al.}}]{Sugiyama:2022}%
  \BibitemOpen
  \bibfield  {author} {\bibinfo {author} {\bibfnamefont {S.}~\bibnamefont
  {{Sugiyama}}}, \bibinfo {author} {\bibfnamefont {M.}~\bibnamefont
  {{Takada}}}, \bibinfo {author} {\bibfnamefont {H.}~\bibnamefont
  {{Miyatake}}}, \bibinfo {author} {\bibfnamefont {T.}~\bibnamefont
  {{Nishimichi}}}, \bibinfo {author} {\bibfnamefont {M.}~\bibnamefont
  {{Shirasaki}}}, \bibinfo {author} {\bibfnamefont {Y.}~\bibnamefont
  {{Kobayashi}}}, \bibinfo {author} {\bibfnamefont {R.}~\bibnamefont
  {{Mandelbaum}}}, \bibinfo {author} {\bibfnamefont {S.}~\bibnamefont
  {{More}}}, \bibinfo {author} {\bibfnamefont {R.}~\bibnamefont {{Takahashi}}},
  \bibinfo {author} {\bibfnamefont {K.}~\bibnamefont {{Osato}}}, \bibinfo
  {author} {\bibfnamefont {M.}~\bibnamefont {{Oguri}}}, \bibinfo {author}
  {\bibfnamefont {J.}~\bibnamefont {{Coupon}}}, \bibinfo {author}
  {\bibfnamefont {C.}~\bibnamefont {{Hikage}}}, \bibinfo {author}
  {\bibfnamefont {B.-C.}\ \bibnamefont {{Hsieh}}}, \bibinfo {author}
  {\bibfnamefont {Y.}~\bibnamefont {{Komiyama}}}, \emph {et~al.},\ }\href
  {https://doi.org/10.1103/PhysRevD.105.123537} {\bibfield  {journal} {\bibinfo
   {journal} {\prd}\ }\textbf {\bibinfo {volume} {105}},\ \bibinfo {eid}
  {123537} (\bibinfo {year} {2022})},\ \Eprint
  {https://arxiv.org/abs/2111.10966} {arXiv:2111.10966 [astro-ph.CO]}
  \BibitemShut {NoStop}%
\bibitem [{\citenamefont {{Miyatake}}\ \emph {et~al.}(2022)\citenamefont
  {{Miyatake}}, \citenamefont {{Sugiyama}}, \citenamefont {{Takada}},
  \citenamefont {{Nishimichi}}, \citenamefont {{Shirasaki}}, \citenamefont
  {{Kobayashi}}, \citenamefont {{Mandelbaum}}, \citenamefont {{More}},
  \citenamefont {{Oguri}}, \citenamefont {{Osato}}, \citenamefont {{Park}},
  \citenamefont {{Takahashi}}, \citenamefont {{Coupon}}, \citenamefont
  {{Hikage}}, \citenamefont {{Hsieh}} \emph {et~al.}}]{Miyatake:2022}%
  \BibitemOpen
  \bibfield  {author} {\bibinfo {author} {\bibfnamefont {H.}~\bibnamefont
  {{Miyatake}}}, \bibinfo {author} {\bibfnamefont {S.}~\bibnamefont
  {{Sugiyama}}}, \bibinfo {author} {\bibfnamefont {M.}~\bibnamefont
  {{Takada}}}, \bibinfo {author} {\bibfnamefont {T.}~\bibnamefont
  {{Nishimichi}}}, \bibinfo {author} {\bibfnamefont {M.}~\bibnamefont
  {{Shirasaki}}}, \bibinfo {author} {\bibfnamefont {Y.}~\bibnamefont
  {{Kobayashi}}}, \bibinfo {author} {\bibfnamefont {R.}~\bibnamefont
  {{Mandelbaum}}}, \bibinfo {author} {\bibfnamefont {S.}~\bibnamefont
  {{More}}}, \bibinfo {author} {\bibfnamefont {M.}~\bibnamefont {{Oguri}}},
  \bibinfo {author} {\bibfnamefont {K.}~\bibnamefont {{Osato}}}, \bibinfo
  {author} {\bibfnamefont {Y.}~\bibnamefont {{Park}}}, \bibinfo {author}
  {\bibfnamefont {R.}~\bibnamefont {{Takahashi}}}, \bibinfo {author}
  {\bibfnamefont {J.}~\bibnamefont {{Coupon}}}, \bibinfo {author}
  {\bibfnamefont {C.}~\bibnamefont {{Hikage}}}, \bibinfo {author}
  {\bibfnamefont {B.-C.}\ \bibnamefont {{Hsieh}}}, \emph {et~al.},\ }\href
  {https://doi.org/10.1103/PhysRevD.106.083520} {\bibfield  {journal} {\bibinfo
   {journal} {\prd}\ }\textbf {\bibinfo {volume} {106}},\ \bibinfo {eid}
  {083520} (\bibinfo {year} {2022})},\ \Eprint
  {https://arxiv.org/abs/2111.02419} {arXiv:2111.02419 [astro-ph.CO]}
  \BibitemShut {NoStop}%
\bibitem [{\citenamefont {{Amon}}\ \emph {et~al.}(2022)\citenamefont {{Amon}},
  \citenamefont {{Gruen}}, \citenamefont {{Troxel}}, \citenamefont
  {{MacCrann}}, \citenamefont {{Dodelson}}, \citenamefont {{Choi}},
  \citenamefont {{Doux}}, \citenamefont {{Secco}}, \citenamefont {{Samuroff}},
  \citenamefont {{Krause}}, \citenamefont {{Cordero}}, \citenamefont {{Myles}},
  \citenamefont {{DeRose}}, \citenamefont {{Wechsler}}, \citenamefont {{Gatti}}
  \emph {et~al.}}]{Amon:2022}%
  \BibitemOpen
  \bibfield  {author} {\bibinfo {author} {\bibfnamefont {A.}~\bibnamefont
  {{Amon}}}, \bibinfo {author} {\bibfnamefont {D.}~\bibnamefont {{Gruen}}},
  \bibinfo {author} {\bibfnamefont {M.~A.}\ \bibnamefont {{Troxel}}}, \bibinfo
  {author} {\bibfnamefont {N.}~\bibnamefont {{MacCrann}}}, \bibinfo {author}
  {\bibfnamefont {S.}~\bibnamefont {{Dodelson}}}, \bibinfo {author}
  {\bibfnamefont {A.}~\bibnamefont {{Choi}}}, \bibinfo {author} {\bibfnamefont
  {C.}~\bibnamefont {{Doux}}}, \bibinfo {author} {\bibfnamefont {L.~F.}\
  \bibnamefont {{Secco}}}, \bibinfo {author} {\bibfnamefont {S.}~\bibnamefont
  {{Samuroff}}}, \bibinfo {author} {\bibfnamefont {E.}~\bibnamefont
  {{Krause}}}, \bibinfo {author} {\bibfnamefont {J.}~\bibnamefont {{Cordero}}},
  \bibinfo {author} {\bibfnamefont {J.}~\bibnamefont {{Myles}}}, \bibinfo
  {author} {\bibfnamefont {J.}~\bibnamefont {{DeRose}}}, \bibinfo {author}
  {\bibfnamefont {R.~H.}\ \bibnamefont {{Wechsler}}}, \bibinfo {author}
  {\bibfnamefont {M.}~\bibnamefont {{Gatti}}}, \emph {et~al.},\ }\href
  {https://doi.org/10.1103/PhysRevD.105.023514} {\bibfield  {journal} {\bibinfo
   {journal} {\prd}\ }\textbf {\bibinfo {volume} {105}},\ \bibinfo {eid}
  {023514} (\bibinfo {year} {2022})},\ \Eprint
  {https://arxiv.org/abs/2105.13543} {arXiv:2105.13543 [astro-ph.CO]}
  \BibitemShut {NoStop}%
\bibitem [{\citenamefont {{DES Collaboration}}\ \emph
  {et~al.}(2021)\citenamefont {{DES Collaboration}}, \citenamefont {{Abbott}},
  \citenamefont {{Aguena}}, \citenamefont {{Alarcon}}, \citenamefont {{Allam}},
  \citenamefont {{Alves}}, \citenamefont {{Amon}}, \citenamefont
  {{Andrade-Oliveira}}, \citenamefont {{Annis}}, \citenamefont {{Avila}},
  \citenamefont {{Bacon}}, \citenamefont {{Baxter}}, \citenamefont {{Bechtol}},
  \citenamefont {{Becker}}, \citenamefont {{Bernstein}} \emph
  {et~al.}}]{DES-Y3}%
  \BibitemOpen
  \bibfield  {author} {\bibinfo {author} {\bibnamefont {{DES Collaboration}}},
  \bibinfo {author} {\bibfnamefont {T.~M.~C.}\ \bibnamefont {{Abbott}}},
  \bibinfo {author} {\bibfnamefont {M.}~\bibnamefont {{Aguena}}}, \bibinfo
  {author} {\bibfnamefont {A.}~\bibnamefont {{Alarcon}}}, \bibinfo {author}
  {\bibfnamefont {S.}~\bibnamefont {{Allam}}}, \bibinfo {author} {\bibfnamefont
  {O.}~\bibnamefont {{Alves}}}, \bibinfo {author} {\bibfnamefont
  {A.}~\bibnamefont {{Amon}}}, \bibinfo {author} {\bibfnamefont
  {F.}~\bibnamefont {{Andrade-Oliveira}}}, \bibinfo {author} {\bibfnamefont
  {J.}~\bibnamefont {{Annis}}}, \bibinfo {author} {\bibfnamefont
  {S.}~\bibnamefont {{Avila}}}, \bibinfo {author} {\bibfnamefont
  {D.}~\bibnamefont {{Bacon}}}, \bibinfo {author} {\bibfnamefont
  {E.}~\bibnamefont {{Baxter}}}, \bibinfo {author} {\bibfnamefont
  {K.}~\bibnamefont {{Bechtol}}}, \bibinfo {author} {\bibfnamefont {M.~R.}\
  \bibnamefont {{Becker}}}, \bibinfo {author} {\bibfnamefont {G.~M.}\
  \bibnamefont {{Bernstein}}}, \emph {et~al.},\ }\href@noop {} {\bibfield
  {journal} {\bibinfo  {journal} {arXiv e-prints}\ ,\ \bibinfo {eid}
  {arXiv:2105.13549}} (\bibinfo {year} {2021})},\ \Eprint
  {https://arxiv.org/abs/2105.13549} {arXiv:2105.13549 [astro-ph.CO]}
  \BibitemShut {NoStop}%
\bibitem [{\citenamefont {{Mo}}\ \emph {et~al.}(2010)\citenamefont {{Mo}},
  \citenamefont {{van den Bosch}},\ and\ \citenamefont
  {{White}}}]{MovdBWhite:2020}%
  \BibitemOpen
  \bibfield  {author} {\bibinfo {author} {\bibfnamefont {H.}~\bibnamefont
  {{Mo}}}, \bibinfo {author} {\bibfnamefont {F.~C.}\ \bibnamefont {{van den
  Bosch}}},\ and\ \bibinfo {author} {\bibfnamefont {S.}~\bibnamefont
  {{White}}},\ }\href@noop {} {\emph {\bibinfo {title} {{Galaxy Formation and
  Evolution}}}}\ (\bibinfo {year} {2010})\BibitemShut {NoStop}%
\bibitem [{\citenamefont {{Bennett}}\ \emph {et~al.}(2013)\citenamefont
  {{Bennett}}, \citenamefont {{Larson}}, \citenamefont {{Weiland}},
  \citenamefont {{Jarosik}}, \citenamefont {{Hinshaw}}, \citenamefont
  {{Odegard}}, \citenamefont {{Smith}}, \citenamefont {{Hill}}, \citenamefont
  {{Gold}}, \citenamefont {{Halpern}}, \citenamefont {{Komatsu}}, \citenamefont
  {{Nolta}}, \citenamefont {{Page}}, \citenamefont {{Spergel}}, \citenamefont
  {{Wollack}} \emph {et~al.}}]{Hinshaw:2013}%
  \BibitemOpen
  \bibfield  {author} {\bibinfo {author} {\bibfnamefont {C.~L.}\ \bibnamefont
  {{Bennett}}}, \bibinfo {author} {\bibfnamefont {D.}~\bibnamefont {{Larson}}},
  \bibinfo {author} {\bibfnamefont {J.~L.}\ \bibnamefont {{Weiland}}}, \bibinfo
  {author} {\bibfnamefont {N.}~\bibnamefont {{Jarosik}}}, \bibinfo {author}
  {\bibfnamefont {G.}~\bibnamefont {{Hinshaw}}}, \bibinfo {author}
  {\bibfnamefont {N.}~\bibnamefont {{Odegard}}}, \bibinfo {author}
  {\bibfnamefont {K.~M.}\ \bibnamefont {{Smith}}}, \bibinfo {author}
  {\bibfnamefont {R.~S.}\ \bibnamefont {{Hill}}}, \bibinfo {author}
  {\bibfnamefont {B.}~\bibnamefont {{Gold}}}, \bibinfo {author} {\bibfnamefont
  {M.}~\bibnamefont {{Halpern}}}, \bibinfo {author} {\bibfnamefont
  {E.}~\bibnamefont {{Komatsu}}}, \bibinfo {author} {\bibfnamefont {M.~R.}\
  \bibnamefont {{Nolta}}}, \bibinfo {author} {\bibfnamefont {L.}~\bibnamefont
  {{Page}}}, \bibinfo {author} {\bibfnamefont {D.~N.}\ \bibnamefont
  {{Spergel}}}, \bibinfo {author} {\bibfnamefont {E.}~\bibnamefont
  {{Wollack}}}, \emph {et~al.},\ }\href
  {https://doi.org/10.1088/0067-0049/208/2/20} {\bibfield  {journal} {\bibinfo
  {journal} {\apjs}\ }\textbf {\bibinfo {volume} {208}},\ \bibinfo {eid} {20}
  (\bibinfo {year} {2013})},\ \Eprint {https://arxiv.org/abs/1212.5225}
  {arXiv:1212.5225 [astro-ph.CO]} \BibitemShut {NoStop}%
\bibitem [{\citenamefont {{de Jong}}\ \emph {et~al.}(2013)\citenamefont {{de
  Jong}}, \citenamefont {{Verdoes Kleijn}}, \citenamefont {{Kuijken}},\ and\
  \citenamefont {{Valentijn}}}]{deJong:2013}%
  \BibitemOpen
  \bibfield  {author} {\bibinfo {author} {\bibfnamefont {J.~T.~A.}\
  \bibnamefont {{de Jong}}}, \bibinfo {author} {\bibfnamefont {G.~A.}\
  \bibnamefont {{Verdoes Kleijn}}}, \bibinfo {author} {\bibfnamefont {K.~H.}\
  \bibnamefont {{Kuijken}}},\ and\ \bibinfo {author} {\bibfnamefont {E.~A.}\
  \bibnamefont {{Valentijn}}},\ }\href
  {https://doi.org/10.1007/s10686-012-9306-1} {\bibfield  {journal} {\bibinfo
  {journal} {Experimental Astronomy}\ }\textbf {\bibinfo {volume} {35}},\
  \bibinfo {pages} {25} (\bibinfo {year} {2013})},\ \Eprint
  {https://arxiv.org/abs/1206.1254} {arXiv:1206.1254 [astro-ph.CO]}
  \BibitemShut {NoStop}%
\bibitem [{Note1()}]{Note1}%
  \BibitemOpen
  \bibinfo {note} {\protect \url
  {https://kids.strw.leidenuniv.nl/}}\BibitemShut {NoStop}%
\bibitem [{\citenamefont {{The Dark Energy Survey
  Collaboration}}(2005)}]{Frieman:2005}%
  \BibitemOpen
  \bibfield  {author} {\bibinfo {author} {\bibnamefont {{The Dark Energy Survey
  Collaboration}}},\ }\href {https://doi.org/10.48550/arXiv.astro-ph/0510346}
  {\bibfield  {journal} {\bibinfo  {journal} {arXiv e-prints}\ ,\ \bibinfo
  {eid} {astro-ph/0510346}} (\bibinfo {year} {2005})},\ \Eprint
  {https://arxiv.org/abs/astro-ph/0510346} {arXiv:astro-ph/0510346 [astro-ph]}
  \BibitemShut {NoStop}%
\bibitem [{Note2()}]{Note2}%
  \BibitemOpen
  \bibinfo {note} {\protect \url
  {https://www.darkenergysurvey.org}}\BibitemShut {NoStop}%
\bibitem [{\citenamefont {{Aihara}}\ \emph
  {et~al.}(2018{\natexlab{a}})\citenamefont {{Aihara}}, \citenamefont
  {{Arimoto}}, \citenamefont {{Armstrong}}, \citenamefont {{Arnouts}},
  \citenamefont {{Bahcall}}, \citenamefont {{Bickerton}}, \citenamefont
  {{Bosch}}, \citenamefont {{Bundy}}, \citenamefont {{Capak}}, \citenamefont
  {{Chan}}, \citenamefont {{Chiba}}, \citenamefont {{Coupon}}, \citenamefont
  {{Egami}}, \citenamefont {{Enoki}}, \citenamefont {{Finet}} \emph
  {et~al.}}]{HSCoverview:17}%
  \BibitemOpen
  \bibfield  {author} {\bibinfo {author} {\bibfnamefont {H.}~\bibnamefont
  {{Aihara}}}, \bibinfo {author} {\bibfnamefont {N.}~\bibnamefont {{Arimoto}}},
  \bibinfo {author} {\bibfnamefont {R.}~\bibnamefont {{Armstrong}}}, \bibinfo
  {author} {\bibfnamefont {S.}~\bibnamefont {{Arnouts}}}, \bibinfo {author}
  {\bibfnamefont {N.~A.}\ \bibnamefont {{Bahcall}}}, \bibinfo {author}
  {\bibfnamefont {S.}~\bibnamefont {{Bickerton}}}, \bibinfo {author}
  {\bibfnamefont {J.}~\bibnamefont {{Bosch}}}, \bibinfo {author} {\bibfnamefont
  {K.}~\bibnamefont {{Bundy}}}, \bibinfo {author} {\bibfnamefont {P.~L.}\
  \bibnamefont {{Capak}}}, \bibinfo {author} {\bibfnamefont {J.~H.~H.}\
  \bibnamefont {{Chan}}}, \bibinfo {author} {\bibfnamefont {M.}~\bibnamefont
  {{Chiba}}}, \bibinfo {author} {\bibfnamefont {J.}~\bibnamefont {{Coupon}}},
  \bibinfo {author} {\bibfnamefont {E.}~\bibnamefont {{Egami}}}, \bibinfo
  {author} {\bibfnamefont {M.}~\bibnamefont {{Enoki}}}, \bibinfo {author}
  {\bibfnamefont {F.}~\bibnamefont {{Finet}}}, \emph {et~al.},\ }\href
  {https://doi.org/10.1093/pasj/psx066} {\bibfield  {journal} {\bibinfo
  {journal} {\pasj}\ }\textbf {\bibinfo {volume} {70}},\ \bibinfo {eid} {S4}
  (\bibinfo {year} {2018}{\natexlab{a}})},\ \Eprint
  {https://arxiv.org/abs/1704.05858} {arXiv:1704.05858 [astro-ph.IM]}
  \BibitemShut {NoStop}%
\bibitem [{Note3()}]{Note3}%
  \BibitemOpen
  \bibinfo {note} {\protect \url {https://hsc.mtk.nao.ac.jp/ssp/}}\BibitemShut
  {NoStop}%
\bibitem [{Gra(2006)}]{GravitationalLensing:SaasFee}%
  \BibitemOpen
  \href@noop {} {\emph {\bibinfo {title} {Saas-Fee Advanced Course 33:
  Gravitational Lensing: Strong, Weak and Micro}}}\ (\bibinfo {year} {2006})\
  \Eprint {https://arxiv.org/abs/astro-ph/0407232} {arXiv:astro-ph/0407232
  [astro-ph]} \BibitemShut {NoStop}%
\bibitem [{\citenamefont {{Tegmark}}\ \emph {et~al.}(2004)\citenamefont
  {{Tegmark}}, \citenamefont {{Blanton}}, \citenamefont {{Strauss}},
  \citenamefont {{Hoyle}}, \citenamefont {{Schlegel}}, \citenamefont
  {{Scoccimarro}}, \citenamefont {{Vogeley}}, \citenamefont {{Weinberg}},
  \citenamefont {{Zehavi}}, \citenamefont {{Berlind}}, \citenamefont
  {{Budavari}}, \citenamefont {{Connolly}}, \citenamefont {{Eisenstein}},
  \citenamefont {{Finkbeiner}}, \citenamefont {{Frieman}} \emph
  {et~al.}}]{Tegmarketal:04}%
  \BibitemOpen
  \bibfield  {author} {\bibinfo {author} {\bibfnamefont {M.}~\bibnamefont
  {{Tegmark}}}, \bibinfo {author} {\bibfnamefont {M.~R.}\ \bibnamefont
  {{Blanton}}}, \bibinfo {author} {\bibfnamefont {M.~A.}\ \bibnamefont
  {{Strauss}}}, \bibinfo {author} {\bibfnamefont {F.}~\bibnamefont {{Hoyle}}},
  \bibinfo {author} {\bibfnamefont {D.}~\bibnamefont {{Schlegel}}}, \bibinfo
  {author} {\bibfnamefont {R.}~\bibnamefont {{Scoccimarro}}}, \bibinfo {author}
  {\bibfnamefont {M.~S.}\ \bibnamefont {{Vogeley}}}, \bibinfo {author}
  {\bibfnamefont {D.~H.}\ \bibnamefont {{Weinberg}}}, \bibinfo {author}
  {\bibfnamefont {I.}~\bibnamefont {{Zehavi}}}, \bibinfo {author}
  {\bibfnamefont {A.}~\bibnamefont {{Berlind}}}, \bibinfo {author}
  {\bibfnamefont {T.}~\bibnamefont {{Budavari}}}, \bibinfo {author}
  {\bibfnamefont {A.}~\bibnamefont {{Connolly}}}, \bibinfo {author}
  {\bibfnamefont {D.~J.}\ \bibnamefont {{Eisenstein}}}, \bibinfo {author}
  {\bibfnamefont {D.}~\bibnamefont {{Finkbeiner}}}, \bibinfo {author}
  {\bibfnamefont {J.~A.}\ \bibnamefont {{Frieman}}}, \emph {et~al.},\ }\href
  {https://doi.org/10.1086/382125} {\bibfield  {journal} {\bibinfo  {journal}
  {\apj}\ }\textbf {\bibinfo {volume} {606}},\ \bibinfo {pages} {702} (\bibinfo
  {year} {2004})},\ \Eprint {https://arxiv.org/abs/arXiv:astro-ph/0310725}
  {arXiv:astro-ph/0310725} \BibitemShut {NoStop}%
\bibitem [{\citenamefont {{Cooray}}\ and\ \citenamefont
  {{Sheth}}(2002)}]{CooraySheth:02}%
  \BibitemOpen
  \bibfield  {author} {\bibinfo {author} {\bibfnamefont {A.}~\bibnamefont
  {{Cooray}}}\ and\ \bibinfo {author} {\bibfnamefont {R.}~\bibnamefont
  {{Sheth}}},\ }\href {https://doi.org/10.1016/S0370-1573(02)00276-4}
  {\bibfield  {journal} {\bibinfo  {journal} {\physrep}\ }\textbf {\bibinfo
  {volume} {372}},\ \bibinfo {pages} {1} (\bibinfo {year} {2002})},\ \Eprint
  {https://arxiv.org/abs/arXiv:astro-ph/0206508} {arXiv:astro-ph/0206508}
  \BibitemShut {NoStop}%
\bibitem [{\citenamefont {{Kaiser}}(1984)}]{Kaiser:1984}%
  \BibitemOpen
  \bibfield  {author} {\bibinfo {author} {\bibfnamefont {N.}~\bibnamefont
  {{Kaiser}}},\ }\href {https://doi.org/10.1086/184341} {\bibfield  {journal}
  {\bibinfo  {journal} {\apjl}\ }\textbf {\bibinfo {volume} {284}},\ \bibinfo
  {pages} {L9} (\bibinfo {year} {1984})}\BibitemShut {NoStop}%
\bibitem [{\citenamefont {{Seljak}}\ \emph {et~al.}(2005)\citenamefont
  {{Seljak}}, \citenamefont {{Makarov}}, \citenamefont {{Mandelbaum}},
  \citenamefont {{Hirata}}, \citenamefont {{Padmanabhan}}, \citenamefont
  {{McDonald}}, \citenamefont {{Blanton}}, \citenamefont {{Tegmark}},
  \citenamefont {{Bahcall}},\ and\ \citenamefont {{Brinkmann}}}]{Seljak:2005}%
  \BibitemOpen
  \bibfield  {author} {\bibinfo {author} {\bibfnamefont {U.}~\bibnamefont
  {{Seljak}}}, \bibinfo {author} {\bibfnamefont {A.}~\bibnamefont {{Makarov}}},
  \bibinfo {author} {\bibfnamefont {R.}~\bibnamefont {{Mandelbaum}}}, \bibinfo
  {author} {\bibfnamefont {C.~M.}\ \bibnamefont {{Hirata}}}, \bibinfo {author}
  {\bibfnamefont {N.}~\bibnamefont {{Padmanabhan}}}, \bibinfo {author}
  {\bibfnamefont {P.}~\bibnamefont {{McDonald}}}, \bibinfo {author}
  {\bibfnamefont {M.~R.}\ \bibnamefont {{Blanton}}}, \bibinfo {author}
  {\bibfnamefont {M.}~\bibnamefont {{Tegmark}}}, \bibinfo {author}
  {\bibfnamefont {N.~A.}\ \bibnamefont {{Bahcall}}},\ and\ \bibinfo {author}
  {\bibfnamefont {J.}~\bibnamefont {{Brinkmann}}},\ }\href
  {https://doi.org/10.1103/PhysRevD.71.043511} {\bibfield  {journal} {\bibinfo
  {journal} {\prd}\ }\textbf {\bibinfo {volume} {71}},\ \bibinfo {eid} {043511}
  (\bibinfo {year} {2005})},\ \Eprint {https://arxiv.org/abs/astro-ph/0406594}
  {arXiv:astro-ph/0406594 [astro-ph]} \BibitemShut {NoStop}%
\bibitem [{\citenamefont {{Tinker}}\ \emph {et~al.}(2010)\citenamefont
  {{Tinker}}, \citenamefont {{Robertson}}, \citenamefont {{Kravtsov}},
  \citenamefont {{Klypin}}, \citenamefont {{Warren}}, \citenamefont {{Yepes}},\
  and\ \citenamefont {{Gottl{\"o}ber}}}]{Tinker:2010}%
  \BibitemOpen
  \bibfield  {author} {\bibinfo {author} {\bibfnamefont {J.~L.}\ \bibnamefont
  {{Tinker}}}, \bibinfo {author} {\bibfnamefont {B.~E.}\ \bibnamefont
  {{Robertson}}}, \bibinfo {author} {\bibfnamefont {A.~V.}\ \bibnamefont
  {{Kravtsov}}}, \bibinfo {author} {\bibfnamefont {A.}~\bibnamefont
  {{Klypin}}}, \bibinfo {author} {\bibfnamefont {M.~S.}\ \bibnamefont
  {{Warren}}}, \bibinfo {author} {\bibfnamefont {G.}~\bibnamefont {{Yepes}}},\
  and\ \bibinfo {author} {\bibfnamefont {S.}~\bibnamefont {{Gottl{\"o}ber}}},\
  }\href {https://doi.org/10.1088/0004-637X/724/2/878} {\bibfield  {journal}
  {\bibinfo  {journal} {\apj}\ }\textbf {\bibinfo {volume} {724}},\ \bibinfo
  {pages} {878} (\bibinfo {year} {2010})},\ \Eprint
  {https://arxiv.org/abs/1001.3162} {arXiv:1001.3162 [astro-ph.CO]}
  \BibitemShut {NoStop}%
\bibitem [{\citenamefont {{Cacciato}}\ \emph {et~al.}(2009)\citenamefont
  {{Cacciato}}, \citenamefont {{van den Bosch}}, \citenamefont {{More}},
  \citenamefont {{Li}}, \citenamefont {{Mo}},\ and\ \citenamefont
  {{Yang}}}]{Cacciato:2009}%
  \BibitemOpen
  \bibfield  {author} {\bibinfo {author} {\bibfnamefont {M.}~\bibnamefont
  {{Cacciato}}}, \bibinfo {author} {\bibfnamefont {F.~C.}\ \bibnamefont {{van
  den Bosch}}}, \bibinfo {author} {\bibfnamefont {S.}~\bibnamefont {{More}}},
  \bibinfo {author} {\bibfnamefont {R.}~\bibnamefont {{Li}}}, \bibinfo {author}
  {\bibfnamefont {H.~J.}\ \bibnamefont {{Mo}}},\ and\ \bibinfo {author}
  {\bibfnamefont {X.}~\bibnamefont {{Yang}}},\ }\href
  {https://doi.org/10.1111/j.1365-2966.2008.14362.x} {\bibfield  {journal}
  {\bibinfo  {journal} {\mnras}\ }\textbf {\bibinfo {volume} {394}},\ \bibinfo
  {pages} {929} (\bibinfo {year} {2009})},\ \Eprint
  {https://arxiv.org/abs/0807.4932} {arXiv:0807.4932 [astro-ph]} \BibitemShut
  {NoStop}%
\bibitem [{\citenamefont {{van den Bosch}}\ \emph {et~al.}(2013)\citenamefont
  {{van den Bosch}}, \citenamefont {{More}}, \citenamefont {{Cacciato}},
  \citenamefont {{Mo}},\ and\ \citenamefont {{Yang}}}]{vandenBosch:2013}%
  \BibitemOpen
  \bibfield  {author} {\bibinfo {author} {\bibfnamefont {F.~C.}\ \bibnamefont
  {{van den Bosch}}}, \bibinfo {author} {\bibfnamefont {S.}~\bibnamefont
  {{More}}}, \bibinfo {author} {\bibfnamefont {M.}~\bibnamefont {{Cacciato}}},
  \bibinfo {author} {\bibfnamefont {H.}~\bibnamefont {{Mo}}},\ and\ \bibinfo
  {author} {\bibfnamefont {X.}~\bibnamefont {{Yang}}},\ }\href
  {https://doi.org/10.1093/mnras/sts006} {\bibfield  {journal} {\bibinfo
  {journal} {\mnras}\ }\textbf {\bibinfo {volume} {430}},\ \bibinfo {pages}
  {725} (\bibinfo {year} {2013})},\ \Eprint {https://arxiv.org/abs/1206.6890}
  {arXiv:1206.6890 [astro-ph.CO]} \BibitemShut {NoStop}%
\bibitem [{\citenamefont {{More}}\ \emph {et~al.}(2013)\citenamefont {{More}},
  \citenamefont {{van den Bosch}}, \citenamefont {{Cacciato}}, \citenamefont
  {{More}}, \citenamefont {{Mo}},\ and\ \citenamefont {{Yang}}}]{More:2013b}%
  \BibitemOpen
  \bibfield  {author} {\bibinfo {author} {\bibfnamefont {S.}~\bibnamefont
  {{More}}}, \bibinfo {author} {\bibfnamefont {F.~C.}\ \bibnamefont {{van den
  Bosch}}}, \bibinfo {author} {\bibfnamefont {M.}~\bibnamefont {{Cacciato}}},
  \bibinfo {author} {\bibfnamefont {A.}~\bibnamefont {{More}}}, \bibinfo
  {author} {\bibfnamefont {H.}~\bibnamefont {{Mo}}},\ and\ \bibinfo {author}
  {\bibfnamefont {X.}~\bibnamefont {{Yang}}},\ }\href
  {https://doi.org/10.1093/mnras/sts697} {\bibfield  {journal} {\bibinfo
  {journal} {\mnras}\ }\textbf {\bibinfo {volume} {430}},\ \bibinfo {pages}
  {747} (\bibinfo {year} {2013})},\ \Eprint {https://arxiv.org/abs/1207.0004}
  {arXiv:1207.0004 [astro-ph.CO]} \BibitemShut {NoStop}%
\bibitem [{\citenamefont {{Cacciato}}\ \emph {et~al.}(2013)\citenamefont
  {{Cacciato}}, \citenamefont {{van den Bosch}}, \citenamefont {{More}},
  \citenamefont {{Mo}},\ and\ \citenamefont {{Yang}}}]{Cacciato:2013}%
  \BibitemOpen
  \bibfield  {author} {\bibinfo {author} {\bibfnamefont {M.}~\bibnamefont
  {{Cacciato}}}, \bibinfo {author} {\bibfnamefont {F.~C.}\ \bibnamefont {{van
  den Bosch}}}, \bibinfo {author} {\bibfnamefont {S.}~\bibnamefont {{More}}},
  \bibinfo {author} {\bibfnamefont {H.}~\bibnamefont {{Mo}}},\ and\ \bibinfo
  {author} {\bibfnamefont {X.}~\bibnamefont {{Yang}}},\ }\href
  {https://doi.org/10.1093/mnras/sts525} {\bibfield  {journal} {\bibinfo
  {journal} {\mnras}\ }\textbf {\bibinfo {volume} {430}},\ \bibinfo {pages}
  {767} (\bibinfo {year} {2013})},\ \Eprint {https://arxiv.org/abs/1207.0503}
  {arXiv:1207.0503 [astro-ph.CO]} \BibitemShut {NoStop}%
\bibitem [{\citenamefont {{More}}\ \emph {et~al.}(2015)\citenamefont {{More}},
  \citenamefont {{Miyatake}}, \citenamefont {{Mandelbaum}}, \citenamefont
  {{Takada}}, \citenamefont {{Spergel}}, \citenamefont {{Brownstein}},\ and\
  \citenamefont {{Schneider}}}]{More:2015}%
  \BibitemOpen
  \bibfield  {author} {\bibinfo {author} {\bibfnamefont {S.}~\bibnamefont
  {{More}}}, \bibinfo {author} {\bibfnamefont {H.}~\bibnamefont {{Miyatake}}},
  \bibinfo {author} {\bibfnamefont {R.}~\bibnamefont {{Mandelbaum}}}, \bibinfo
  {author} {\bibfnamefont {M.}~\bibnamefont {{Takada}}}, \bibinfo {author}
  {\bibfnamefont {D.~N.}\ \bibnamefont {{Spergel}}}, \bibinfo {author}
  {\bibfnamefont {J.~R.}\ \bibnamefont {{Brownstein}}},\ and\ \bibinfo {author}
  {\bibfnamefont {D.~P.}\ \bibnamefont {{Schneider}}},\ }\href
  {https://doi.org/10.1088/0004-637X/806/1/2} {\bibfield  {journal} {\bibinfo
  {journal} {\apj}\ }\textbf {\bibinfo {volume} {806}},\ \bibinfo {eid} {2}
  (\bibinfo {year} {2015})},\ \Eprint {https://arxiv.org/abs/1407.1856}
  {arXiv:1407.1856 [astro-ph.CO]} \BibitemShut {NoStop}%
\bibitem [{\citenamefont {{Mandelbaum}}\ \emph {et~al.}(2013)\citenamefont
  {{Mandelbaum}}, \citenamefont {{Slosar}}, \citenamefont {{Baldauf}},
  \citenamefont {{Seljak}}, \citenamefont {{Hirata}}, \citenamefont
  {{Nakajima}}, \citenamefont {{Reyes}},\ and\ \citenamefont
  {{Smith}}}]{Mandelbaumetal:13}%
  \BibitemOpen
  \bibfield  {author} {\bibinfo {author} {\bibfnamefont {R.}~\bibnamefont
  {{Mandelbaum}}}, \bibinfo {author} {\bibfnamefont {A.}~\bibnamefont
  {{Slosar}}}, \bibinfo {author} {\bibfnamefont {T.}~\bibnamefont {{Baldauf}}},
  \bibinfo {author} {\bibfnamefont {U.}~\bibnamefont {{Seljak}}}, \bibinfo
  {author} {\bibfnamefont {C.~M.}\ \bibnamefont {{Hirata}}}, \bibinfo {author}
  {\bibfnamefont {R.}~\bibnamefont {{Nakajima}}}, \bibinfo {author}
  {\bibfnamefont {R.}~\bibnamefont {{Reyes}}},\ and\ \bibinfo {author}
  {\bibfnamefont {R.~E.}\ \bibnamefont {{Smith}}},\ }\href
  {https://doi.org/10.1093/mnras/stt572} {\bibfield  {journal} {\bibinfo
  {journal} {\mnras}\ }\textbf {\bibinfo {volume} {432}},\ \bibinfo {pages}
  {1544} (\bibinfo {year} {2013})},\ \Eprint {https://arxiv.org/abs/1207.1120}
  {arXiv:1207.1120} \BibitemShut {NoStop}%
\bibitem [{\citenamefont {Miyatake}\ \emph {et~al.}(2023)\citenamefont
  {Miyatake}, \citenamefont {Sugiyama}, \citenamefont {Takada}, \citenamefont
  {Nishimichi}, \citenamefont {chong Li}, \citenamefont {Shirasaki},
  \citenamefont {More}, \citenamefont {Kobayashi}, \citenamefont {Nishizawa},
  \citenamefont {Rau}, \citenamefont {Zhang}, \citenamefont {Takahashi},
  \citenamefont {Dalal}, \citenamefont {Mandelbaum}, \citenamefont {Strauss}
  \emph {et~al.}}]{Miyatake_hscy3}%
  \BibitemOpen
  \bibfield  {author} {\bibinfo {author} {\bibfnamefont {H.}~\bibnamefont
  {Miyatake}}, \bibinfo {author} {\bibfnamefont {S.}~\bibnamefont {Sugiyama}},
  \bibinfo {author} {\bibfnamefont {M.}~\bibnamefont {Takada}}, \bibinfo
  {author} {\bibfnamefont {T.}~\bibnamefont {Nishimichi}}, \bibinfo {author}
  {\bibfnamefont {X.}~\bibnamefont {chong Li}}, \bibinfo {author}
  {\bibfnamefont {M.}~\bibnamefont {Shirasaki}}, \bibinfo {author}
  {\bibfnamefont {S.}~\bibnamefont {More}}, \bibinfo {author} {\bibfnamefont
  {Y.}~\bibnamefont {Kobayashi}}, \bibinfo {author} {\bibfnamefont {A.~J.}\
  \bibnamefont {Nishizawa}}, \bibinfo {author} {\bibfnamefont {M.~M.}\
  \bibnamefont {Rau}}, \bibinfo {author} {\bibfnamefont {T.}~\bibnamefont
  {Zhang}}, \bibinfo {author} {\bibfnamefont {R.}~\bibnamefont {Takahashi}},
  \bibinfo {author} {\bibfnamefont {R.}~\bibnamefont {Dalal}}, \bibinfo
  {author} {\bibfnamefont {R.}~\bibnamefont {Mandelbaum}}, \bibinfo {author}
  {\bibfnamefont {M.~A.}\ \bibnamefont {Strauss}}, \emph {et~al.},\ }\href@noop
  {} {\bibfield  {journal} {\bibinfo  {journal} {arXiv e-prints}\ } (\bibinfo
  {year} {2023})},\ \Eprint {https://arxiv.org/abs/2304.00704}
  {arXiv:2304.00704 [astro-ph.CO]} \BibitemShut {NoStop}%
\bibitem [{\citenamefont {Sugiyama}\ \emph {et~al.}(2023)\citenamefont
  {Sugiyama}, \citenamefont {Miyatake}, \citenamefont {More}, \citenamefont
  {Li}, \citenamefont {Shirasaki}, \citenamefont {Takada}, \citenamefont
  {Kobayashi}, \citenamefont {Takahashi}, \citenamefont {Nishimichi},
  \citenamefont {Nishizawa}, \citenamefont {Rau}, \citenamefont {Zhang},
  \citenamefont {Dalal}, \citenamefont {Mandelbaum}, \citenamefont {Strauss}
  \emph {et~al.}}]{Sugiyama_hscy3}%
  \BibitemOpen
  \bibfield  {author} {\bibinfo {author} {\bibfnamefont {S.}~\bibnamefont
  {Sugiyama}}, \bibinfo {author} {\bibfnamefont {H.}~\bibnamefont {Miyatake}},
  \bibinfo {author} {\bibfnamefont {S.}~\bibnamefont {More}}, \bibinfo {author}
  {\bibfnamefont {X.}~\bibnamefont {Li}}, \bibinfo {author} {\bibfnamefont
  {M.}~\bibnamefont {Shirasaki}}, \bibinfo {author} {\bibfnamefont
  {M.}~\bibnamefont {Takada}}, \bibinfo {author} {\bibfnamefont
  {Y.}~\bibnamefont {Kobayashi}}, \bibinfo {author} {\bibfnamefont
  {R.}~\bibnamefont {Takahashi}}, \bibinfo {author} {\bibfnamefont
  {T.}~\bibnamefont {Nishimichi}}, \bibinfo {author} {\bibfnamefont {A.~J.}\
  \bibnamefont {Nishizawa}}, \bibinfo {author} {\bibfnamefont {M.~M.}\
  \bibnamefont {Rau}}, \bibinfo {author} {\bibfnamefont {T.}~\bibnamefont
  {Zhang}}, \bibinfo {author} {\bibfnamefont {R.}~\bibnamefont {Dalal}},
  \bibinfo {author} {\bibfnamefont {R.}~\bibnamefont {Mandelbaum}}, \bibinfo
  {author} {\bibfnamefont {M.~A.}\ \bibnamefont {Strauss}}, \emph {et~al.},\
  }\href@noop {} {\bibfield  {journal} {\bibinfo  {journal} {arXiv e-prints}\ }
  (\bibinfo {year} {2023})},\ \Eprint {https://arxiv.org/abs/2304.00705}
  {arXiv:2304.00705 [astro-ph.CO]} \BibitemShut {NoStop}%
\bibitem [{\citenamefont {Li}\ \emph {et~al.}(2023)\citenamefont {Li},
  \citenamefont {Zhang}, \citenamefont {Sugiyama}, \citenamefont {Dalal},
  \citenamefont {Rau}, \citenamefont {Mandelbaum}, \citenamefont {Takada},
  \citenamefont {More}, \citenamefont {Strauss}, \citenamefont {Miyatake},
  \citenamefont {Shirasaki}, \citenamefont {Hamana}, \citenamefont {Oguri},
  \citenamefont {Luo}, \citenamefont {Nishizawa} \emph {et~al.}}]{Li_hscy3}%
  \BibitemOpen
  \bibfield  {author} {\bibinfo {author} {\bibfnamefont {X.}~\bibnamefont
  {Li}}, \bibinfo {author} {\bibfnamefont {T.}~\bibnamefont {Zhang}}, \bibinfo
  {author} {\bibfnamefont {S.}~\bibnamefont {Sugiyama}}, \bibinfo {author}
  {\bibfnamefont {R.}~\bibnamefont {Dalal}}, \bibinfo {author} {\bibfnamefont
  {M.~M.}\ \bibnamefont {Rau}}, \bibinfo {author} {\bibfnamefont
  {R.}~\bibnamefont {Mandelbaum}}, \bibinfo {author} {\bibfnamefont
  {M.}~\bibnamefont {Takada}}, \bibinfo {author} {\bibfnamefont
  {S.}~\bibnamefont {More}}, \bibinfo {author} {\bibfnamefont {M.~A.}\
  \bibnamefont {Strauss}}, \bibinfo {author} {\bibfnamefont {H.}~\bibnamefont
  {Miyatake}}, \bibinfo {author} {\bibfnamefont {M.}~\bibnamefont {Shirasaki}},
  \bibinfo {author} {\bibfnamefont {T.}~\bibnamefont {Hamana}}, \bibinfo
  {author} {\bibfnamefont {M.}~\bibnamefont {Oguri}}, \bibinfo {author}
  {\bibfnamefont {W.}~\bibnamefont {Luo}}, \bibinfo {author} {\bibfnamefont
  {A.~J.}\ \bibnamefont {Nishizawa}}, \emph {et~al.},\ }\href@noop {}
  {\bibfield  {journal} {\bibinfo  {journal} {arXiv e-prints}\ } (\bibinfo
  {year} {2023})},\ \Eprint {https://arxiv.org/abs/2304.00702}
  {arXiv:2304.00702 [astro-ph.CO]} \BibitemShut {NoStop}%
\bibitem [{\citenamefont {Dalal}\ \emph {et~al.}(2023)\citenamefont {Dalal},
  \citenamefont {Li}, \citenamefont {Nicola}, \citenamefont {Zuntz},
  \citenamefont {Strauss}, \citenamefont {Sugiyama}, \citenamefont {Zhang},
  \citenamefont {Rau}, \citenamefont {Mandelbaum}, \citenamefont {Takada},
  \citenamefont {More}, \citenamefont {Miyatake}, \citenamefont {Kannawadi},
  \citenamefont {Shirasaki}, \citenamefont {Taniguchi} \emph
  {et~al.}}]{Dalal_hscy3}%
  \BibitemOpen
  \bibfield  {author} {\bibinfo {author} {\bibfnamefont {R.}~\bibnamefont
  {Dalal}}, \bibinfo {author} {\bibfnamefont {X.}~\bibnamefont {Li}}, \bibinfo
  {author} {\bibfnamefont {A.}~\bibnamefont {Nicola}}, \bibinfo {author}
  {\bibfnamefont {J.}~\bibnamefont {Zuntz}}, \bibinfo {author} {\bibfnamefont
  {M.~A.}\ \bibnamefont {Strauss}}, \bibinfo {author} {\bibfnamefont
  {S.}~\bibnamefont {Sugiyama}}, \bibinfo {author} {\bibfnamefont
  {T.}~\bibnamefont {Zhang}}, \bibinfo {author} {\bibfnamefont {M.~M.}\
  \bibnamefont {Rau}}, \bibinfo {author} {\bibfnamefont {R.}~\bibnamefont
  {Mandelbaum}}, \bibinfo {author} {\bibfnamefont {M.}~\bibnamefont {Takada}},
  \bibinfo {author} {\bibfnamefont {S.}~\bibnamefont {More}}, \bibinfo {author}
  {\bibfnamefont {H.}~\bibnamefont {Miyatake}}, \bibinfo {author}
  {\bibfnamefont {A.}~\bibnamefont {Kannawadi}}, \bibinfo {author}
  {\bibfnamefont {M.}~\bibnamefont {Shirasaki}}, \bibinfo {author}
  {\bibfnamefont {T.}~\bibnamefont {Taniguchi}}, \emph {et~al.},\ }\href@noop
  {} {\bibfield  {journal} {\bibinfo  {journal} {arXiv e-prints}\ } (\bibinfo
  {year} {2023})},\ \Eprint {https://arxiv.org/abs/2304.00701}
  {arXiv:2304.00701 [astro-ph.CO]} \BibitemShut {NoStop}%
\bibitem [{\citenamefont {{Takahashi}}\ \emph {et~al.}(2017)\citenamefont
  {{Takahashi}}, \citenamefont {{Hamana}}, \citenamefont {{Shirasaki}},
  \citenamefont {{Namikawa}}, \citenamefont {{Nishimichi}}, \citenamefont
  {{Osato}},\ and\ \citenamefont {{Shiroyama}}}]{2017ApJ...850...24T}%
  \BibitemOpen
  \bibfield  {author} {\bibinfo {author} {\bibfnamefont {R.}~\bibnamefont
  {{Takahashi}}}, \bibinfo {author} {\bibfnamefont {T.}~\bibnamefont
  {{Hamana}}}, \bibinfo {author} {\bibfnamefont {M.}~\bibnamefont
  {{Shirasaki}}}, \bibinfo {author} {\bibfnamefont {T.}~\bibnamefont
  {{Namikawa}}}, \bibinfo {author} {\bibfnamefont {T.}~\bibnamefont
  {{Nishimichi}}}, \bibinfo {author} {\bibfnamefont {K.}~\bibnamefont
  {{Osato}}},\ and\ \bibinfo {author} {\bibfnamefont {K.}~\bibnamefont
  {{Shiroyama}}},\ }\href {https://doi.org/10.3847/1538-4357/aa943d} {\bibfield
   {journal} {\bibinfo  {journal} {\apj}\ }\textbf {\bibinfo {volume} {850}},\
  \bibinfo {eid} {24} (\bibinfo {year} {2017})},\ \Eprint
  {https://arxiv.org/abs/1706.01472} {arXiv:1706.01472 [astro-ph.CO]}
  \BibitemShut {NoStop}%
\bibitem [{\citenamefont {{Miyazaki}}\ \emph {et~al.}(2018)\citenamefont
  {{Miyazaki}}, \citenamefont {{Komiyama}}, \citenamefont {{Kawanomoto}},
  \citenamefont {{Doi}}, \citenamefont {{Furusawa}}, \citenamefont {{Hamana}},
  \citenamefont {{Hayashi}}, \citenamefont {{Ikeda}}, \citenamefont {{Kamata}},
  \citenamefont {{Karoji}}, \citenamefont {{Koike}}, \citenamefont
  {{Kurakami}}, \citenamefont {{Miyama}}, \citenamefont {{Morokuma}},
  \citenamefont {{Nakata}}, \citenamefont {{Namikawa}}, \citenamefont
  {{Nakaya}}, \citenamefont {{Nariai}}, \citenamefont {{Obuchi}}, \citenamefont
  {{Oishi}}, \citenamefont {{Okada}}, \citenamefont {{Okura}} \emph
  {et~al.}}]{2018PASJ...70S...1M}%
  \BibitemOpen
  \bibfield  {author} {\bibinfo {author} {\bibfnamefont {S.}~\bibnamefont
  {{Miyazaki}}}, \bibinfo {author} {\bibfnamefont {Y.}~\bibnamefont
  {{Komiyama}}}, \bibinfo {author} {\bibfnamefont {S.}~\bibnamefont
  {{Kawanomoto}}}, \bibinfo {author} {\bibfnamefont {Y.}~\bibnamefont {{Doi}}},
  \bibinfo {author} {\bibfnamefont {H.}~\bibnamefont {{Furusawa}}}, \bibinfo
  {author} {\bibfnamefont {T.}~\bibnamefont {{Hamana}}}, \bibinfo {author}
  {\bibfnamefont {Y.}~\bibnamefont {{Hayashi}}}, \bibinfo {author}
  {\bibfnamefont {H.}~\bibnamefont {{Ikeda}}}, \bibinfo {author} {\bibfnamefont
  {Y.}~\bibnamefont {{Kamata}}}, \bibinfo {author} {\bibfnamefont
  {H.}~\bibnamefont {{Karoji}}}, \bibinfo {author} {\bibfnamefont
  {M.}~\bibnamefont {{Koike}}}, \bibinfo {author} {\bibfnamefont
  {T.}~\bibnamefont {{Kurakami}}}, \bibinfo {author} {\bibfnamefont
  {S.}~\bibnamefont {{Miyama}}}, \bibinfo {author} {\bibfnamefont
  {T.}~\bibnamefont {{Morokuma}}}, \bibinfo {author} {\bibfnamefont
  {F.}~\bibnamefont {{Nakata}}}, \bibinfo {author} {\bibfnamefont
  {K.}~\bibnamefont {{Namikawa}}}, \bibinfo {author} {\bibfnamefont
  {H.}~\bibnamefont {{Nakaya}}}, \bibinfo {author} {\bibfnamefont
  {K.}~\bibnamefont {{Nariai}}}, \bibinfo {author} {\bibfnamefont
  {Y.}~\bibnamefont {{Obuchi}}}, \bibinfo {author} {\bibfnamefont
  {Y.}~\bibnamefont {{Oishi}}}, \bibinfo {author} {\bibfnamefont
  {N.}~\bibnamefont {{Okada}}}, \bibinfo {author} {\bibfnamefont
  {Y.}~\bibnamefont {{Okura}}}, \emph {et~al.},\ }\href
  {https://doi.org/10.1093/pasj/psx063} {\bibfield  {journal} {\bibinfo
  {journal} {\pasj}\ }\textbf {\bibinfo {volume} {70}},\ \bibinfo {eid} {S1}
  (\bibinfo {year} {2018})}\BibitemShut {NoStop}%
\bibitem [{\citenamefont {{Komiyama}}\ \emph {et~al.}(2018)\citenamefont
  {{Komiyama}}, \citenamefont {{Obuchi}}, \citenamefont {{Nakaya}},
  \citenamefont {{Kamata}}, \citenamefont {{Kawanomoto}}, \citenamefont
  {{Utsumi}}, \citenamefont {{Miyazaki}}, \citenamefont {{Uraguchi}},
  \citenamefont {{Furusawa}}, \citenamefont {{Morokuma}}, \citenamefont
  {{Uchida}}, \citenamefont {{Miyatake}}, \citenamefont {{Mineo}},
  \citenamefont {{Fujimori}}, \citenamefont {{Aihara}}, \citenamefont
  {{Karoji}}, \citenamefont {{Gunn}},\ and\ \citenamefont
  {{Wang}}}]{2018PASJ...70S...2K}%
  \BibitemOpen
  \bibfield  {author} {\bibinfo {author} {\bibfnamefont {Y.}~\bibnamefont
  {{Komiyama}}}, \bibinfo {author} {\bibfnamefont {Y.}~\bibnamefont
  {{Obuchi}}}, \bibinfo {author} {\bibfnamefont {H.}~\bibnamefont {{Nakaya}}},
  \bibinfo {author} {\bibfnamefont {Y.}~\bibnamefont {{Kamata}}}, \bibinfo
  {author} {\bibfnamefont {S.}~\bibnamefont {{Kawanomoto}}}, \bibinfo {author}
  {\bibfnamefont {Y.}~\bibnamefont {{Utsumi}}}, \bibinfo {author}
  {\bibfnamefont {S.}~\bibnamefont {{Miyazaki}}}, \bibinfo {author}
  {\bibfnamefont {F.}~\bibnamefont {{Uraguchi}}}, \bibinfo {author}
  {\bibfnamefont {H.}~\bibnamefont {{Furusawa}}}, \bibinfo {author}
  {\bibfnamefont {T.}~\bibnamefont {{Morokuma}}}, \bibinfo {author}
  {\bibfnamefont {T.}~\bibnamefont {{Uchida}}}, \bibinfo {author}
  {\bibfnamefont {H.}~\bibnamefont {{Miyatake}}}, \bibinfo {author}
  {\bibfnamefont {S.}~\bibnamefont {{Mineo}}}, \bibinfo {author} {\bibfnamefont
  {H.}~\bibnamefont {{Fujimori}}}, \bibinfo {author} {\bibfnamefont
  {H.}~\bibnamefont {{Aihara}}}, \bibinfo {author} {\bibfnamefont
  {H.}~\bibnamefont {{Karoji}}}, \bibinfo {author} {\bibfnamefont {J.~E.}\
  \bibnamefont {{Gunn}}},\ and\ \bibinfo {author} {\bibfnamefont {S.-Y.}\
  \bibnamefont {{Wang}}},\ }\href {https://doi.org/10.1093/pasj/psx069}
  {\bibfield  {journal} {\bibinfo  {journal} {\pasj}\ }\textbf {\bibinfo
  {volume} {70}},\ \bibinfo {eid} {S2} (\bibinfo {year} {2018})}\BibitemShut
  {NoStop}%
\bibitem [{\citenamefont {{Furusawa}}\ \emph {et~al.}(2018)\citenamefont
  {{Furusawa}}, \citenamefont {{Koike}}, \citenamefont {{Takata}},
  \citenamefont {{Okura}}, \citenamefont {{Miyatake}}, \citenamefont
  {{Lupton}}, \citenamefont {{Bickerton}}, \citenamefont {{Price}},
  \citenamefont {{Bosch}}, \citenamefont {{Yasuda}}, \citenamefont {{Mineo}},
  \citenamefont {{Yamada}}, \citenamefont {{Miyazaki}}, \citenamefont
  {{Nakata}}, \citenamefont {{Koshida}}, \citenamefont {{Komiyama}},
  \citenamefont {{Utsumi}}, \citenamefont {{Kawanomoto}}, \citenamefont
  {{Jeschke}}, \citenamefont {{Noumaru}}, \citenamefont {{Schubert}},
  \citenamefont {{Iwata}} \emph {et~al.}}]{2018PASJ...70S...3F}%
  \BibitemOpen
  \bibfield  {author} {\bibinfo {author} {\bibfnamefont {H.}~\bibnamefont
  {{Furusawa}}}, \bibinfo {author} {\bibfnamefont {M.}~\bibnamefont {{Koike}}},
  \bibinfo {author} {\bibfnamefont {T.}~\bibnamefont {{Takata}}}, \bibinfo
  {author} {\bibfnamefont {Y.}~\bibnamefont {{Okura}}}, \bibinfo {author}
  {\bibfnamefont {H.}~\bibnamefont {{Miyatake}}}, \bibinfo {author}
  {\bibfnamefont {R.~H.}\ \bibnamefont {{Lupton}}}, \bibinfo {author}
  {\bibfnamefont {S.}~\bibnamefont {{Bickerton}}}, \bibinfo {author}
  {\bibfnamefont {P.~A.}\ \bibnamefont {{Price}}}, \bibinfo {author}
  {\bibfnamefont {J.}~\bibnamefont {{Bosch}}}, \bibinfo {author} {\bibfnamefont
  {N.}~\bibnamefont {{Yasuda}}}, \bibinfo {author} {\bibfnamefont
  {S.}~\bibnamefont {{Mineo}}}, \bibinfo {author} {\bibfnamefont
  {Y.}~\bibnamefont {{Yamada}}}, \bibinfo {author} {\bibfnamefont
  {S.}~\bibnamefont {{Miyazaki}}}, \bibinfo {author} {\bibfnamefont
  {F.}~\bibnamefont {{Nakata}}}, \bibinfo {author} {\bibfnamefont
  {S.}~\bibnamefont {{Koshida}}}, \bibinfo {author} {\bibfnamefont
  {Y.}~\bibnamefont {{Komiyama}}}, \bibinfo {author} {\bibfnamefont
  {Y.}~\bibnamefont {{Utsumi}}}, \bibinfo {author} {\bibfnamefont
  {S.}~\bibnamefont {{Kawanomoto}}}, \bibinfo {author} {\bibfnamefont
  {E.}~\bibnamefont {{Jeschke}}}, \bibinfo {author} {\bibfnamefont
  {J.}~\bibnamefont {{Noumaru}}}, \bibinfo {author} {\bibfnamefont
  {K.}~\bibnamefont {{Schubert}}}, \bibinfo {author} {\bibfnamefont
  {I.}~\bibnamefont {{Iwata}}}, \emph {et~al.},\ }\href
  {https://doi.org/10.1093/pasj/psx079} {\bibfield  {journal} {\bibinfo
  {journal} {\pasj}\ }\textbf {\bibinfo {volume} {70}},\ \bibinfo {eid} {S3}
  (\bibinfo {year} {2018})}\BibitemShut {NoStop}%
\bibitem [{\citenamefont {{Kawanomoto}}\ \emph {et~al.}(2018)\citenamefont
  {{Kawanomoto}}, \citenamefont {{Uraguchi}}, \citenamefont {{Komiyama}},
  \citenamefont {{Miyazaki}}, \citenamefont {{Furusawa}}, \citenamefont
  {{Finet}}, \citenamefont {{Hattori}}, \citenamefont {{Wang}}, \citenamefont
  {{Yasuda}},\ and\ \citenamefont {{Suzuki}}}]{2018PASJ...70...66K}%
  \BibitemOpen
  \bibfield  {author} {\bibinfo {author} {\bibfnamefont {S.}~\bibnamefont
  {{Kawanomoto}}}, \bibinfo {author} {\bibfnamefont {F.}~\bibnamefont
  {{Uraguchi}}}, \bibinfo {author} {\bibfnamefont {Y.}~\bibnamefont
  {{Komiyama}}}, \bibinfo {author} {\bibfnamefont {S.}~\bibnamefont
  {{Miyazaki}}}, \bibinfo {author} {\bibfnamefont {H.}~\bibnamefont
  {{Furusawa}}}, \bibinfo {author} {\bibfnamefont {F.}~\bibnamefont {{Finet}}},
  \bibinfo {author} {\bibfnamefont {T.}~\bibnamefont {{Hattori}}}, \bibinfo
  {author} {\bibfnamefont {S.-Y.}\ \bibnamefont {{Wang}}}, \bibinfo {author}
  {\bibfnamefont {N.}~\bibnamefont {{Yasuda}}},\ and\ \bibinfo {author}
  {\bibfnamefont {N.}~\bibnamefont {{Suzuki}}},\ }\href
  {https://doi.org/10.1093/pasj/psy056} {\bibfield  {journal} {\bibinfo
  {journal} {\pasj}\ }\textbf {\bibinfo {volume} {70}},\ \bibinfo {eid} {66}
  (\bibinfo {year} {2018})}\BibitemShut {NoStop}%
\bibitem [{\citenamefont {{Bosch}}\ \emph {et~al.}(2018)\citenamefont
  {{Bosch}}, \citenamefont {{Armstrong}}, \citenamefont {{Bickerton}},
  \citenamefont {{Furusawa}}, \citenamefont {{Ikeda}}, \citenamefont {{Koike}},
  \citenamefont {{Lupton}}, \citenamefont {{Mineo}}, \citenamefont {{Price}},
  \citenamefont {{Takata}}, \citenamefont {{Tanaka}}, \citenamefont {{Yasuda}},
  \citenamefont {{AlSayyad}}, \citenamefont {{Becker}}, \citenamefont
  {{Coulton}} \emph {et~al.}}]{2018PASJ...70S...5B}%
  \BibitemOpen
  \bibfield  {author} {\bibinfo {author} {\bibfnamefont {J.}~\bibnamefont
  {{Bosch}}}, \bibinfo {author} {\bibfnamefont {R.}~\bibnamefont
  {{Armstrong}}}, \bibinfo {author} {\bibfnamefont {S.}~\bibnamefont
  {{Bickerton}}}, \bibinfo {author} {\bibfnamefont {H.}~\bibnamefont
  {{Furusawa}}}, \bibinfo {author} {\bibfnamefont {H.}~\bibnamefont {{Ikeda}}},
  \bibinfo {author} {\bibfnamefont {M.}~\bibnamefont {{Koike}}}, \bibinfo
  {author} {\bibfnamefont {R.}~\bibnamefont {{Lupton}}}, \bibinfo {author}
  {\bibfnamefont {S.}~\bibnamefont {{Mineo}}}, \bibinfo {author} {\bibfnamefont
  {P.}~\bibnamefont {{Price}}}, \bibinfo {author} {\bibfnamefont
  {T.}~\bibnamefont {{Takata}}}, \bibinfo {author} {\bibfnamefont
  {M.}~\bibnamefont {{Tanaka}}}, \bibinfo {author} {\bibfnamefont
  {N.}~\bibnamefont {{Yasuda}}}, \bibinfo {author} {\bibfnamefont
  {Y.}~\bibnamefont {{AlSayyad}}}, \bibinfo {author} {\bibfnamefont {A.~C.}\
  \bibnamefont {{Becker}}}, \bibinfo {author} {\bibfnamefont {W.}~\bibnamefont
  {{Coulton}}}, \emph {et~al.},\ }\href {https://doi.org/10.1093/pasj/psx080}
  {\bibfield  {journal} {\bibinfo  {journal} {\pasj}\ }\textbf {\bibinfo
  {volume} {70}},\ \bibinfo {eid} {S5} (\bibinfo {year} {2018})},\ \Eprint
  {https://arxiv.org/abs/1705.06766} {arXiv:1705.06766 [astro-ph.IM]}
  \BibitemShut {NoStop}%
\bibitem [{\citenamefont {{Aihara}}\ \emph
  {et~al.}(2018{\natexlab{b}})\citenamefont {{Aihara}}, \citenamefont
  {{Armstrong}}, \citenamefont {{Bickerton}}, \citenamefont {{Bosch}},
  \citenamefont {{Coupon}}, \citenamefont {{Furusawa}}, \citenamefont
  {{Hayashi}}, \citenamefont {{Ikeda}}, \citenamefont {{Kamata}}, \citenamefont
  {{Karoji}}, \citenamefont {{Kawanomoto}}, \citenamefont {{Koike}},
  \citenamefont {{Komiyama}}, \citenamefont {{Lang}}, \citenamefont {{Lupton}}
  \emph {et~al.}}]{HSCDR1:17}%
  \BibitemOpen
  \bibfield  {author} {\bibinfo {author} {\bibfnamefont {H.}~\bibnamefont
  {{Aihara}}}, \bibinfo {author} {\bibfnamefont {R.}~\bibnamefont
  {{Armstrong}}}, \bibinfo {author} {\bibfnamefont {S.}~\bibnamefont
  {{Bickerton}}}, \bibinfo {author} {\bibfnamefont {J.}~\bibnamefont
  {{Bosch}}}, \bibinfo {author} {\bibfnamefont {J.}~\bibnamefont {{Coupon}}},
  \bibinfo {author} {\bibfnamefont {H.}~\bibnamefont {{Furusawa}}}, \bibinfo
  {author} {\bibfnamefont {Y.}~\bibnamefont {{Hayashi}}}, \bibinfo {author}
  {\bibfnamefont {H.}~\bibnamefont {{Ikeda}}}, \bibinfo {author} {\bibfnamefont
  {Y.}~\bibnamefont {{Kamata}}}, \bibinfo {author} {\bibfnamefont
  {H.}~\bibnamefont {{Karoji}}}, \bibinfo {author} {\bibfnamefont
  {S.}~\bibnamefont {{Kawanomoto}}}, \bibinfo {author} {\bibfnamefont
  {M.}~\bibnamefont {{Koike}}}, \bibinfo {author} {\bibfnamefont
  {Y.}~\bibnamefont {{Komiyama}}}, \bibinfo {author} {\bibfnamefont
  {D.}~\bibnamefont {{Lang}}}, \bibinfo {author} {\bibfnamefont {R.~H.}\
  \bibnamefont {{Lupton}}}, \emph {et~al.},\ }\href
  {https://doi.org/10.1093/pasj/psx081} {\bibfield  {journal} {\bibinfo
  {journal} {\pasj}\ }\textbf {\bibinfo {volume} {70}},\ \bibinfo {eid} {S8}
  (\bibinfo {year} {2018}{\natexlab{b}})},\ \Eprint
  {https://arxiv.org/abs/1702.08449} {arXiv:1702.08449 [astro-ph.IM]}
  \BibitemShut {NoStop}%
\bibitem [{\citenamefont {{Aihara}}\ \emph {et~al.}(2019)\citenamefont
  {{Aihara}}, \citenamefont {{AlSayyad}}, \citenamefont {{Ando}}, \citenamefont
  {{Armstrong}}, \citenamefont {{Bosch}}, \citenamefont {{Egami}},
  \citenamefont {{Furusawa}}, \citenamefont {{Furusawa}}, \citenamefont
  {{Goulding}}, \citenamefont {{Harikane}}, \citenamefont {{Hikage}},
  \citenamefont {{Ho}}, \citenamefont {{Hsieh}}, \citenamefont {{Huang}},
  \citenamefont {{Ikeda}} \emph {et~al.}}]{HSCDR2:2019}%
  \BibitemOpen
  \bibfield  {author} {\bibinfo {author} {\bibfnamefont {H.}~\bibnamefont
  {{Aihara}}}, \bibinfo {author} {\bibfnamefont {Y.}~\bibnamefont
  {{AlSayyad}}}, \bibinfo {author} {\bibfnamefont {M.}~\bibnamefont {{Ando}}},
  \bibinfo {author} {\bibfnamefont {R.}~\bibnamefont {{Armstrong}}}, \bibinfo
  {author} {\bibfnamefont {J.}~\bibnamefont {{Bosch}}}, \bibinfo {author}
  {\bibfnamefont {E.}~\bibnamefont {{Egami}}}, \bibinfo {author} {\bibfnamefont
  {H.}~\bibnamefont {{Furusawa}}}, \bibinfo {author} {\bibfnamefont
  {J.}~\bibnamefont {{Furusawa}}}, \bibinfo {author} {\bibfnamefont
  {A.}~\bibnamefont {{Goulding}}}, \bibinfo {author} {\bibfnamefont
  {Y.}~\bibnamefont {{Harikane}}}, \bibinfo {author} {\bibfnamefont
  {C.}~\bibnamefont {{Hikage}}}, \bibinfo {author} {\bibfnamefont {P.~T.~P.}\
  \bibnamefont {{Ho}}}, \bibinfo {author} {\bibfnamefont {B.-C.}\ \bibnamefont
  {{Hsieh}}}, \bibinfo {author} {\bibfnamefont {S.}~\bibnamefont {{Huang}}},
  \bibinfo {author} {\bibfnamefont {H.}~\bibnamefont {{Ikeda}}}, \emph
  {et~al.},\ }\href {https://doi.org/10.1093/pasj/psz103} {\bibfield  {journal}
  {\bibinfo  {journal} {\pasj}\ }\textbf {\bibinfo {volume} {71}},\ \bibinfo
  {eid} {114} (\bibinfo {year} {2019})},\ \Eprint
  {https://arxiv.org/abs/1905.12221} {arXiv:1905.12221 [astro-ph.IM]}
  \BibitemShut {NoStop}%
\bibitem [{\citenamefont {{Aihara}}\ \emph {et~al.}(2022)\citenamefont
  {{Aihara}}, \citenamefont {{AlSayyad}}, \citenamefont {{Ando}}, \citenamefont
  {{Armstrong}}, \citenamefont {{Bosch}}, \citenamefont {{Egami}},
  \citenamefont {{Furusawa}}, \citenamefont {{Furusawa}}, \citenamefont
  {{Harasawa}}, \citenamefont {{Harikane}}, \citenamefont {{Hsieh}},
  \citenamefont {{Ikeda}}, \citenamefont {{Ito}}, \citenamefont {{Iwata}},
  \citenamefont {{Kodama}} \emph {et~al.}}]{HSCDR3:2022}%
  \BibitemOpen
  \bibfield  {author} {\bibinfo {author} {\bibfnamefont {H.}~\bibnamefont
  {{Aihara}}}, \bibinfo {author} {\bibfnamefont {Y.}~\bibnamefont
  {{AlSayyad}}}, \bibinfo {author} {\bibfnamefont {M.}~\bibnamefont {{Ando}}},
  \bibinfo {author} {\bibfnamefont {R.}~\bibnamefont {{Armstrong}}}, \bibinfo
  {author} {\bibfnamefont {J.}~\bibnamefont {{Bosch}}}, \bibinfo {author}
  {\bibfnamefont {E.}~\bibnamefont {{Egami}}}, \bibinfo {author} {\bibfnamefont
  {H.}~\bibnamefont {{Furusawa}}}, \bibinfo {author} {\bibfnamefont
  {J.}~\bibnamefont {{Furusawa}}}, \bibinfo {author} {\bibfnamefont
  {S.}~\bibnamefont {{Harasawa}}}, \bibinfo {author} {\bibfnamefont
  {Y.}~\bibnamefont {{Harikane}}}, \bibinfo {author} {\bibfnamefont {B.-C.}\
  \bibnamefont {{Hsieh}}}, \bibinfo {author} {\bibfnamefont {H.}~\bibnamefont
  {{Ikeda}}}, \bibinfo {author} {\bibfnamefont {K.}~\bibnamefont {{Ito}}},
  \bibinfo {author} {\bibfnamefont {I.}~\bibnamefont {{Iwata}}}, \bibinfo
  {author} {\bibfnamefont {T.}~\bibnamefont {{Kodama}}}, \emph {et~al.},\
  }\href {https://doi.org/10.1093/pasj/psab122} {\bibfield  {journal} {\bibinfo
   {journal} {\pasj}\ }\textbf {\bibinfo {volume} {74}},\ \bibinfo {pages}
  {247} (\bibinfo {year} {2022})},\ \Eprint {https://arxiv.org/abs/2108.13045}
  {arXiv:2108.13045 [astro-ph.IM]} \BibitemShut {NoStop}%
\bibitem [{\citenamefont {{Li}}\ \emph {et~al.}(2021)\citenamefont {{Li}},
  \citenamefont {{Miyatake}}, \citenamefont {{Luo}}, \citenamefont {{More}},
  \citenamefont {{Oguri}}, \citenamefont {{Hamana}}, \citenamefont
  {{Mandelbaum}}, \citenamefont {{Shirasaki}}, \citenamefont {{Takada}},
  \citenamefont {{Armstrong}}, \citenamefont {{Kannawadi}}, \citenamefont
  {{Takita}}, \citenamefont {{Miyazaki}}, \citenamefont {{Nishizawa}},
  \citenamefont {{Plazas Malag{\'o}n}} \emph {et~al.}}]{Li2021}%
  \BibitemOpen
  \bibfield  {author} {\bibinfo {author} {\bibfnamefont {X.}~\bibnamefont
  {{Li}}}, \bibinfo {author} {\bibfnamefont {H.}~\bibnamefont {{Miyatake}}},
  \bibinfo {author} {\bibfnamefont {W.}~\bibnamefont {{Luo}}}, \bibinfo
  {author} {\bibfnamefont {S.}~\bibnamefont {{More}}}, \bibinfo {author}
  {\bibfnamefont {M.}~\bibnamefont {{Oguri}}}, \bibinfo {author} {\bibfnamefont
  {T.}~\bibnamefont {{Hamana}}}, \bibinfo {author} {\bibfnamefont
  {R.}~\bibnamefont {{Mandelbaum}}}, \bibinfo {author} {\bibfnamefont
  {M.}~\bibnamefont {{Shirasaki}}}, \bibinfo {author} {\bibfnamefont
  {M.}~\bibnamefont {{Takada}}}, \bibinfo {author} {\bibfnamefont
  {R.}~\bibnamefont {{Armstrong}}}, \bibinfo {author} {\bibfnamefont
  {A.}~\bibnamefont {{Kannawadi}}}, \bibinfo {author} {\bibfnamefont
  {S.}~\bibnamefont {{Takita}}}, \bibinfo {author} {\bibfnamefont
  {S.}~\bibnamefont {{Miyazaki}}}, \bibinfo {author} {\bibfnamefont {A.~J.}\
  \bibnamefont {{Nishizawa}}}, \bibinfo {author} {\bibfnamefont {A.~A.}\
  \bibnamefont {{Plazas Malag{\'o}n}}}, \emph {et~al.},\ }\href@noop {}
  {\bibfield  {journal} {\bibinfo  {journal} {arXiv e-prints}\ ,\ \bibinfo
  {eid} {arXiv:2107.00136}} (\bibinfo {year} {2021})},\ \Eprint
  {https://arxiv.org/abs/2107.00136} {arXiv:2107.00136 [astro-ph.CO]}
  \BibitemShut {NoStop}%
\bibitem [{\citenamefont {{Mandelbaum}}\ \emph
  {et~al.}(2018{\natexlab{a}})\citenamefont {{Mandelbaum}}, \citenamefont
  {{Miyatake}}, \citenamefont {{Hamana}}, \citenamefont {{Oguri}},
  \citenamefont {{Simet}}, \citenamefont {{Armstrong}}, \citenamefont
  {{Bosch}}, \citenamefont {{Murata}}, \citenamefont {{Lanusse}}, \citenamefont
  {{Leauthaud}}, \citenamefont {{Coupon}}, \citenamefont {{More}},
  \citenamefont {{Takada}}, \citenamefont {{Miyazaki}}, \citenamefont
  {{Speagle}} \emph {et~al.}}]{HSCDR1_shear:17}%
  \BibitemOpen
  \bibfield  {author} {\bibinfo {author} {\bibfnamefont {R.}~\bibnamefont
  {{Mandelbaum}}}, \bibinfo {author} {\bibfnamefont {H.}~\bibnamefont
  {{Miyatake}}}, \bibinfo {author} {\bibfnamefont {T.}~\bibnamefont
  {{Hamana}}}, \bibinfo {author} {\bibfnamefont {M.}~\bibnamefont {{Oguri}}},
  \bibinfo {author} {\bibfnamefont {M.}~\bibnamefont {{Simet}}}, \bibinfo
  {author} {\bibfnamefont {R.}~\bibnamefont {{Armstrong}}}, \bibinfo {author}
  {\bibfnamefont {J.}~\bibnamefont {{Bosch}}}, \bibinfo {author} {\bibfnamefont
  {R.}~\bibnamefont {{Murata}}}, \bibinfo {author} {\bibfnamefont
  {F.}~\bibnamefont {{Lanusse}}}, \bibinfo {author} {\bibfnamefont
  {A.}~\bibnamefont {{Leauthaud}}}, \bibinfo {author} {\bibfnamefont
  {J.}~\bibnamefont {{Coupon}}}, \bibinfo {author} {\bibfnamefont
  {S.}~\bibnamefont {{More}}}, \bibinfo {author} {\bibfnamefont
  {M.}~\bibnamefont {{Takada}}}, \bibinfo {author} {\bibfnamefont
  {S.}~\bibnamefont {{Miyazaki}}}, \bibinfo {author} {\bibfnamefont {J.~S.}\
  \bibnamefont {{Speagle}}}, \emph {et~al.},\ }\href
  {https://doi.org/10.1093/pasj/psx130} {\bibfield  {journal} {\bibinfo
  {journal} {\pasj}\ }\textbf {\bibinfo {volume} {70}},\ \bibinfo {eid} {S25}
  (\bibinfo {year} {2018}{\natexlab{a}})},\ \Eprint
  {https://arxiv.org/abs/1705.06745} {arXiv:1705.06745 [astro-ph.CO]}
  \BibitemShut {NoStop}%
\bibitem [{\citenamefont {{Mandelbaum}}\ \emph
  {et~al.}(2018{\natexlab{b}})\citenamefont {{Mandelbaum}}, \citenamefont
  {{Lanusse}}, \citenamefont {{Leauthaud}}, \citenamefont {{Armstrong}},
  \citenamefont {{Simet}}, \citenamefont {{Miyatake}}, \citenamefont
  {{Meyers}}, \citenamefont {{Bosch}}, \citenamefont {{Murata}}, \citenamefont
  {{Miyazaki}},\ and\ \citenamefont {{Tanaka}}}]{2018MNRAS.481.3170M}%
  \BibitemOpen
  \bibfield  {author} {\bibinfo {author} {\bibfnamefont {R.}~\bibnamefont
  {{Mandelbaum}}}, \bibinfo {author} {\bibfnamefont {F.}~\bibnamefont
  {{Lanusse}}}, \bibinfo {author} {\bibfnamefont {A.}~\bibnamefont
  {{Leauthaud}}}, \bibinfo {author} {\bibfnamefont {R.}~\bibnamefont
  {{Armstrong}}}, \bibinfo {author} {\bibfnamefont {M.}~\bibnamefont
  {{Simet}}}, \bibinfo {author} {\bibfnamefont {H.}~\bibnamefont {{Miyatake}}},
  \bibinfo {author} {\bibfnamefont {J.~E.}\ \bibnamefont {{Meyers}}}, \bibinfo
  {author} {\bibfnamefont {J.}~\bibnamefont {{Bosch}}}, \bibinfo {author}
  {\bibfnamefont {R.}~\bibnamefont {{Murata}}}, \bibinfo {author}
  {\bibfnamefont {S.}~\bibnamefont {{Miyazaki}}},\ and\ \bibinfo {author}
  {\bibfnamefont {M.}~\bibnamefont {{Tanaka}}},\ }\href
  {https://doi.org/10.1093/mnras/sty2420} {\bibfield  {journal} {\bibinfo
  {journal} {\mnras}\ }\textbf {\bibinfo {volume} {481}},\ \bibinfo {pages}
  {3170} (\bibinfo {year} {2018}{\natexlab{b}})},\ \Eprint
  {https://arxiv.org/abs/1710.00885} {arXiv:1710.00885 [astro-ph.CO]}
  \BibitemShut {NoStop}%
\bibitem [{\citenamefont {{Nishizawa}}\ \emph {et~al.}(2020)\citenamefont
  {{Nishizawa}}, \citenamefont {{Hsieh}}, \citenamefont {{Tanaka}},\ and\
  \citenamefont {{Takata}}}]{2020arXiv200301511N}%
  \BibitemOpen
  \bibfield  {author} {\bibinfo {author} {\bibfnamefont {A.~J.}\ \bibnamefont
  {{Nishizawa}}}, \bibinfo {author} {\bibfnamefont {B.-C.}\ \bibnamefont
  {{Hsieh}}}, \bibinfo {author} {\bibfnamefont {M.}~\bibnamefont {{Tanaka}}},\
  and\ \bibinfo {author} {\bibfnamefont {T.}~\bibnamefont {{Takata}}},\
  }\href@noop {} {\bibfield  {journal} {\bibinfo  {journal} {arXiv e-prints}\
  ,\ \bibinfo {eid} {arXiv:2003.01511}} (\bibinfo {year} {2020})},\ \Eprint
  {https://arxiv.org/abs/2003.01511} {arXiv:2003.01511 [astro-ph.GA]}
  \BibitemShut {NoStop}%
\bibitem [{Note4()}]{Note4}%
  \BibitemOpen
  \bibinfo {note} {However, the photo-$z$ catalog released for PDR2 is
  different from the catalog we use in this analysis. We will make the shape
  catalog, with the photo-$z$ information, publicly available after the
  cosmology papers are published.}\BibitemShut {Stop}%
\bibitem [{Note5()}]{Note5}%
  \BibitemOpen
  \bibinfo {note} {Self-organizing map is yet another way of characterizing the
  photometric redshift distribution similar to {\protect \sc DEmPz}~/{\protect
  \sc DNNz}~~methods. The differences between these methods lie in the machine
  learning architecture used for the calibration and inference. Please see
  \protect \citet {Rau:2023} for detailed discussion about the differences in
  the calibration strategies adopted in the current surveys.}\BibitemShut
  {Stop}%
\bibitem [{\citenamefont {{Oguri}}\ and\ \citenamefont
  {{Takada}}(2011)}]{OguriTakada:11}%
  \BibitemOpen
  \bibfield  {author} {\bibinfo {author} {\bibfnamefont {M.}~\bibnamefont
  {{Oguri}}}\ and\ \bibinfo {author} {\bibfnamefont {M.}~\bibnamefont
  {{Takada}}},\ }\href {https://doi.org/10.1103/PhysRevD.83.023008} {\bibfield
  {journal} {\bibinfo  {journal} {\prd}\ }\textbf {\bibinfo {volume} {83}},\
  \bibinfo {pages} {023008} (\bibinfo {year} {2011})},\ \Eprint
  {https://arxiv.org/abs/1010.0744} {arXiv:1010.0744 [astro-ph.CO]}
  \BibitemShut {NoStop}%
\bibitem [{Note6()}]{Note6}%
  \BibitemOpen
  \bibinfo {note} {Although a cut that retains only 24 percent of source
  galaxies is quite severe, in these initial round of cosmological analyses, we
  stick to a single source bin at redshifts larger than any of our lens samples
  to conservatively constrain any systematics in the redshift distribution of
  the source galaxies and avoid any source galaxies that may be physically
  correlated with our source galaxies.}\BibitemShut {Stop}%
\bibitem [{\citenamefont {{Rau}}\ \emph {et~al.}(2022)\citenamefont {{Rau}},
  \citenamefont {{Dalal}}, \citenamefont {{Zhang}}, \citenamefont {{Li}},
  \citenamefont {{Nishizawa}}, \citenamefont {{More}}, \citenamefont
  {{Mandelbaum}}, \citenamefont {{Strauss}},\ and\ \citenamefont
  {{Takada}}}]{Rau:2023}%
  \BibitemOpen
  \bibfield  {author} {\bibinfo {author} {\bibfnamefont {M.~M.}\ \bibnamefont
  {{Rau}}}, \bibinfo {author} {\bibfnamefont {R.}~\bibnamefont {{Dalal}}},
  \bibinfo {author} {\bibfnamefont {T.}~\bibnamefont {{Zhang}}}, \bibinfo
  {author} {\bibfnamefont {X.}~\bibnamefont {{Li}}}, \bibinfo {author}
  {\bibfnamefont {A.~J.}\ \bibnamefont {{Nishizawa}}}, \bibinfo {author}
  {\bibfnamefont {S.}~\bibnamefont {{More}}}, \bibinfo {author} {\bibfnamefont
  {R.}~\bibnamefont {{Mandelbaum}}}, \bibinfo {author} {\bibfnamefont {M.~A.}\
  \bibnamefont {{Strauss}}},\ and\ \bibinfo {author} {\bibfnamefont
  {M.}~\bibnamefont {{Takada}}},\ }\href
  {https://doi.org/10.48550/arXiv.2211.16516} {\bibfield  {journal} {\bibinfo
  {journal} {arXiv e-prints}\ ,\ \bibinfo {eid} {arXiv:2211.16516}} (\bibinfo
  {year} {2022})},\ \Eprint {https://arxiv.org/abs/2211.16516}
  {arXiv:2211.16516 [astro-ph.CO]} \BibitemShut {NoStop}%
\bibitem [{\citenamefont {{Oguri}}(2014)}]{2014MNRAS.444..147O}%
  \BibitemOpen
  \bibfield  {author} {\bibinfo {author} {\bibfnamefont {M.}~\bibnamefont
  {{Oguri}}},\ }\href {https://doi.org/10.1093/mnras/stu1446} {\bibfield
  {journal} {\bibinfo  {journal} {\mnras}\ }\textbf {\bibinfo {volume} {444}},\
  \bibinfo {pages} {147} (\bibinfo {year} {2014})},\ \Eprint
  {https://arxiv.org/abs/1407.4693} {arXiv:1407.4693 [astro-ph.CO]}
  \BibitemShut {NoStop}%
\bibitem [{\citenamefont {{Oguri}}\ \emph {et~al.}(2018)\citenamefont
  {{Oguri}}, \citenamefont {{Lin}}, \citenamefont {{Lin}}, \citenamefont
  {{Nishizawa}}, \citenamefont {{More}}, \citenamefont {{More}}, \citenamefont
  {{Hsieh}}, \citenamefont {{Medezinski}}, \citenamefont {{Miyatake}},
  \citenamefont {{Jian}}, \citenamefont {{Lin}}, \citenamefont {{Takada}},
  \citenamefont {{Okabe}}, \citenamefont {{Speagle}}, \citenamefont {{Coupon}},
  \citenamefont {{Leauthaud}}, \citenamefont {{Lupton}}, \citenamefont
  {{Miyazaki}}, \citenamefont {{Price}}, \citenamefont {{Tanaka}},
  \citenamefont {{Chiu}}, \citenamefont {{Komiyama}} \emph
  {et~al.}}]{2018PASJ...70S..20O}%
  \BibitemOpen
  \bibfield  {author} {\bibinfo {author} {\bibfnamefont {M.}~\bibnamefont
  {{Oguri}}}, \bibinfo {author} {\bibfnamefont {Y.-T.}\ \bibnamefont {{Lin}}},
  \bibinfo {author} {\bibfnamefont {S.-C.}\ \bibnamefont {{Lin}}}, \bibinfo
  {author} {\bibfnamefont {A.~J.}\ \bibnamefont {{Nishizawa}}}, \bibinfo
  {author} {\bibfnamefont {A.}~\bibnamefont {{More}}}, \bibinfo {author}
  {\bibfnamefont {S.}~\bibnamefont {{More}}}, \bibinfo {author} {\bibfnamefont
  {B.-C.}\ \bibnamefont {{Hsieh}}}, \bibinfo {author} {\bibfnamefont
  {E.}~\bibnamefont {{Medezinski}}}, \bibinfo {author} {\bibfnamefont
  {H.}~\bibnamefont {{Miyatake}}}, \bibinfo {author} {\bibfnamefont {H.-Y.}\
  \bibnamefont {{Jian}}}, \bibinfo {author} {\bibfnamefont {L.}~\bibnamefont
  {{Lin}}}, \bibinfo {author} {\bibfnamefont {M.}~\bibnamefont {{Takada}}},
  \bibinfo {author} {\bibfnamefont {N.}~\bibnamefont {{Okabe}}}, \bibinfo
  {author} {\bibfnamefont {J.~S.}\ \bibnamefont {{Speagle}}}, \bibinfo {author}
  {\bibfnamefont {J.}~\bibnamefont {{Coupon}}}, \bibinfo {author}
  {\bibfnamefont {A.}~\bibnamefont {{Leauthaud}}}, \bibinfo {author}
  {\bibfnamefont {R.~H.}\ \bibnamefont {{Lupton}}}, \bibinfo {author}
  {\bibfnamefont {S.}~\bibnamefont {{Miyazaki}}}, \bibinfo {author}
  {\bibfnamefont {P.~A.}\ \bibnamefont {{Price}}}, \bibinfo {author}
  {\bibfnamefont {M.}~\bibnamefont {{Tanaka}}}, \bibinfo {author}
  {\bibfnamefont {I.~N.}\ \bibnamefont {{Chiu}}}, \bibinfo {author}
  {\bibfnamefont {Y.}~\bibnamefont {{Komiyama}}}, \emph {et~al.},\ }\href
  {https://doi.org/10.1093/pasj/psx042} {\bibfield  {journal} {\bibinfo
  {journal} {\pasj}\ }\textbf {\bibinfo {volume} {70}},\ \bibinfo {eid} {S20}
  (\bibinfo {year} {2018})},\ \Eprint {https://arxiv.org/abs/1701.00818}
  {arXiv:1701.00818 [astro-ph.CO]} \BibitemShut {NoStop}%
\bibitem [{\citenamefont {{Hikage}}\ \emph
  {et~al.}(2019{\natexlab{b}})\citenamefont {{Hikage}}, \citenamefont
  {{Oguri}}, \citenamefont {{Hamana}}, \citenamefont {{More}}, \citenamefont
  {{Mandelbaum}}, \citenamefont {{Takada}}, \citenamefont {{K{\"o}hlinger}},
  \citenamefont {{Miyatake}}, \citenamefont {{Nishizawa}}, \citenamefont
  {{Aihara}}, \citenamefont {{Armstrong}}, \citenamefont {{Bosch}},
  \citenamefont {{Coupon}}, \citenamefont {{Ducout}}, \citenamefont {{Ho}},
  \citenamefont {{Hsieh}}, \citenamefont {{Komiyama}}, \citenamefont
  {{Lanusse}}, \citenamefont {{Leauthaud}}, \citenamefont {{Lupton}},
  \citenamefont {{Medezinski}}, \citenamefont {{Mineo}}, \citenamefont
  {{Miyama}} \emph {et~al.}}]{2019PASJ...71...43H}%
  \BibitemOpen
  \bibfield  {author} {\bibinfo {author} {\bibfnamefont {C.}~\bibnamefont
  {{Hikage}}}, \bibinfo {author} {\bibfnamefont {M.}~\bibnamefont {{Oguri}}},
  \bibinfo {author} {\bibfnamefont {T.}~\bibnamefont {{Hamana}}}, \bibinfo
  {author} {\bibfnamefont {S.}~\bibnamefont {{More}}}, \bibinfo {author}
  {\bibfnamefont {R.}~\bibnamefont {{Mandelbaum}}}, \bibinfo {author}
  {\bibfnamefont {M.}~\bibnamefont {{Takada}}}, \bibinfo {author}
  {\bibfnamefont {F.}~\bibnamefont {{K{\"o}hlinger}}}, \bibinfo {author}
  {\bibfnamefont {H.}~\bibnamefont {{Miyatake}}}, \bibinfo {author}
  {\bibfnamefont {A.~J.}\ \bibnamefont {{Nishizawa}}}, \bibinfo {author}
  {\bibfnamefont {H.}~\bibnamefont {{Aihara}}}, \bibinfo {author}
  {\bibfnamefont {R.}~\bibnamefont {{Armstrong}}}, \bibinfo {author}
  {\bibfnamefont {J.}~\bibnamefont {{Bosch}}}, \bibinfo {author} {\bibfnamefont
  {J.}~\bibnamefont {{Coupon}}}, \bibinfo {author} {\bibfnamefont
  {A.}~\bibnamefont {{Ducout}}}, \bibinfo {author} {\bibfnamefont
  {P.}~\bibnamefont {{Ho}}}, \bibinfo {author} {\bibfnamefont {B.-C.}\
  \bibnamefont {{Hsieh}}}, \bibinfo {author} {\bibfnamefont {Y.}~\bibnamefont
  {{Komiyama}}}, \bibinfo {author} {\bibfnamefont {F.}~\bibnamefont
  {{Lanusse}}}, \bibinfo {author} {\bibfnamefont {A.}~\bibnamefont
  {{Leauthaud}}}, \bibinfo {author} {\bibfnamefont {R.~H.}\ \bibnamefont
  {{Lupton}}}, \bibinfo {author} {\bibfnamefont {E.}~\bibnamefont
  {{Medezinski}}}, \bibinfo {author} {\bibfnamefont {S.}~\bibnamefont
  {{Mineo}}}, \bibinfo {author} {\bibfnamefont {S.}~\bibnamefont {{Miyama}}},
  \emph {et~al.},\ }\href {https://doi.org/10.1093/pasj/psz010} {\bibfield
  {journal} {\bibinfo  {journal} {\pasj}\ }\textbf {\bibinfo {volume} {71}},\
  \bibinfo {eid} {43} (\bibinfo {year} {2019}{\natexlab{b}})},\ \Eprint
  {https://arxiv.org/abs/1809.09148} {arXiv:1809.09148 [astro-ph.CO]}
  \BibitemShut {NoStop}%
\bibitem [{Note7()}]{Note7}%
  \BibitemOpen
  \bibinfo {note} {\protect \url {https://www.sdss.org/dr11/}}\BibitemShut
  {NoStop}%
\bibitem [{\citenamefont {{Alam}}\ \emph {et~al.}(2015)\citenamefont {{Alam}},
  \citenamefont {{Albareti}}, \citenamefont {{Allende Prieto}}, \citenamefont
  {{Anders}}, \citenamefont {{Anderson}}, \citenamefont {{Anderton}},
  \citenamefont {{Andrews}}, \citenamefont {{Armengaud}}, \citenamefont
  {{Aubourg}}, \citenamefont {{Bailey}}, \citenamefont {{Basu}}, \citenamefont
  {{Bautista}}, \citenamefont {{Beaton}}, \citenamefont {{Beers}},
  \citenamefont {{Bender}} \emph {et~al.}}]{Alam:2015}%
  \BibitemOpen
  \bibfield  {author} {\bibinfo {author} {\bibfnamefont {S.}~\bibnamefont
  {{Alam}}}, \bibinfo {author} {\bibfnamefont {F.~D.}\ \bibnamefont
  {{Albareti}}}, \bibinfo {author} {\bibfnamefont {C.}~\bibnamefont {{Allende
  Prieto}}}, \bibinfo {author} {\bibfnamefont {F.}~\bibnamefont {{Anders}}},
  \bibinfo {author} {\bibfnamefont {S.~F.}\ \bibnamefont {{Anderson}}},
  \bibinfo {author} {\bibfnamefont {T.}~\bibnamefont {{Anderton}}}, \bibinfo
  {author} {\bibfnamefont {B.~H.}\ \bibnamefont {{Andrews}}}, \bibinfo {author}
  {\bibfnamefont {E.}~\bibnamefont {{Armengaud}}}, \bibinfo {author}
  {\bibfnamefont {{\'E}.}~\bibnamefont {{Aubourg}}}, \bibinfo {author}
  {\bibfnamefont {S.}~\bibnamefont {{Bailey}}}, \bibinfo {author}
  {\bibfnamefont {S.}~\bibnamefont {{Basu}}}, \bibinfo {author} {\bibfnamefont
  {J.~E.}\ \bibnamefont {{Bautista}}}, \bibinfo {author} {\bibfnamefont
  {R.~L.}\ \bibnamefont {{Beaton}}}, \bibinfo {author} {\bibfnamefont {T.~C.}\
  \bibnamefont {{Beers}}}, \bibinfo {author} {\bibfnamefont {C.~F.}\
  \bibnamefont {{Bender}}}, \emph {et~al.},\ }\href
  {https://doi.org/10.1088/0067-0049/219/1/12} {\bibfield  {journal} {\bibinfo
  {journal} {\apjs}\ }\textbf {\bibinfo {volume} {219}},\ \bibinfo {eid} {12}
  (\bibinfo {year} {2015})},\ \Eprint {https://arxiv.org/abs/1501.00963}
  {arXiv:1501.00963 [astro-ph.IM]} \BibitemShut {NoStop}%
\bibitem [{\citenamefont {{Dawson}}\ \emph {et~al.}(2013)\citenamefont
  {{Dawson}}, \citenamefont {{Schlegel}}, \citenamefont {{Ahn}}, \citenamefont
  {{Anderson}}, \citenamefont {{Aubourg}}, \citenamefont {{Bailey}},
  \citenamefont {{Barkhouser}}, \citenamefont {{Bautista}}, \citenamefont
  {{Beifiori}}, \citenamefont {{Berlind}}, \citenamefont {{Bhardwaj}},
  \citenamefont {{Bizyaev}}, \citenamefont {{Blake}}, \citenamefont
  {{Blanton}}, \citenamefont {{Blomqvist}}, \citenamefont {{Bolton}},
  \citenamefont {{Borde}}, \citenamefont {{Bovy}}, \citenamefont {{Brandt}},
  \citenamefont {{Brewington}}, \citenamefont {{Brinkmann}} \emph
  {et~al.}}]{2013AJ....145...10D}%
  \BibitemOpen
  \bibfield  {author} {\bibinfo {author} {\bibfnamefont {K.~S.}\ \bibnamefont
  {{Dawson}}}, \bibinfo {author} {\bibfnamefont {D.~J.}\ \bibnamefont
  {{Schlegel}}}, \bibinfo {author} {\bibfnamefont {C.~P.}\ \bibnamefont
  {{Ahn}}}, \bibinfo {author} {\bibfnamefont {S.~F.}\ \bibnamefont
  {{Anderson}}}, \bibinfo {author} {\bibfnamefont {{\'E}.}~\bibnamefont
  {{Aubourg}}}, \bibinfo {author} {\bibfnamefont {S.}~\bibnamefont {{Bailey}}},
  \bibinfo {author} {\bibfnamefont {R.~H.}\ \bibnamefont {{Barkhouser}}},
  \bibinfo {author} {\bibfnamefont {J.~E.}\ \bibnamefont {{Bautista}}},
  \bibinfo {author} {\bibfnamefont {A.~r.}\ \bibnamefont {{Beifiori}}},
  \bibinfo {author} {\bibfnamefont {A.~A.}\ \bibnamefont {{Berlind}}}, \bibinfo
  {author} {\bibfnamefont {V.}~\bibnamefont {{Bhardwaj}}}, \bibinfo {author}
  {\bibfnamefont {D.}~\bibnamefont {{Bizyaev}}}, \bibinfo {author}
  {\bibfnamefont {C.~H.}\ \bibnamefont {{Blake}}}, \bibinfo {author}
  {\bibfnamefont {M.~R.}\ \bibnamefont {{Blanton}}}, \bibinfo {author}
  {\bibfnamefont {M.}~\bibnamefont {{Blomqvist}}}, \bibinfo {author}
  {\bibfnamefont {A.~S.}\ \bibnamefont {{Bolton}}}, \bibinfo {author}
  {\bibfnamefont {A.}~\bibnamefont {{Borde}}}, \bibinfo {author} {\bibfnamefont
  {J.}~\bibnamefont {{Bovy}}}, \bibinfo {author} {\bibfnamefont {W.~N.}\
  \bibnamefont {{Brandt}}}, \bibinfo {author} {\bibfnamefont {H.}~\bibnamefont
  {{Brewington}}}, \bibinfo {author} {\bibfnamefont {J.}~\bibnamefont
  {{Brinkmann}}}, \emph {et~al.},\ }\href
  {https://doi.org/10.1088/0004-6256/145/1/10} {\bibfield  {journal} {\bibinfo
  {journal} {\aj}\ }\textbf {\bibinfo {volume} {145}},\ \bibinfo {eid} {10}
  (\bibinfo {year} {2013})},\ \Eprint {https://arxiv.org/abs/1208.0022}
  {arXiv:1208.0022 [astro-ph.CO]} \BibitemShut {NoStop}%
\bibitem [{\citenamefont {{Miyatake}}\ \emph {et~al.}(2020)\citenamefont
  {{Miyatake}}, \citenamefont {{Kobayashi}}, \citenamefont {{Takada}},
  \citenamefont {{Nishimichi}}, \citenamefont {{Shirasaki}}, \citenamefont
  {{Sugiyama}}, \citenamefont {{Takahashi}}, \citenamefont {{Osato}},
  \citenamefont {{More}},\ and\ \citenamefont {{Park}}}]{2021arXiv210100113M}%
  \BibitemOpen
  \bibfield  {author} {\bibinfo {author} {\bibfnamefont {H.}~\bibnamefont
  {{Miyatake}}}, \bibinfo {author} {\bibfnamefont {Y.}~\bibnamefont
  {{Kobayashi}}}, \bibinfo {author} {\bibfnamefont {M.}~\bibnamefont
  {{Takada}}}, \bibinfo {author} {\bibfnamefont {T.}~\bibnamefont
  {{Nishimichi}}}, \bibinfo {author} {\bibfnamefont {M.}~\bibnamefont
  {{Shirasaki}}}, \bibinfo {author} {\bibfnamefont {S.}~\bibnamefont
  {{Sugiyama}}}, \bibinfo {author} {\bibfnamefont {R.}~\bibnamefont
  {{Takahashi}}}, \bibinfo {author} {\bibfnamefont {K.}~\bibnamefont
  {{Osato}}}, \bibinfo {author} {\bibfnamefont {S.}~\bibnamefont {{More}}},\
  and\ \bibinfo {author} {\bibfnamefont {Y.}~\bibnamefont {{Park}}},\
  }\href@noop {} {\bibfield  {journal} {\bibinfo  {journal} {arXiv e-prints}\
  ,\ \bibinfo {eid} {arXiv:2101.00113}} (\bibinfo {year} {2020})},\ \Eprint
  {https://arxiv.org/abs/2101.00113} {arXiv:2101.00113 [astro-ph.CO]}
  \BibitemShut {NoStop}%
\bibitem [{\citenamefont {{Sugiyama}}\ \emph {et~al.}(2021)\citenamefont
  {{Sugiyama}}, \citenamefont {{Takada}}, \citenamefont {{Miyatake}},
  \citenamefont {{Nishimichi}}, \citenamefont {{Shirasaki}}, \citenamefont
  {{Kobayashi}}, \citenamefont {{More}}, \citenamefont {{Takahashi}},
  \citenamefont {{Osato}}, \citenamefont {{Oguri}}, \citenamefont {{Coupon}},
  \citenamefont {{Hikage}}, \citenamefont {{Hsieh}}, \citenamefont
  {{Komiyama}}, \citenamefont {{Leauthaud}} \emph {et~al.}}]{Sugiyama:2021}%
  \BibitemOpen
  \bibfield  {author} {\bibinfo {author} {\bibfnamefont {S.}~\bibnamefont
  {{Sugiyama}}}, \bibinfo {author} {\bibfnamefont {M.}~\bibnamefont
  {{Takada}}}, \bibinfo {author} {\bibfnamefont {H.}~\bibnamefont
  {{Miyatake}}}, \bibinfo {author} {\bibfnamefont {T.}~\bibnamefont
  {{Nishimichi}}}, \bibinfo {author} {\bibfnamefont {M.}~\bibnamefont
  {{Shirasaki}}}, \bibinfo {author} {\bibfnamefont {Y.}~\bibnamefont
  {{Kobayashi}}}, \bibinfo {author} {\bibfnamefont {S.}~\bibnamefont {{More}}},
  \bibinfo {author} {\bibfnamefont {R.}~\bibnamefont {{Takahashi}}}, \bibinfo
  {author} {\bibfnamefont {K.}~\bibnamefont {{Osato}}}, \bibinfo {author}
  {\bibfnamefont {M.}~\bibnamefont {{Oguri}}}, \bibinfo {author} {\bibfnamefont
  {J.}~\bibnamefont {{Coupon}}}, \bibinfo {author} {\bibfnamefont
  {C.}~\bibnamefont {{Hikage}}}, \bibinfo {author} {\bibfnamefont {B.-C.}\
  \bibnamefont {{Hsieh}}}, \bibinfo {author} {\bibfnamefont {Y.}~\bibnamefont
  {{Komiyama}}}, \bibinfo {author} {\bibfnamefont {A.}~\bibnamefont
  {{Leauthaud}}}, \emph {et~al.},\ }\href@noop {} {\bibfield  {journal}
  {\bibinfo  {journal} {arXiv e-prints}\ ,\ \bibinfo {eid} {arXiv:2111.10966}}
  (\bibinfo {year} {2021})},\ \Eprint {https://arxiv.org/abs/2111.10966}
  {arXiv:2111.10966 [astro-ph.CO]} \BibitemShut {NoStop}%
\bibitem [{\citenamefont {{Miyatake}}\ \emph {et~al.}(2015)\citenamefont
  {{Miyatake}}, \citenamefont {{More}}, \citenamefont {{Mandelbaum}},
  \citenamefont {{Takada}}, \citenamefont {{Spergel}}, \citenamefont {{Kneib}},
  \citenamefont {{Schneider}}, \citenamefont {{Brinkmann}},\ and\ \citenamefont
  {{Brownstein}}}]{Miyatakeetal:15}%
  \BibitemOpen
  \bibfield  {author} {\bibinfo {author} {\bibfnamefont {H.}~\bibnamefont
  {{Miyatake}}}, \bibinfo {author} {\bibfnamefont {S.}~\bibnamefont {{More}}},
  \bibinfo {author} {\bibfnamefont {R.}~\bibnamefont {{Mandelbaum}}}, \bibinfo
  {author} {\bibfnamefont {M.}~\bibnamefont {{Takada}}}, \bibinfo {author}
  {\bibfnamefont {D.~N.}\ \bibnamefont {{Spergel}}}, \bibinfo {author}
  {\bibfnamefont {J.-P.}\ \bibnamefont {{Kneib}}}, \bibinfo {author}
  {\bibfnamefont {D.~P.}\ \bibnamefont {{Schneider}}}, \bibinfo {author}
  {\bibfnamefont {J.}~\bibnamefont {{Brinkmann}}},\ and\ \bibinfo {author}
  {\bibfnamefont {J.~R.}\ \bibnamefont {{Brownstein}}},\ }\href
  {https://doi.org/10.1088/0004-637X/806/1/1} {\bibfield  {journal} {\bibinfo
  {journal} {\apj}\ }\textbf {\bibinfo {volume} {806}},\ \bibinfo {eid} {1}
  (\bibinfo {year} {2015})},\ \Eprint {https://arxiv.org/abs/1311.1480}
  {arXiv:1311.1480} \BibitemShut {NoStop}%
\bibitem [{\citenamefont {{Abazajian}}\ \emph {et~al.}(2009)\citenamefont
  {{Abazajian}}, \citenamefont {{Adelman-McCarthy}}, \citenamefont
  {{Ag{\"u}eros}}, \citenamefont {{Allam}}, \citenamefont {{Allende Prieto}},
  \citenamefont {{An}}, \citenamefont {{Anderson}}, \citenamefont {{Anderson}},
  \citenamefont {{Annis}}, \citenamefont {{Bahcall}}, \citenamefont
  {{Bailer-Jones}}, \citenamefont {{Barentine}}, \citenamefont {{Bassett}},
  \citenamefont {{Becker}}, \citenamefont {{Beers}} \emph
  {et~al.}}]{2009ApJS..182..543A}%
  \BibitemOpen
  \bibfield  {author} {\bibinfo {author} {\bibfnamefont {K.~N.}\ \bibnamefont
  {{Abazajian}}}, \bibinfo {author} {\bibfnamefont {J.~K.}\ \bibnamefont
  {{Adelman-McCarthy}}}, \bibinfo {author} {\bibfnamefont {M.~A.}\ \bibnamefont
  {{Ag{\"u}eros}}}, \bibinfo {author} {\bibfnamefont {S.~S.}\ \bibnamefont
  {{Allam}}}, \bibinfo {author} {\bibfnamefont {C.}~\bibnamefont {{Allende
  Prieto}}}, \bibinfo {author} {\bibfnamefont {D.}~\bibnamefont {{An}}},
  \bibinfo {author} {\bibfnamefont {K.~S.~J.}\ \bibnamefont {{Anderson}}},
  \bibinfo {author} {\bibfnamefont {S.~F.}\ \bibnamefont {{Anderson}}},
  \bibinfo {author} {\bibfnamefont {J.}~\bibnamefont {{Annis}}}, \bibinfo
  {author} {\bibfnamefont {N.~A.}\ \bibnamefont {{Bahcall}}}, \bibinfo {author}
  {\bibfnamefont {C.~A.~L.}\ \bibnamefont {{Bailer-Jones}}}, \bibinfo {author}
  {\bibfnamefont {J.~C.}\ \bibnamefont {{Barentine}}}, \bibinfo {author}
  {\bibfnamefont {B.~A.}\ \bibnamefont {{Bassett}}}, \bibinfo {author}
  {\bibfnamefont {A.~C.}\ \bibnamefont {{Becker}}}, \bibinfo {author}
  {\bibfnamefont {T.~C.}\ \bibnamefont {{Beers}}}, \emph {et~al.},\ }\href
  {https://doi.org/10.1088/0067-0049/182/2/543} {\bibfield  {journal} {\bibinfo
   {journal} {\apjs}\ }\textbf {\bibinfo {volume} {182}},\ \bibinfo {pages}
  {543} (\bibinfo {year} {2009})},\ \Eprint {https://arxiv.org/abs/0812.0649}
  {arXiv:0812.0649 [astro-ph]} \BibitemShut {NoStop}%
\bibitem [{\citenamefont {{Gunn}}\ \emph {et~al.}(2006)\citenamefont {{Gunn}},
  \citenamefont {{Siegmund}}, \citenamefont {{Mannery}}, \citenamefont
  {{Owen}}, \citenamefont {{Hull}}, \citenamefont {{Leger}}, \citenamefont
  {{Carey}}, \citenamefont {{Knapp}}, \citenamefont {{York}}, \citenamefont
  {{Boroski}}, \citenamefont {{Kent}}, \citenamefont {{Lupton}}, \citenamefont
  {{Rockosi}}, \citenamefont {{Evans}}, \citenamefont {{Waddell}} \emph
  {et~al.}}]{2006AJ....131.2332G}%
  \BibitemOpen
  \bibfield  {author} {\bibinfo {author} {\bibfnamefont {J.~E.}\ \bibnamefont
  {{Gunn}}}, \bibinfo {author} {\bibfnamefont {W.~A.}\ \bibnamefont
  {{Siegmund}}}, \bibinfo {author} {\bibfnamefont {E.~J.}\ \bibnamefont
  {{Mannery}}}, \bibinfo {author} {\bibfnamefont {R.~E.}\ \bibnamefont
  {{Owen}}}, \bibinfo {author} {\bibfnamefont {C.~L.}\ \bibnamefont {{Hull}}},
  \bibinfo {author} {\bibfnamefont {R.~F.}\ \bibnamefont {{Leger}}}, \bibinfo
  {author} {\bibfnamefont {L.~N.}\ \bibnamefont {{Carey}}}, \bibinfo {author}
  {\bibfnamefont {G.~R.}\ \bibnamefont {{Knapp}}}, \bibinfo {author}
  {\bibfnamefont {D.~G.}\ \bibnamefont {{York}}}, \bibinfo {author}
  {\bibfnamefont {W.~N.}\ \bibnamefont {{Boroski}}}, \bibinfo {author}
  {\bibfnamefont {S.~M.}\ \bibnamefont {{Kent}}}, \bibinfo {author}
  {\bibfnamefont {R.~H.}\ \bibnamefont {{Lupton}}}, \bibinfo {author}
  {\bibfnamefont {C.~M.}\ \bibnamefont {{Rockosi}}}, \bibinfo {author}
  {\bibfnamefont {M.~L.}\ \bibnamefont {{Evans}}}, \bibinfo {author}
  {\bibfnamefont {P.}~\bibnamefont {{Waddell}}}, \emph {et~al.},\ }\href
  {https://doi.org/10.1086/500975} {\bibfield  {journal} {\bibinfo  {journal}
  {\aj}\ }\textbf {\bibinfo {volume} {131}},\ \bibinfo {pages} {2332} (\bibinfo
  {year} {2006})},\ \Eprint {https://arxiv.org/abs/astro-ph/0602326}
  {arXiv:astro-ph/0602326 [astro-ph]} \BibitemShut {NoStop}%
\bibitem [{\citenamefont {{Fukugita}}\ \emph {et~al.}(1996)\citenamefont
  {{Fukugita}}, \citenamefont {{Ichikawa}}, \citenamefont {{Gunn}},
  \citenamefont {{Doi}}, \citenamefont {{Shimasaku}},\ and\ \citenamefont
  {{Schneider}}}]{1996AJ....111.1748F}%
  \BibitemOpen
  \bibfield  {author} {\bibinfo {author} {\bibfnamefont {M.}~\bibnamefont
  {{Fukugita}}}, \bibinfo {author} {\bibfnamefont {T.}~\bibnamefont
  {{Ichikawa}}}, \bibinfo {author} {\bibfnamefont {J.~E.}\ \bibnamefont
  {{Gunn}}}, \bibinfo {author} {\bibfnamefont {M.}~\bibnamefont {{Doi}}},
  \bibinfo {author} {\bibfnamefont {K.}~\bibnamefont {{Shimasaku}}},\ and\
  \bibinfo {author} {\bibfnamefont {D.~P.}\ \bibnamefont {{Schneider}}},\
  }\href {https://doi.org/10.1086/117915} {\bibfield  {journal} {\bibinfo
  {journal} {\aj}\ }\textbf {\bibinfo {volume} {111}},\ \bibinfo {pages} {1748}
  (\bibinfo {year} {1996})}\BibitemShut {NoStop}%
\bibitem [{\citenamefont {{Smith}}\ \emph {et~al.}(2002)\citenamefont
  {{Smith}}, \citenamefont {{Tucker}}, \citenamefont {{Kent}}, \citenamefont
  {{Richmond}}, \citenamefont {{Fukugita}}, \citenamefont {{Ichikawa}},
  \citenamefont {{Ichikawa}}, \citenamefont {{Jorgensen}}, \citenamefont
  {{Uomoto}}, \citenamefont {{Gunn}}, \citenamefont {{Hamabe}}, \citenamefont
  {{Watanabe}}, \citenamefont {{Tolea}}, \citenamefont {{Henden}},
  \citenamefont {{Annis}} \emph {et~al.}}]{2002AJ....123.2121S}%
  \BibitemOpen
  \bibfield  {author} {\bibinfo {author} {\bibfnamefont {J.~A.}\ \bibnamefont
  {{Smith}}}, \bibinfo {author} {\bibfnamefont {D.~L.}\ \bibnamefont
  {{Tucker}}}, \bibinfo {author} {\bibfnamefont {S.}~\bibnamefont {{Kent}}},
  \bibinfo {author} {\bibfnamefont {M.~W.}\ \bibnamefont {{Richmond}}},
  \bibinfo {author} {\bibfnamefont {M.}~\bibnamefont {{Fukugita}}}, \bibinfo
  {author} {\bibfnamefont {T.}~\bibnamefont {{Ichikawa}}}, \bibinfo {author}
  {\bibfnamefont {S.-i.}\ \bibnamefont {{Ichikawa}}}, \bibinfo {author}
  {\bibfnamefont {A.~M.}\ \bibnamefont {{Jorgensen}}}, \bibinfo {author}
  {\bibfnamefont {A.}~\bibnamefont {{Uomoto}}}, \bibinfo {author}
  {\bibfnamefont {J.~E.}\ \bibnamefont {{Gunn}}}, \bibinfo {author}
  {\bibfnamefont {M.}~\bibnamefont {{Hamabe}}}, \bibinfo {author}
  {\bibfnamefont {M.}~\bibnamefont {{Watanabe}}}, \bibinfo {author}
  {\bibfnamefont {A.}~\bibnamefont {{Tolea}}}, \bibinfo {author} {\bibfnamefont
  {A.}~\bibnamefont {{Henden}}}, \bibinfo {author} {\bibfnamefont
  {J.}~\bibnamefont {{Annis}}}, \emph {et~al.},\ }\href
  {https://doi.org/10.1086/339311} {\bibfield  {journal} {\bibinfo  {journal}
  {\aj}\ }\textbf {\bibinfo {volume} {123}},\ \bibinfo {pages} {2121} (\bibinfo
  {year} {2002})},\ \Eprint {https://arxiv.org/abs/astro-ph/0201143}
  {arXiv:astro-ph/0201143 [astro-ph]} \BibitemShut {NoStop}%
\bibitem [{\citenamefont {{Doi}}\ \emph {et~al.}(2010)\citenamefont {{Doi}},
  \citenamefont {{Tanaka}}, \citenamefont {{Fukugita}}, \citenamefont {{Gunn}},
  \citenamefont {{Yasuda}}, \citenamefont {{Ivezi{\'c}}}, \citenamefont
  {{Brinkmann}}, \citenamefont {{de Haars}}, \citenamefont {{Kleinman}},
  \citenamefont {{Krzesinski}},\ and\ \citenamefont {{French
  Leger}}}]{2010AJ....139.1628D}%
  \BibitemOpen
  \bibfield  {author} {\bibinfo {author} {\bibfnamefont {M.}~\bibnamefont
  {{Doi}}}, \bibinfo {author} {\bibfnamefont {M.}~\bibnamefont {{Tanaka}}},
  \bibinfo {author} {\bibfnamefont {M.}~\bibnamefont {{Fukugita}}}, \bibinfo
  {author} {\bibfnamefont {J.~E.}\ \bibnamefont {{Gunn}}}, \bibinfo {author}
  {\bibfnamefont {N.}~\bibnamefont {{Yasuda}}}, \bibinfo {author}
  {\bibfnamefont {{\v{Z}}.}~\bibnamefont {{Ivezi{\'c}}}}, \bibinfo {author}
  {\bibfnamefont {J.}~\bibnamefont {{Brinkmann}}}, \bibinfo {author}
  {\bibfnamefont {E.}~\bibnamefont {{de Haars}}}, \bibinfo {author}
  {\bibfnamefont {S.~J.}\ \bibnamefont {{Kleinman}}}, \bibinfo {author}
  {\bibfnamefont {J.}~\bibnamefont {{Krzesinski}}},\ and\ \bibinfo {author}
  {\bibfnamefont {R.}~\bibnamefont {{French Leger}}},\ }\href
  {https://doi.org/10.1088/0004-6256/139/4/1628} {\bibfield  {journal}
  {\bibinfo  {journal} {\aj}\ }\textbf {\bibinfo {volume} {139}},\ \bibinfo
  {pages} {1628} (\bibinfo {year} {2010})},\ \Eprint
  {https://arxiv.org/abs/1002.3701} {arXiv:1002.3701 [astro-ph.IM]}
  \BibitemShut {NoStop}%
\bibitem [{\citenamefont {{Eisenstein}}\ \emph {et~al.}(2011)\citenamefont
  {{Eisenstein}}, \citenamefont {{Weinberg}}, \citenamefont {{Agol}},
  \citenamefont {{Aihara}}, \citenamefont {{Allende Prieto}}, \citenamefont
  {{Anderson}}, \citenamefont {{Arns}}, \citenamefont {{Aubourg}},
  \citenamefont {{Bailey}}, \citenamefont {{Balbinot}}, \citenamefont
  {{Barkhouser}}, \citenamefont {{Beers}}, \citenamefont {{Berlind}},
  \citenamefont {{Bickerton}}, \citenamefont {{Bizyaev}} \emph
  {et~al.}}]{2011AJ....142...72E}%
  \BibitemOpen
  \bibfield  {author} {\bibinfo {author} {\bibfnamefont {D.~J.}\ \bibnamefont
  {{Eisenstein}}}, \bibinfo {author} {\bibfnamefont {D.~H.}\ \bibnamefont
  {{Weinberg}}}, \bibinfo {author} {\bibfnamefont {E.}~\bibnamefont {{Agol}}},
  \bibinfo {author} {\bibfnamefont {H.}~\bibnamefont {{Aihara}}}, \bibinfo
  {author} {\bibfnamefont {C.}~\bibnamefont {{Allende Prieto}}}, \bibinfo
  {author} {\bibfnamefont {S.~F.}\ \bibnamefont {{Anderson}}}, \bibinfo
  {author} {\bibfnamefont {J.~A.}\ \bibnamefont {{Arns}}}, \bibinfo {author}
  {\bibfnamefont {{\'E}.}~\bibnamefont {{Aubourg}}}, \bibinfo {author}
  {\bibfnamefont {S.}~\bibnamefont {{Bailey}}}, \bibinfo {author}
  {\bibfnamefont {E.}~\bibnamefont {{Balbinot}}}, \bibinfo {author}
  {\bibfnamefont {R.}~\bibnamefont {{Barkhouser}}}, \bibinfo {author}
  {\bibfnamefont {T.~C.}\ \bibnamefont {{Beers}}}, \bibinfo {author}
  {\bibfnamefont {A.~A.}\ \bibnamefont {{Berlind}}}, \bibinfo {author}
  {\bibfnamefont {S.~J.}\ \bibnamefont {{Bickerton}}}, \bibinfo {author}
  {\bibfnamefont {D.}~\bibnamefont {{Bizyaev}}}, \emph {et~al.},\ }\href
  {https://doi.org/10.1088/0004-6256/142/3/72} {\bibfield  {journal} {\bibinfo
  {journal} {\aj}\ }\textbf {\bibinfo {volume} {142}},\ \bibinfo {eid} {72}
  (\bibinfo {year} {2011})},\ \Eprint {https://arxiv.org/abs/1101.1529}
  {arXiv:1101.1529 [astro-ph.IM]} \BibitemShut {NoStop}%
\bibitem [{\citenamefont {{Ahn}}\ \emph {et~al.}(2012)\citenamefont {{Ahn}},
  \citenamefont {{Alexandroff}}, \citenamefont {{Allende Prieto}},
  \citenamefont {{Anderson}}, \citenamefont {{Anderton}}, \citenamefont
  {{Andrews}}, \citenamefont {{Aubourg}}, \citenamefont {{Bailey}},
  \citenamefont {{Balbinot}}, \citenamefont {{Barnes}}, \citenamefont
  {{Bautista}}, \citenamefont {{Beers}}, \citenamefont {{Beifiori}},
  \citenamefont {{Berlind}}, \citenamefont {{Bhardwaj}} \emph
  {et~al.}}]{2012ApJS..203...21A}%
  \BibitemOpen
  \bibfield  {author} {\bibinfo {author} {\bibfnamefont {C.~P.}\ \bibnamefont
  {{Ahn}}}, \bibinfo {author} {\bibfnamefont {R.}~\bibnamefont
  {{Alexandroff}}}, \bibinfo {author} {\bibfnamefont {C.}~\bibnamefont
  {{Allende Prieto}}}, \bibinfo {author} {\bibfnamefont {S.~F.}\ \bibnamefont
  {{Anderson}}}, \bibinfo {author} {\bibfnamefont {T.}~\bibnamefont
  {{Anderton}}}, \bibinfo {author} {\bibfnamefont {B.~H.}\ \bibnamefont
  {{Andrews}}}, \bibinfo {author} {\bibfnamefont {{\'E}.}~\bibnamefont
  {{Aubourg}}}, \bibinfo {author} {\bibfnamefont {S.}~\bibnamefont {{Bailey}}},
  \bibinfo {author} {\bibfnamefont {E.}~\bibnamefont {{Balbinot}}}, \bibinfo
  {author} {\bibfnamefont {R.}~\bibnamefont {{Barnes}}}, \bibinfo {author}
  {\bibfnamefont {J.}~\bibnamefont {{Bautista}}}, \bibinfo {author}
  {\bibfnamefont {T.~C.}\ \bibnamefont {{Beers}}}, \bibinfo {author}
  {\bibfnamefont {A.}~\bibnamefont {{Beifiori}}}, \bibinfo {author}
  {\bibfnamefont {A.~A.}\ \bibnamefont {{Berlind}}}, \bibinfo {author}
  {\bibfnamefont {V.}~\bibnamefont {{Bhardwaj}}}, \emph {et~al.},\ }\href
  {https://doi.org/10.1088/0067-0049/203/2/21} {\bibfield  {journal} {\bibinfo
  {journal} {\apjs}\ }\textbf {\bibinfo {volume} {203}},\ \bibinfo {eid} {21}
  (\bibinfo {year} {2012})},\ \Eprint {https://arxiv.org/abs/1207.7137}
  {arXiv:1207.7137 [astro-ph.IM]} \BibitemShut {NoStop}%
\bibitem [{\citenamefont {{Aihara}}\ \emph {et~al.}(2011)\citenamefont
  {{Aihara}}, \citenamefont {{Allende Prieto}}, \citenamefont {{An}},
  \citenamefont {{Anderson}}, \citenamefont {{Aubourg}}, \citenamefont
  {{Balbinot}}, \citenamefont {{Beers}}, \citenamefont {{Berlind}},
  \citenamefont {{Bickerton}}, \citenamefont {{Bizyaev}}, \citenamefont
  {{Blanton}}, \citenamefont {{Bochanski}}, \citenamefont {{Bolton}},
  \citenamefont {{Bovy}}, \citenamefont {{Brandt}} \emph
  {et~al.}}]{2011ApJS..193...29A}%
  \BibitemOpen
  \bibfield  {author} {\bibinfo {author} {\bibfnamefont {H.}~\bibnamefont
  {{Aihara}}}, \bibinfo {author} {\bibfnamefont {C.}~\bibnamefont {{Allende
  Prieto}}}, \bibinfo {author} {\bibfnamefont {D.}~\bibnamefont {{An}}},
  \bibinfo {author} {\bibfnamefont {S.~F.}\ \bibnamefont {{Anderson}}},
  \bibinfo {author} {\bibfnamefont {{\'E}.}~\bibnamefont {{Aubourg}}}, \bibinfo
  {author} {\bibfnamefont {E.}~\bibnamefont {{Balbinot}}}, \bibinfo {author}
  {\bibfnamefont {T.~C.}\ \bibnamefont {{Beers}}}, \bibinfo {author}
  {\bibfnamefont {A.~A.}\ \bibnamefont {{Berlind}}}, \bibinfo {author}
  {\bibfnamefont {S.~J.}\ \bibnamefont {{Bickerton}}}, \bibinfo {author}
  {\bibfnamefont {D.}~\bibnamefont {{Bizyaev}}}, \bibinfo {author}
  {\bibfnamefont {M.~R.}\ \bibnamefont {{Blanton}}}, \bibinfo {author}
  {\bibfnamefont {J.~J.}\ \bibnamefont {{Bochanski}}}, \bibinfo {author}
  {\bibfnamefont {A.~S.}\ \bibnamefont {{Bolton}}}, \bibinfo {author}
  {\bibfnamefont {J.}~\bibnamefont {{Bovy}}}, \bibinfo {author} {\bibfnamefont
  {W.~N.}\ \bibnamefont {{Brandt}}}, \emph {et~al.},\ }\href
  {https://doi.org/10.1088/0067-0049/193/2/29} {\bibfield  {journal} {\bibinfo
  {journal} {\apjs}\ }\textbf {\bibinfo {volume} {193}},\ \bibinfo {eid} {29}
  (\bibinfo {year} {2011})},\ \Eprint {https://arxiv.org/abs/1101.1559}
  {arXiv:1101.1559 [astro-ph.IM]} \BibitemShut {NoStop}%
\bibitem [{\citenamefont {{Lupton}}\ \emph {et~al.}(2001)\citenamefont
  {{Lupton}}, \citenamefont {{Gunn}}, \citenamefont {{Ivezi{\'c}}},
  \citenamefont {{Knapp}},\ and\ \citenamefont {{Kent}}}]{2001ASPC..238..269L}%
  \BibitemOpen
  \bibfield  {author} {\bibinfo {author} {\bibfnamefont {R.}~\bibnamefont
  {{Lupton}}}, \bibinfo {author} {\bibfnamefont {J.~E.}\ \bibnamefont
  {{Gunn}}}, \bibinfo {author} {\bibfnamefont {Z.}~\bibnamefont
  {{Ivezi{\'c}}}}, \bibinfo {author} {\bibfnamefont {G.~R.}\ \bibnamefont
  {{Knapp}}},\ and\ \bibinfo {author} {\bibfnamefont {S.}~\bibnamefont
  {{Kent}}},\ }in\ \href@noop {} {\emph {\bibinfo {booktitle} {Astronomical
  Data Analysis Software and Systems X}}},\ \bibinfo {series} {Astronomical
  Society of the Pacific Conference Series}, Vol.\ \bibinfo {volume} {238},\
  \bibinfo {editor} {edited by\ \bibinfo {editor} {\bibfnamefont
  {J.}~\bibnamefont {{Harnden}}, \bibfnamefont {F.~R.}}, \bibinfo {editor}
  {\bibfnamefont {F.~A.}\ \bibnamefont {{Primini}}},\ and\ \bibinfo {editor}
  {\bibfnamefont {H.~E.}\ \bibnamefont {{Payne}}}}\ (\bibinfo {year} {2001})\
  p.\ \bibinfo {pages} {269},\ \Eprint {https://arxiv.org/abs/astro-ph/0101420}
  {arXiv:astro-ph/0101420 [astro-ph]} \BibitemShut {NoStop}%
\bibitem [{\citenamefont {{Pier}}\ \emph {et~al.}(2003)\citenamefont {{Pier}},
  \citenamefont {{Munn}}, \citenamefont {{Hindsley}}, \citenamefont
  {{Hennessy}}, \citenamefont {{Kent}}, \citenamefont {{Lupton}},\ and\
  \citenamefont {{Ivezi{\'c}}}}]{2003AJ....125.1559P}%
  \BibitemOpen
  \bibfield  {author} {\bibinfo {author} {\bibfnamefont {J.~R.}\ \bibnamefont
  {{Pier}}}, \bibinfo {author} {\bibfnamefont {J.~A.}\ \bibnamefont {{Munn}}},
  \bibinfo {author} {\bibfnamefont {R.~B.}\ \bibnamefont {{Hindsley}}},
  \bibinfo {author} {\bibfnamefont {G.~S.}\ \bibnamefont {{Hennessy}}},
  \bibinfo {author} {\bibfnamefont {S.~M.}\ \bibnamefont {{Kent}}}, \bibinfo
  {author} {\bibfnamefont {R.~H.}\ \bibnamefont {{Lupton}}},\ and\ \bibinfo
  {author} {\bibfnamefont {{\v{Z}}.}~\bibnamefont {{Ivezi{\'c}}}},\ }\href
  {https://doi.org/10.1086/346138} {\bibfield  {journal} {\bibinfo  {journal}
  {\aj}\ }\textbf {\bibinfo {volume} {125}},\ \bibinfo {pages} {1559} (\bibinfo
  {year} {2003})},\ \Eprint {https://arxiv.org/abs/astro-ph/0211375}
  {arXiv:astro-ph/0211375 [astro-ph]} \BibitemShut {NoStop}%
\bibitem [{\citenamefont {{Padmanabhan}}\ \emph {et~al.}(2008)\citenamefont
  {{Padmanabhan}}, \citenamefont {{Schlegel}}, \citenamefont {{Finkbeiner}},
  \citenamefont {{Barentine}}, \citenamefont {{Blanton}}, \citenamefont
  {{Brewington}}, \citenamefont {{Gunn}}, \citenamefont {{Harvanek}},
  \citenamefont {{Hogg}}, \citenamefont {{Ivezi{\'c}}}, \citenamefont
  {{Johnston}}, \citenamefont {{Kent}}, \citenamefont {{Kleinman}},
  \citenamefont {{Knapp}}, \citenamefont {{Krzesinski}} \emph
  {et~al.}}]{2008ApJ...674.1217P}%
  \BibitemOpen
  \bibfield  {author} {\bibinfo {author} {\bibfnamefont {N.}~\bibnamefont
  {{Padmanabhan}}}, \bibinfo {author} {\bibfnamefont {D.~J.}\ \bibnamefont
  {{Schlegel}}}, \bibinfo {author} {\bibfnamefont {D.~P.}\ \bibnamefont
  {{Finkbeiner}}}, \bibinfo {author} {\bibfnamefont {J.~C.}\ \bibnamefont
  {{Barentine}}}, \bibinfo {author} {\bibfnamefont {M.~R.}\ \bibnamefont
  {{Blanton}}}, \bibinfo {author} {\bibfnamefont {H.~J.}\ \bibnamefont
  {{Brewington}}}, \bibinfo {author} {\bibfnamefont {J.~E.}\ \bibnamefont
  {{Gunn}}}, \bibinfo {author} {\bibfnamefont {M.}~\bibnamefont {{Harvanek}}},
  \bibinfo {author} {\bibfnamefont {D.~W.}\ \bibnamefont {{Hogg}}}, \bibinfo
  {author} {\bibfnamefont {{\v{Z}}.}~\bibnamefont {{Ivezi{\'c}}}}, \bibinfo
  {author} {\bibfnamefont {D.}~\bibnamefont {{Johnston}}}, \bibinfo {author}
  {\bibfnamefont {S.~M.}\ \bibnamefont {{Kent}}}, \bibinfo {author}
  {\bibfnamefont {S.~J.}\ \bibnamefont {{Kleinman}}}, \bibinfo {author}
  {\bibfnamefont {G.~R.}\ \bibnamefont {{Knapp}}}, \bibinfo {author}
  {\bibfnamefont {J.}~\bibnamefont {{Krzesinski}}}, \emph {et~al.},\ }\href
  {https://doi.org/10.1086/524677} {\bibfield  {journal} {\bibinfo  {journal}
  {\apj}\ }\textbf {\bibinfo {volume} {674}},\ \bibinfo {pages} {1217}
  (\bibinfo {year} {2008})},\ \Eprint {https://arxiv.org/abs/astro-ph/0703454}
  {arXiv:astro-ph/0703454 [astro-ph]} \BibitemShut {NoStop}%
\bibitem [{\citenamefont {{Schlegel}}\ \emph {et~al.}(1998)\citenamefont
  {{Schlegel}}, \citenamefont {{Finkbeiner}},\ and\ \citenamefont
  {{Davis}}}]{1998ApJ...500..525S}%
  \BibitemOpen
  \bibfield  {author} {\bibinfo {author} {\bibfnamefont {D.~J.}\ \bibnamefont
  {{Schlegel}}}, \bibinfo {author} {\bibfnamefont {D.~P.}\ \bibnamefont
  {{Finkbeiner}}},\ and\ \bibinfo {author} {\bibfnamefont {M.}~\bibnamefont
  {{Davis}}},\ }\href {https://doi.org/10.1086/305772} {\bibfield  {journal}
  {\bibinfo  {journal} {\apj}\ }\textbf {\bibinfo {volume} {500}},\ \bibinfo
  {pages} {525} (\bibinfo {year} {1998})},\ \Eprint
  {https://arxiv.org/abs/astro-ph/9710327} {arXiv:astro-ph/9710327 [astro-ph]}
  \BibitemShut {NoStop}%
\bibitem [{\citenamefont {{Bolton}}\ \emph {et~al.}(2012)\citenamefont
  {{Bolton}}, \citenamefont {{Schlegel}}, \citenamefont {{Aubourg}},
  \citenamefont {{Bailey}}, \citenamefont {{Bhardwaj}}, \citenamefont
  {{Brownstein}}, \citenamefont {{Burles}}, \citenamefont {{Chen}},
  \citenamefont {{Dawson}}, \citenamefont {{Eisenstein}}, \citenamefont
  {{Gunn}}, \citenamefont {{Knapp}}, \citenamefont {{Loomis}}, \citenamefont
  {{Lupton}}, \citenamefont {{Maraston}} \emph {et~al.}}]{2012AJ....144..144B}%
  \BibitemOpen
  \bibfield  {author} {\bibinfo {author} {\bibfnamefont {A.~S.}\ \bibnamefont
  {{Bolton}}}, \bibinfo {author} {\bibfnamefont {D.~J.}\ \bibnamefont
  {{Schlegel}}}, \bibinfo {author} {\bibfnamefont {{\'E}.}~\bibnamefont
  {{Aubourg}}}, \bibinfo {author} {\bibfnamefont {S.}~\bibnamefont {{Bailey}}},
  \bibinfo {author} {\bibfnamefont {V.}~\bibnamefont {{Bhardwaj}}}, \bibinfo
  {author} {\bibfnamefont {J.~R.}\ \bibnamefont {{Brownstein}}}, \bibinfo
  {author} {\bibfnamefont {S.}~\bibnamefont {{Burles}}}, \bibinfo {author}
  {\bibfnamefont {Y.-M.}\ \bibnamefont {{Chen}}}, \bibinfo {author}
  {\bibfnamefont {K.}~\bibnamefont {{Dawson}}}, \bibinfo {author}
  {\bibfnamefont {D.~J.}\ \bibnamefont {{Eisenstein}}}, \bibinfo {author}
  {\bibfnamefont {J.~E.}\ \bibnamefont {{Gunn}}}, \bibinfo {author}
  {\bibfnamefont {G.~R.}\ \bibnamefont {{Knapp}}}, \bibinfo {author}
  {\bibfnamefont {C.~P.}\ \bibnamefont {{Loomis}}}, \bibinfo {author}
  {\bibfnamefont {R.~H.}\ \bibnamefont {{Lupton}}}, \bibinfo {author}
  {\bibfnamefont {C.}~\bibnamefont {{Maraston}}}, \emph {et~al.},\ }\href
  {https://doi.org/10.1088/0004-6256/144/5/144} {\bibfield  {journal} {\bibinfo
   {journal} {\aj}\ }\textbf {\bibinfo {volume} {144}},\ \bibinfo {eid} {144}
  (\bibinfo {year} {2012})},\ \Eprint {https://arxiv.org/abs/1207.7326}
  {arXiv:1207.7326 [astro-ph.CO]} \BibitemShut {NoStop}%
\bibitem [{\citenamefont {{Reid}}\ \emph {et~al.}(2016)\citenamefont {{Reid}},
  \citenamefont {{Ho}}, \citenamefont {{Padmanabhan}}, \citenamefont
  {{Percival}}, \citenamefont {{Tinker}}, \citenamefont {{Tojeiro}},
  \citenamefont {{White}}, \citenamefont {{Eisenstein}}, \citenamefont
  {{Maraston}}, \citenamefont {{Ross}}, \citenamefont {{S{\'a}nchez}},
  \citenamefont {{Schlegel}}, \citenamefont {{Sheldon}}, \citenamefont
  {{Strauss}}, \citenamefont {{Thomas}} \emph {et~al.}}]{Reid:2016}%
  \BibitemOpen
  \bibfield  {author} {\bibinfo {author} {\bibfnamefont {B.}~\bibnamefont
  {{Reid}}}, \bibinfo {author} {\bibfnamefont {S.}~\bibnamefont {{Ho}}},
  \bibinfo {author} {\bibfnamefont {N.}~\bibnamefont {{Padmanabhan}}}, \bibinfo
  {author} {\bibfnamefont {W.~J.}\ \bibnamefont {{Percival}}}, \bibinfo
  {author} {\bibfnamefont {J.}~\bibnamefont {{Tinker}}}, \bibinfo {author}
  {\bibfnamefont {R.}~\bibnamefont {{Tojeiro}}}, \bibinfo {author}
  {\bibfnamefont {M.}~\bibnamefont {{White}}}, \bibinfo {author} {\bibfnamefont
  {D.~J.}\ \bibnamefont {{Eisenstein}}}, \bibinfo {author} {\bibfnamefont
  {C.}~\bibnamefont {{Maraston}}}, \bibinfo {author} {\bibfnamefont {A.~J.}\
  \bibnamefont {{Ross}}}, \bibinfo {author} {\bibfnamefont {A.~G.}\
  \bibnamefont {{S{\'a}nchez}}}, \bibinfo {author} {\bibfnamefont
  {D.}~\bibnamefont {{Schlegel}}}, \bibinfo {author} {\bibfnamefont
  {E.}~\bibnamefont {{Sheldon}}}, \bibinfo {author} {\bibfnamefont {M.~A.}\
  \bibnamefont {{Strauss}}}, \bibinfo {author} {\bibfnamefont {D.}~\bibnamefont
  {{Thomas}}}, \emph {et~al.},\ }\href {https://doi.org/10.1093/mnras/stv2382}
  {\bibfield  {journal} {\bibinfo  {journal} {\mnras}\ }\textbf {\bibinfo
  {volume} {455}},\ \bibinfo {pages} {1553} (\bibinfo {year} {2016})},\ \Eprint
  {https://arxiv.org/abs/1509.06529} {arXiv:1509.06529 [astro-ph.CO]}
  \BibitemShut {NoStop}%
\bibitem [{\citenamefont {{Anderson}}\ \emph {et~al.}(2014)\citenamefont
  {{Anderson}}, \citenamefont {{Aubourg}}, \citenamefont {{Bailey}},
  \citenamefont {{Beutler}}, \citenamefont {{Bhardwaj}}, \citenamefont
  {{Blanton}}, \citenamefont {{Bolton}}, \citenamefont {{Brinkmann}},
  \citenamefont {{Brownstein}}, \citenamefont {{Burden}}, \citenamefont
  {{Chuang}}, \citenamefont {{Cuesta}}, \citenamefont {{Dawson}}, \citenamefont
  {{Eisenstein}}, \citenamefont {{Escoffier}} \emph
  {et~al.}}]{2014MNRAS.441...24A}%
  \BibitemOpen
  \bibfield  {author} {\bibinfo {author} {\bibfnamefont {L.}~\bibnamefont
  {{Anderson}}}, \bibinfo {author} {\bibfnamefont {{\'E}.}~\bibnamefont
  {{Aubourg}}}, \bibinfo {author} {\bibfnamefont {S.}~\bibnamefont {{Bailey}}},
  \bibinfo {author} {\bibfnamefont {F.}~\bibnamefont {{Beutler}}}, \bibinfo
  {author} {\bibfnamefont {V.}~\bibnamefont {{Bhardwaj}}}, \bibinfo {author}
  {\bibfnamefont {M.}~\bibnamefont {{Blanton}}}, \bibinfo {author}
  {\bibfnamefont {A.~S.}\ \bibnamefont {{Bolton}}}, \bibinfo {author}
  {\bibfnamefont {J.}~\bibnamefont {{Brinkmann}}}, \bibinfo {author}
  {\bibfnamefont {J.~R.}\ \bibnamefont {{Brownstein}}}, \bibinfo {author}
  {\bibfnamefont {A.}~\bibnamefont {{Burden}}}, \bibinfo {author}
  {\bibfnamefont {C.-H.}\ \bibnamefont {{Chuang}}}, \bibinfo {author}
  {\bibfnamefont {A.~J.}\ \bibnamefont {{Cuesta}}}, \bibinfo {author}
  {\bibfnamefont {K.~S.}\ \bibnamefont {{Dawson}}}, \bibinfo {author}
  {\bibfnamefont {D.~J.}\ \bibnamefont {{Eisenstein}}}, \bibinfo {author}
  {\bibfnamefont {S.}~\bibnamefont {{Escoffier}}}, \emph {et~al.},\ }\href
  {https://doi.org/10.1093/mnras/stu523} {\bibfield  {journal} {\bibinfo
  {journal} {\mnras}\ }\textbf {\bibinfo {volume} {441}},\ \bibinfo {pages}
  {24} (\bibinfo {year} {2014})},\ \Eprint {https://arxiv.org/abs/1312.4877}
  {arXiv:1312.4877 [astro-ph.CO]} \BibitemShut {NoStop}%
\bibitem [{\citenamefont {{Guo}}\ \emph {et~al.}(2012)\citenamefont {{Guo}},
  \citenamefont {{Zehavi}},\ and\ \citenamefont {{Zheng}}}]{Guo:2012}%
  \BibitemOpen
  \bibfield  {author} {\bibinfo {author} {\bibfnamefont {H.}~\bibnamefont
  {{Guo}}}, \bibinfo {author} {\bibfnamefont {I.}~\bibnamefont {{Zehavi}}},\
  and\ \bibinfo {author} {\bibfnamefont {Z.}~\bibnamefont {{Zheng}}},\ }\href
  {https://doi.org/10.1088/0004-637X/756/2/127} {\bibfield  {journal} {\bibinfo
   {journal} {\apj}\ }\textbf {\bibinfo {volume} {756}},\ \bibinfo {eid} {127}
  (\bibinfo {year} {2012})},\ \Eprint {https://arxiv.org/abs/1111.6598}
  {arXiv:1111.6598 [astro-ph.CO]} \BibitemShut {NoStop}%
\bibitem [{\citenamefont {{Wake}}\ \emph {et~al.}(2006)\citenamefont {{Wake}},
  \citenamefont {{Nichol}}, \citenamefont {{Eisenstein}}, \citenamefont
  {{Loveday}}, \citenamefont {{Edge}}, \citenamefont {{Cannon}}, \citenamefont
  {{Smail}}, \citenamefont {{Schneider}}, \citenamefont {{Scranton}},
  \citenamefont {{Carson}}, \citenamefont {{Ross}}, \citenamefont {{Brunner}},
  \citenamefont {{Colless}}, \citenamefont {{Couch}}, \citenamefont {{Croom}}
  \emph {et~al.}}]{2006MNRAS.372..537W}%
  \BibitemOpen
  \bibfield  {author} {\bibinfo {author} {\bibfnamefont {D.~A.}\ \bibnamefont
  {{Wake}}}, \bibinfo {author} {\bibfnamefont {R.~C.}\ \bibnamefont
  {{Nichol}}}, \bibinfo {author} {\bibfnamefont {D.~J.}\ \bibnamefont
  {{Eisenstein}}}, \bibinfo {author} {\bibfnamefont {J.}~\bibnamefont
  {{Loveday}}}, \bibinfo {author} {\bibfnamefont {A.~C.}\ \bibnamefont
  {{Edge}}}, \bibinfo {author} {\bibfnamefont {R.}~\bibnamefont {{Cannon}}},
  \bibinfo {author} {\bibfnamefont {I.}~\bibnamefont {{Smail}}}, \bibinfo
  {author} {\bibfnamefont {D.~P.}\ \bibnamefont {{Schneider}}}, \bibinfo
  {author} {\bibfnamefont {R.}~\bibnamefont {{Scranton}}}, \bibinfo {author}
  {\bibfnamefont {D.}~\bibnamefont {{Carson}}}, \bibinfo {author}
  {\bibfnamefont {N.~P.}\ \bibnamefont {{Ross}}}, \bibinfo {author}
  {\bibfnamefont {R.~J.}\ \bibnamefont {{Brunner}}}, \bibinfo {author}
  {\bibfnamefont {M.}~\bibnamefont {{Colless}}}, \bibinfo {author}
  {\bibfnamefont {W.~J.}\ \bibnamefont {{Couch}}}, \bibinfo {author}
  {\bibfnamefont {S.~M.}\ \bibnamefont {{Croom}}}, \emph {et~al.},\ }\href
  {https://doi.org/10.1111/j.1365-2966.2006.10831.x} {\bibfield  {journal}
  {\bibinfo  {journal} {\mnras}\ }\textbf {\bibinfo {volume} {372}},\ \bibinfo
  {pages} {537} (\bibinfo {year} {2006})},\ \Eprint
  {https://arxiv.org/abs/astro-ph/0607629} {arXiv:astro-ph/0607629 [astro-ph]}
  \BibitemShut {NoStop}%
\bibitem [{\citenamefont {{Bruzual}}\ and\ \citenamefont
  {{Charlot}}(2003)}]{Bruzual_Charlot:2003}%
  \BibitemOpen
  \bibfield  {author} {\bibinfo {author} {\bibfnamefont {G.}~\bibnamefont
  {{Bruzual}}}\ and\ \bibinfo {author} {\bibfnamefont {S.}~\bibnamefont
  {{Charlot}}},\ }\href {https://doi.org/10.1046/j.1365-8711.2003.06897.x}
  {\bibfield  {journal} {\bibinfo  {journal} {\mnras}\ }\textbf {\bibinfo
  {volume} {344}},\ \bibinfo {pages} {1000} (\bibinfo {year} {2003})},\ \Eprint
  {https://arxiv.org/abs/astro-ph/0309134} {arXiv:astro-ph/0309134 [astro-ph]}
  \BibitemShut {NoStop}%
\bibitem [{\citenamefont {{Ross}}\ \emph {et~al.}(2012)\citenamefont {{Ross}},
  \citenamefont {{Percival}}, \citenamefont {{S{\'a}nchez}}, \citenamefont
  {{Samushia}}, \citenamefont {{Ho}}, \citenamefont {{Kazin}}, \citenamefont
  {{Manera}}, \citenamefont {{Reid}}, \citenamefont {{White}}, \citenamefont
  {{Tojeiro}}, \citenamefont {{McBride}}, \citenamefont {{Xu}}, \citenamefont
  {{Wake}}, \citenamefont {{Strauss}}, \citenamefont {{Montesano}} \emph
  {et~al.}}]{Ross:2012}%
  \BibitemOpen
  \bibfield  {author} {\bibinfo {author} {\bibfnamefont {A.~J.}\ \bibnamefont
  {{Ross}}}, \bibinfo {author} {\bibfnamefont {W.~J.}\ \bibnamefont
  {{Percival}}}, \bibinfo {author} {\bibfnamefont {A.~G.}\ \bibnamefont
  {{S{\'a}nchez}}}, \bibinfo {author} {\bibfnamefont {L.}~\bibnamefont
  {{Samushia}}}, \bibinfo {author} {\bibfnamefont {S.}~\bibnamefont {{Ho}}},
  \bibinfo {author} {\bibfnamefont {E.}~\bibnamefont {{Kazin}}}, \bibinfo
  {author} {\bibfnamefont {M.}~\bibnamefont {{Manera}}}, \bibinfo {author}
  {\bibfnamefont {B.}~\bibnamefont {{Reid}}}, \bibinfo {author} {\bibfnamefont
  {M.}~\bibnamefont {{White}}}, \bibinfo {author} {\bibfnamefont
  {R.}~\bibnamefont {{Tojeiro}}}, \bibinfo {author} {\bibfnamefont {C.~K.}\
  \bibnamefont {{McBride}}}, \bibinfo {author} {\bibfnamefont {X.}~\bibnamefont
  {{Xu}}}, \bibinfo {author} {\bibfnamefont {D.~A.}\ \bibnamefont {{Wake}}},
  \bibinfo {author} {\bibfnamefont {M.~A.}\ \bibnamefont {{Strauss}}}, \bibinfo
  {author} {\bibfnamefont {F.}~\bibnamefont {{Montesano}}}, \emph {et~al.},\
  }\href {https://doi.org/10.1111/j.1365-2966.2012.21235.x} {\bibfield
  {journal} {\bibinfo  {journal} {\mnras}\ }\textbf {\bibinfo {volume} {424}},\
  \bibinfo {pages} {564} (\bibinfo {year} {2012})},\ \Eprint
  {https://arxiv.org/abs/1203.6499} {arXiv:1203.6499 [astro-ph.CO]}
  \BibitemShut {NoStop}%
\bibitem [{Note8()}]{Note8}%
  \BibitemOpen
  \bibinfo {note} {The gap in the redshift distributions between LOWZ and CMASS
  samples is due to our use of the SDSS DR11 parent sample. Using the latest
  DR12 catalog would have allowed us to add in a few more galaxies at
  intermediate redshifts. However, that would have required us to modify the
  BOSS large scale structure sample to include the fiber collided and redshift
  failure galaxies. Since this procedure was already performed for DR11, we
  decided to stick with this sample.}\BibitemShut {Stop}%
\bibitem [{\citenamefont {{Hinshaw}}\ \emph {et~al.}(2013)\citenamefont
  {{Hinshaw}}, \citenamefont {{Larson}}, \citenamefont {{Komatsu}},
  \citenamefont {{Spergel}}, \citenamefont {{Bennett}}, \citenamefont
  {{Dunkley}}, \citenamefont {{Nolta}}, \citenamefont {{Halpern}},
  \citenamefont {{Hill}}, \citenamefont {{Odegard}}, \citenamefont {{Page}},
  \citenamefont {{Smith}}, \citenamefont {{Weiland}}, \citenamefont {{Gold}},
  \citenamefont {{Jarosik}} \emph {et~al.}}]{WMAP9}%
  \BibitemOpen
  \bibfield  {author} {\bibinfo {author} {\bibfnamefont {G.}~\bibnamefont
  {{Hinshaw}}}, \bibinfo {author} {\bibfnamefont {D.}~\bibnamefont {{Larson}}},
  \bibinfo {author} {\bibfnamefont {E.}~\bibnamefont {{Komatsu}}}, \bibinfo
  {author} {\bibfnamefont {D.~N.}\ \bibnamefont {{Spergel}}}, \bibinfo {author}
  {\bibfnamefont {C.~L.}\ \bibnamefont {{Bennett}}}, \bibinfo {author}
  {\bibfnamefont {J.}~\bibnamefont {{Dunkley}}}, \bibinfo {author}
  {\bibfnamefont {M.~R.}\ \bibnamefont {{Nolta}}}, \bibinfo {author}
  {\bibfnamefont {M.}~\bibnamefont {{Halpern}}}, \bibinfo {author}
  {\bibfnamefont {R.~S.}\ \bibnamefont {{Hill}}}, \bibinfo {author}
  {\bibfnamefont {N.}~\bibnamefont {{Odegard}}}, \bibinfo {author}
  {\bibfnamefont {L.}~\bibnamefont {{Page}}}, \bibinfo {author} {\bibfnamefont
  {K.~M.}\ \bibnamefont {{Smith}}}, \bibinfo {author} {\bibfnamefont {J.~L.}\
  \bibnamefont {{Weiland}}}, \bibinfo {author} {\bibfnamefont {B.}~\bibnamefont
  {{Gold}}}, \bibinfo {author} {\bibfnamefont {N.}~\bibnamefont {{Jarosik}}},
  \emph {et~al.},\ }\href {https://doi.org/10.1088/0067-0049/208/2/19}
  {\bibfield  {journal} {\bibinfo  {journal} {\apjs}\ }\textbf {\bibinfo
  {volume} {208}},\ \bibinfo {eid} {19} (\bibinfo {year} {2013})},\ \Eprint
  {https://arxiv.org/abs/1212.5226} {arXiv:1212.5226 [astro-ph.CO]}
  \BibitemShut {NoStop}%
\bibitem [{\citenamefont {{Shirasaki}}\ \emph {et~al.}(2019)\citenamefont
  {{Shirasaki}}, \citenamefont {{Hamana}}, \citenamefont {{Takada}},
  \citenamefont {{Takahashi}},\ and\ \citenamefont
  {{Miyatake}}}]{2019MNRAS.486...52S}%
  \BibitemOpen
  \bibfield  {author} {\bibinfo {author} {\bibfnamefont {M.}~\bibnamefont
  {{Shirasaki}}}, \bibinfo {author} {\bibfnamefont {T.}~\bibnamefont
  {{Hamana}}}, \bibinfo {author} {\bibfnamefont {M.}~\bibnamefont {{Takada}}},
  \bibinfo {author} {\bibfnamefont {R.}~\bibnamefont {{Takahashi}}},\ and\
  \bibinfo {author} {\bibfnamefont {H.}~\bibnamefont {{Miyatake}}},\ }\href
  {https://doi.org/10.1093/mnras/stz791} {\bibfield  {journal} {\bibinfo
  {journal} {\mnras}\ }\textbf {\bibinfo {volume} {486}},\ \bibinfo {pages}
  {52} (\bibinfo {year} {2019})},\ \Eprint {https://arxiv.org/abs/1901.09488}
  {arXiv:1901.09488 [astro-ph.CO]} \BibitemShut {NoStop}%
\bibitem [{\citenamefont {{Navarro}}\ \emph {et~al.}(1997)\citenamefont
  {{Navarro}}, \citenamefont {{Frenk}},\ and\ \citenamefont
  {{White}}}]{Navarroetal:97}%
  \BibitemOpen
  \bibfield  {author} {\bibinfo {author} {\bibfnamefont {J.~F.}\ \bibnamefont
  {{Navarro}}}, \bibinfo {author} {\bibfnamefont {C.~S.}\ \bibnamefont
  {{Frenk}}},\ and\ \bibinfo {author} {\bibfnamefont {S.~D.~M.}\ \bibnamefont
  {{White}}},\ }\href {https://doi.org/10.1086/304888} {\bibfield  {journal}
  {\bibinfo  {journal} {\apj}\ }\textbf {\bibinfo {volume} {490}},\ \bibinfo
  {pages} {493} (\bibinfo {year} {1997})},\ \Eprint
  {https://arxiv.org/abs/arXiv:astro-ph/9611107} {arXiv:astro-ph/9611107}
  \BibitemShut {NoStop}%
\bibitem [{\citenamefont {{Behroozi}}\ \emph {et~al.}(2013)\citenamefont
  {{Behroozi}}, \citenamefont {{Wechsler}},\ and\ \citenamefont
  {{Wu}}}]{Behroozi:2013}%
  \BibitemOpen
  \bibfield  {author} {\bibinfo {author} {\bibfnamefont {P.~S.}\ \bibnamefont
  {{Behroozi}}}, \bibinfo {author} {\bibfnamefont {R.~H.}\ \bibnamefont
  {{Wechsler}}},\ and\ \bibinfo {author} {\bibfnamefont {H.-Y.}\ \bibnamefont
  {{Wu}}},\ }\href {https://doi.org/10.1088/0004-637X/762/2/109} {\bibfield
  {journal} {\bibinfo  {journal} {\apj}\ }\textbf {\bibinfo {volume} {762}},\
  \bibinfo {eid} {109} (\bibinfo {year} {2013})},\ \Eprint
  {https://arxiv.org/abs/1110.4372} {arXiv:1110.4372 [astro-ph.CO]}
  \BibitemShut {NoStop}%
\bibitem [{Note9()}]{Note9}%
  \BibitemOpen
  \bibinfo {note} {The original code was written in C++ by Matthew B Kennel,
  the first author of this paper has carried out some bug fixes, memory leaks
  and optimizations.}\BibitemShut {Stop}%
\bibitem [{\citenamefont {{Hikage}}\ \emph {et~al.}(2011)\citenamefont
  {{Hikage}}, \citenamefont {{Takada}}, \citenamefont {{Hamana}},\ and\
  \citenamefont {{Spergel}}}]{Hikageetal:11}%
  \BibitemOpen
  \bibfield  {author} {\bibinfo {author} {\bibfnamefont {C.}~\bibnamefont
  {{Hikage}}}, \bibinfo {author} {\bibfnamefont {M.}~\bibnamefont {{Takada}}},
  \bibinfo {author} {\bibfnamefont {T.}~\bibnamefont {{Hamana}}},\ and\
  \bibinfo {author} {\bibfnamefont {D.}~\bibnamefont {{Spergel}}},\ }\href
  {https://doi.org/10.1111/j.1365-2966.2010.17886.x} {\bibfield  {journal}
  {\bibinfo  {journal} {\mnras}\ }\textbf {\bibinfo {volume} {412}},\ \bibinfo
  {pages} {65} (\bibinfo {year} {2011})},\ \Eprint
  {https://arxiv.org/abs/1004.3542} {arXiv:1004.3542 [astro-ph.CO]}
  \BibitemShut {NoStop}%
\bibitem [{\citenamefont {Troxel}\ \emph {et~al.}(2018)\citenamefont {Troxel}
  \emph {et~al.}}]{Troxel:2018}%
  \BibitemOpen
  \bibfield  {author} {\bibinfo {author} {\bibfnamefont {M.~A.}\ \bibnamefont
  {Troxel}} \emph {et~al.} (\bibinfo {collaboration} {DES}),\ }\href
  {https://doi.org/10.1103/PhysRevD.98.043528} {\bibfield  {journal} {\bibinfo
  {journal} {Phys. Rev. D}\ }\textbf {\bibinfo {volume} {98}},\ \bibinfo
  {pages} {043528} (\bibinfo {year} {2018})},\ \Eprint
  {https://arxiv.org/abs/1708.01538} {arXiv:1708.01538 [astro-ph.CO]}
  \BibitemShut {NoStop}%
\bibitem [{\citenamefont {{Zhang}}\ \emph {et~al.}(2022)\citenamefont
  {{Zhang}}, \citenamefont {{Li}}, \citenamefont {{Dalal}}, \citenamefont
  {{Mandelbaum}}, \citenamefont {{Strauss}}, \citenamefont {{Kannawadi}},
  \citenamefont {{Miyatake}}, \citenamefont {{Nicola}}, \citenamefont {{Plazas
  Malag{\'o}n}}, \citenamefont {{Shirasaki}}, \citenamefont {{Sugiyama}},\ and\
  \citenamefont {{Takada}}}]{TQ:2023}%
  \BibitemOpen
  \bibfield  {author} {\bibinfo {author} {\bibfnamefont {T.}~\bibnamefont
  {{Zhang}}}, \bibinfo {author} {\bibfnamefont {X.}~\bibnamefont {{Li}}},
  \bibinfo {author} {\bibfnamefont {R.}~\bibnamefont {{Dalal}}}, \bibinfo
  {author} {\bibfnamefont {R.}~\bibnamefont {{Mandelbaum}}}, \bibinfo {author}
  {\bibfnamefont {M.~A.}\ \bibnamefont {{Strauss}}}, \bibinfo {author}
  {\bibfnamefont {A.}~\bibnamefont {{Kannawadi}}}, \bibinfo {author}
  {\bibfnamefont {H.}~\bibnamefont {{Miyatake}}}, \bibinfo {author}
  {\bibfnamefont {A.}~\bibnamefont {{Nicola}}}, \bibinfo {author}
  {\bibfnamefont {A.~A.}\ \bibnamefont {{Plazas Malag{\'o}n}}}, \bibinfo
  {author} {\bibfnamefont {M.}~\bibnamefont {{Shirasaki}}}, \bibinfo {author}
  {\bibfnamefont {S.}~\bibnamefont {{Sugiyama}}},\ and\ \bibinfo {author}
  {\bibfnamefont {M.}~\bibnamefont {{Takada}}},\ }\href
  {https://doi.org/10.48550/arXiv.2212.03257} {\bibfield  {journal} {\bibinfo
  {journal} {arXiv e-prints}\ ,\ \bibinfo {eid} {arXiv:2212.03257}} (\bibinfo
  {year} {2022})},\ \Eprint {https://arxiv.org/abs/2212.03257}
  {arXiv:2212.03257 [astro-ph.CO]} \BibitemShut {NoStop}%
\bibitem [{\citenamefont {{Hartlap}}\ \emph {et~al.}(2007)\citenamefont
  {{Hartlap}}, \citenamefont {{Simon}},\ and\ \citenamefont
  {{Schneider}}}]{2007A&A...464..399H}%
  \BibitemOpen
  \bibfield  {author} {\bibinfo {author} {\bibfnamefont {J.}~\bibnamefont
  {{Hartlap}}}, \bibinfo {author} {\bibfnamefont {P.}~\bibnamefont {{Simon}}},\
  and\ \bibinfo {author} {\bibfnamefont {P.}~\bibnamefont {{Schneider}}},\
  }\href {https://doi.org/10.1051/0004-6361:20066170} {\bibfield  {journal}
  {\bibinfo  {journal} {\aap}\ }\textbf {\bibinfo {volume} {464}},\ \bibinfo
  {pages} {399} (\bibinfo {year} {2007})},\ \Eprint
  {https://arxiv.org/abs/astro-ph/0608064} {arXiv:astro-ph/0608064 [astro-ph]}
  \BibitemShut {NoStop}%
\bibitem [{Note10()}]{Note10}%
  \BibitemOpen
  \bibinfo {note} {Note that unlike cosmic shear covariance, there is no mixed
  term between large structure and shape noise in the galaxy-galaxy lensing
  measurements.}\BibitemShut {Stop}%
\bibitem [{\citenamefont {{Hamana}}\ \emph
  {et~al.}(2020{\natexlab{b}})\citenamefont {{Hamana}}, \citenamefont
  {{Shirasaki}}, \citenamefont {{Miyazaki}}, \citenamefont {{Hikage}},
  \citenamefont {{Oguri}}, \citenamefont {{More}}, \citenamefont {{Armstrong}},
  \citenamefont {{Leauthaud}}, \citenamefont {{Mandelbaum}}, \citenamefont
  {{Miyatake}}, \citenamefont {{Nishizawa}}, \citenamefont {{Simet}},
  \citenamefont {{Takada}}, \citenamefont {{Aihara}}, \citenamefont {{Bosch}},
  \citenamefont {{Komiyama}}, \citenamefont {{Lupton}}, \citenamefont
  {{Murayama}}, \citenamefont {{Strauss}},\ and\ \citenamefont
  {{Tanaka}}}]{2020PASJ...72...16H}%
  \BibitemOpen
  \bibfield  {author} {\bibinfo {author} {\bibfnamefont {T.}~\bibnamefont
  {{Hamana}}}, \bibinfo {author} {\bibfnamefont {M.}~\bibnamefont
  {{Shirasaki}}}, \bibinfo {author} {\bibfnamefont {S.}~\bibnamefont
  {{Miyazaki}}}, \bibinfo {author} {\bibfnamefont {C.}~\bibnamefont
  {{Hikage}}}, \bibinfo {author} {\bibfnamefont {M.}~\bibnamefont {{Oguri}}},
  \bibinfo {author} {\bibfnamefont {S.}~\bibnamefont {{More}}}, \bibinfo
  {author} {\bibfnamefont {R.}~\bibnamefont {{Armstrong}}}, \bibinfo {author}
  {\bibfnamefont {A.}~\bibnamefont {{Leauthaud}}}, \bibinfo {author}
  {\bibfnamefont {R.}~\bibnamefont {{Mandelbaum}}}, \bibinfo {author}
  {\bibfnamefont {H.}~\bibnamefont {{Miyatake}}}, \bibinfo {author}
  {\bibfnamefont {A.~J.}\ \bibnamefont {{Nishizawa}}}, \bibinfo {author}
  {\bibfnamefont {M.}~\bibnamefont {{Simet}}}, \bibinfo {author} {\bibfnamefont
  {M.}~\bibnamefont {{Takada}}}, \bibinfo {author} {\bibfnamefont
  {H.}~\bibnamefont {{Aihara}}}, \bibinfo {author} {\bibfnamefont
  {J.}~\bibnamefont {{Bosch}}}, \bibinfo {author} {\bibfnamefont
  {Y.}~\bibnamefont {{Komiyama}}}, \bibinfo {author} {\bibfnamefont
  {R.}~\bibnamefont {{Lupton}}}, \bibinfo {author} {\bibfnamefont
  {H.}~\bibnamefont {{Murayama}}}, \bibinfo {author} {\bibfnamefont {M.~A.}\
  \bibnamefont {{Strauss}}},\ and\ \bibinfo {author} {\bibfnamefont
  {M.}~\bibnamefont {{Tanaka}}},\ }\href {https://doi.org/10.1093/pasj/psz138}
  {\bibfield  {journal} {\bibinfo  {journal} {\pasj}\ }\textbf {\bibinfo
  {volume} {72}},\ \bibinfo {eid} {16} (\bibinfo {year}
  {2020}{\natexlab{b}})},\ \Eprint {https://arxiv.org/abs/1906.06041}
  {arXiv:1906.06041 [astro-ph.CO]} \BibitemShut {NoStop}%
\bibitem [{Note11()}]{Note11}%
  \BibitemOpen
  \bibinfo {note} {One may think that the factor of $1.08$ is not enough
  because the blue contour does not cover the green contour in Fig.~\ref
  {fig:psf-alpha-beta-contour}. Depending on the flags, {\protect \sc i\protect
  \_calib\protect \_psf\protect \_used=True} or {False}, we not only use
  different coefficients but also different $\xi ^{\protect \rm pp,pq,qq}$, and
  the difference is not so significant in predicted $\protect \hat \xi
  ^{\protect \rm psf}$ than how it appears in Fig.~\ref
  {fig:psf-alpha-beta-contour}}\BibitemShut {NoStop}%
\end{thebibliography}%

\appendix
\section{Systematic differences in the inferred redshift distribution}
\label{sec:nz_differences_app}

As shown in Fig.~\ref{fig:dndz}, we see differences in the inferred redshift
distribution of our source sample depending upon whether we use the redshift
PDFs from \dempz or \dnnz. We estimate the potential biases in the measured
weak lensing signals due to these systematic uncertainties.

For the galaxy-galaxy lensing signal this systematic bias can be quantified by
computing the average ratio of the critical surface density of the source
sample (described in Section~\ref{sec:data_shape_catalog}) by utilizing the
inferred redshifts from either of the two estimates. We compute this ratio by
using the median redshifts of each of our lens samples (described in
Section~\ref{sec:data_lens_catalog}). We find that for the LOWZ sample the
critical surface density based on the redshifts inferred based on \dempz is
higher by 1.7 percent compared to those inferred based on \dnnz, while this
difference grows to 3.3 and 4.2 percent for the CMASS1 and CMASS2 subsamples.

For the cosmic shear signal, we compute the theoretical estimate of $\xi+$ and
$\xi-$ for the source sample for the Planck cosmological model using either of
the two inferred redshift distributions. For the cosmic shear signal, we also
find that the signals based on the redshifts inferred from \dempz is predicted
to be about 6 percent higher than that from \dnnz.

We would like to have a simple way to parameterize these differences and
marginalize over them in our cosmology analysis. We find that the average
redshift of the source sample based on the inferred redshift distribution from
\dempz is higher than that of \dnnz by $\Delta \bar z$ of $0.04$. If we simply
shift the inferred redshift distribution from \dempz lower by the same amount
while maintaining the shape of the redshift distribution, we can account for
almost all of the differences. When such a shift is included, the ratios of the
values of the critical surface density for each of the lens subsamples is
reduced to sub-percent levels ($<0.75$ percent), while the cosmic shear signals
also agree at the sub percent level. We recommend the use of such $\Delta z$ in
the systematic uncertainty analyses for cosmological inference. This is
particularly important because we do not have any clustering measurements which
can constrain the redshift PDFs at high redshifts for our source galaxy sample.

As will be shown in our companion paper Miyatake \etal \cite{Miyatake_hscy3}, using the self
calibration technique of \citet[][]{OguriTakada:11}, there are hints that
support a value of $\Delta z$ of the order $-0.06$ (albeit at low
significance), indicating that the true mean redshifts may possibly be even
higher than that suggested by \dempz.

\section{The variation of the signal in the redshift bin with BOSS full sample}

\begin{figure*}
    \centering
    \includegraphics[width=2\columnwidth]{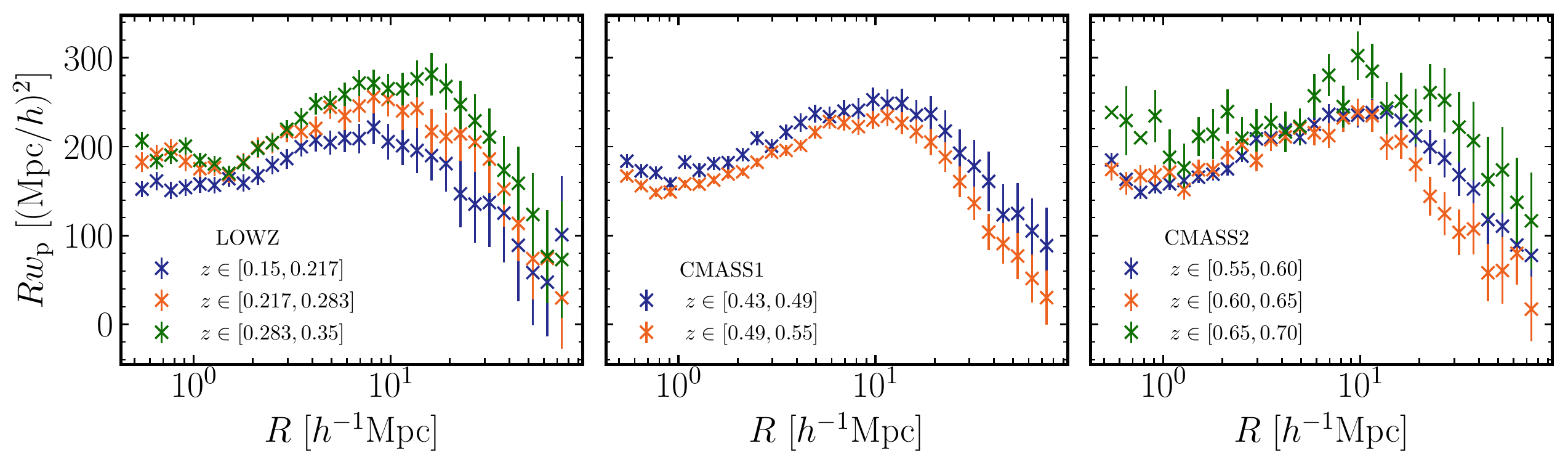}
    \caption{Similar plot as Fig.~\ref{fig:clustering_signal_finer_bin}, but we use the BOSS full sample for the clustering measurements rather than the luminosity limited sample, which is the fiducial lens sample used in this paper.}
    \label{fig:clustering_signal_finer_bin_full_sample}
\end{figure*}

In this section, we present the variation of the clustering signal with the
BOSS full sample, i.e the flux limited sample. We used all the galaxies in the
SDSS DR11 BOSS catalogs, rather than applying the
luminosity cuts used in the main part of this paper. Using this BOSS full
sample, we tested the variation of the clustering signal in each redshift in a
similar way as Section~\ref{subsec:clustering-systematics}: we divide the
galaxies into three or two subsamples in each redshift bin and measure the
clustering signal in each subsample.
Fig.~\ref{fig:clustering_signal_finer_bin_full_sample} shows the variation of
the clustering signal in each redshift bin using the BOSS full sample.
Comparing to Fig.~\ref{fig:clustering_signal_finer_bin}, we can find that the
flux limited sample shows the stronger variation, indicating a stronger
evolution of the galaxy property as a function of redshift. Therefore, the
luminosity limited sample has almost the same galaxy property, and we can model
the galaxy physics with a single set of the galaxy-related parameters, e.g. the
HOD parameters or galaxy bias parameter, without accounting for the redshift
evolution of the galaxy-related parameters in the model in each redshift bin.

\section{Measurements with \dnnz and \mizuki}
\label{sec:dempz_mizuki_meas_app}

\begin{figure*}
    \centering
    \includegraphics[width=2\columnwidth]{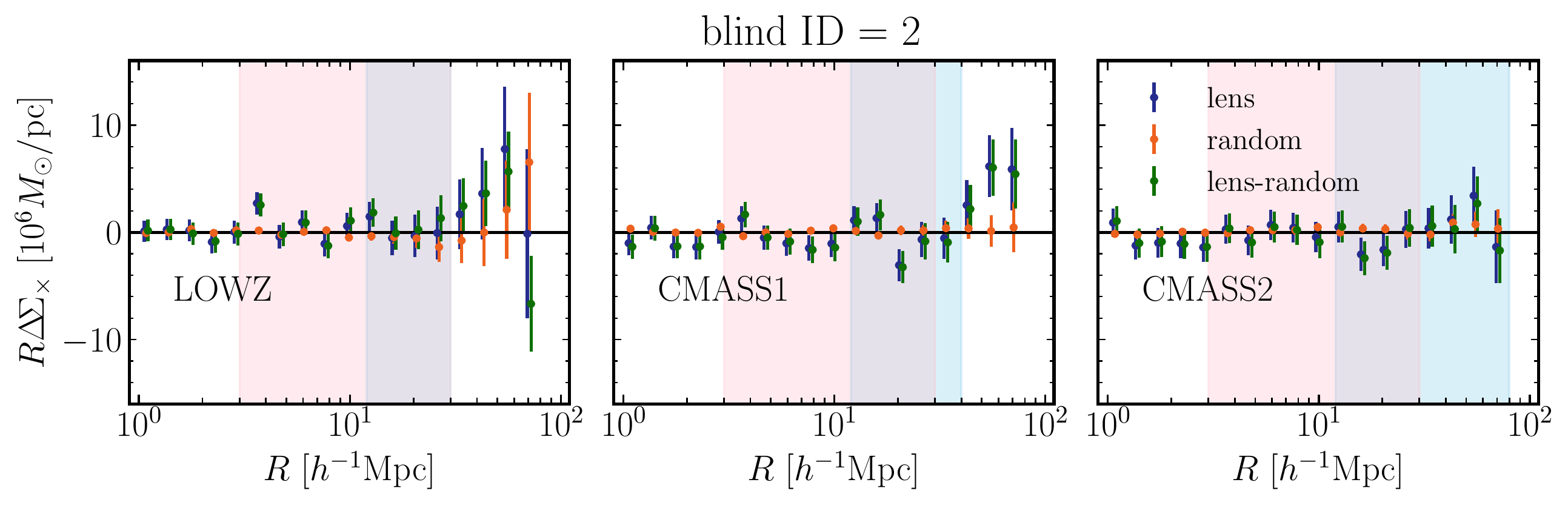}
    \includegraphics[width=2\columnwidth]{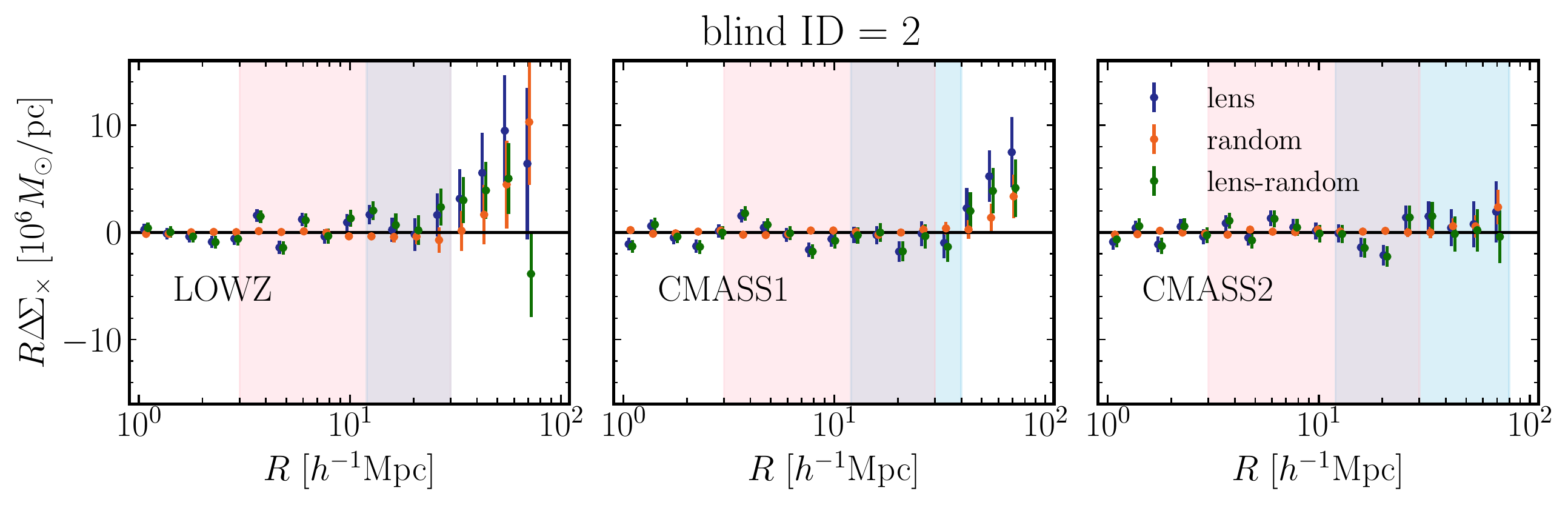}
    \caption{Similar plot as Fig.~\ref{fig:dsigmacross}, but using \dnnz and \mizuki for sample selection.}
    \label{fig:dsigmacross-dnnz-mizuki}
\end{figure*}

In this appendix, we present the weak lensing measurements and the systematics
tests but for the source samples selected using \dnnz and \mizuki,
respectively. The upper and lower panels of
Fig.~\ref{fig:dsigmacross-dnnz-mizuki} present the cross signal measured for
the three subsamples of lenses we use for our measurements for the two
different source samples. The same comments apply as were observed for the
\dempz based source sample, as written in Sec.\ref{subsec:gglens-meas}. We
apply the same scale cuts and show the $\chi^2$ and the corresponding p-values
in Table~\ref{tab:dsigmacross} for the large and small-scale analyses,
respectively.

\begin{figure*}
    \centering
    \includegraphics[width=2\columnwidth]{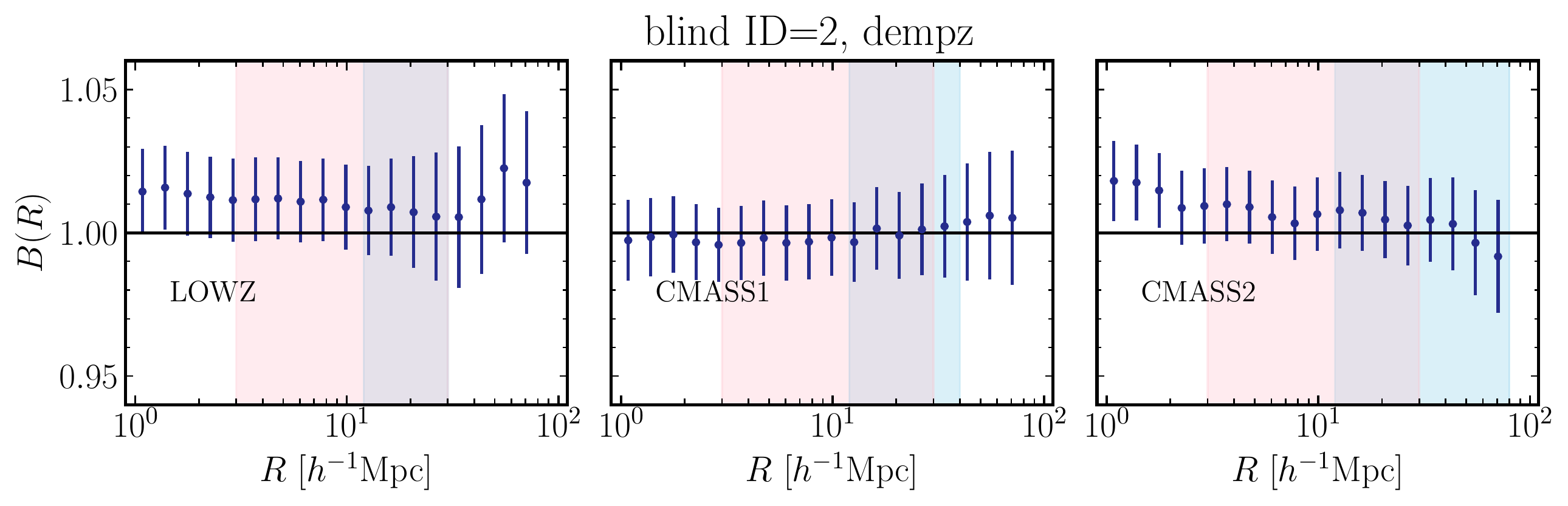}
    \includegraphics[width=2\columnwidth]{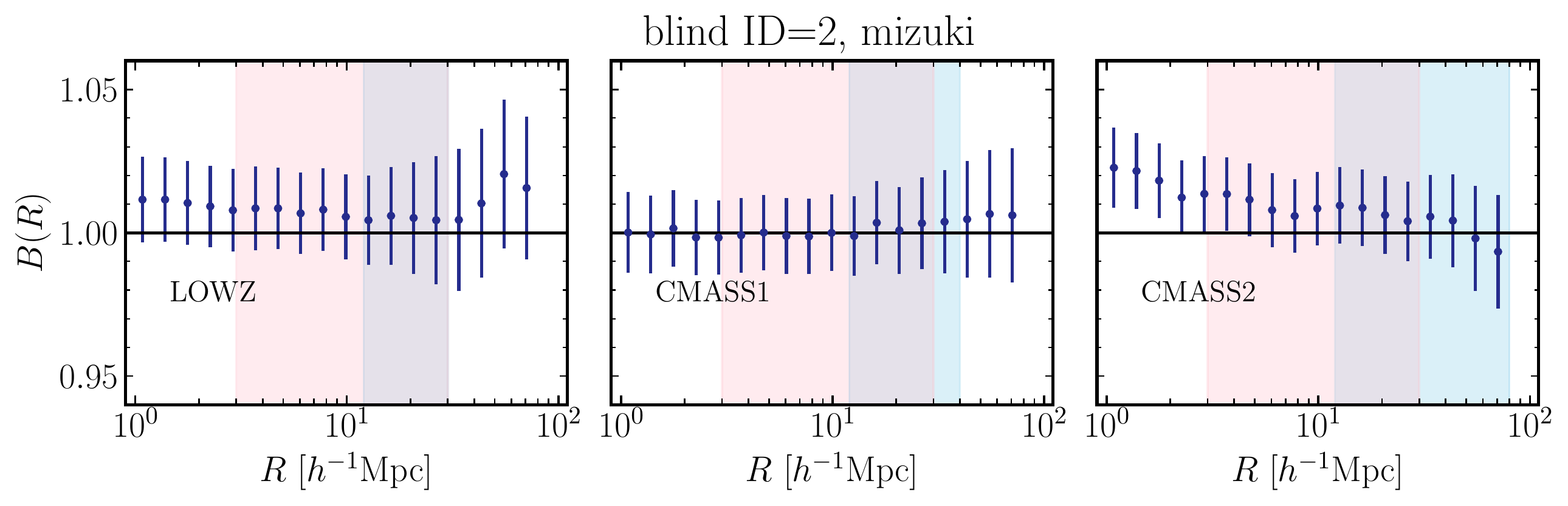}
    \caption{Similar plot as Fig.~\ref{fig:boost}, but using \dnnz and \mizuki for sample selection.}
    \label{fig:boost-dnnz-mizuki}
\end{figure*}

We also compute the boost factors for the corresponding lens subsamples to
estimate the contamination of the signal due to galaxies physically associated
with the lens sample. These are shown in Fig~\ref{fig:boost-dnnz-mizuki}. The
deviation of these signals from unity and the values of the corresponding
$\chi^2$ per degree of freedom and the p-values, on scales that we will use for
our cosmological analyses are listed in Table~\ref{tab:boost}.

\begin{figure*}
    \centering
    \includegraphics[width=1.5\columnwidth]{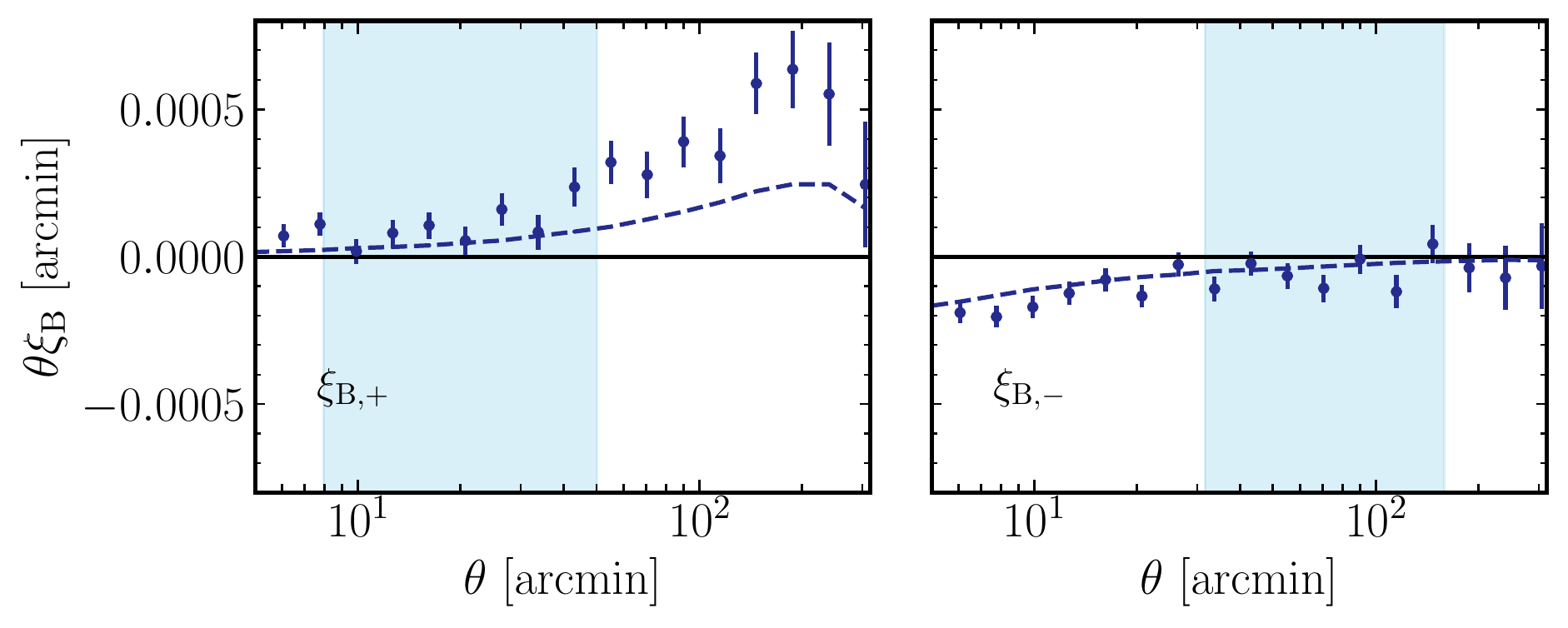}
    \includegraphics[width=1.5\columnwidth]{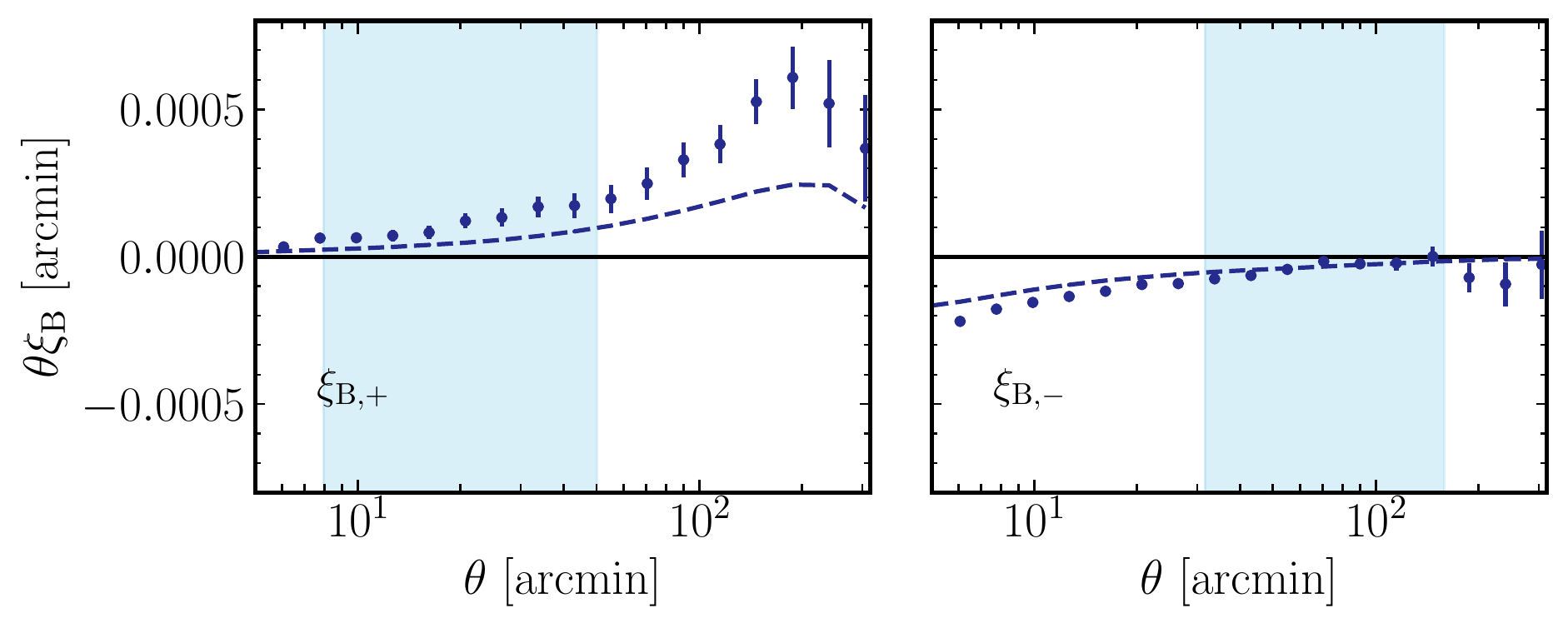}
    \caption{Systematics test of cosmic shear signal, i.e. $B$-mode test for the \dnnz and the \mizuki for sample selections.}
    \label{fig:Bmode-dnnz-mizuki}
\end{figure*}

The B-modes from the cosmic shear measurements for the source subsamples
selected by the two methods are shown in Figs.~\ref{fig:Bmode-dnnz-mizuki}. In
the scales that we consider for our analyses, the B-mode signals are consistent
with zero, as in the fiducial analysis.

The galaxy-galaxy lensing measurements, the cosmic shear measurements as well
as the redshift distributions for the source samples will be presented in
electronic form on the data repository.

\section{A problematic region in GAMA09H field}
\label{sec:gama09h_bmodes_app}

\begin{figure*}
    \centering
    \includegraphics[width=1.5\columnwidth]{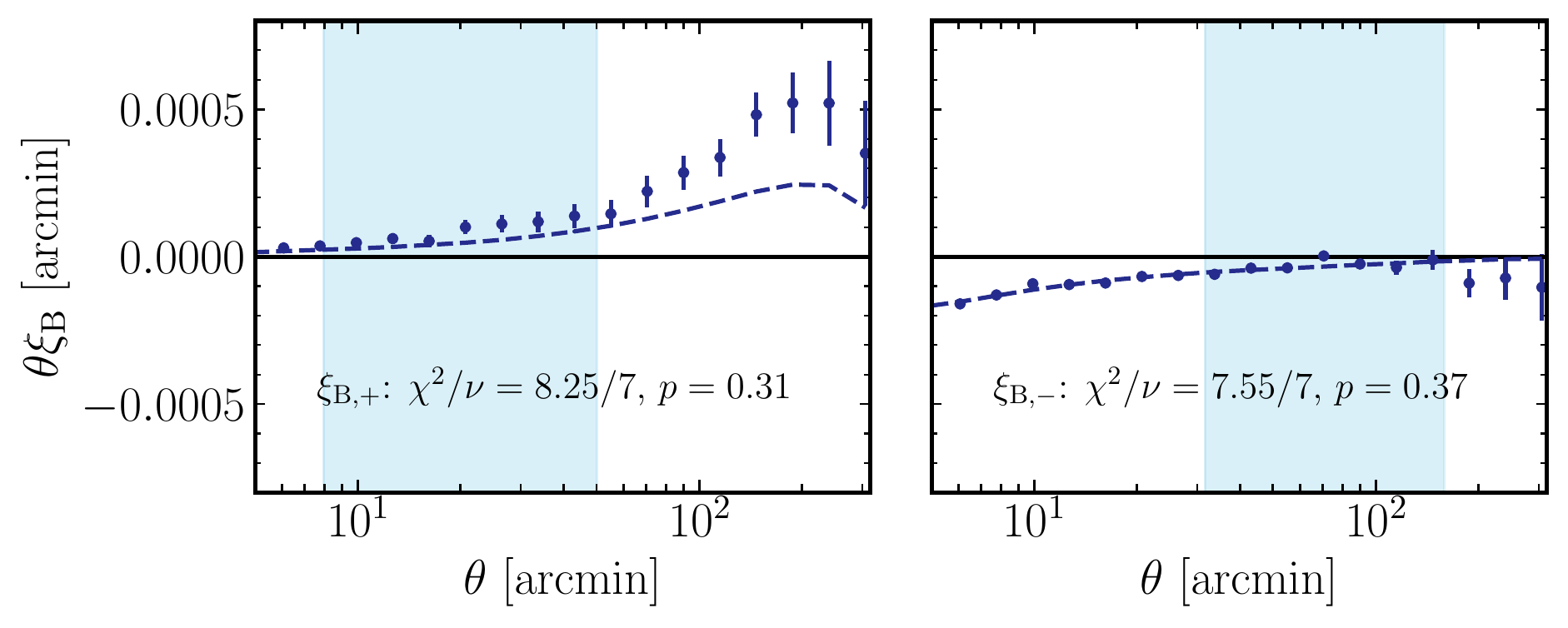}
    \caption{Cosmic shear $B$ mode test with the problematic region in GAMA09H field.}
    \label{fig:xiB-G09H20deg2}
\end{figure*}

We have computed the cosmic shear signal using source galaxies selected from
\dempz from the entire full depth full color region of the HSC-Y3 shape
catalog. In Fig.~\ref{fig:xiB-G09H20deg2}, we present the B-mode signals
measured based on $\xi_+$ and $\xi_-$ and compare it to the mean expectation
from mock shape catalogs. The p-value for the B-modes computed from $xi_+$ show
somewhat low values of $0.07$, although it is not troublingly low. However,
when we had performed this analysis previously with a sample based on \dnnz
selection, we have noticed unacceptably low p-values, driven partly by the
larger amplitude of the B-mode signal from $\xi_+$, and the erratic behaviour
of the B-mode signal from $\xi_-$. By performing a field-by-field analysis with
the \dnnz sample, we were able to track down such odd behaviour to one of the
subfields within HSC, namely GAMA09H. We were able to identify a $20$ sq degree
region bounded by right ascension between $132.5$ and $140.0$ deg and
declination between $1.6$ and $4.7$ deg. We have not been able to entirely
track down the cause of this issue yet, but empirically this excluded region
includes an area which was observed in some of the best seeing conditions.
Therefore, to be on the safe side, we carry out all our measurements of the
galaxy-galaxy lensing signal and the cosmic shear signal by excluding that
particular patch of the sky.

\section{Analytic expression of magnification bias covariance}
\label{sec:cov-mag}
In this section, we derive the analytic expression of magnification bias effect
on the cross-covariance between galaxy-galaxy lensing signal and cosmic shear
signal. We consider the galaxy-galaxy lensing of the lens at the representative
redshift $z_{\rm l}$ and the source sample at source bin $z_{\rm s_0}$, and the
cosmic shear signal of the source samples at source bin $z_{\rm s_1}$ and
$z_{\rm s_2}$. Considering only the Gaussian terms, the cross-covariance is
expressed as
\begin{widetext}
\begin{align}
    &\hspace{-2em}{\rm Cov}[\hat\dSigma(R_n, z_l, z_{{\rm s_0}}), 
    \hat\xi_{\pm}(\theta_m, z_{{\rm s_1}}, z_{{\rm s_2}})] \nonumber\\
    = & \Sigma_{\rm cr}(z_{\rm l}, z_{{\rm s_0}})\frac{1}{\Omega_{\rm s}}
    \int\frac{\ell{\rm d}\ell}{2\pi} J_{2}\left(\ell\frac{R_n}{\chi_l}\right)J_{0/4}(\ell\theta_m) \nonumber\\
    &\times\left[
    C_{g\kappa_{\rm s}}(\ell;z_{\rm l},z_{{\rm s_1}}) 
    \left(C_{\kappa_{\rm s}\kappa_{\rm s}}(\ell; z_{{\rm s_0}}, z_{{\rm s_2}})+\frac{\sigma_\epsilon^2}{\bar{n}_{\rm s_0}}\delta_{\rm s_0, s_2}^{\rm K}\right)
    +
    C_{g\kappa_{\rm s}}(\ell;z_{\rm l},z_{{\rm s_2}}) 
    \left(C_{\kappa_{\rm s}\kappa_{\rm s}}(\ell; z_{{\rm s_0}}, z_{{\rm s_1}})+\frac{\sigma_\epsilon^2}{\bar{n}_{\rm s_0}}\delta_{\rm s_0, s_1}^{\rm K}\right)
    \right.\nonumber\\
    &\hspace{2em}+2(\alpha_{\rm mag, l}-1)\left(
    C_{\kappa_{\rm l}\kappa_{\rm s}}(\ell; z_{\rm l}, z_{{\rm s_1}})
    \left(C_{\kappa_{\rm s}\kappa_{\rm s}}(\ell; z_{{\rm s_0}}, z_{{\rm s_2}})+\frac{\sigma_\epsilon^2}{\bar{n}_{\rm s_0}}\delta_{\rm s_0, s_2}^{\rm K}\right)
    +\left.
    C_{\kappa_{\rm l}\kappa_{\rm s}}(\ell; z_{\rm l}, z_{{\rm s_2}})
    \left(C_{\kappa_{\rm s}\kappa_{\rm s}}(\ell; z_{{\rm s_0}}, z_{{\rm s_1}})+\frac{\sigma_\epsilon^2}{\bar{n}_{\rm s_0}}\delta_{\rm s_0, s_1}^{\rm K}\right)
    \right)\right].
\end{align}
Here the angular correlation functions $C_{\rm XY}$ are defined in Appendix A
of \citet{Sugiyama:2021}. $\sigma^2_\epsilon/\bar{n}_{\rm s_0}$ is the shape
noise term of source sample in redshift bin $z_{\rm s_0}$, and $\delta^{\rm
K}_{X, Y}$ is the Kronecker delta. $R_n$ and $\theta_m$ are the $m$-th and
$n$-th angular bin of galaxy-galaxy lensing and cosmic shear respectively.
$J_n$ is the $n$-th order Bessel function, and $J_0$ and $J_4$ are for $\xi_+$
and $\xi_-$ respectively. $\alpha_{\rm mag, l}$ is the magnification bias
parameter of lens sample $z_{\rm l}$. The last two terms are the contribution
of the magnification bias effect. In this paper, we have single source sample,
i.e. $z_{\rm s_0}=z_{\rm s_1}=z_{\rm s_2}\equiv z_{\rm s}$, and hence the
contribution from magnification bias effect to be added on the covariance
estimated from mock measurements in Sections \ref{subsec:gglens-meas} and
\ref{subsec:cosmicshear-meas} is 
\begin{align}
    &\hspace{-2em}\delta{\rm Cov}[\Delta\Sigma(R_n, z_l, z_{\rm s}), \xi_{\pm}(\theta_m, z_{\rm s}, z_{\rm s})]\nonumber\\
    &= 2\Sigma_{\rm cr}(z_{\rm l}, z_{\rm s})\frac{1}{\Omega_{\rm s}}
    \int\frac{\ell{\rm d}\ell}{2\pi} J_{2}\left(\ell\frac{R_n}{\chi_l}\right)J_{0/4}(\ell\theta_m)
    2(\alpha_{\rm l}-1)
    C_{\kappa_{\rm l}\kappa_{\rm s}}(\ell; z_{\rm l}, z_{\rm s})
    \left(C_{\kappa_{\rm s}\kappa_{\rm s}}(\ell; z_{\rm s}, z_{\rm s})+\frac{\sigma_\epsilon^2}{\bar{n}_{\rm s}}\right).
\end{align}
\end{widetext}

\end{document}